\documentclass[10pt,twocolumn,letterpaper]{article}
\usepackage[pagenumbers]{cvpr}
\usepackage[toc,page,header]{appendix}
\usepackage{minitoc}

\usepackage{graphicx}
\usepackage{amsmath}
\usepackage{amssymb}
\usepackage{booktabs}
\usepackage{hhline}
\usepackage{titling}
\usepackage{dblfloatfix}
\usepackage[pagebackref,breaklinks,colorlinks]{hyperref}
\usepackage{subcaption}
\usepackage{pifont}
\usepackage[dvipsnames]{xcolor}
\definecolor{limegreen}{HTML}{32CD32}
\newcommand{\cmark}{\textcolor{limegreen}{\ding{51}}}%
\newcommand{\xmark}{\textcolor{red}{\ding{55}}}%
\usepackage{arydshln}
\usepackage{multirow}
\usepackage{multicol}
\usepackage{xcolor}
\usepackage[export]{adjustbox}

\def\confName{CVPR}
\date{}

\newcommand{\sketch}{\mathcal{S}}
\newcommand{\image}{\mathcal{I}}
\newcommand{\renderer}{\mathcal{R}}

\newcommand{\widthabs}{0.23}
\newcommand{\widthrescomp}{0.15}
\newcommand{\widthresnet}{0.1}
\newcommand{\widthteapot}{0.11}
\newcommand{\widthattn}{0.15}
\newcommand{\widthseed}{0.2}
\newcommand{\widthmany}{0.09}

\newcommand{\widthdeformfaces}{0.14}
\newcommand{\widthbackground}{0.13}
\newcommand{\widthvit}{0.1}
\newcommand{\widthmiss}{0.24}
\newcommand{\widthface}{0.17}
\newcommand{\widthfaceour}{0.11}

\title{CLIPasso: Semantically-Aware Object Sketching}
\author{
Yael Vinker$^{2,1}$
\and
Ehsan Pajouheshgar$^{1}$
\and
Jessica Y. Bo$^{1}$
\and
Roman Christian Bachmann$^{1}$
\and
Amit Haim Bermano$^{2}$
\and
Daniel Cohen-Or$^{2}$
\and
Amir Zamir$^{1}$
\and
Ariel Shamir$^{3}$
\vspace{2mm}
\and $^{1}$Swiss Federal Institute of Technology (EPFL) \and $^{2}$Tel-Aviv University \and $^{3}$Reichman University
\vspace{1mm}
\and
\small\url{https://clipasso.github.io/clipasso/}
}

\begin{document}
\doparttoc 
\faketableofcontents 

\twocolumn[{
\vspace{-11mm}
	\maketitle
	\vspace{-4em}
	\renewcommand\twocolumn[1][]{#1}
	\begin{center}
		\centering
		\includegraphics[width=0.96\textwidth]{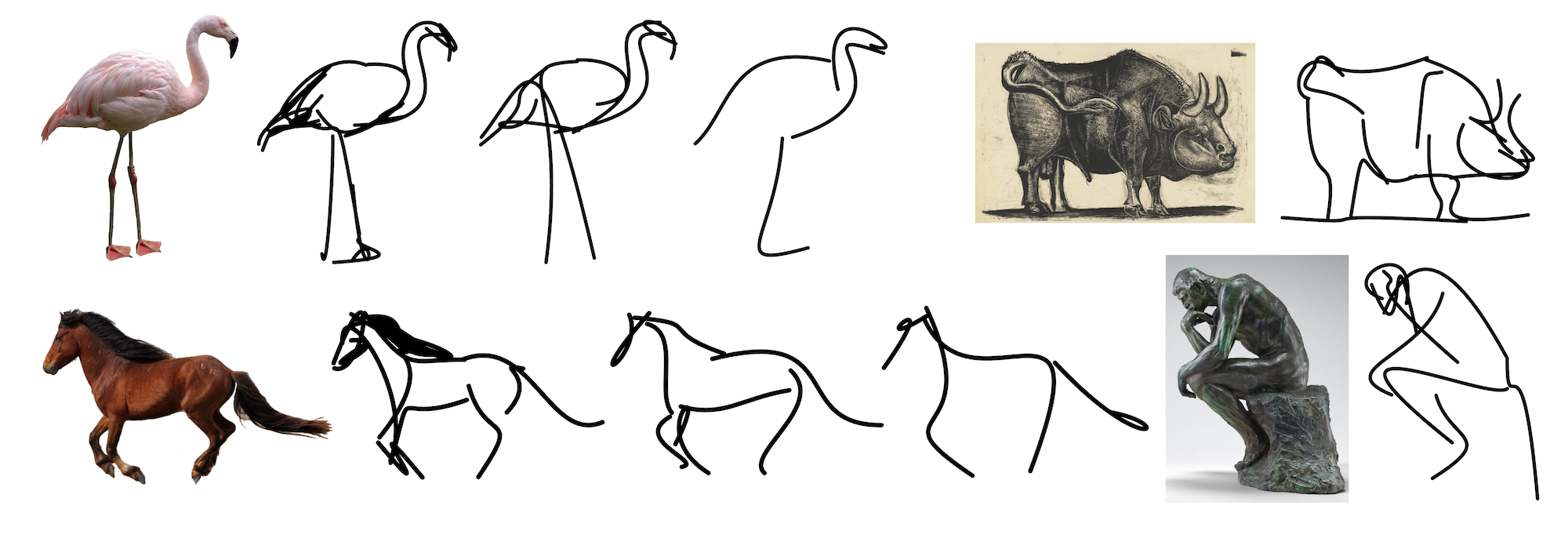}
		\vspace{-0.5em}
		\captionof{figure}{Our work converts an image of an object to a sketch, allowing for varying levels of abstraction, while preserving its key visual features. Even with a very minimal representation (the rightmost flamingo and horse are drawn with only a few strokes), one can recognize both the semantics and the structure of the subject depicted.}
		\label{fig:teaser}
	\end{center}
}]

\begin{abstract}
    Abstraction is at the heart of sketching due to the simple and minimal nature of line drawings.
    Abstraction entails identifying the essential visual properties of an object or scene, which requires semantic understanding and prior knowledge of high-level concepts.
    Abstract depictions are therefore challenging for artists, and even more so for machines.
    We present CLIPasso, an object sketching method that can achieve different levels of abstraction, guided by geometric and semantic simplifications.
    While sketch generation methods often rely on explicit sketch datasets for training, we utilize the remarkable ability of CLIP (Contrastive-Language-Image-Pretraining) to distill semantic concepts from sketches and images alike.
    We define a sketch as a set of Bézier curves and use a differentiable rasterizer to optimize the parameters of the curves directly with respect to a CLIP-based perceptual loss.
    The abstraction degree is controlled by varying the number of strokes.
    The generated sketches demonstrate multiple levels of abstraction while maintaining recognizability, underlying structure, and essential visual components of the subject drawn.
\end{abstract}

\section{Introduction}
\label{sec:introduction}
\begin{figure}[ht]
  \centering
  \includegraphics[width=0.9\linewidth]{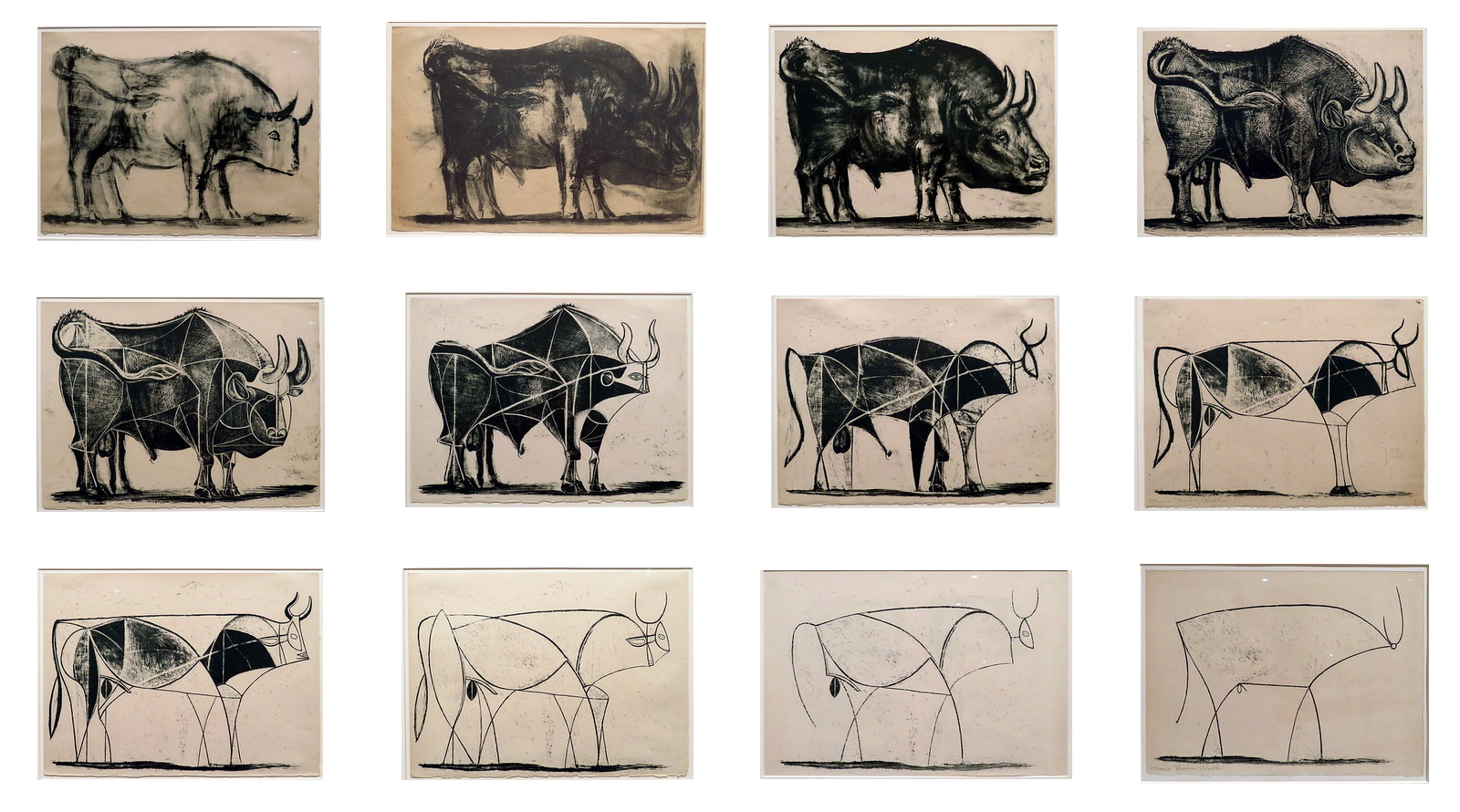}
   
  \caption{"Le Taureau" by Picasso — note how the abstraction process is achieved by gradually \emph{removing} elements while the bull's essence is preserved. }
  \vspace{-2mm}
 \label{fig:picasso}

\end{figure}

Free-hand sketching is a valuable visual tool for expressing ideas, concepts, and actions \cite{Fan2018CommonOR, Hertzmann2020WhyDL, xu2020deep, Gryaditskaya2019OpenSketchAR}. As sketches consist of only strokes, and often only a limited number of strokes, the process of \emph{abstraction} is central to sketching. 
An artist must make representational decisions to choose key visual features of the subject drawn to capture the relevant information she wishes to express, while omitting (many) others~\cite{genesis_errors}. 
For example, in the famous "Le Taureau" series (Figure \ref{fig:picasso}), Picasso depicts the progressive abstraction of a bull. In this series of lithographs, the artist transforms a bull from a concrete, fully rendered, anatomical drawing, into a sketch composition of a few lines that still manages to capture the essence of the bull.

In this paper, we pose the question — can computer renderings imitate such a process of sketching abstraction, converting a photograph from a concrete depiction to an abstract one?

Today, machines can render realistic sketches simply by applying mathematical and geometric operations to an input photograph \cite{Winnemller2012XDoGAE, canny1986computational}. However, creating abstractions is more difficult for machines to achieve. The abstraction process suggests that the artist selects visual features that capture the underlying structure and semantic meaning of the object or scene, to produce a minimal, yet descriptive rendering. 
This demands semantic understanding of the subject, which is more complex than applying simple geometric operations to the image. To fill this semantic gap, we use CLIP \cite{clip}, a neural network trained on various styles of images paired with text. CLIP is exceptional at encoding the semantic meaning of visual depictions, regardless of their style \cite{goh2021multimodal}.

Previous works that attempt replicating human-like sketching often use sketch datasets of the desired level of abstraction to guide the form and style of the generated sketch \cite{Berger2013, Deformable_Stroke, Deep-Sketch-Abstraction}. While this data-driven approach can certainly imitate the final rendering of human artwork, it requires the existence and availability of relevant datasets, and it restricts the output style to match this data. 
In contrast, we present an optimization-based photo-to-sketch generation technique that achieves different levels of abstraction without requiring an explicit sketch dataset. 

Our method uses the CLIP image encoder to guide the process of converting a photograph to an abstract sketch. CLIP encoding provides the semantic understanding of the concept depicted, while the photograph itself provides the geometric grounding of the sketch to the concrete subject.

Our sketches are defined using a set of thin, black strokes (Bézier curves) placed on a white background, and the level of abstraction is dictated by the number of strokes used.
Given the target image to be drawn, we use a differentiable rasterizer \cite{diffvg} to directly optimize the strokes’ parameters (control points positions) with respect to a CLIP-based loss.
We combine the final and intermediate activations of a pre-trained CLIP model to achieve both geometric and semantic simplifications.
For improved robustness, we propose a saliency-guided initialization process, based on the local attention maps of a pretrained vision transformer model.

The resulting sketches demonstrate a combination of the semantic and visual features that capture the essence of the input object, while still being minimal and providing good category and instance level object recognition clues.

\begin{figure}
\centering
\begin{tabular}{@{\hskip2pt}c@{\hskip2pt}c@{\hskip2pt}c@{\hskip2pt}c}
    \includegraphics[width=\widthabs\linewidth]{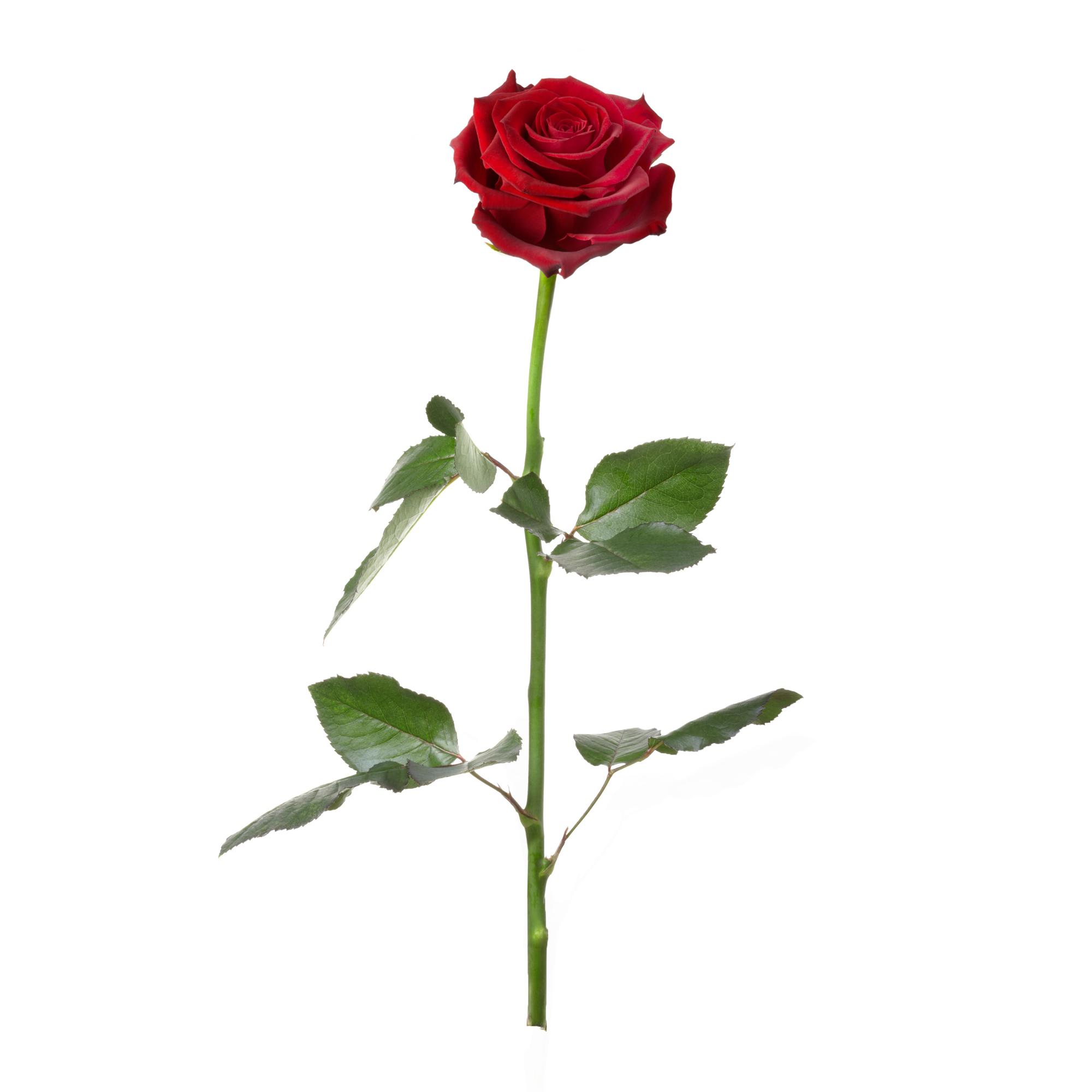} &
    \includegraphics[width=\widthabs\linewidth]{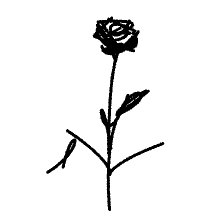} &
    \includegraphics[width=\widthabs\linewidth]{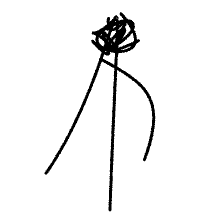} &
    \includegraphics[width=\widthabs\linewidth]{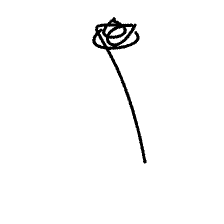} \\
    
    \midrule
    \includegraphics[width=\widthabs\linewidth]{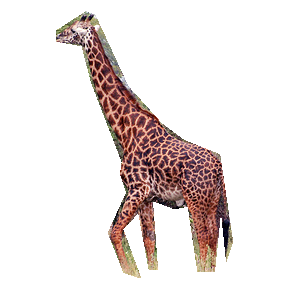} &
    \includegraphics[width=\widthabs\linewidth]{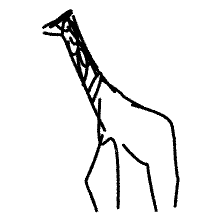} &
    \includegraphics[width=\widthabs\linewidth]{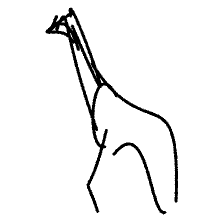} &
    \includegraphics[width=\widthabs\linewidth]{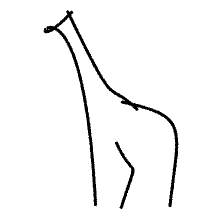} \\

    \midrule
    \includegraphics[width=\widthabs\linewidth]{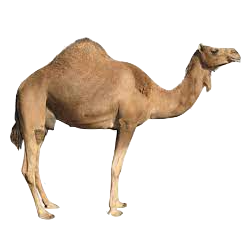} &
    \includegraphics[width=\widthabs\linewidth]{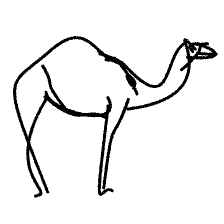} &
    \includegraphics[width=\widthabs\linewidth]{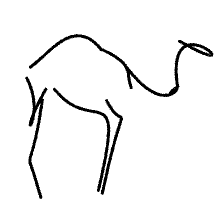} &
    \includegraphics[width=\widthabs\linewidth]{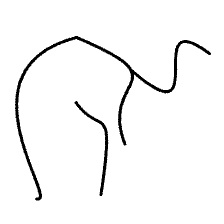} \\

    \midrule
    \includegraphics[width=\widthabs\linewidth]{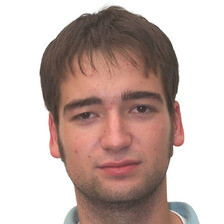} &
  \includegraphics[width=\widthabs\linewidth]{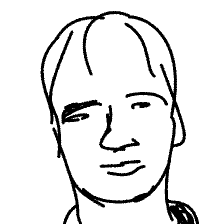} &
  \includegraphics[width=\widthabs\linewidth]{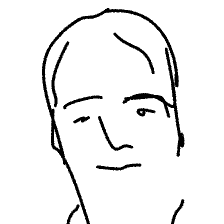} &
  \includegraphics[width=\widthabs\linewidth]{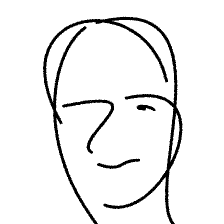} \\

\end{tabular}
 \caption{Different levels of abstraction produced by our method. Left to right: input images and increased level of abstraction, achieved by reducing the number of strokes used by half each column.}
\label{fig:abstraction_levels}
\end{figure}

\section{Related Work}
\label{sec:related}

\begin{figure}[ht]
    \centering
    \includegraphics[width=1\linewidth]{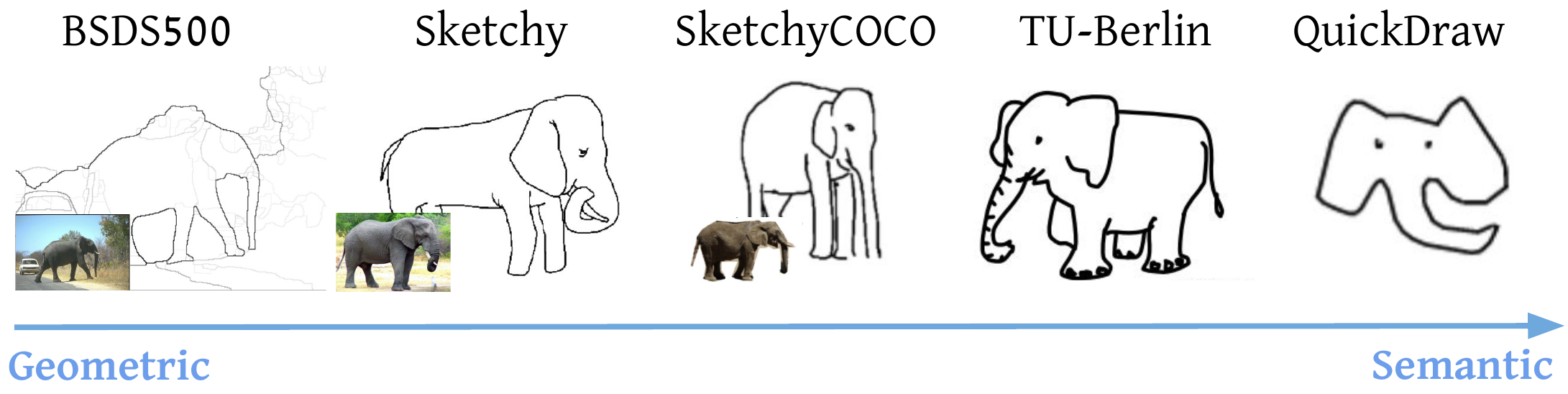}
    \caption{Variations in style and abstraction among sketch datasets — examples are arranged from left to right by the degree of abstraction: from edge-based to category-based sketches. For datasets that have image references, the input image is placed alongside the sketch; otherwise, the input is just the category label (e.g., ``elephant'').}
    \label{fig:datasets}
\end{figure}

Unlike edge-map extraction methods \cite{Winnemller2012XDoGAE, canny1986computational} which are purely based on geometry, free-hand sketch generation aims to produce sketches that are abstract in terms of structure and semantic interpretation so as to mimic a human-like style.
This high-level goal varies among different works, as there are many styles and levels of abstraction that can be produced. Consequently, existing works tend to choose the desired output style based on a given dataset: from highly abstract — guided only by a category-based text prompt \cite{SketchRNN}, 
to more concrete \cite{BSDS500data}, which is guided by contour detection. 
Figure \ref{fig:datasets} illustrates this spectrum.
While methods that rely on sketch datasets are limited to the abstraction levels present, our method is optimization based. Hence, it is capable of producing multiple levels of abstraction without relying on the existence of suitable sketch datasets or requiring a lengthy new training phase.

\newcommand\RotText[1]{\rotatebox{90}{\parbox{2cm}{\centering#1}}}

\begin{table}[ht!]
\caption{Comparison of sketch synthesis algorithms. \textbf{(A)} Is not restricted to categories from training dataset, \textbf{(B)} Can produce different levels of abstractions, \textbf{(C)} Is not limited to abstractions in the dataset, \textbf{(D)} Can produce  vector sketches, \textbf{(E)} Can produce a sequential sketch \textbf{(F)} Is not directly relying on the edge map.}
\resizebox{\linewidth}{!}{%
\begin{tabular}{ccccccc}
\toprule
\begin{tabular}[c]{@{}c@{}}{Method}\end{tabular} &
\begin{tabular}[c]{@{}c@{}}{A}\end{tabular} &
\begin{tabular}[c]{@{}c@{}}{B}\end{tabular} 
&\begin{tabular}[c]{@{}c@{}}{C}\end{tabular} & \begin{tabular}[c]{@{}c@{}}{D}\end{tabular} & \begin{tabular}[c]{@{}c@{}}{E}\end{tabular} & \begin{tabular}[c]{@{}c@{}}{F}\end{tabular} \\

\midrule 
Berger et al. \cite{Berger2013} & \xmark & \cmark & \xmark & \xmark & \cmark & \xmark \\

Li et al. \cite{Deformable_Stroke} &\xmark & \xmark & \xmark & \cmark & \xmark & \xmark \\

Muhammad et al. \cite{Deep-Sketch-Abstraction} &\xmark & \cmark & \cmark & \cmark & \cmark & \xmark \\

Song et al. \cite{song2018learning} &\xmark & \xmark & \xmark & \cmark & \cmark & \cmark \\

Li et al. \cite{li2019photosketching} &\cmark & \xmark & \xmark & \xmark & \xmark & \cmark \\

Kampelm{\"{u}}hler and Pinz \cite{human-like-sketches} & \xmark & \xmark & \xmark & \xmark & \xmark & \cmark \\

Qi and Su et al. \cite{Qi2021SketchLatticeLR} & \xmark & \cmark & \cmark & \cmark & \cmark & \xmark \\
\midrule
\textbf{Ours} & \cmark & \cmark & \cmark & \cmark & \xmark & \cmark \\
\bottomrule
\end{tabular}
}

\label{tab:sketch_synth_comparison}
\end{table}

We provide a brief review of existing relevant photo-sketch synthesis works, which all rely on sketch-specific datasets.
Table \ref{tab:sketch_synth_comparison} summarizes the high-level characteristics that differentiate these methods.

\textbf{Photo-Sketch Synthesis}
Early methods learn explicit models to synthesize facial sketches \cite{2001faces, Berger2013}. To generalise to categories beyond faces, Li et al. \cite{Deformable_Stroke} learn a deformable stroke model based on perceptual grouping.

In the deep learning era, it is intuitive to think of photo-sketch generation as a domain translation task.
However, the highly sparse and abstract nature of sketches introduces challenges for trivial methods \cite{pix2pix, pix2pixHD} to adhere to the sketch domain, and therefore sketch-specific adjustments must be made. 

Song et al.  \cite{song2018learning} propose a hybrid supervised-unsupervised multi-task learning approach with a shortcut cycle consistency constraint. 
Li et al. \cite{li2019photosketching} present a learning-based contour generation algorithm to resolve the diversity of the human drawings in the dataset.
Kampelmuhler and Pinz \cite{human-like-sketches} propose an encoder-decoder architecture, where the loss is guided by a pretrained sketch classifier network.
Qi and Su et al. \cite{Qi2021SketchLatticeLR} propose a lattice representation for sketches, employing LSTM and graph models to generate a vector sketch from points sampled from the edge map. The density of points determines the abstraction level of the sketch.

\textbf{Vector Graphics}
There is a substantial literature on stroke-based rendering, contour visualization, and feature line rendering, summarized in the surveys by Hertzmann \cite{Hertzmann2003-survey}, and by Bénard and Hertzmann \cite{Bnard2019LineDF}.
Vector representations are widely used for a variety of sketching tasks and applications, employing a number of deep learning models including RNN \cite{SketchRNN}, BERT \cite{Lin2020SketchBERTLS}, Transformers \cite{Bhunia2020PixelorAC, Ribeiro2020SketchformerTR}, CNNs \cite{Chen2017Sketchpix2seqAM} GANs \cite{Varshaneya2021TeachingGT} and reinforcement learning algorithms \cite{Zhou2018LearningTS, spiralpp, Ganin2018SynthesizingPF}.
The recent development of differentiable rendering algorithms \cite{Zheng2019StrokeNetAN, Mihai2021DifferentiableDA, diffvg} makes it possible to manipulate or synthesize vector content by using raster-based loss functions. We use the method of Li et al. \cite{diffvg}, as it can handle a wide range of curves and strokes, including Bézier curves.

\textbf{Sketches Abstraction}
Only two previous works propose a unified model to produce sketches of a given image at different levels of abstraction.
Berger et al. \cite{Berger2013} collected a dataset of portraits drawn by professional artists at different levels of abstraction.
For each artist, a library of strokes is created indexed by shape, curvature, and length, and these are used to replace curves extracted from the image edge map.
Their method is limited to portraits and requires a new dataset for each level of abstraction.

Muhammad et al. \cite{Deep-Sketch-Abstraction} propose a stroke-level sketch abstraction model. A reinforcement learning agent is trained to select which strokes can be removed from an edge map representation of the input image without affecting its recognizability.
The recognition signal is provided by a sketch classifier trained on 9 classes from the QuickDraw dataset \cite{SketchRNN}, and hence to operate on new classes, a fine-tuning stage is required.

\begin{figure*}[t]
  \centering
  \includegraphics[width=0.9\linewidth]{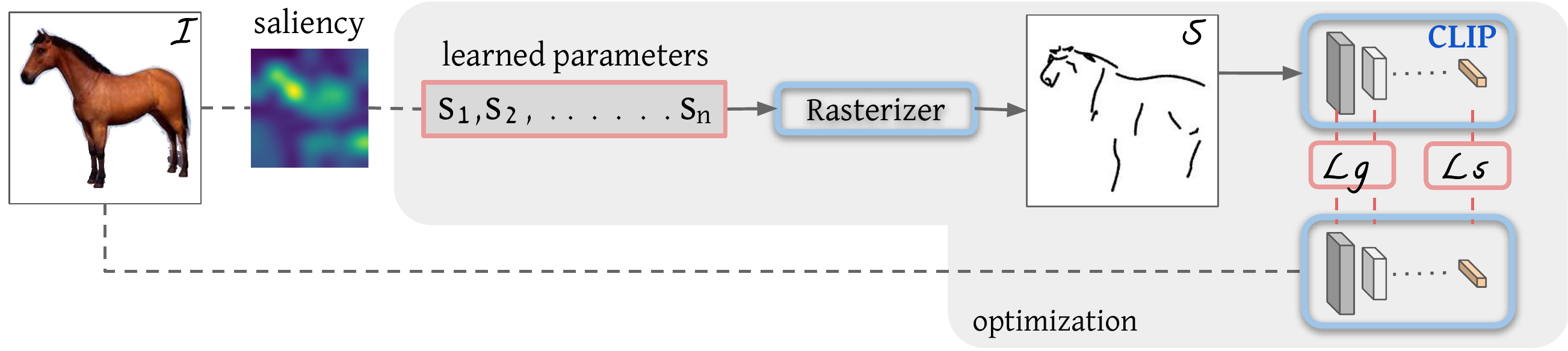}
  \caption{Method overview – Given a target image $\image$, and the number of strokes $n$, a saliency map is used as the distribution to sample the initial strokes locations $\{s_1, .. s_n\}$. A differentiable rasterizer $\renderer$ is used to create a resterized sketch $\sketch$. Both the sketch and the image are fed into a pretrained CLIP model to evaluate the geometric distance $L_g$ and semantic distance $L_s$ between the two. The loss is backpropagated through $\renderer$ to optimize the strokes parameters until convergence.
  The learned parameters and loss terms are highlighted in red, while the blue components are frozen during the entire optimization process, solid arrows are used to mark the backpropagation path.
  }
\label{fig:pipeline}
\end{figure*}

\textbf{CLIP-based Image Abstraction} 
CLIP \cite{clip} is a neural network trained on 400 million image-text pairs collected from the internet with the objective of creating a joint latent space using contrastive learning. 
Being trained on a wide variety of image domains along with lingual concepts, CLIP models are found to be very useful for a wide range of zero-shot tasks.
The most relevant works within our context are by Frans et al.~\cite{CLIPDraw} (CLIPDraw), and Tian and Ha \cite{Evolution-Strategies-for-Creativity}.
CLIPDraw optimizes a set of random Bezier curves to create a drawing that maximizes the CLIP similarity for a given text prompt. Likewise, we also use a differentiable rasterizer \cite{diffvg} and a CLIP-based loss. However, while CLIPDraw is purely text-driven, we allow control over the output appearance, conditioned on the input image. For this purpose, we introduce a new geometric loss term and a saliency-guided initialization procedure.

Tian and Ha \cite{Evolution-Strategies-for-Creativity} employ evolutionary algorithms combined with CLIP, to produce creative abstract concepts represented by colored triangles guided by text or shape. Their results are limited to either fully semantic (using CLIP's text encoder) or entirely geometric (using L2), whereas we are able to integrate both.

\section{Method}
\label{sec:method}

We define a sketch as a set of $n$ black strokes $\{s_1, .. s_n\}$ placed on a white background. We use a two-dimensional Bézier curve with four control points $s_i = \{p_i^j\}_{j=1}^4 = \{(x_i,y_i)^j\}_{j=1}^4$ to represent each stroke. For simplicity, we only optimize the position of control points and choose to keep the degree, width, and opacity of the strokes fixed. However, these parameters can later be used to achieve variations in style (see Figure~\ref{fig:n_control_points}).
The parameters of the strokes are fed to a differentiable rasterizer $\renderer$, which forms the rasterized sketch $\sketch = \renderer(\{p_1^j\}_{j=1}^4, ... \{p_n^j\}_{j=1}^4) = \renderer(s_1, .. s_n)$. As is often conventional \cite{Evolution-Strategies-for-Creativity, spiralpp, Deep-Sketch-Abstraction}, we vary the number of strokes $n$ to create different levels of abstraction. 

An overview of our method can be seen in Figure \ref{fig:pipeline}.
Given a target image $\image$ of the desired subject, our goal is to synthesize the corresponding sketch $\sketch$ while maintaining both the semantic and geometric attributes of the subject.
We begin by extracting the salient regions of the input image to define the initial locations of the strokes.
Next, in each step of the optimization we feed the stroke parameters to a differentiable rasterizer $\renderer$ to produce the rasterized sketch. The resulting sketch, as well as the original image are then fed into CLIP to define a CLIP-based perceptual loss. We back-propagate the loss through the 
differentiable rasterizer and update the strokes' control points directly at each step until convergence of the loss function.

\subsection{Loss Function}
\label{sec:loss}
\begin{figure}[ht]
    \centering
    \begin{tabular}{@{\hskip2pt}c@{\hskip2pt}c@{\hskip2pt}:c@{\hskip2pt}c@{\hskip2pt}c}\\
    \toprule
    Input & XDoG & L2 & LPIPS & Ours \\
        \hline  
        \includegraphics[width=0.18\linewidth]{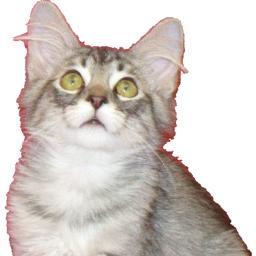} &
        \includegraphics[width=0.18\linewidth]{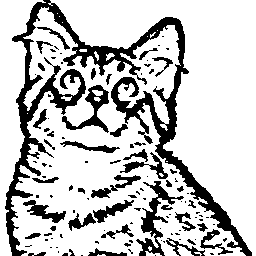}&
        \includegraphics[width=0.18\linewidth]{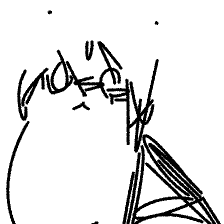} &
        \includegraphics[width=0.18\linewidth]{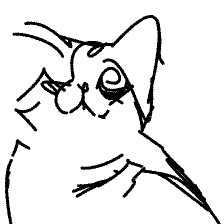} &
        \includegraphics[width=0.18\linewidth]{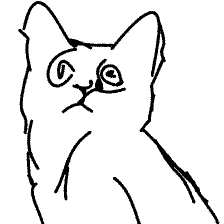}
    \end{tabular}
    \caption{Loss functions comparison — we optimize the strokes by minimizing different losses: L2 loss simply encourages the filling of colored pixels, LPIPS is more semantically aware, but the resulting sketch is still close to the edge map (see the XDog edges for comparison). In contrast, our CLIP-based loss allows better semantic depiction while preserving the morphology of the subject. }
    \label{fig:loss_compare}
\end{figure}

As sketches are highly sparse and abstract, pixel-wise metrics are not sufficient to measure the distance between a sketch and an image. Additionally, even though perceptual losses such as LPIPS \cite{Zhang2018TheUE} can encode semantic information from images, they may not be suitable to encode abstract sketches, as illustrated in Figure \ref{fig:loss_compare} (for further analysis, please refer to the supplementary material).
One solution is to train task-specific encoders to learn a shared embedding space of images and sketches under which the distance between the two modalities can be computed~\cite{human-like-sketches, song2018learning}. This approach depends on the availability of such datasets, and requires additional effort for training the models.

Instead, we utilize the pretrained image encoder model of CLIP, which was trained on various image modalities so that it can encode information from both natural images and sketches without the need for further training.
CLIP encodes high-level semantic attributes in the last layer since it was trained on both images and text.
We therefore define the distance between the embeddings of the sketch $CLIP(\renderer(\{s_i\}_{i=1}^n)$ and image $CLIP(\image)$ as:

\begin{equation}
\label{eq:semnticloss}
    L_{semantic} = dist\big(CLIP(\image),  CLIP(\renderer(\{s_i\}_{i=1}^n)\big),
\end{equation}

where $dist(x,y)=1 - \frac{x\cdot y}{||x||\cdot ||y||}$ is the cosine distance.
However, the final encoding of the network is agnostic to low-level spatial features such as pose and structure.
To measure the geometric similarity between the image and the sketch, and consequently, allow some control over the appearance of the output, we compute the $L2$ distance between intermediate level activations of CLIP:
\begin{equation}
\label{eq:geometric}
    L_{geometric} =\ \sum_{l}{\big{\|}\ CLIP_{l}(\image) - \    CLIP_{l}(\renderer(\{s_i\}_{i=1}^n))\ \big{\|}_2^2},
\end{equation}

where $CLIP_{l}$ is the $CLIP$ encoder activation at layer $l$. Specifically, we use layers 3 and 4 of the ResNet101 CLIP model.
The final objective of the optimization is then defined as:
\begin{equation}
\label{eq:finalloss}
\min_{\{s_i\}_{i=1}^n} {L_{geometry} + w_{s} \cdot L_{semantic}},
\end{equation}
with $w_s = 0.1$.
We analyze the contribution of different layers and weights, as well as the results of using different CLIP models in the supplementary material.

\subsection{Optimization}
\label{sec:optimization}

\begin{figure}[h]
\centering
\begin{tabular}{@{\hskip2pt}c@{\hskip2pt}c@{\hskip2pt}c@{\hskip2pt}c@{\hskip2pt}c@{\hskip2pt}c}
    \midrule
    Input & 0 & 100 & 300 & 1000 & 2000 \\
    
    \midrule
    \includegraphics[width=0.14\linewidth]{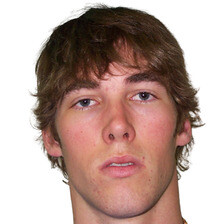} &
    \includegraphics[width=0.14\linewidth]{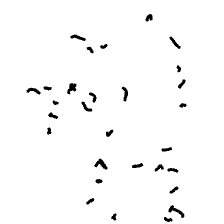} &
    \includegraphics[width=0.14\linewidth]{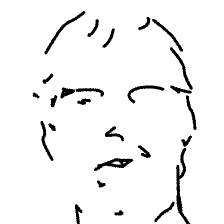} &
    \includegraphics[width=0.14\linewidth]{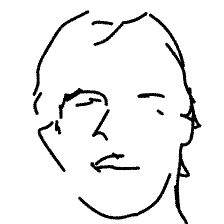} &
    \includegraphics[width=0.14\linewidth]{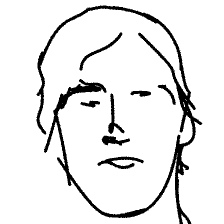} &
    \includegraphics[width=0.14\linewidth]{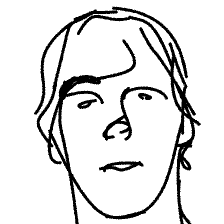}\\
\end{tabular}
\caption{The sketch appearance throughout the optimization iterations.}
\label{fig:optimization_stages}
\end{figure}

Our goal is to optimize the set of parameters 
$\{s_i\}_{i=1}^{n} = \{\{p_i^j\}_{j=1}^4\}_{i=1}^n$
to define a sketch that closely resembles the target image $\image$ in terms of both geometry and semantics.
At each step of the optimization, we use off-the-shelf gradient-based solving technique to compute the gradients of the loss with respect to the strokes' parameters $\{s_i\}_{i=1}^n$.
We follow the same data augmentation scheme suggested in CLIPDraw \cite{CLIPDraw} and augment both the sketch and the target image before feed-forwarding into CLIP. 
These augmentations prevent the generation of adversarial sketches, which minimize the objective but are not meaningful to humans.
We repeat this process until convergence, when the optimization error does not change significantly (taking typically $\sim2000$ iterations).
Figure \ref{fig:optimization_stages} illustrates the progression of the generated sketch as the optimization evolves.

\subsection{Strokes Initialization}
\label{sec:initialisation}
\begin{figure}
    \centering
    \begin{subfigure}[b]{0.99\linewidth}
        \begin{tabular}{@{\hskip2pt}c@{\hskip2pt}c@{\hskip2pt}c@{\hskip2pt}c@{\hskip2pt}c}
            \midrule
            Input & Attention & Distribution & Proposed & Random \\
            \midrule
            \includegraphics[width=0.18\linewidth]{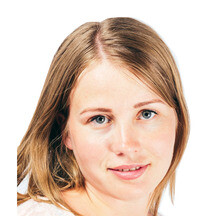} &
            \includegraphics[width=0.18\linewidth]{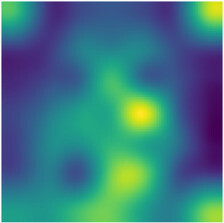} &
            \includegraphics[width=0.18\linewidth]{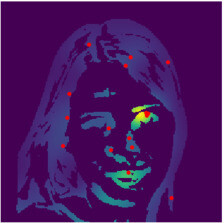} &
            \includegraphics[width=0.18\linewidth]{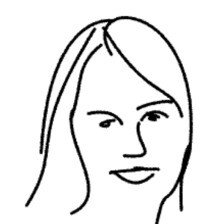} &
            \includegraphics[width=0.18\linewidth]{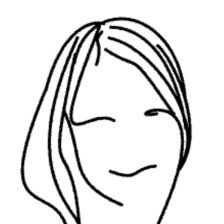}
        \end{tabular}\subcaption{Saliency-guided initialization }
        \label{fig:initialisation_random_comp}
    \end{subfigure}
    \\[3ex]
    \begin{subfigure}[b]{0.99\linewidth}
        \begin{tabular}{@{\hskip2pt}c@{\hskip2pt}c@{\hskip2pt}c@{\hskip2pt}c@{\hskip5pt}c@{\hskip2pt}c@{\hskip2pt}c@{\hskip2pt}c@{\hskip2pt}c}
        \toprule
            \includegraphics[width=0.11\linewidth]{figs/abstraction_levels/input/camel/camel.png} &
            \includegraphics[width=0.11\linewidth,cfbox=cyan 0.5pt 0.5pt]{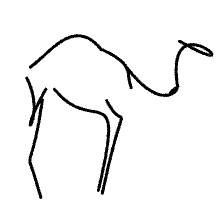} &
            \includegraphics[width=0.11\linewidth]{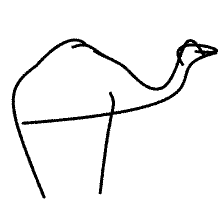} &
            \includegraphics[width=0.11\linewidth]{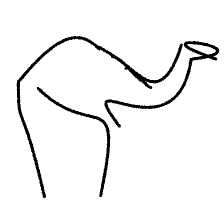} & &
            
            \includegraphics[width=0.11\linewidth]{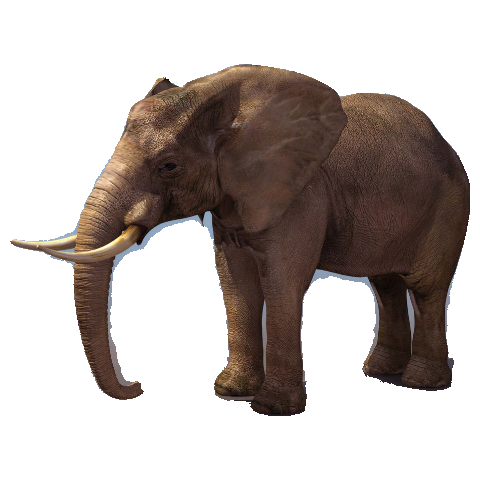} &
            \includegraphics[width=0.11\linewidth,cfbox=cyan 0.5pt 0.5pt]{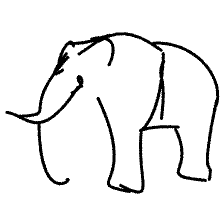} &
            \includegraphics[width=0.11\linewidth]{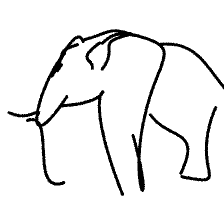} &
            \includegraphics[width=0.11\linewidth]{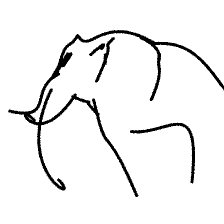} \\
            \bottomrule
        \end{tabular}
        \subcaption{Automatic selection procedure of the final sketch}
        \label{fig:different_seeds}
    \end{subfigure}
    \caption{Strokes Initialization. (a) Left to right: input, the saliency map produced from CLIP ViT activations, final distribution map (adjusted to adhere to image edges) with sampled initial stroke locations (in red), the sketch produced using the proposed initialization procedure, and the sketch produced when using random initialization. (b) Results of three different initializations with the same number of strokes. The sketches marked in blue produced the lowest loss value, and would thus be used as the final output.}
    \label{fig:initialisation_examples}
\end{figure}

Our objective function is highly non-convex. Therefore, the optimization process is susceptible to the initialization (i.e., the initial location of the strokes). 
This is especially significant at higher levels of abstraction – where very few strokes must be wisely placed to emphasize semantic components.
For example, in Figure \ref{fig:initialisation_random_comp}, the sketches in the last two columns were produced using the same number of strokes, however, in the ``Random'' initialization case, more strokes were devoted to the hair while the eyes, nose, and mouth are more salient and critical features of the face.

To improve convergence towards semantic depictions, we place the initial strokes based on the salient regions of the target image. We use the pretrained vision transformer~\cite{Kolesnikov2021ViT} ViT-B/32 model of CLIP, that performs global context modeling using self-attention between patches of a given image to capture meaningful features.
We use the recent transformer interpretability method by Chefer et al. \cite{Chefer_2021_ICCV} to extract a relevancy map from the self-attention heads, without any text supervision.

Our final distribution map is produced by multiplying the relevancy map with the edge map of the image extracted using XDoG \cite{Winnemller2012XDoGAE}, followed by a softmax normalization. 
XDoG is used to strengthen the morphological positioning of the strokes, motivated by the hypothesis that edges are effective in predicting where people draw lines \cite{Hertzmann2021TheRO}.

Figure \ref{fig:initialisation_random_comp} illustrates this procedure.
It can be seen that our saliency-based initialization contributes significantly to the quality of the final sketch compared to random initialization. 

This sampling-based approach also lends itself to providing variability in the results. In all our examples we typically use 3 initializations, and automatically choose the one that yields the lowest loss (see Figure \ref{fig:different_seeds}).
We further analyze the initialization procedure and variability in the supplementary material.

\begin{figure}[b]
    \centering
    \includegraphics[width=1\linewidth]{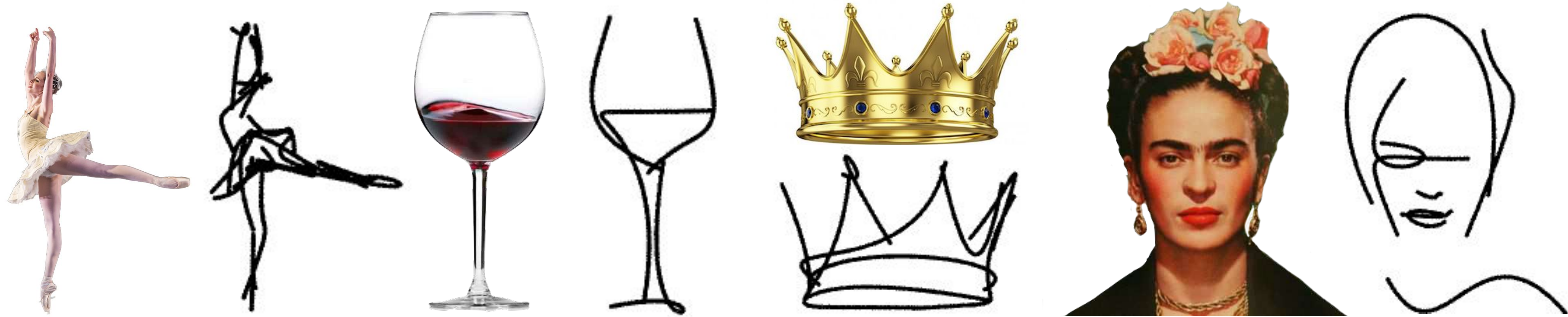}
    
    \caption{Sketches produced by our method for infrequent categories.}
    \label{fig:robust}
\end{figure}

\section{Results}
\label{sec:results}
Section \ref{sec:qualitative} provide qualitative evaluations.
In section \ref{sec:comparison} we compare our method with existing image-to-sketch methods, which were all trained on sketch-specific datasets.
In Section \ref{sec:quantitative}, we supply a quantitative evaluation of our method's ability to produce recognizable sketches testing both category and instance recognition.
For images with background, we use an automatic method (U2-Net \cite{Qin_2020_PR}) to mask out their background.
We provide further analysis of our method, extra results, and extended comparison with other methods in the supplemental file.

\subsection{Qualitative Evaluation}
\label{sec:qualitative}
Our approach is different from conventional sketching methods in that it does not utilize a sketch dataset for training, rather it is optimized under the guidance of CLIP. Thus, our method is not limited to specific categories observed during training, as no category definition was introduced at any stage. This makes our method robust to various inputs, as shown in Figures~\ref{fig:teaser} and \ref{fig:robust}.

In Figures \ref{fig:teaser} and ~\ref{fig:abstraction_levels} we demonstrate the ability of our method to produce sketches at different levels of abstraction.
As the number of strokes decreases, the task of minimizing the loss becomes more challenging, forcing the strokes to capture the essence of the object. For example, in the abstraction process of the flamingo in Figure \ref{fig:teaser}, the transition from 16 to 4 strokes led to the removal of details such as the eyes, feathers, and feet, while maintaining the important visual features such as the general pose, the neck and legs which are iconic characteristics of a flamingo. 

\begin{figure}
    \begin{subfigure}[b]{0.62\linewidth}
            \begin{tabular}{@{\hskip2pt}c@{\hskip2pt}c@{\hskip2pt}c@{\hskip2pt}c}
            Input & 4 cp & 3 cp & 2 cp  \\
            \includegraphics[width=0.23\linewidth]{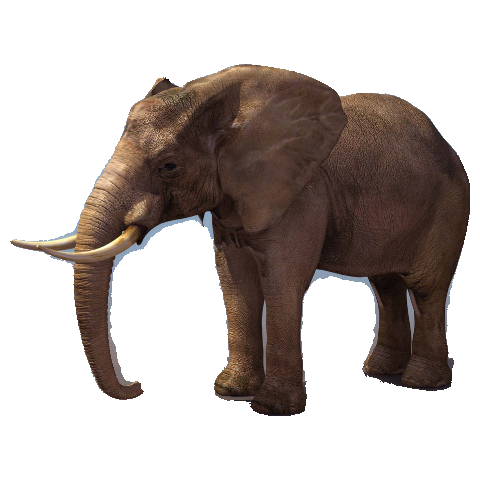} &
            \includegraphics[width=0.23\linewidth]{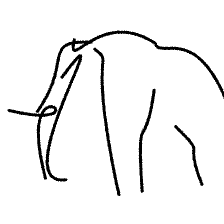} & 
            \includegraphics[width=0.23\linewidth]{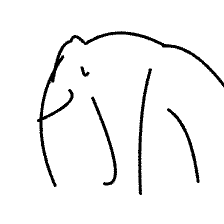} &
            \includegraphics[width=0.23\linewidth]{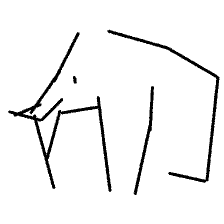}
            \end{tabular}
             \subcaption{Changing the degree of the curves}\label{fig:n_control_points}
        \end{subfigure}
        \begin{subfigure}[b]{0.37\linewidth}
            \includegraphics[width=0.45\linewidth]{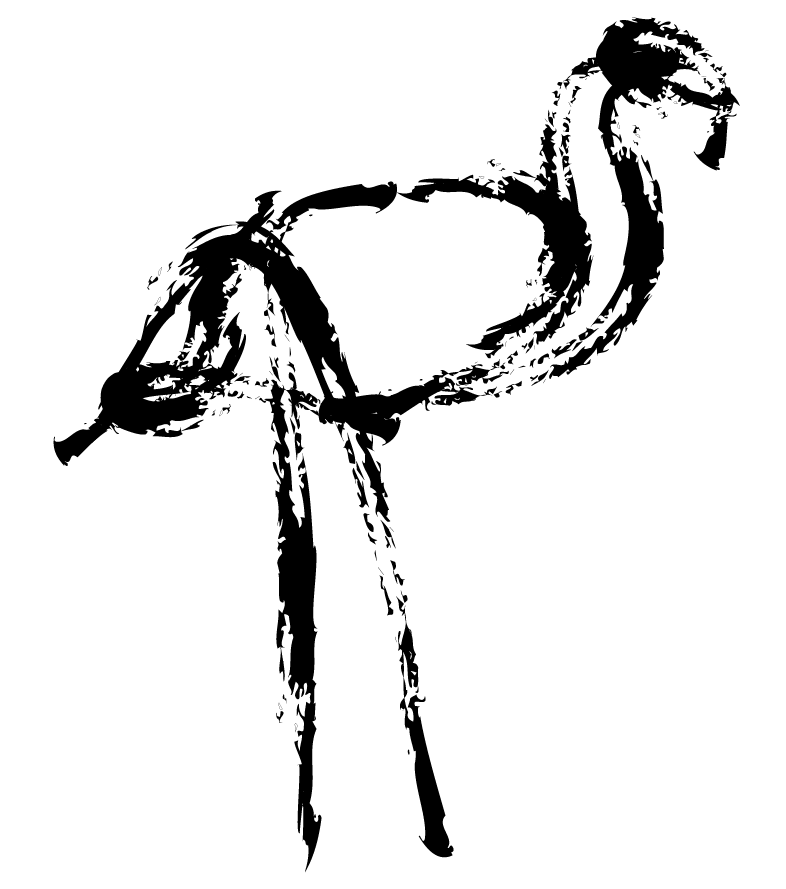}
            \includegraphics[width=0.52\linewidth]{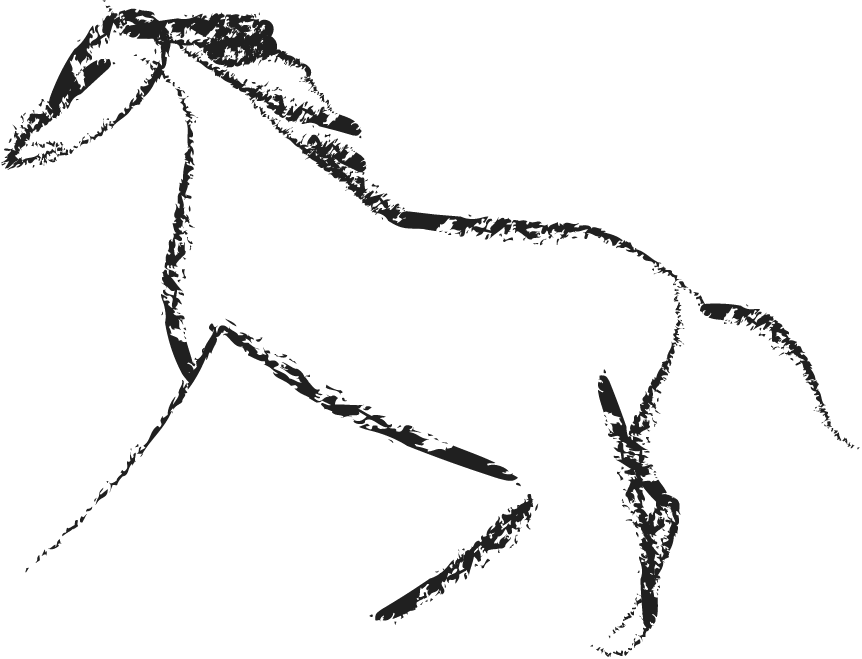}
            \subcaption{Editing the brush style on SVGs}
            \label{fig:pencil_style}
        \end{subfigure}
        
    \caption{Changing sketch style. (a) From left to right are the results produced by our method when using Bézier curves with 4, 3, and 2 control points (cp), respectively. We can see how this affects the style of the output sketch. (b) Using Adobe Illustrator, horse — pencil feather, flamingo — dry brush. }
    \label{fig:style_change}
\end{figure}

Besides changing the number of strokes, different sketch styles can be achieved by varying the degree of the strokes (Figure~\ref{fig:n_control_points}) or using a brush style on top of the vector strokes (Figure \ref{fig:pencil_style}).

\subsection{Comparison with Existing Methods}
\label{sec:comparison}
\begin{figure}
    \centering
    \includegraphics[width=\linewidth]{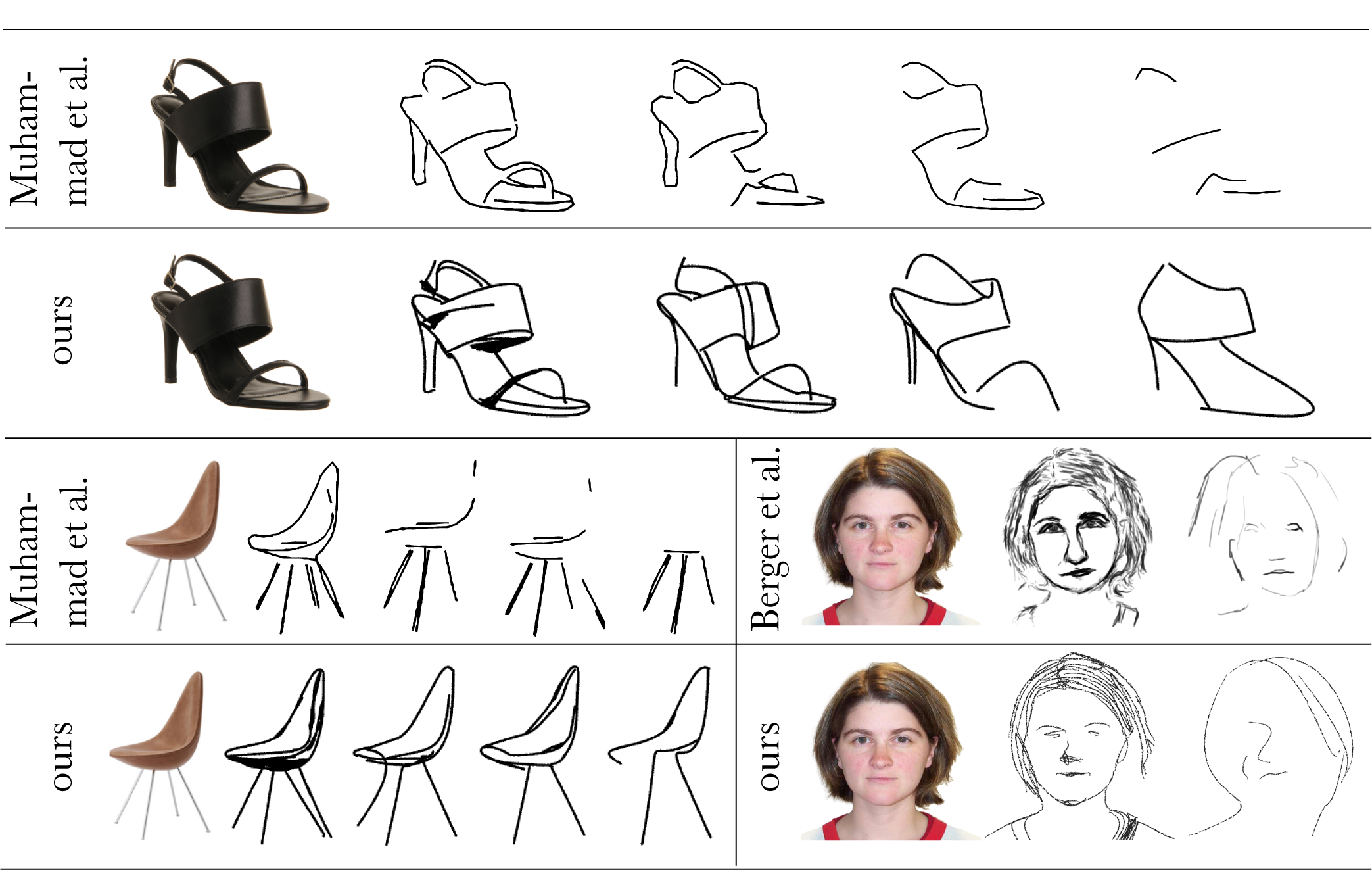}
    \caption{Levels of Abstraction Comparison — in the top and left part are comparisons to Muhammad et al. \cite{Deep-Sketch-Abstraction}. The leftmost column shows the input image, and the next four columns show different levels of abstraction. For the shoe and chair our results were produced using 32, 16, 8, and 4 curves (from left to right).
    In the right bottom part is a comparison to Berger et al. \cite{Berger2013}, we use 64 and 8 strokes to generate our sketches.}
    \label{fig:comp_shoes}
\end{figure}

\paragraph{Sketches with different levels of abstraction.}
Only a few works have attempted to sketch objects at different levels of abstraction. In Figure \ref{fig:comp_shoes} we compare with Muhammad et al. \cite{Deep-Sketch-Abstraction} and Berger et al. \cite{Berger2013}.
The results by Muhammad et al. demonstrate four levels of abstraction on two simple inputs — a shoe and a chair (in the absence of their code, the results were taken directly from the paper).
We produce sketches at four levels of abstraction using 32, 16, 8, and 4 strokes.
The sketches by Muhammad et al. are coherent with the geometry of the image; but to achieve higher levels of abstraction, they only remove strokes from the generated sketch without changing the remaining ones. This can result in losing class-level recognizability at higher levels of abstraction (rightmost sketches). 
Such an approach is sub-optimal, since a better arrangement may be possible for fewer strokes.
Our method successfully produces a recognizable rendition of the subject while preserving its geometry, even in the challenging 4-stroke case (rightmost sketch). 

In the right bottom part of Figure \ref{fig:comp_shoes} we compare with the method of Berger et al. \cite{Berger2013}. Their results were provided by the authors and demonstrate two levels of abstraction generated based on the style of a particular artist. 
We use 64 and 8 strokes, respectively, to achieve two comparable levels of abstraction and place a pencil style on top of the generated sketch to better fit the artist's style. As can be seen, our approach is more geometrically coherent while still allowing abstraction. Their results fit better to a specific style, but can only work with faces and are limited to the dataset gathered.

\begin{figure}
\centering
\begin{tabular}{@{\hskip1pt}c|@{\hskip1pt}c@{\hskip1pt}c@{\hskip1pt}c|@{\hskip1pt}c@{\hskip1pt}c}
     Input & \multicolumn{3}{c}{Data-Driven} & \multicolumn{2}{|c}{Ours} \\
     & A & B & C & 16s & 8s \\
    \hline
    \includegraphics[trim={0 1.5cm 0  1.5cm},clip,width=\widthrescomp\linewidth]{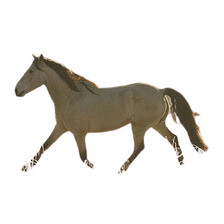} &
    \includegraphics[trim={0 1cm 0  1cm},clip,width=\widthrescomp\linewidth]{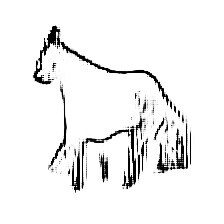} &
    \includegraphics[width=\widthrescomp\linewidth]{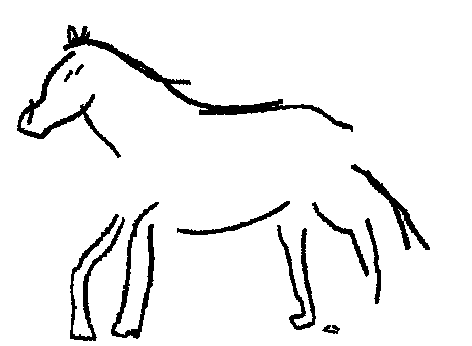} &
    \includegraphics[trim={0 1.5cm 0  1.5cm},clip,width=\widthrescomp\linewidth]{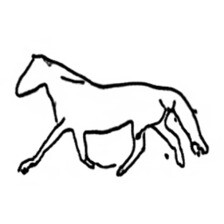} &
    \includegraphics[trim={0 1cm 0  1cm},clip,width=\widthrescomp\linewidth]{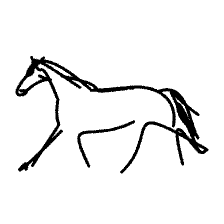} &
    \includegraphics[trim={0 1cm 0  1cm},clip,width=\widthrescomp\linewidth]{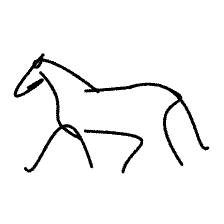} \\
   
    \includegraphics[trim={0 1.5cm 0  1.5cm},clip,width=\widthrescomp\linewidth]{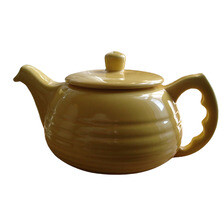} &
    \includegraphics[trim={0 1cm 0  1cm},clip,width=\widthrescomp\linewidth]{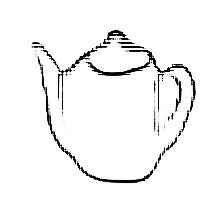} &
    \raisebox{.2\height}{\includegraphics[width=\widthrescomp\linewidth]{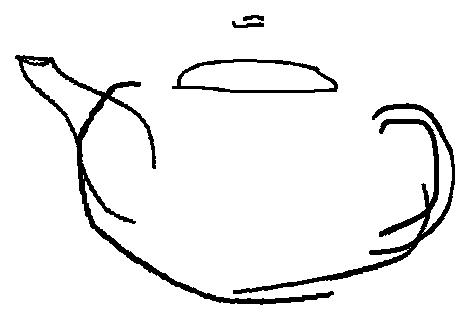}} &
    \includegraphics[trim={0 1.5cm 0  2cm},clip,width=\widthrescomp\linewidth]{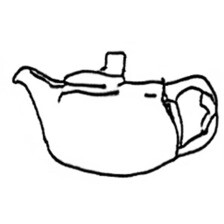} &
    \includegraphics[trim={0 1cm 0  1cm},clip,width=\widthrescomp\linewidth]{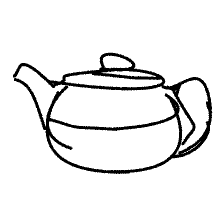} &
    \includegraphics[trim={0 1cm 0  1cm},clip,width=\widthrescomp\linewidth]{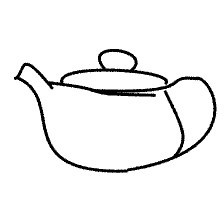} \\
    
    \includegraphics[trim={0 2cm 0  2cm},clip,width=\widthrescomp\linewidth]{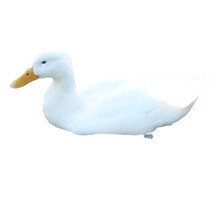} &
    \raisebox{.1\height}{\includegraphics[trim={0.5cm 0 0.5cm  0},clip,width=\widthrescomp\linewidth, height=0.1\linewidth]{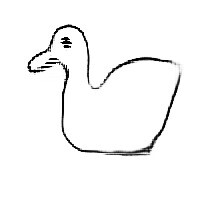}} &
    \raisebox{.1\height}{\includegraphics[width=\widthrescomp\linewidth, height=0.1\linewidth]{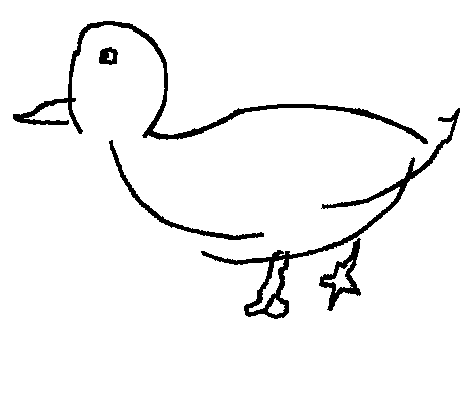}} &
    \includegraphics[trim={0 2cm 0  2cm},clip,width=\widthrescomp\linewidth]{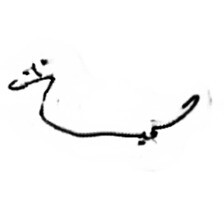} &
    \includegraphics[trim={0 1cm 0  1cm},clip,width=\widthrescomp\linewidth]{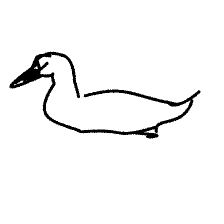} &
    \includegraphics[trim={0 1cm 0  1cm},clip,width=\widthrescomp\linewidth]{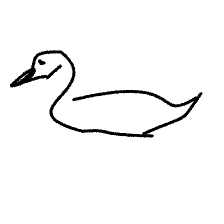}\\
    
     \midrule
      & D & E &  & 16 & 4 \\
    \includegraphics[trim={0 1cm 0  1cm},clip,width=\widthrescomp\linewidth]{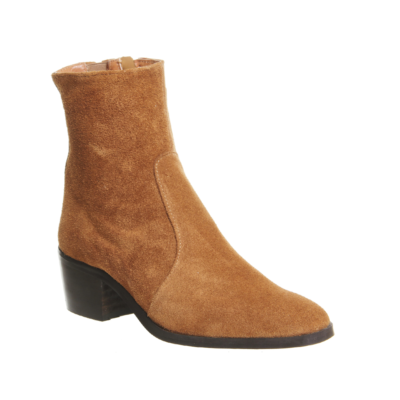} & 
    \includegraphics[trim={0 1cm 0  1cm},clip,width=\widthrescomp\linewidth]{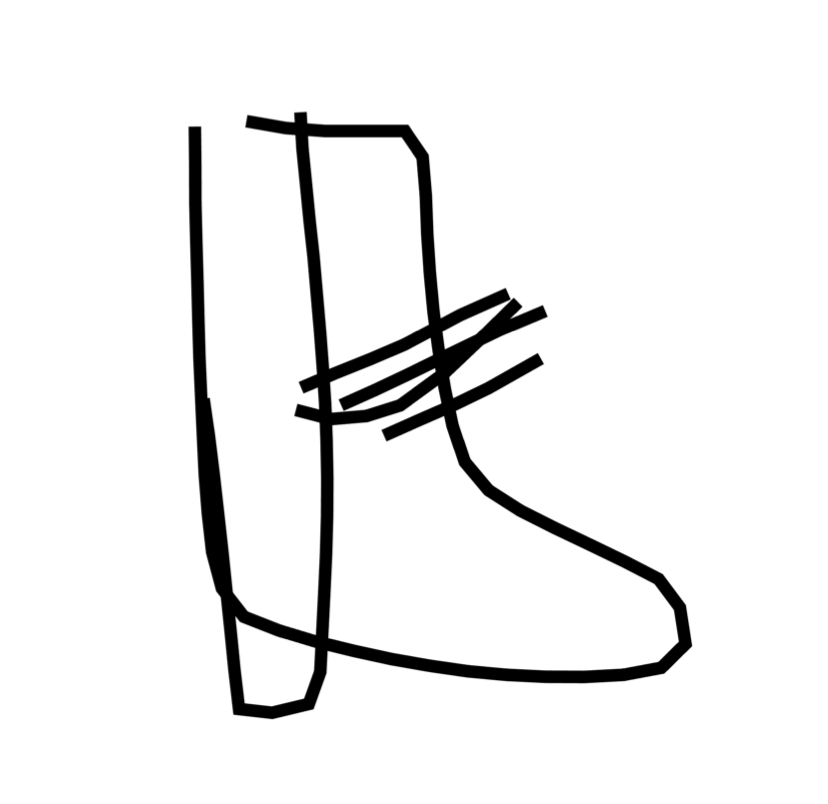} & 
    \includegraphics[trim={0 1cm 0  1cm},clip,width=\widthrescomp\linewidth]{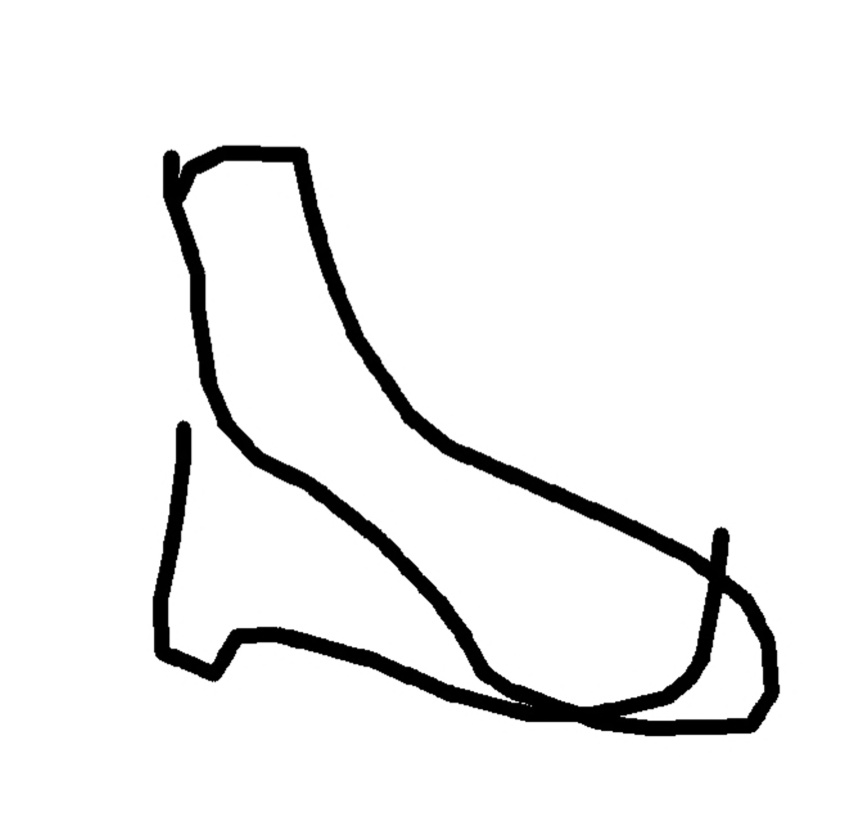} &
    &
    \includegraphics[trim={0 1cm 0  1cm},clip,width=\widthrescomp\linewidth]{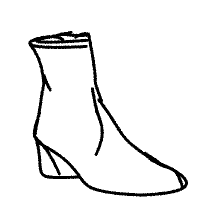} &
    \includegraphics[trim={0 1cm 0  1cm},clip,width=\widthrescomp\linewidth]{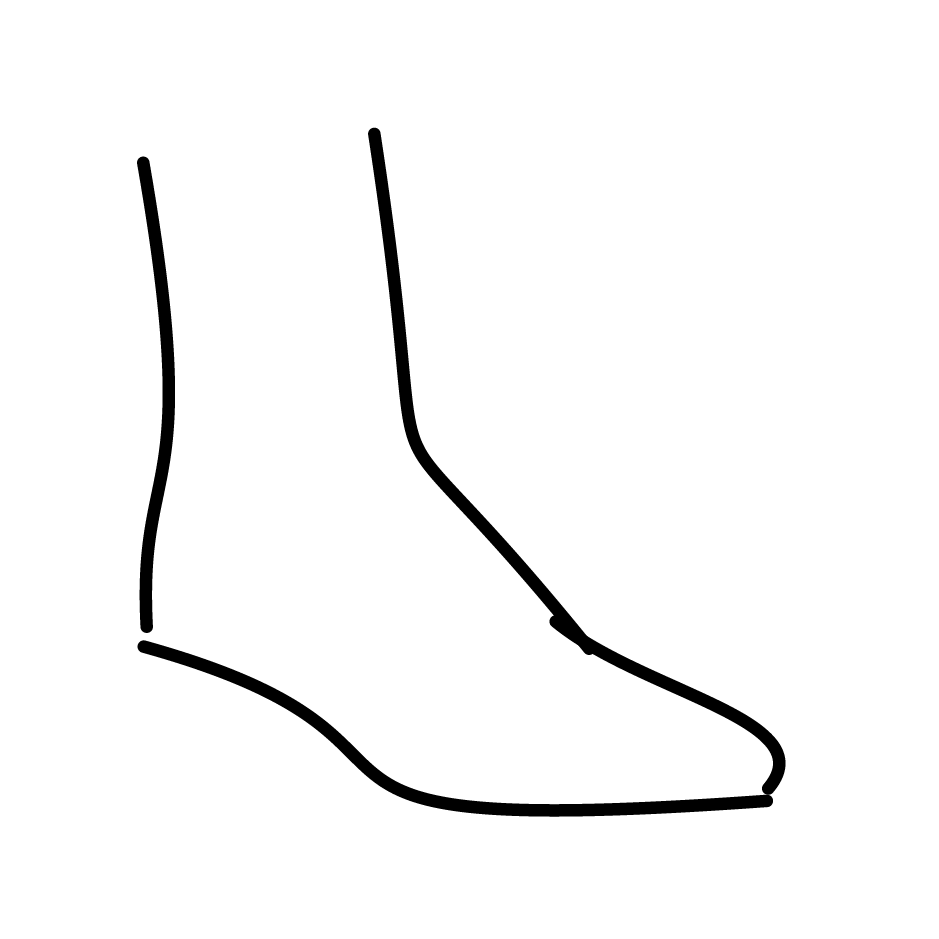}

\end{tabular}
 \caption{Comparison to Existing Image-to-Sketch Works — the leftmost column shows the input images. The methods presented are (A) Kampelmühler and Pinz \cite{human-like-sketches}, (B) Li et al. \cite{Deformable_Stroke}, (C) Li et al. \cite{li2019photosketching} (D) Song et al. \cite{song2018learning}, (E) SketchLattice \cite{Qi2021SketchLatticeLR}.}
\label{fig:comp1}
\end{figure}

\paragraph{Photo-Sketch Synthesis.}
In Figure \ref{fig:comp1} we present a comparison with the five works outlined in Table \ref{tab:sketch_synth_comparison}. 
The results by Kampelmühler and Pinz \cite{human-like-sketches} (A), Li et al. \cite{Deformable_Stroke} (B) and Li et al. \cite{li2019photosketching} (C) were generated based on the authors' implementation and best practice.
Due to the lack of a publicly available implementation of SketchLattice \cite{Qi2021SketchLatticeLR} (E), their results are taken directly from the paper.
We present the sketches of Song et al. \cite{song2018learning} (D) on shoe images, since their method only works with shoes and chairs. 

Each of these methods define a specific objective which influences their dataset selection and final output style.
Li et al. \cite{li2019photosketching} (C) aim for boundary-like drawings, and indeed, geometric coherence is achieved with the input image, capturing salient outlines.
The other methods are designed to produce human-like sketches of non-experts, and indeed, the synthesized sketches exhibit a ``doodle-like'' style. Furthermore, the methods learn highly abstract concepts (such as highlighting the eyes) while maintaining some relation to the geometry of the input object.

Each method accomplishes their respective objective, but we wish to emphasize the particular benefits of our method.
First, all of the above methods are sketch-data dependent, meaning they can only be used with the style and level of abstraction observed during training. With our framework, we can handle images of all categories and produce sketches of various levels of abstraction simply by changing the number of strokes. Although every method can be retrained with new datasets, this is neither convenient nor practical, and depends on the availability of such datasets.
Second, while each method leans towards a more semantic or a more geometric sketching style, our method can provide both. For example, our method did not produce a perfect alignment of the legs of the horse, as in (C), but it captured the movement of the horse in a minimal way.

\begin{figure}
\centering
    \begin{tabular}{ccc}
    \toprule
         Input & CLIPDraw & Ours  \\
         \midrule
         \includegraphics[width=0.2\linewidth]{figs/comp_objects/input/men/face_7_edge.jpg} &
         \includegraphics[width=0.2\linewidth]{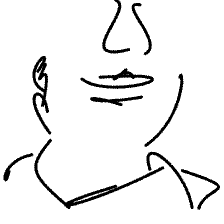} &
         \includegraphics[width=0.2\linewidth]{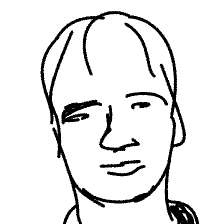} \\
         
         \includegraphics[width=0.2\linewidth]{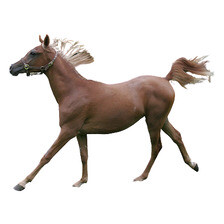} &
        \includegraphics[width=0.2\linewidth]{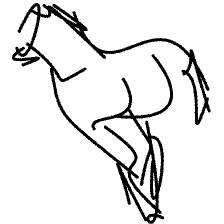} &
        \includegraphics[width=0.2\linewidth]{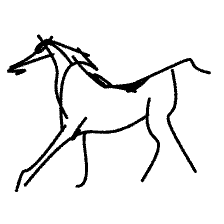}
    \end{tabular}
    \vspace{-2mm}
    \caption{Comparison to CLIPDraw \cite{CLIPDraw}.
    All sketches were produced using 16 strokes.}
    \label{fig:clipdraw_comp}
\end{figure}

In Figure \ref{fig:clipdraw_comp} we provide a comparison with CLIPDraw \cite{CLIPDraw}. The text input for CLIPDraw is replaced with the target image. This was made possible since CLIP encodes both text and images to the same latent space. To provide a comparable visualization, we constrain the output primitives of CLIPDraw in the same manner as we defined our strokes.
As can be seen, although the parts of the subject can be recognizable using CLIPDraw, since there is no geometric grounding to the image, the overall structure is destroyed. Further comparisons of CLIPDraw incorporating text and color can be found in the supplemental file.

\subsection{Quantitative Evaluation}
\label{sec:quantitative}
We conduct a user study with 121 participants to assess both the category-level and instance-level recognizability of the sketches generated by our method at different levels of abstraction.
Additionally, similar to previous image-to-sketch works \cite{human-like-sketches, song2018learning, Deep-Sketch-Abstraction}, we also use pretrained classifier networks to evaluate the category-level recognizability of the sketches generated by our method. 

\paragraph{User-Study}
We choose five popular animal classes from the SketchyCOCO dataset \cite{gao2020sketchycoco} and randomly sample five images per class.
We synthesize sketches at four levels of abstraction for each image, with 4, 8, 16, and 32 strokes. We compare the recognition rates with the two recent photo-sketch synthesis methods by Kampelmühler and Pinz \cite{human-like-sketches} and Li et al. \cite{li2019photosketching}. The sketches were produced with one level of abstraction, suited to the capabilities of the methods. 

Each participant was presented with randomly selected sketches of a single method. To examine category-level recognition, participants were asked to choose the correct category text description alongside four confound categories and the option 'None'. For the instance-level recognition experiment, the distractors are images from the same object category. 
Table \ref{tab:user_study_animal} shows the average recognition rates attained from the user-study.
Both the category-level and instance-level recognizability are inversely correlated with the level of abstraction. At four strokes, we can see that our sketches are hardly recognizable at the category level (36\%), which illustrates a ``breaking point'' of our method. At eight strokes and above, both instance and class level rates are high. Li et al. \cite{li2019photosketching} demonstrate full recognition at the instance level; this is understandable given that the sketches are contour-based and not abstract. At 16 and 32 strokes, we achieve comparable rates, and even with the high level of abstraction when using only eight strokes, we achieve 95\% instance level recognizability. The sketches by Kampelmühler and Pinz are more abstract, which explains their low accuracy at the instance and class level.

\paragraph{Sketch classifier}
Two different classifiers are used for evaluating the synthesized sketches; a ResNet34 classifier from Kampelm{\"{u}}hler and Pinz \cite{human-like-sketches} trained on the Sketchy-Database \cite{Sketchy-Database} with 125 categories, and a CLIP ViT-B/32 zero-shot classifier using text prompts defined as "A sketch of a(n) \textit{class-name}". Note that this is not the CLIP model we use for training.
Table \ref{tab:classification_res} compares the sketch-classifier recognition accuracy of our sketches to that of Kampelm{\"{u}}hler and Pinz \cite{human-like-sketches} and Li et al. \cite{li2019photosketching} based upon 200 randomly selected images of 10 categories from the SketchyCOCO dataset \cite{gao2020sketchycoco}.
The recognition accuracy on human sketches from the SketchyCOCO dataset is also calculated as a baseline.
The method by Kampelm{\"{u}}hler and Pinz achieves the highest scores for ResNet34 classifier, possibly since they use the same model and dataset during training.
Despite the distribution differences, our method still achieves good recognition rates under this classifier.
With CLIP classifier, our method achieves very high accuracy levels of 78\% with 16 strokes and 91\% with 32 strokes. For more details and analysis please refer to the supplementary material.

\begin{table}
  \caption{User study results – average recognition rates. (A) Kampelmühler and Pinz \cite{human-like-sketches}, (B) Li et al. 
\cite{li2019photosketching}.}
  \label{tab:user_study_animal}
  \begin{tabular}{@{\hskip5pt}c@{\hskip5pt}c@{\hskip5pt}c@{\hskip5pt}c@{\hskip5pt}c@{\hskip5pt}c@{\hskip5pt}c}
    \toprule
     & A & B & \begin{tabular}[c]{@{}c@{}}Ours \\ 4s\end{tabular} & \begin{tabular}[c]{@{}c@{}}Ours \\ 8s\end{tabular} & \begin{tabular}[c]{@{}c@{}}Ours \\ 16s\end{tabular} & \begin{tabular}[c]{@{}c@{}}Ours \\ 32s\end{tabular} \\
    \midrule
    \begin{tabular}[c]{@{}c@{}}Category-\\ Level\end{tabular} & 
    \begin{tabular}[c]{@{}c@{}}65\% \\ $\pm
    2\%$\end{tabular} & 
    \begin{tabular}[c]{@{}c@{}}96.9\% \\ $\pm 0.7\%$\end{tabular} & 
    
    \begin{tabular}[c]{@{}c@{}} 36\%    \\ $\pm 3\%$\end{tabular} & 
    \begin{tabular}[c]{@{}c@{}} 87\% \\ $\pm 2\%$\end{tabular} & 
    \begin{tabular}[c]{@{}c@{}} 97.9\% \\ $\pm 0.8\%$\end{tabular} & 
    \begin{tabular}[c]{@{}c@{}} 99.3\% \\ $\pm 0.5\%$\end{tabular} \\
    \midrule
    
    \begin{tabular}[c]{@{}c@{}}Instance-\\ Level\end{tabular} & 
    \begin{tabular}[c]{@{}c@{}}65\% \\ $\pm 2\%$\end{tabular} & 
    \begin{tabular}[c]{@{}c@{}}99.1\% \\ $\pm 0.4\%$\end{tabular} & 
    
    \begin{tabular}[c]{@{}c@{}}72\% \\ $\pm 3\%$\end{tabular} & 
    \begin{tabular}[c]{@{}c@{}}95\% \\ $\pm 1\%$\end{tabular} & 
    \begin{tabular}[c]{@{}c@{}}96\% \\ $\pm 1\%$\end{tabular} & 
    \begin{tabular}[c]{@{}c@{}}97\% \\ $\pm 1\%$\end{tabular} \\
    
  \bottomrule
\end{tabular}
\end{table}

\begin{table}
\caption{Top-1 and Top-3 sketch recognition accuracy computed with ResNet34 and CLIP ViT-B/32 on 200 sketches from 10 categories. (A) Kampelmühler and Pinz \cite{human-like-sketches}, (B) Li et al. 
\cite{li2019photosketching}}
\label{tab:classification_res}
\centering
\begin{tabular}{@{\hskip3pt}c@{\hskip3pt}c|c:cccc}
\toprule
                        Classifier  &      & \begin{tabular}[c]{@{}c@{}}Human \\ Sketches\end{tabular} & A & B & \begin{tabular}[c]{@{}c@{}}Ours \\ 16s\end{tabular} & \begin{tabular}[c]{@{}c@{}}Ours \\ 32s\end{tabular} \\ 
\hline
\multirow{2}{*}{ResNet34} & Top1 & 98\%  & \textbf{\textbf{67\%}}  & 61\%   & 54\%  &  63\%  \\
                          & Top3 & 99\%  & \textbf{\textbf{82\%}}  & 78\%   & 75\%  &  77\%  \\ 
\hline
\multirow{2}{*}{\begin{tabular}[c]{@{}c@{}}CLIP\\ ViT-B/32\end{tabular}}  & Top1 & 75\%  & 49\%  & 60\%  & \textbf{\textbf{78\%}} & 91\%  \\
                                & Top3 & 93\%  & 65\%  & 77\%  & \textbf{\textbf{93\%}} & 97\%  \\
\hline
\end{tabular}
\end{table}

\section{Limitations}
For images with background, our method's performance is reduced at higher abstraction levels.
This limitation can be addressed by using an automatic mask. However, a potential development would be to include such a remedial term within the loss function.
In addition, our sketches are not created sequentially and all strokes are optimized simultaneously, which differs from the conventional way of sketching.
Furthermore, the number of strokes must be determined in advance to achieve the desired level of abstraction. Another possible extension could be to make this a learned parameter, as different images might require different numbers of strokes to reach similar levels of abstraction.

\section{Conclusions}
We presented a method for photo-sketch synthesis, producing sketches with different levels of abstraction, without the need to train on specific sketch datasets.
Our method can generalize to various categories and cope with challenging levels of abstraction, while maintaining the semantic visual clues that allow for instance-level and class-level recognition.

{\small
\bibliographystyle{ieee_fullname}
\bibliography{bibliography}
}
\clearpage

\title{CLIPasso: Semantically-Aware Object Sketching \\
Supplementary Material}
\author{
Yael Vinker$^{2,1}$
\and
Ehsan Pajouheshgar$^{1}$
\and
Jessica Y. Bo$^{1}$
\and
Roman Christian Bachmann$^{1}$
\and
Amit Haim Bermano$^{2}$
\and
Daniel Cohen-Or$^{2}$
\and
Amir Zamir$^{1}$
\and
Ariel Shamir$^{3}$
\vspace{2mm}
\and $^{1}$Swiss Federal Institute of Technology (EPFL) \and $^{2}$Tel-Aviv University \and $^{3}$Reichman University
}

\maketitle

\appendix
\addcontentsline{toc}{section}{} 
\part{} 
\vspace{-3em}
\parttoc 
\section{Implementation Details}
We implement our method in PyTorch \cite{NEURIPS2019_9015} utilizing the "diffvg" differentiable rasterizer implementation by \cite{diffvg}.

\textbf{Initialization details}: Using the ViT-32/B CLIP~\cite{clip} model, we extract the salient regions of an image by averaging all of the attention heads of each self-attention layer, resulting in 12 attention maps. Then, we average these attention maps and take the attention between the final class embedding and all 49 patches to get our relevancy map.
We multiply the relevancy map with the edge map extracted using XDoG~\cite{Winnemller2012XDoGAE}, and then use this final attention map to sample the locations of the initial strokes.
We first sample $n$ positions for the first control point of each curve using the map and then randomly sample the next three control points of each Bezier curve within a small radius (0.05) of the first point.

\textbf{Augmentation details}: Following the implementation of CLIPDraw \cite{CLIPDraw}, we apply random affine augmentations to both the sketch and target images before passing them as inputs to CLIP. The transformations we use are RandomPerspective and RandomResizedCrop. These augmentations make the optimization less prone to adversarial samples and improve the quality of the generated sketch.

\textbf{Optimization details}: We use Adam optimizer with a learning rate set to 1. We evaluate the output sketch every 10 iterations. Evaluation is done by computing the loss without random augmentations. We repeat the optimization process until convergence (when the difference in loss between two successive evaluations is less than 0.00001), this typically takes around 2000 iterations.
As noted in the paper, the final result is highly sensitive to the initialization, therefore, we run the process in parallel with three different seeds and select the final sketch which has the lowest evaluation loss.
It takes 6 minutes to run 2000 iterations on a single Tesla V100 GPU, however, after 100 iterations it is already possible to get a recognizable sketch for most images.
In the case of images with backgrounds, it is possible to use an automatic tool such as U2-Net for background removal, and then apply the proposed method.

\section{Dealing With Background}
As stated in the paper, our method is most suitable for input images that contain only the object to be drawn without background.
Generally, methods for object sketching that can cope with background are trained in a supervised manner to overcome this challenge, i.e. the training dataset included pairs of image with background and the corresponding sketch without background \cite{Sketchy-Database}. As a result, these models are encouraged to omit the background.
The fact that we perform zero-shot generation implies that we do not rely on the characteristics of a particular sketches dataset to constrain the appearance of the output sketch. Consequently, our method includes consideration of the background. Moreover, the geometric loss term encodes relatively shallow features, making it less semantic than the final fully connected layer. This also contributes to the inclusion of background details in the final sketch results.

Even though the task is challenging, our model can still handle images with background, but with reduced performance.

\begin{figure}
\centering
\begin{tabular}{@{\hskip2pt}c@{\hskip2pt}c@{\hskip2pt}c@{\hskip2pt}c@{\hskip2pt}c@{\hskip2pt}c@{\hskip2pt}c}
    \midrule
    Input & Attention & 32 & 16 & 8 & Mask & 16 \\
    
    \midrule
    \includegraphics[width=\widthbackground\linewidth]{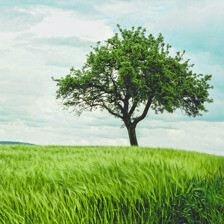} &
    \includegraphics[width=\widthbackground\linewidth]{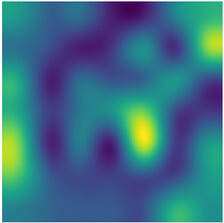} &
    \includegraphics[width=\widthbackground\linewidth]{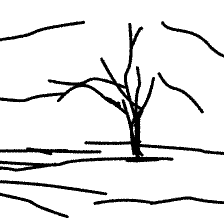} &
    \includegraphics[width=\widthbackground\linewidth]{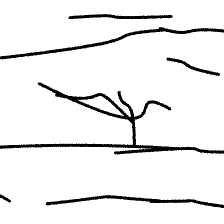} &
    \includegraphics[width=\widthbackground\linewidth]{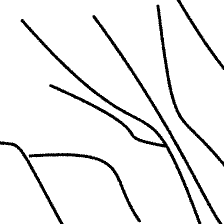} &
    \includegraphics[width=\widthbackground\linewidth]{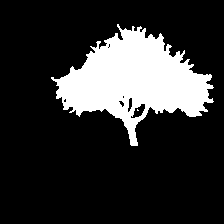} &
    \includegraphics[width=\widthbackground\linewidth]{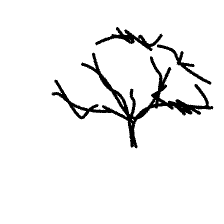} \\
    
    \midrule
    \includegraphics[width=\widthbackground\linewidth]{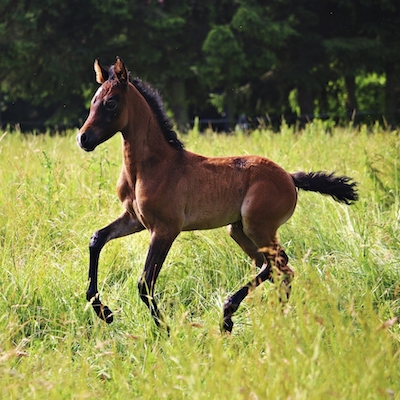} &
    \includegraphics[width=\widthbackground\linewidth]{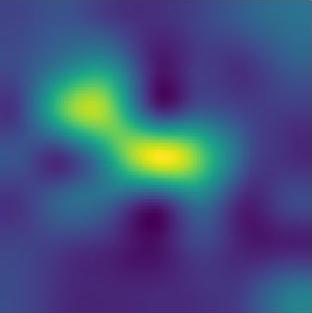} &
    \includegraphics[width=\widthbackground\linewidth]{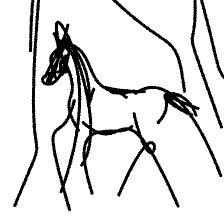} &
    \includegraphics[width=\widthbackground\linewidth]{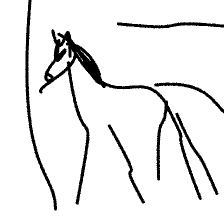} &
    \includegraphics[width=\widthbackground\linewidth]{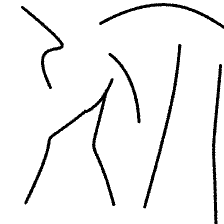} &
    \includegraphics[width=\widthbackground\linewidth]{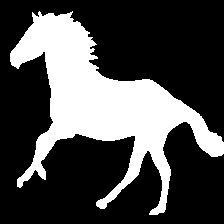} &
    \includegraphics[width=\widthbackground\linewidth]{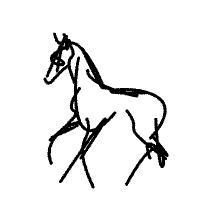} \\
    
    \midrule
    \includegraphics[width=\widthbackground\linewidth]{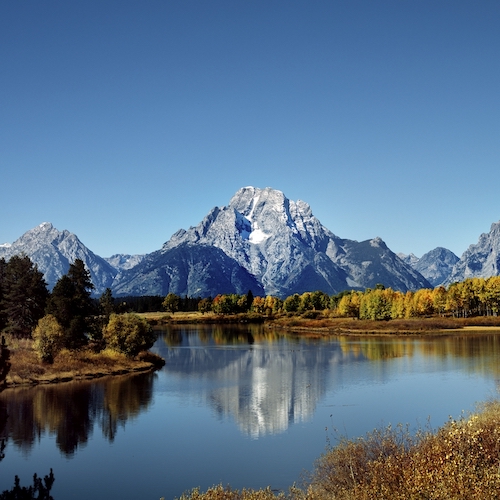} &
    \includegraphics[width=\widthbackground\linewidth]{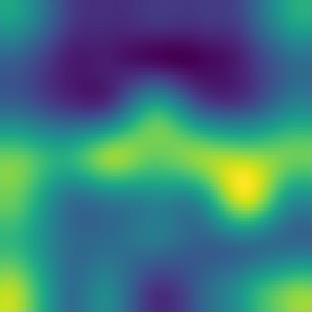} &
    \includegraphics[width=\widthbackground\linewidth]{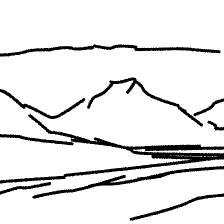} &
    \includegraphics[width=\widthbackground\linewidth]{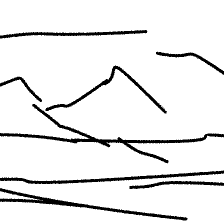} &
    \includegraphics[width=\widthbackground\linewidth]{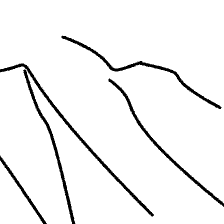} &
    \includegraphics[width=\widthbackground\linewidth]{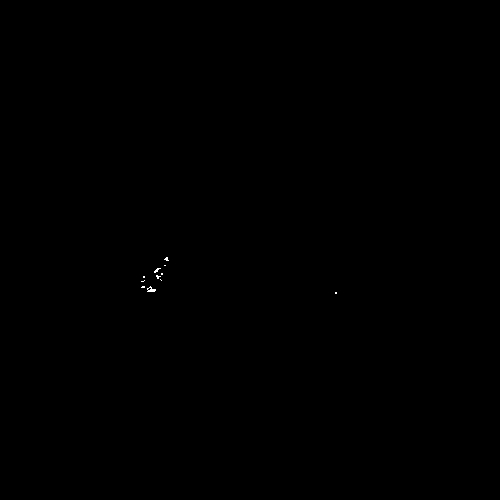} &
    \includegraphics[width=\widthbackground\linewidth]{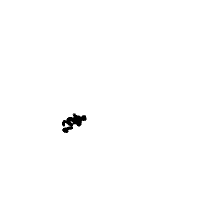} \\
    
    \midrule
    \includegraphics[width=\widthbackground\linewidth]{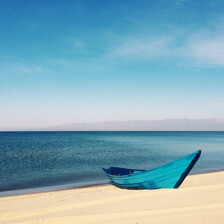} &
    \includegraphics[width=\widthbackground\linewidth]{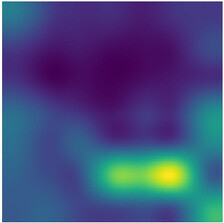} &
    \includegraphics[width=\widthbackground\linewidth]{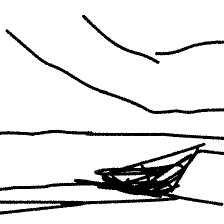} &
    \includegraphics[width=\widthbackground\linewidth]{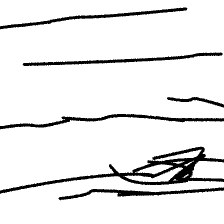} &
    \includegraphics[width=\widthbackground\linewidth]{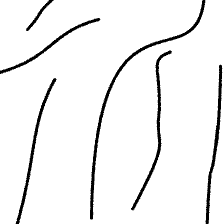} &
    \includegraphics[width=\widthbackground\linewidth]{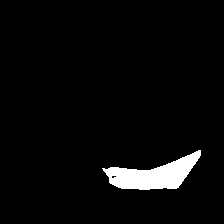} &
    \includegraphics[width=\widthbackground\linewidth]{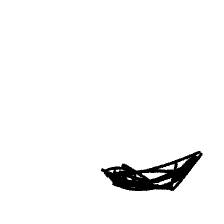} \\
    
\end{tabular}
 \caption{Background Analysis - the first column shows the input images with background that were used for our optimization, the second column shows the extracted attention map. Columns 3, 4, and 5 show the output sketches produced by our method, using 32, 16, and 8 strokes respectively. Column 6 shows the mask produced by pretrained U2-Net, and column 7 shows the results of our method when applied on the masked images.}
\label{fig:backgroundanalysis}
\end{figure}

In Figure \ref{fig:backgroundanalysis} we demonstrate two use cases in which a background is present -- (1) images that contain a salient object to be drawn (for example, the tree and horse in rows 1 and 2), and (2) images of natural landscapes in which the background plays an important role and cannot be ignored, such as the mountains and beach in columns 3 and 4.
As our method is designed for object sketching, the solution for case number (1) is straightforward - we simply use an automatic tool (such as U2-Net \cite{Qin_2020_PR}) for masking the salient object in the scene. On the rightmost column are the sketches produced after applying the mask to the input image. It can be seen that the generated sketches for images with a salient object are satisfactory. 
Regarding use case number (2), although our method is not designed to handle such input sources, we can see that it nonetheless produces pleasing results when applied with 32 and 16 strokes (columns 3, 4). The method, however, failed to produce reasonable drawings for higher levels of abstraction (8 strokes) as seen in column 5. 

\section{User Study}
This section provides additional details and analysis on the user study conducted on the sketches produced by our method.

\begin{figure}[h]
\centering
\begin{tabular}{@{\hskip2pt}c@{\hskip2pt}c@{\hskip2pt}c@{\hskip2pt}c@{\hskip2pt}c}
    \includegraphics[width=0.18\linewidth]{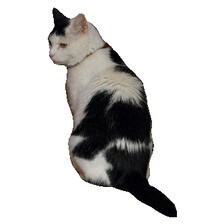} &
    \includegraphics[width=0.18\linewidth]{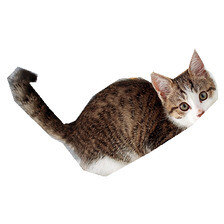} &
    \includegraphics[width=0.18\linewidth]{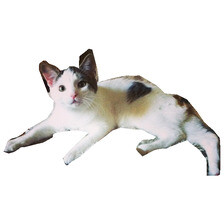} &
    \includegraphics[width=0.18\linewidth]{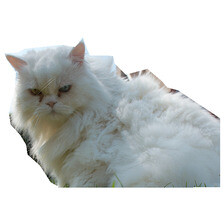} &
    \includegraphics[width=0.18\linewidth]{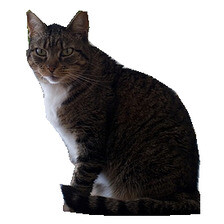} \\
    \midrule
    
    \includegraphics[width=0.18\linewidth]{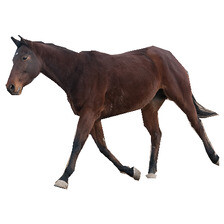} &
    \includegraphics[width=0.18\linewidth]{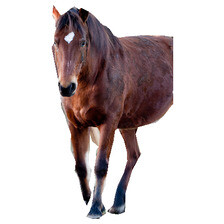} &
    \includegraphics[width=0.18\linewidth]{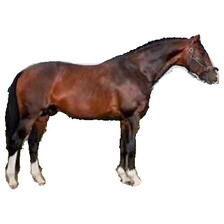} &
    \includegraphics[width=0.18\linewidth]{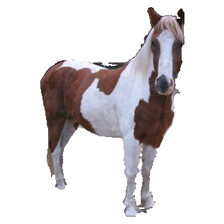} &
    \includegraphics[width=0.18\linewidth]{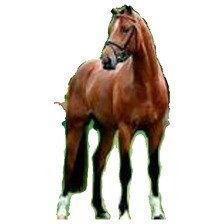} \\
    \midrule
    \includegraphics[width=0.18\linewidth]{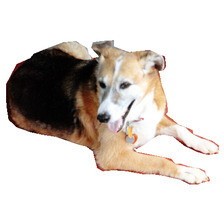} &
    \includegraphics[width=0.18\linewidth]{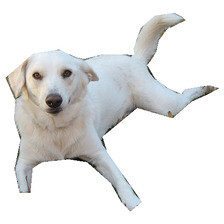} &
    \includegraphics[width=0.18\linewidth]{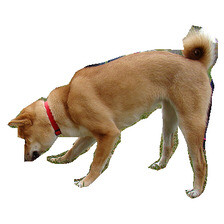} &
    \includegraphics[width=0.18\linewidth]{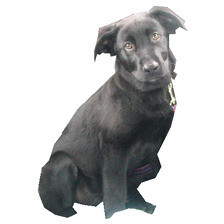} &
    \includegraphics[width=0.18\linewidth]{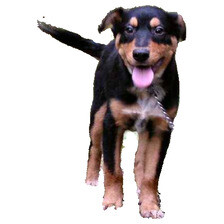} \\
    \midrule
    \includegraphics[width=0.18\linewidth]{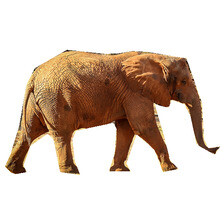} &
    \includegraphics[width=0.18\linewidth]{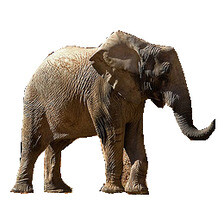} &
    \includegraphics[width=0.18\linewidth]{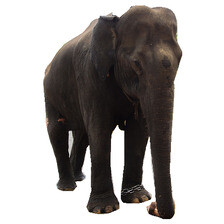} &
   \includegraphics[width=0.18\linewidth]{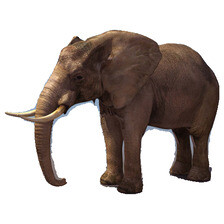} &
   \includegraphics[width=0.18\linewidth]{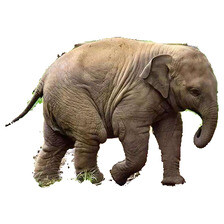} \\
    
    \midrule
    \includegraphics[width=0.18\linewidth]{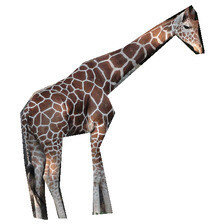} &
    \includegraphics[width=0.18\linewidth]{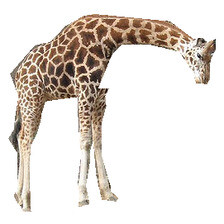} &
    \includegraphics[width=0.18\linewidth]{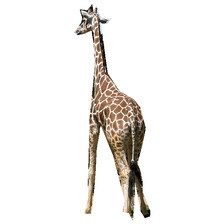} &
    \includegraphics[width=0.18\linewidth]{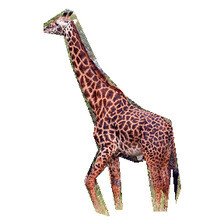} &
    \includegraphics[width=0.18\linewidth]{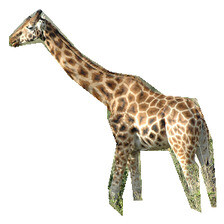} \\
\end{tabular}
 \caption{The randomly sampled animals images used for the user study.}
 \vspace{-2em}
\label{fig:user_study_input_animals}
\end{figure}

\begin{figure}[h]
\centering
\begin{tabular}{@{\hskip2pt}c@{\hskip2pt}c@{\hskip2pt}c@{\hskip2pt}c@{\hskip2pt}c}
    
    \midrule
    \includegraphics[width=0.18\linewidth]{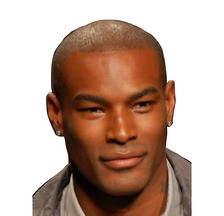} &
    \includegraphics[width=0.18\linewidth]{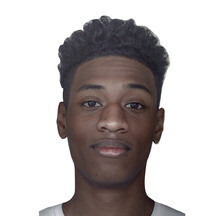} &
    \includegraphics[width=0.18\linewidth]{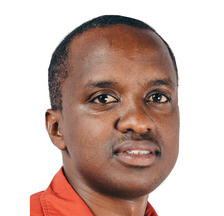} &
    \includegraphics[width=0.18\linewidth]{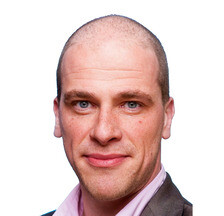} &
    \includegraphics[width=0.18\linewidth]{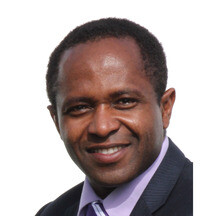} \\
    
    \midrule
    \includegraphics[width=0.18\linewidth]{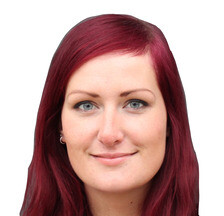} &
    \includegraphics[width=0.18\linewidth]{figs/user_study/input_images/women/02-19879809258_3de6bc082b_o.jpg} &
    \includegraphics[width=0.18\linewidth]{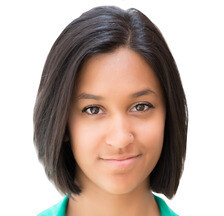} &
    \includegraphics[width=0.18\linewidth]{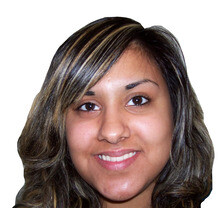} &
    \includegraphics[width=0.18\linewidth]{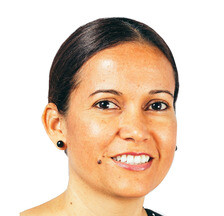} \\

\end{tabular}
 \caption{The randomly sampled face images used for the user study.}
\label{fig:user_study_input_face}
\end{figure}

\begin{figure}[h]
\centering
\begin{tabular}{@{\hskip2pt}c@{\hskip2pt}c@{\hskip2pt}c@{\hskip2pt}c@{\hskip2pt}c}
    \midrule
     & 32  & 16 & 8 & 4 \\
    \midrule
    \rotatebox{90}{\parbox[t]{0.2\linewidth}{\hspace*{\fill}Cat\hspace*{\fill}}} &
    \includegraphics[width=0.2\linewidth]{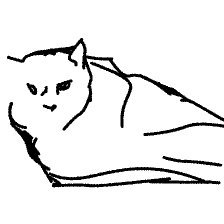} &
    \includegraphics[width=0.2\linewidth]{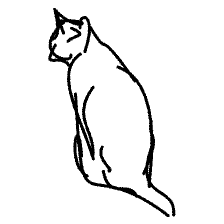} &
    \includegraphics[width=0.2\linewidth]{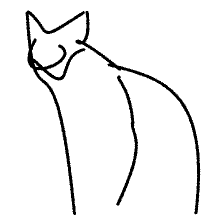} &
    \includegraphics[width=0.2\linewidth]{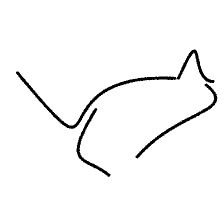} \\
    \rotatebox{90}{\parbox[t]{0.2\linewidth}{\hspace*{\fill}Horse\hspace*{\fill}}} &
    \includegraphics[width=0.2\linewidth]{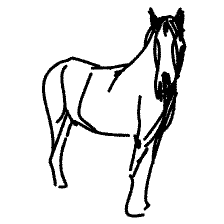} &
    \includegraphics[width=0.2\linewidth]{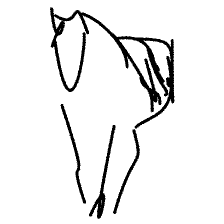} &
    \includegraphics[width=0.2\linewidth]{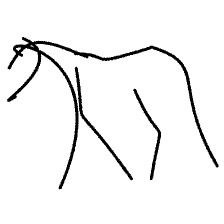} &
    \includegraphics[width=0.2\linewidth]{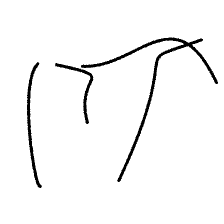} \\
    \rotatebox{90}{\parbox[t]{0.2\linewidth}{\hspace*{\fill}Dog\hspace*{\fill}}} &
    \includegraphics[width=0.2\linewidth]{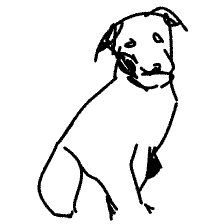} &
    \includegraphics[width=0.2\linewidth]{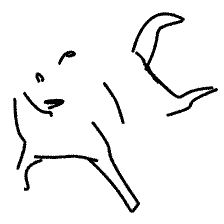} &
    \includegraphics[width=0.2\linewidth]{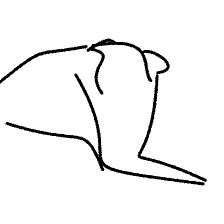} &
    \includegraphics[width=0.2\linewidth]{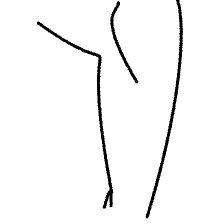} \\
    \rotatebox{90}{\parbox[t]{0.2\linewidth}{\hspace*{\fill}Elephant\hspace*{\fill}}} &
    \includegraphics[width=0.2\linewidth]{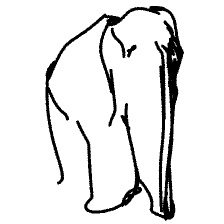} &
    \includegraphics[width=0.2\linewidth]{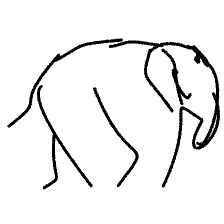} &
    \includegraphics[width=0.2\linewidth]{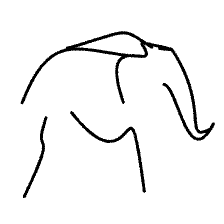} &
    \includegraphics[width=0.2\linewidth]{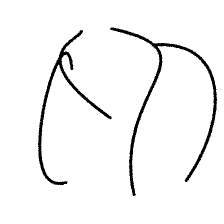} \\
    \rotatebox{90}{\parbox[t]{0.2\linewidth}{\hspace*{\fill}Giraffe\hspace*{\fill}}} &
    \includegraphics[width=0.2\linewidth]{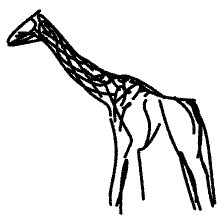} &
    \includegraphics[width=0.2\linewidth]{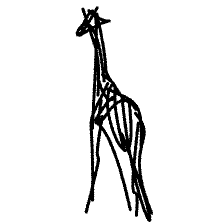} &
    \includegraphics[width=0.2\linewidth]{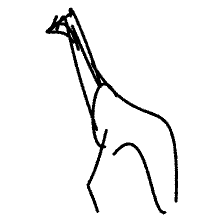} &
    \includegraphics[width=0.2\linewidth]{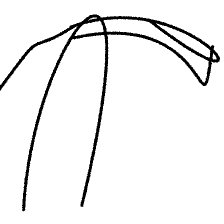} \\
\end{tabular}
 \caption{Sample animal sketches produced by our method, that were used in the user study.}
\label{fig:user_study_samples}
\end{figure}

\begin{figure}[h]
\centering
\begin{tabular}{@{\hskip2pt}c@{\hskip2pt}c@{\hskip2pt}c@{\hskip2pt}c@{\hskip2pt}c}
    \midrule
     & 32  & 16 & 8 & 4 \\
    \midrule
    \rotatebox{90}{\parbox[t]{0.2\linewidth}{\hspace*{\fill}Man\hspace*{\fill}}} &
    \includegraphics[width=0.2\linewidth]{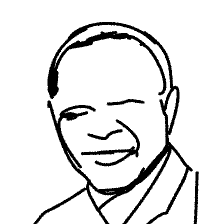} &
    \includegraphics[width=0.2\linewidth]{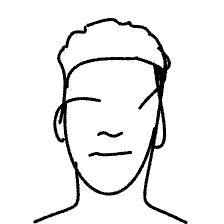} &
    \includegraphics[width=0.2\linewidth]{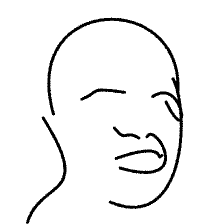} &
    \includegraphics[width=0.2\linewidth]{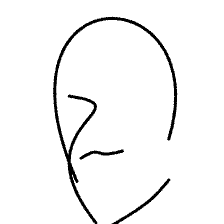} \\
    
    \rotatebox{90}{\parbox[t]{0.2\linewidth}{\hspace*{\fill}Woman\hspace*{\fill}}} &
    \includegraphics[width=0.2\linewidth]{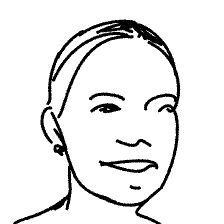} &
    \includegraphics[width=0.2\linewidth]{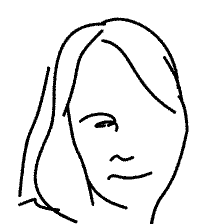} &
    \includegraphics[width=0.2\linewidth]{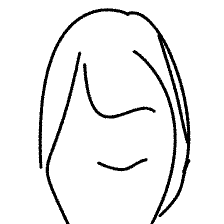} &
    \includegraphics[width=0.2\linewidth]{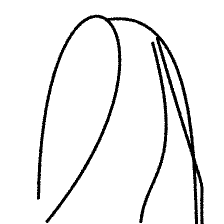} \\
    
\end{tabular}
 \caption{Sample face sketches produced by our method that were used in
the user study.}
\label{fig:user_study_samples_face}
\end{figure}

We select five animal classes for the user study (cat, dog, elephant, giraffe, and horse), as animals typically have distinctive pictorial representations but also share visual features that would make class-level differentiation a non-trivial task. 
As was stated in the main paper, we use the SketchyCOCO dataset \cite{gao2020sketchycoco}, and randomly sample 5 images for each class, see Figure \ref{fig:user_study_input_animals} for reference. For the instance-level recognition test, we also include an additional class of human faces from the \cite{Rosin2017BenchmarkingNR} dataset, and sample five portraits of men and five of women (see Figure \ref{fig:user_study_input_face}). The 'face' samples were not included in the main paper's analyses.
In total, we generate 100 sketches of animals and 40 sketches of human faces.
We randomly split the sketches to 5 questionnaires, each questionnaire with 20 questions for the category-level recognition experiment (five classes at four levels of abstraction) and 28 questions for the instance-level recognition experiment (five animal classes, male faces, and female faces at four levels of abstraction).
Figures \ref{fig:user_study_samples} and \ref{fig:user_study_samples_face} demonstrate the output sketches used in one of the questionnaires.

We surveyed 121 people in total. Out of 121, 60 people were assigned to evaluate our sketches, 38 to the sketches by Kampelmühler and Pinz \cite{human-like-sketches}, and 24 to the sketches by Li et al. \cite{li2019photosketching}. Participants evaluating our sketches were allowed to randomly select the set of questions they received, with the ultimate split being 13 people receiving version one, 15 for version two, 8 for version three, 15 for version four, and 9 for version five. The weight of each response contributed equally to the overall accuracy computation, regardless of the version. It is possible that certain versions of the survey had higher difficulty than others, but the variations are not significant. In the following section we provide the detailed results of the user-study on our sketches. 

\subsection{Category-Level Recognition}
\begin{figure}[h]
    \centering
    \begin{subfigure}{\linewidth}
        \includegraphics[width=\linewidth]{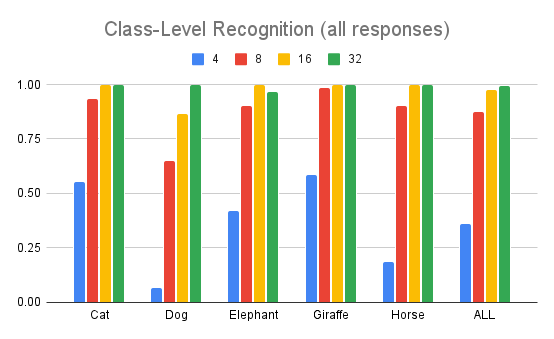}
    \end{subfigure}
    
    \begin{subfigure}{\linewidth}
        \includegraphics[width=\linewidth]{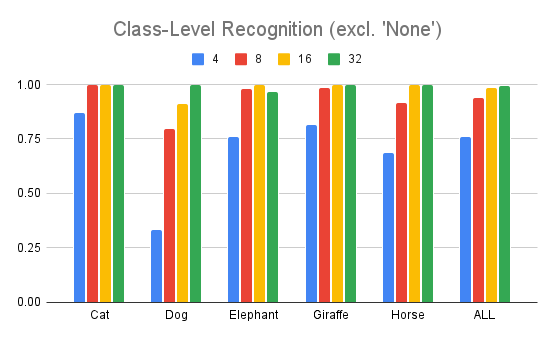}
    \end{subfigure}
    
    \caption{Class level recognizability user study results.}
    \label{fig:user_study_class}
\end{figure}

Figure \ref{fig:user_study_class} shows the per-class recognition accuracy at each abstraction level. The color of the bar graph represents the number of strokes, for example blue represents 4-stroke sketches. In the main paper, we reported the accuracy averaged across all classes to be 36\%. In the class-by-class breakdown, we observe that 'dog' and 'horse' classes received below average accuracies, while 'cat' and 'giraffe' had much better performance. One theory to support this is that cats and giraffes contain more unambiguous discriminative features in their pictorial representations --- namely, triangular ears for the former, and a long neck for the latter. These are features that even amateur artists would capture while sketching lightweight representations, and evidently they are well captured by our method through the saliency-based initialization and the CLIP-based loss.

Another interesting observation we made is in the confidence of the answers. The baseline performance of randomly guessing between the five animal classes would result in an accuracy of 20\%. However, since the sixth choice is 'None', participants are able to defer their guess if they are uncertain of the answer. In highly abstract sketches with 4 strokes, 53\% of the answers are 'None', but this fell to 7\% in 8-stroke sketches, and virtually none in 16- and 32-stroke sketches, indicating that added stroke details reduced uncertainty significantly. If we exclude 'None' answers (as in, leaving only the confident guesses), then the recognition accuracy of 4-stroke sketches is improved from 36\% to 76\%. This shows that while high abstraction may introduce uncertainty, there is still enough semantic and/or geometric information to represent the object class for the study participants who were relatively confident in answering. 

\begin{figure}[h]
    \centering
    \begin{subfigure}{\linewidth}
        \includegraphics[width=\linewidth]{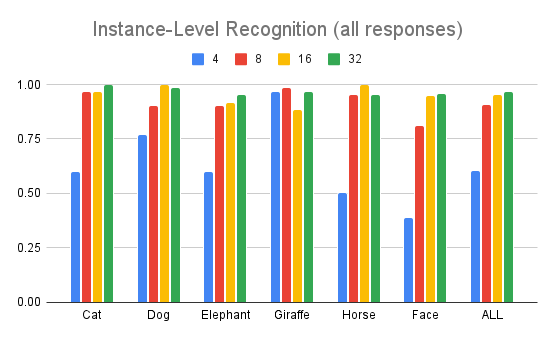}
    \end{subfigure}
    
    \begin{subfigure}{\linewidth}
        \includegraphics[width=\linewidth]{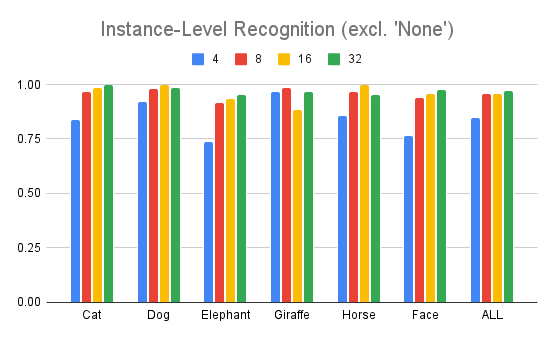}
    \end{subfigure}
    
    \caption{Instance level recognizability user study results.}
    \label{fig:user_study_instance}
\end{figure}

\subsection{Instance-Level Recognition}
Figure \ref{fig:user_study_instance} shows the instance-level recognition for each class and abstraction level. Included in this experiment is the 'face' class, which depicts portrait-style human faces of various demographics\footnote{Notice that the face class was not included in the results reported in the main paper because one of the methods was only limited to the categories in the Sketchy Database}. As the goal is to identify the input image of a sketch given four additional confounder images of the same class, instance-level recognition relies more on the geometric feature matching between the sketch and the image. Semantic and class-level discriminative features are less important since these features would be shared across all options. Thus the instance-level recognizability is more determined by the geometric fidelity of the sketch to the original image.

The results show that in comparison to the class recognition test, the 4-stroke sketches yielded much better recognizability and less uncertainty. The reason is likely because geometric features are easy to match between the input image and the sketch. It should be noted that the variance across classes for 4-stroke sketches is quite high, from 39\% on 'face' to 97\% on 'giraffe', and the average accuracy across classes being 60\%. The reason for the lower accuracy of the 'face' class is likely due to the less-distinctive poses of the subjects, as all faces had the similar orientations and sizes, so the participants had to rely more on the detailed features of the face for instance recognition. The proportion of 'None' answers is 40\% for 4 strokes, and the accuracy discounting 'None' answers is 85\%.  
The increase in overall recognizability from 4 strokes (60\%) to 32 strokes (97\%) is evident but less substantial than improvements seen in the class recognition task, likely due to visually similar poses of the subjects of the confounder images.

\section{Initialization Analysis}
\subsection{Initialization Ablation}
This section provides a visual analysis demonstrating the sensitivity to initialization.

\begin{figure}[h!]
\centering
\begin{tabular}{@{\hskip2pt}c@{\hskip2pt}c@{\hskip2pt}c@{\hskip2pt}c}
    \midrule
    Input & Seed a & Seed b & Seed c \\
    \midrule
    \includegraphics[width=\widthseed\linewidth]{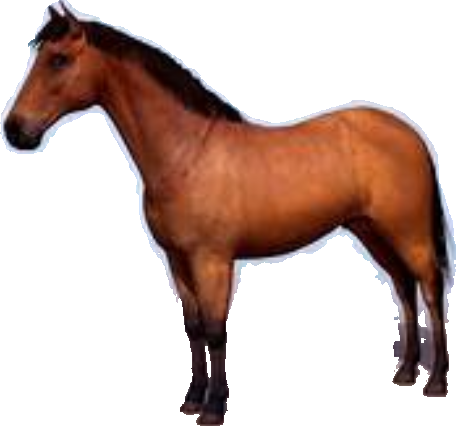} &
    \includegraphics[width=\widthseed\linewidth]{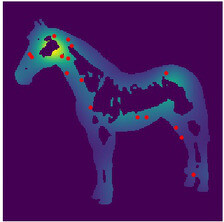} &
    \includegraphics[width=\widthseed\linewidth]{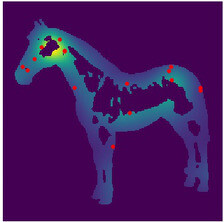} &
    \includegraphics[width=\widthseed\linewidth]{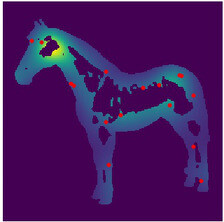}\\
    & 
    \includegraphics[width=\widthseed\linewidth]{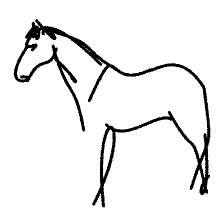} &
    \includegraphics[width=\widthseed\linewidth]{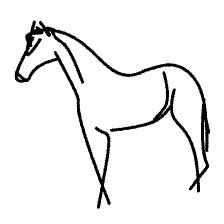} &
    \includegraphics[width=\widthseed\linewidth]{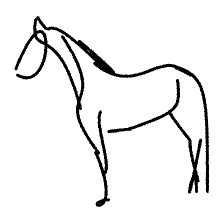} \\
    
    \midrule
    \includegraphics[width=\widthseed\linewidth]{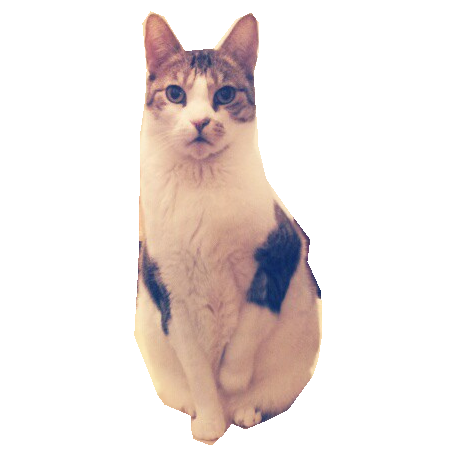} &
    \includegraphics[width=\widthseed\linewidth]{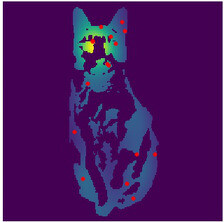} &
    \includegraphics[width=\widthseed\linewidth]{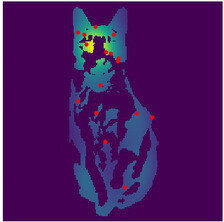} &
    \includegraphics[width=\widthseed\linewidth]{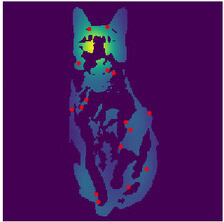} \\
    & \includegraphics[width=\widthseed\linewidth]{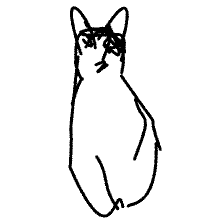} &
    \includegraphics[width=\widthseed\linewidth]{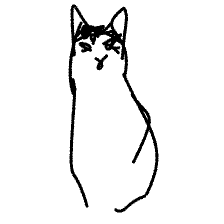} &
    \includegraphics[width=\widthseed\linewidth]{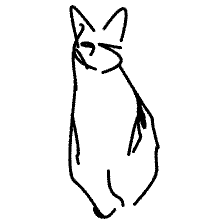} \\
    
    \midrule
    \includegraphics[width=\widthseed\linewidth]{figs/abstraction_levels/input/rose/rose1.jpeg} &
    \includegraphics[width=\widthseed\linewidth]{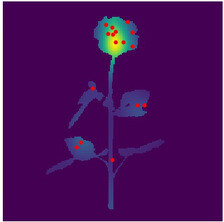} &
    \includegraphics[width=\widthseed\linewidth]{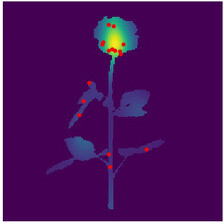} &
    \includegraphics[width=\widthseed\linewidth]{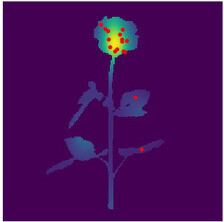} \\
    & \includegraphics[width=\widthseed\linewidth]{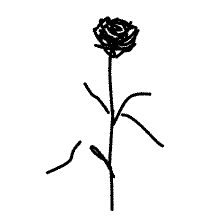} &
    \includegraphics[width=\widthseed\linewidth]{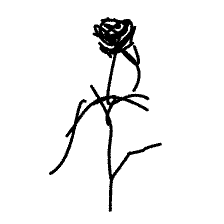} &
    \includegraphics[width=\widthseed\linewidth]{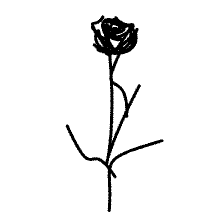} \\
    
    \midrule
    \includegraphics[width=\widthseed\linewidth]{figs/abstraction_levels/input/giraffe/596728.png} &
    \includegraphics[width=\widthseed\linewidth]{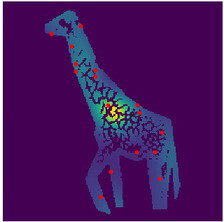} &
    \includegraphics[width=\widthseed\linewidth]{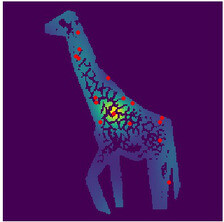} &
    \includegraphics[width=\widthseed\linewidth]{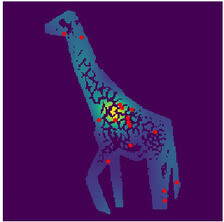}\\
    
    & \includegraphics[width=\widthseed\linewidth]{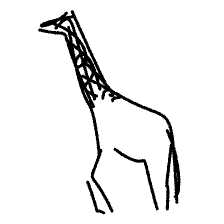} &
    \includegraphics[width=\widthseed\linewidth]{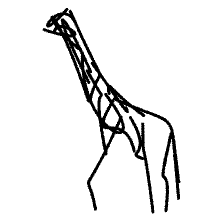} &
    \includegraphics[width=\widthseed\linewidth]{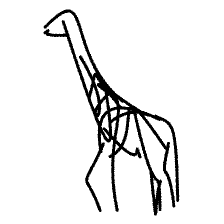} \\
 
\end{tabular}
 \caption{Different Seeds — sorted by final loss score from left to right (best to worst).}
\label{fig:different_seeds_}
\end{figure}

\begin{figure}
\centering
\begin{tabular}{@{\hskip2pt}c@{\hskip2pt}c@{\hskip2pt}c@{\hskip2pt}c@{\hskip2pt}c@{\hskip2pt}c}
    \midrule
     & Input & Saliency & \begin{tabular}[c]{@{}c@{}}Points\\ Sampled\end{tabular} & \begin{tabular}[c]{@{}c@{}}Initial\\ Strokes\end{tabular} & Output \\
    
    \midrule
    \rotatebox{90}{\parbox[t]{\widthattn\linewidth}{\hspace*{\fill}\begin{tabular}[c]{@{}c@{}}XDoG\\ \& ViT\end{tabular}\hspace*{\fill}}} &
    \includegraphics[width=\widthattn\linewidth]{figs/initialisation_comp/cat/xdog_clip/cat_10544.jpg} &
    \includegraphics[width=\widthattn\linewidth]{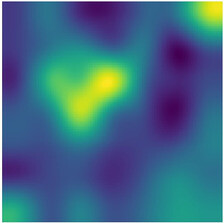} &
    \includegraphics[width=\widthattn\linewidth]{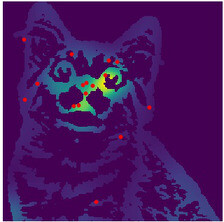} &
    \includegraphics[width=\widthattn\linewidth]{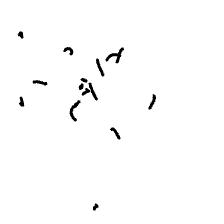} &
    \includegraphics[width=\widthattn\linewidth]{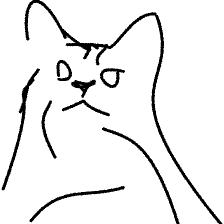} \\
    
    \midrule
    \rotatebox{90}{\parbox[t]{\widthattn\linewidth}{\hspace*{\fill}ViT\hspace*{\fill}}} &
    \includegraphics[width=\widthattn\linewidth]{figs/initialisation_comp/cat/xdog_clip/cat_10544.jpg} &
    \includegraphics[width=\widthattn\linewidth]{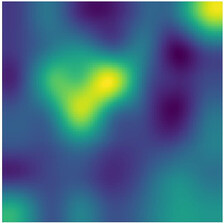} &
    \includegraphics[width=\widthattn\linewidth]{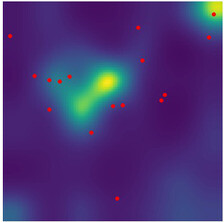} &
    \includegraphics[width=\widthattn\linewidth]{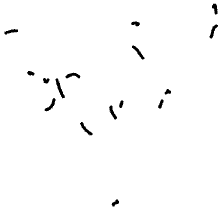} &
    \includegraphics[width=\widthattn\linewidth]{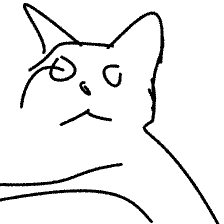} \\
    
    \midrule
    \rotatebox{90}{\parbox[t]{\widthattn\linewidth}{\hspace*{\fill}Random\hspace*{\fill}}} &
    \includegraphics[width=\widthattn\linewidth]{figs/initialisation_comp/cat/xdog_clip/cat_10544.jpg} &
     &
     &
     \includegraphics[width=\widthattn\linewidth]{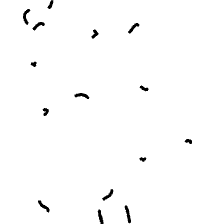} &
    \includegraphics[width=\widthattn\linewidth]{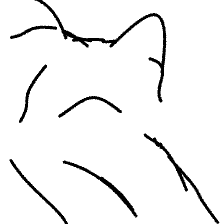}
    
\end{tabular}
 \caption{Comparison of different initialization approaches.}
\label{fig:init_analysis_cat}
\end{figure}

\begin{figure}
    \centering
    \begin{subfigure}{\linewidth}
        \includegraphics[width=\linewidth]{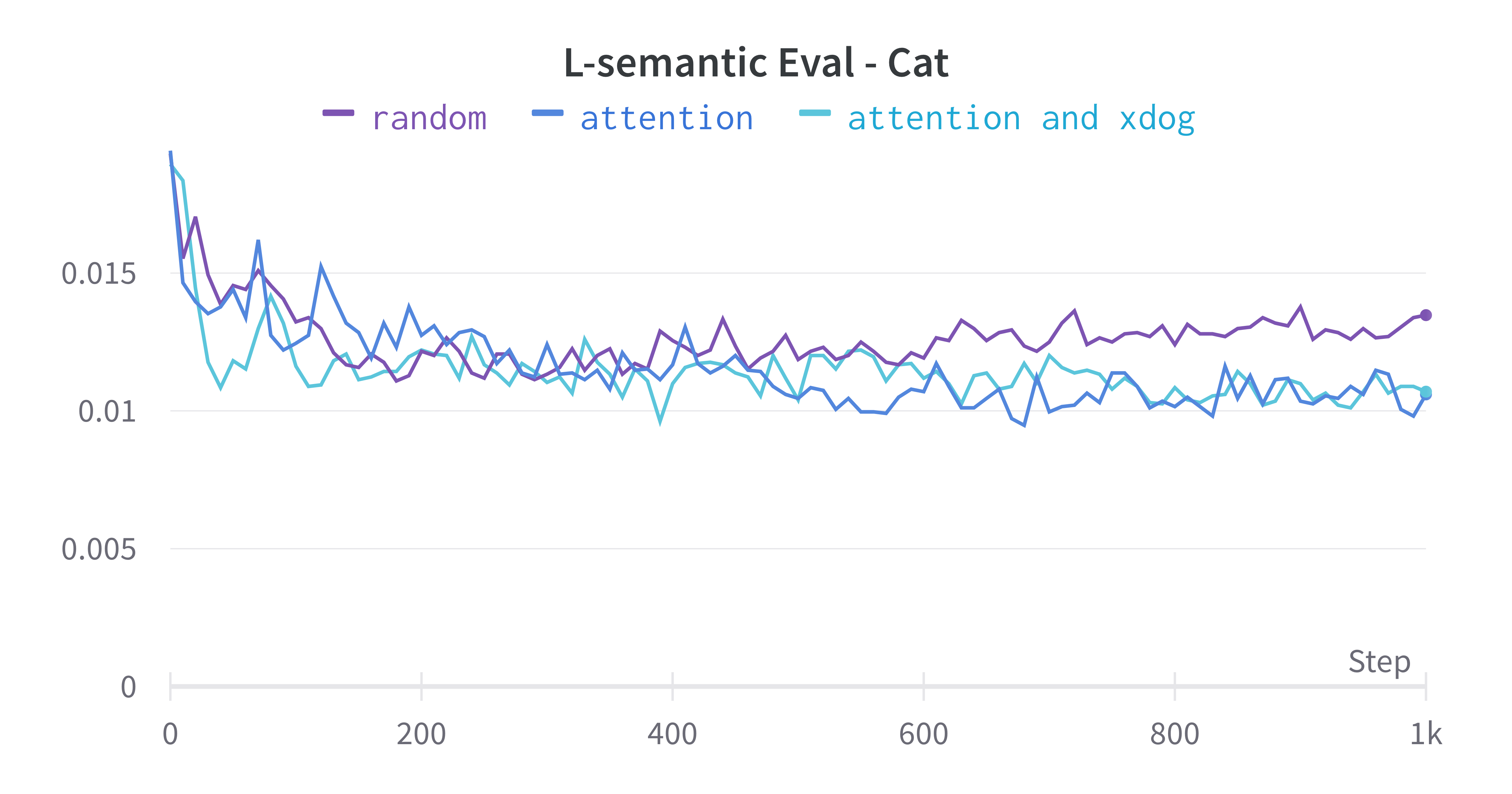}
    \end{subfigure}
    
    \begin{subfigure}{\linewidth}
        \includegraphics[width=\linewidth]{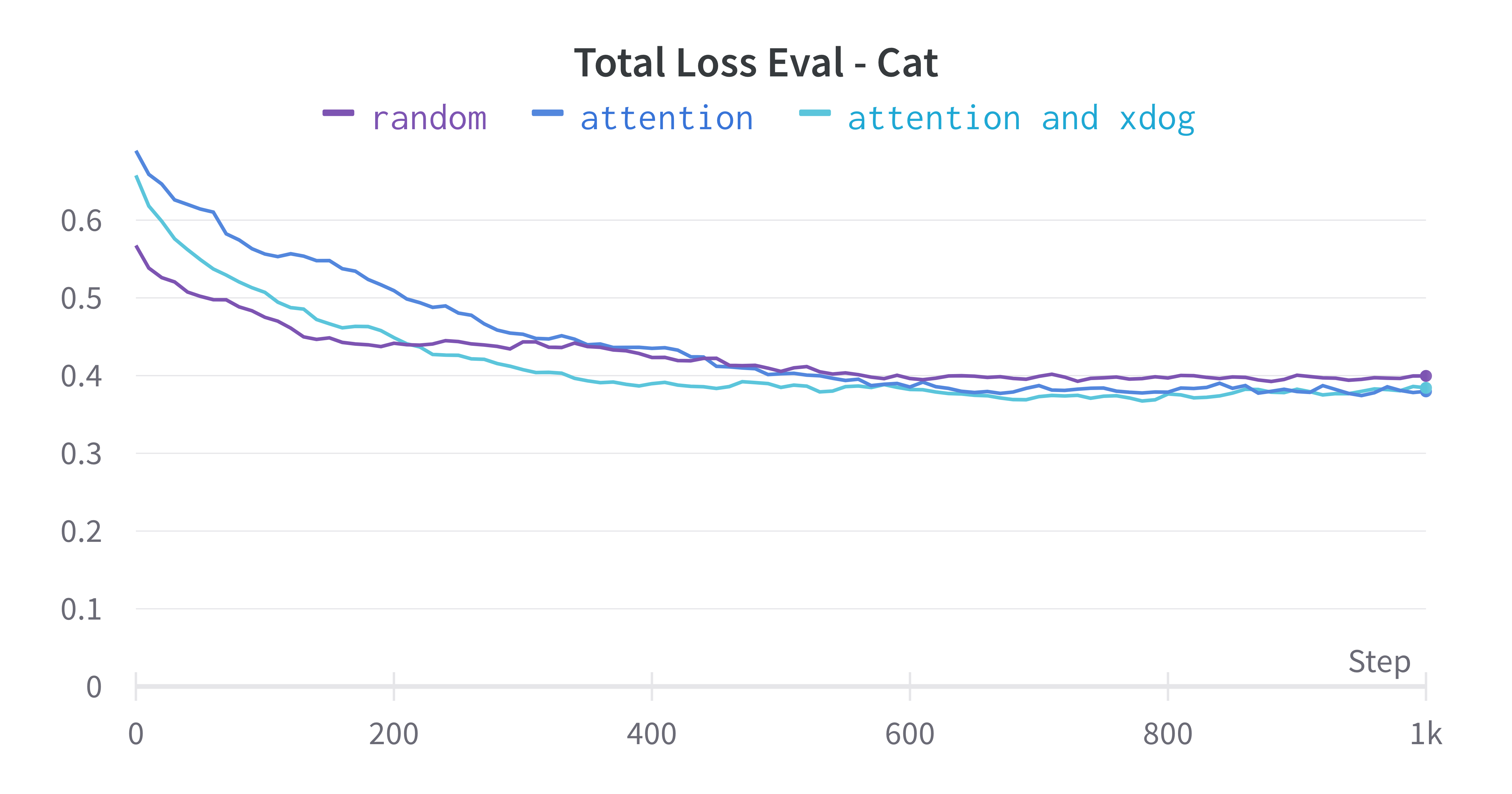}
    \end{subfigure}
    
    \caption{Convergence comparison with different initialization approaches.}
    \label{fig:init_cat_conv}
\end{figure}

\begin{figure}
\centering
\begin{tabular}{@{\hskip2pt}c@{\hskip2pt}c@{\hskip2pt}c@{\hskip2pt}c@{\hskip2pt}c@{\hskip2pt}c}
    \midrule
    & Input & Saliency & \begin{tabular}[c]{@{}c@{}}Points\\ Sampled\end{tabular} & \begin{tabular}[c]{@{}c@{}}Initial\\ Strokes\end{tabular} & Output \\
    
    \midrule
    \rotatebox{90}{\parbox[t]{\widthattn\linewidth}{\hspace*{\fill}\begin{tabular}[c]{@{}c@{}}XDoG\\ \& ViT\end{tabular}\hspace*{\fill}}} &
    \includegraphics[width=\widthattn\linewidth]{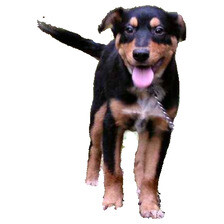} &
    \includegraphics[width=\widthattn\linewidth]{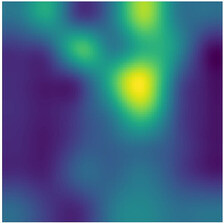} &
    \includegraphics[width=\widthattn\linewidth]{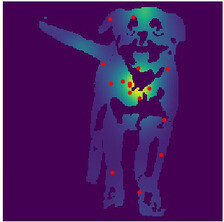} &
    \includegraphics[width=\widthattn\linewidth]{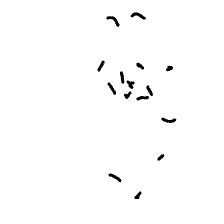} &
    \includegraphics[width=\widthattn\linewidth]{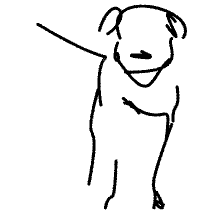} \\
    
    \midrule
    \rotatebox{90}{\parbox[t]{\widthattn\linewidth}{\hspace*{\fill}ViT\hspace*{\fill}}} &
    \includegraphics[width=\widthattn\linewidth]{figs/initialisation_comp/dog_update/dog.10346.jpg} &
     \includegraphics[width=\widthattn\linewidth]{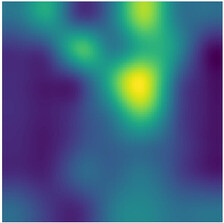} &
    \includegraphics[width=\widthattn\linewidth]{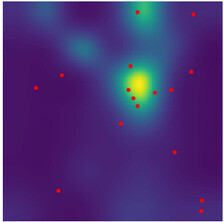} &
    \includegraphics[width=\widthattn\linewidth]{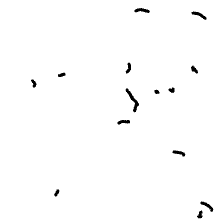} &
    \includegraphics[width=\widthattn\linewidth]{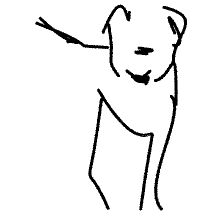} \\
    
    \midrule
    \rotatebox{90}{\parbox[t]{\widthattn\linewidth}{\hspace*{\fill}Random\hspace*{\fill}}} &
    \includegraphics[width=\widthattn\linewidth]{figs/initialisation_comp/dog_update/dog.10346.jpg} &
     &
     &
    \includegraphics[width=\widthattn\linewidth]{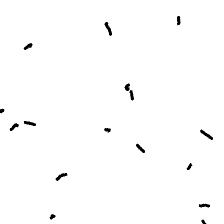} &
    \includegraphics[width=\widthattn\linewidth]{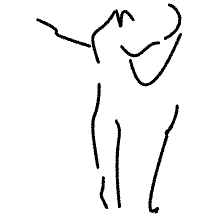} \\
    
\end{tabular}
 \caption{Comparison of different initialization approaches.}
\label{fig:init_analysis_dog}
\end{figure}

\begin{figure}
    \centering
    \begin{subfigure}{\linewidth}
        \includegraphics[width=\linewidth]{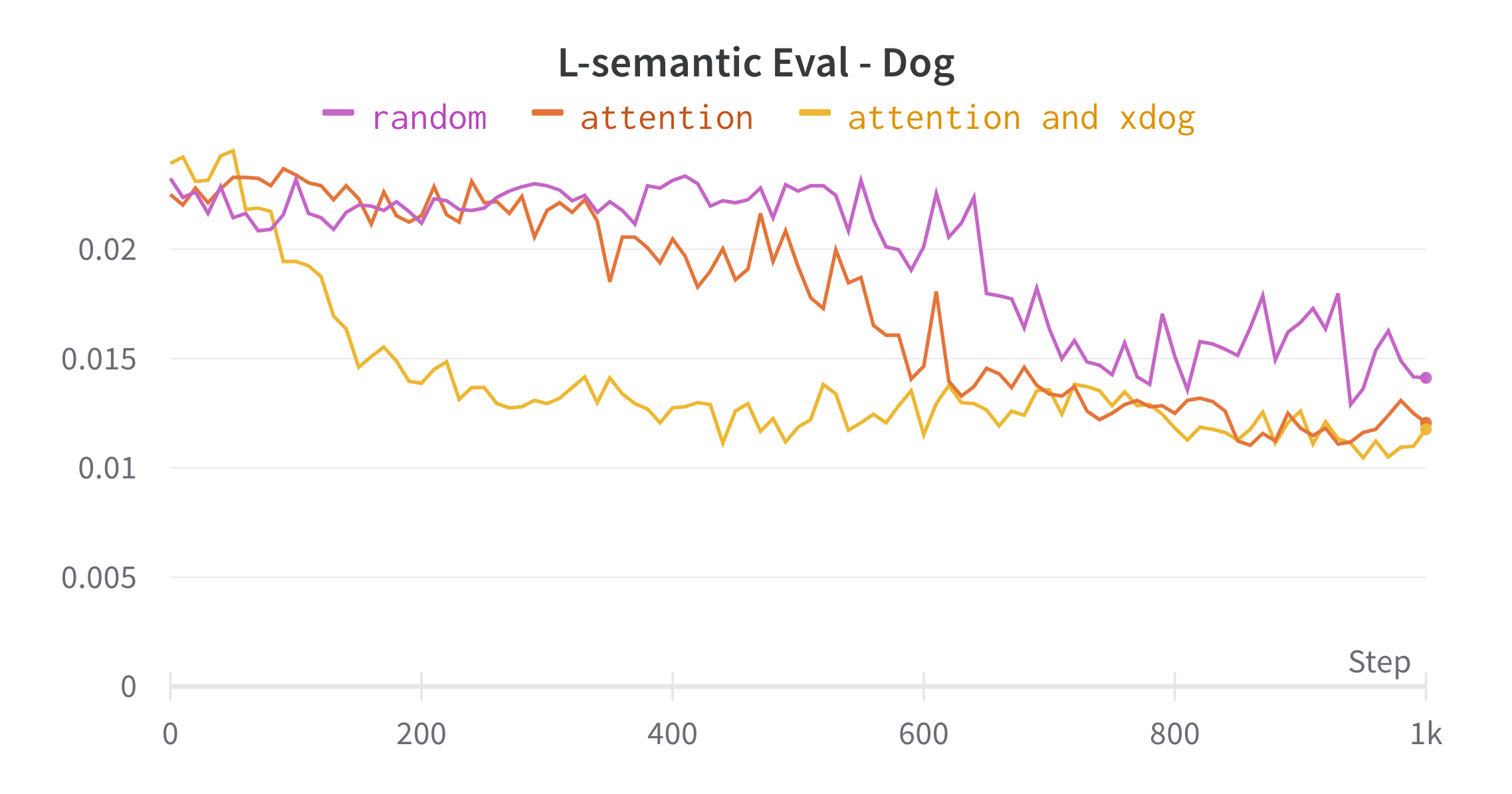}
    \end{subfigure}
    
    \begin{subfigure}{\linewidth}
        \includegraphics[width=\linewidth]{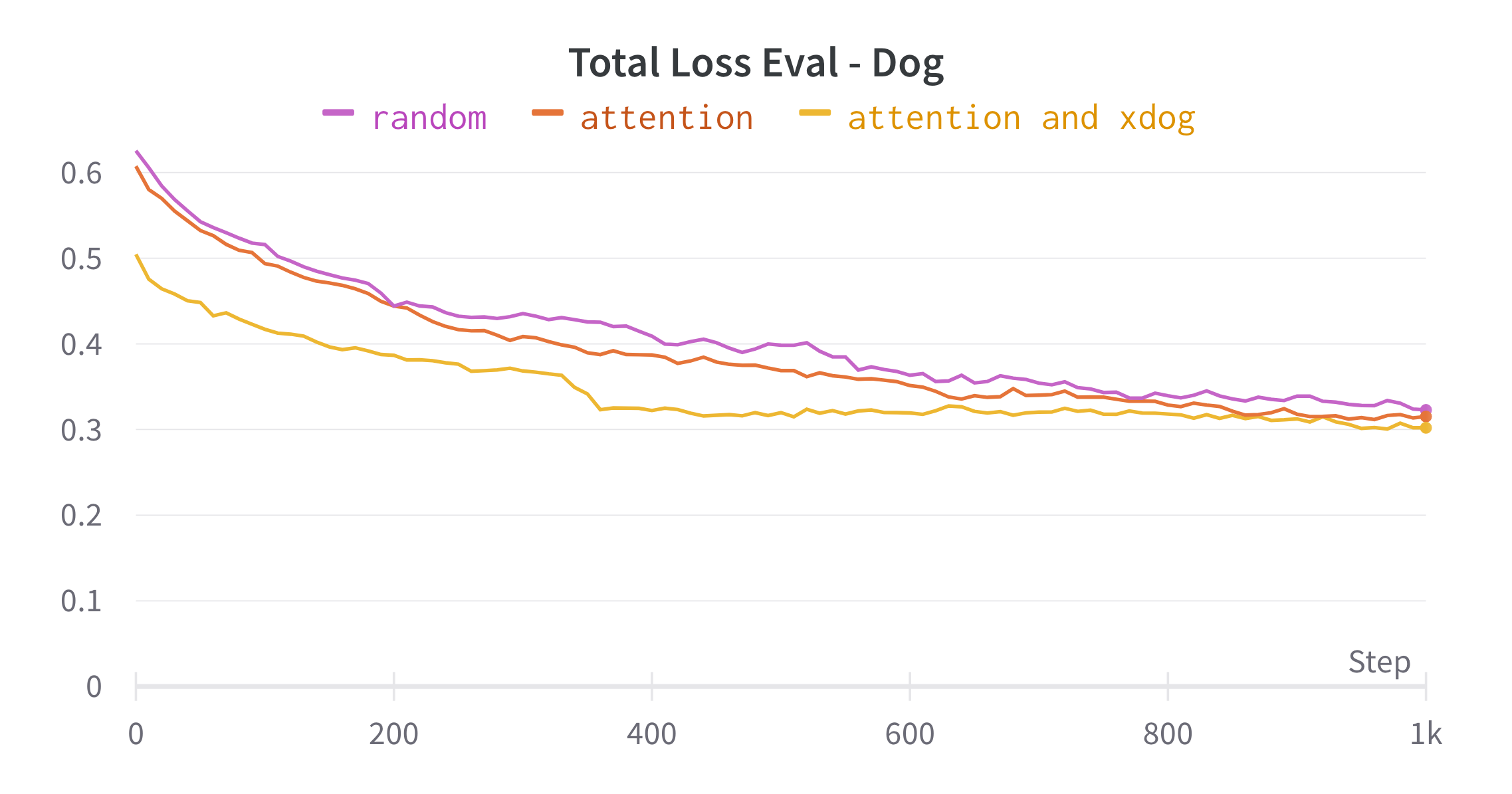}
    \end{subfigure}
    
    \caption{Convergence comparison with different initialization approaches.}
    \label{fig:init_dog_conv}
\end{figure}

\begin{figure}
\centering
\begin{tabular}{@{\hskip2pt}c@{\hskip2pt}c@{\hskip2pt}c@{\hskip2pt}c@{\hskip2pt}c@{\hskip2pt}c}
    \midrule
     & Input & Saliency & \begin{tabular}[c]{@{}c@{}}Points\\ Sampled\end{tabular} & \begin{tabular}[c]{@{}c@{}}Initial\\ Strokes\end{tabular} & Output \\
    
    \midrule
    \rotatebox{90}{\parbox[t]{\widthattn\linewidth}{\hspace*{\fill}\begin{tabular}[c]{@{}c@{}}XDoG\\ \& ViT\end{tabular}\hspace*{\fill}}} &
    \includegraphics[width=\widthattn\linewidth]{figs/user_study/input_images/women/02-19879809258_3de6bc082b_o.jpg} &
    \includegraphics[width=\widthattn\linewidth]{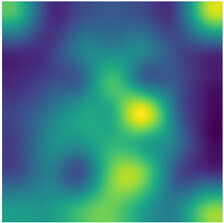} &
    \includegraphics[width=\widthattn\linewidth]{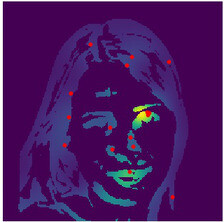} &
    \includegraphics[width=\widthattn\linewidth]{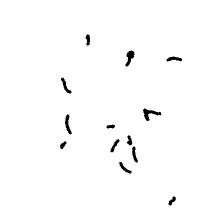} &
    \includegraphics[width=\widthattn\linewidth]{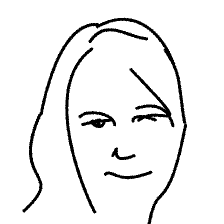} \\
    
    \midrule
    \rotatebox{90}{\parbox[t]{\widthattn\linewidth}{\hspace*{\fill}ViT\hspace*{\fill}}} &
    \includegraphics[width=\widthattn\linewidth]{figs/user_study/input_images/women/02-19879809258_3de6bc082b_o.jpg} &
     \includegraphics[width=\widthattn\linewidth]{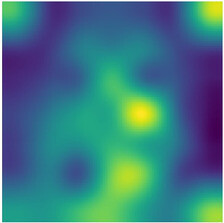} &
    \includegraphics[width=\widthattn\linewidth]{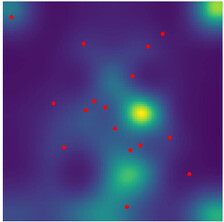} &
    \includegraphics[width=\widthattn\linewidth]{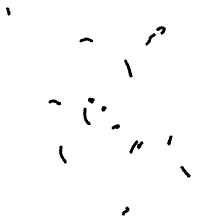} &
    \includegraphics[width=\widthattn\linewidth]{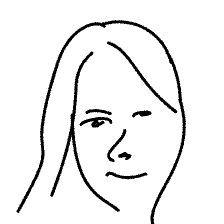} \\
    
    \midrule
    
    \rotatebox{90}{\parbox[t]{\widthattn\linewidth}{\hspace*{\fill}Random\hspace*{\fill}}} &
    \includegraphics[width=\widthattn\linewidth]{figs/user_study/input_images/women/02-19879809258_3de6bc082b_o.jpg} &
     &
     &
    \includegraphics[width=\widthattn\linewidth]{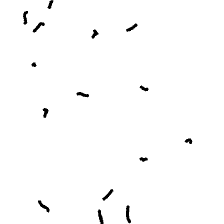} &
    \includegraphics[width=\widthattn\linewidth]{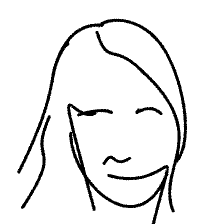}\\
    
\end{tabular}
 \caption{Comparison of different initialization approaches.}
\label{fig:init_analysis_woman}
\end{figure}

\begin{figure}[ht]
    \centering
    \begin{subfigure}{\linewidth}
        \includegraphics[width=\linewidth]{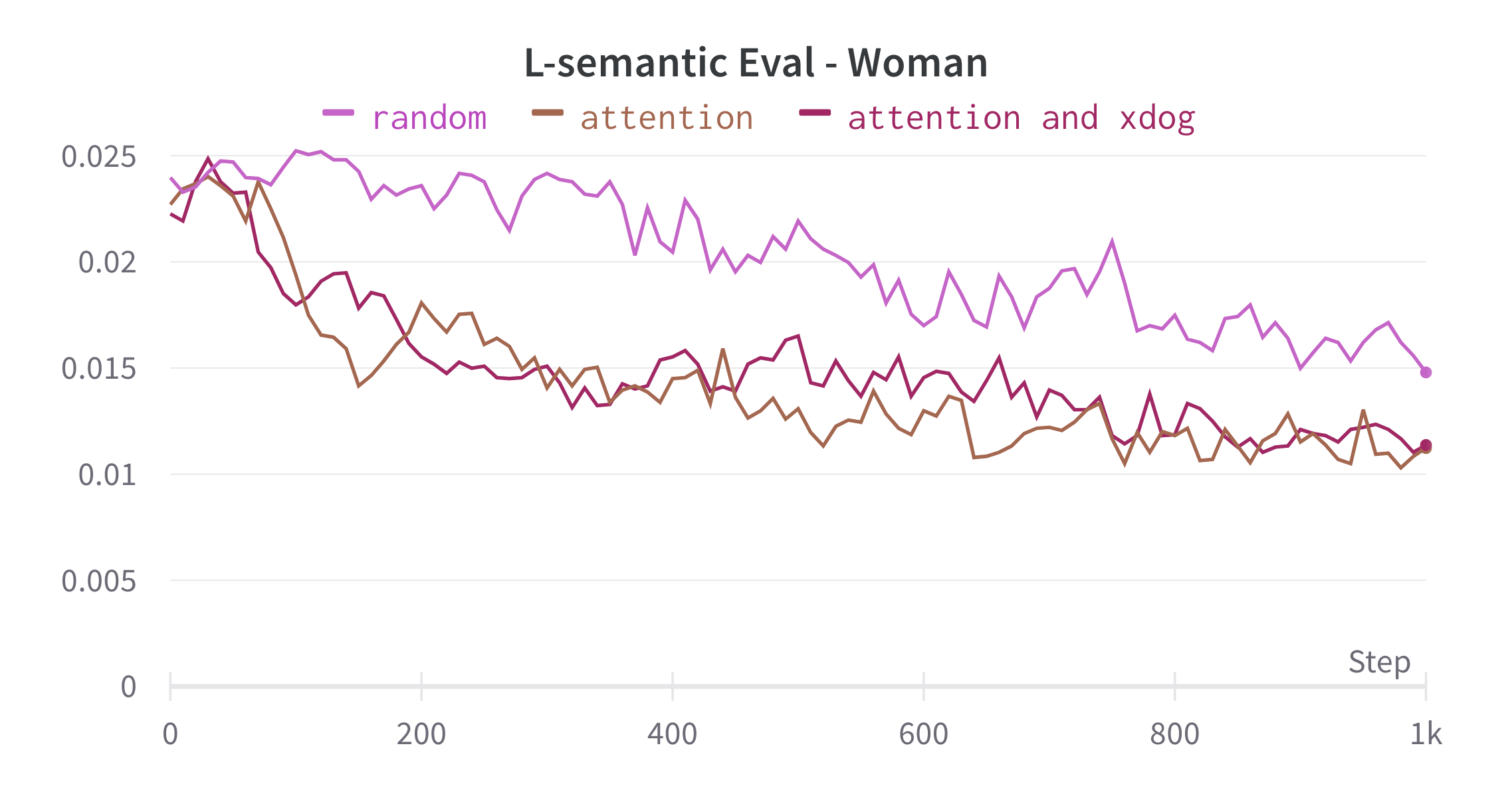}
    \end{subfigure}
    
    \begin{subfigure}{\linewidth}
        \includegraphics[width=\linewidth]{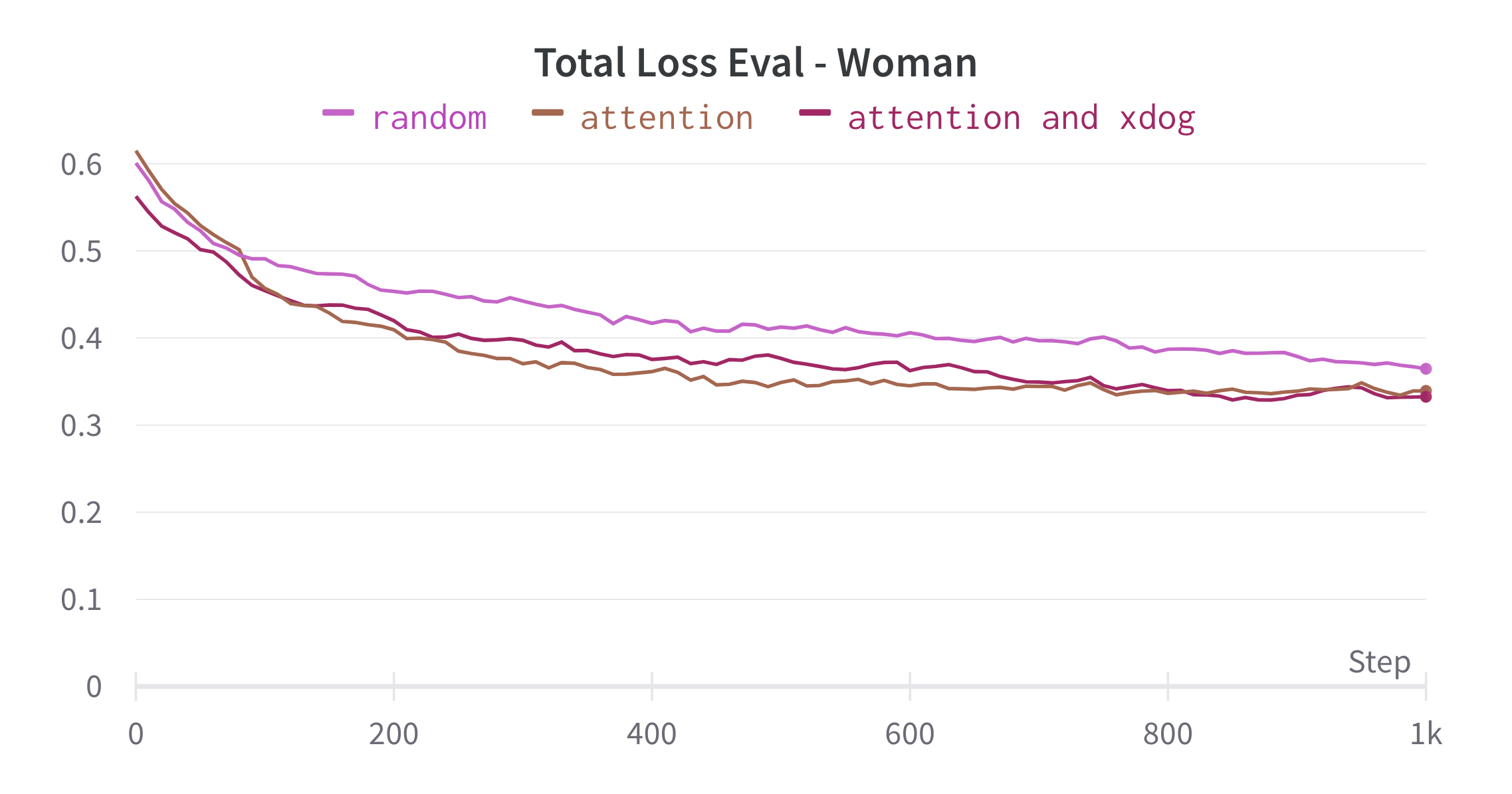}
    \end{subfigure}
    
    \caption{Convergence comparison with different initialization approaches.}
    \label{fig:init_woman_conv}
\end{figure}

In Figure \ref{fig:different_seeds_} we provide additional examples showing how multiple sketches of the same level of abstraction can be produced from a single input object by changing the random seed used. The red dots on each saliency map indicate the initial strokes' location sampled for three different random seeds.
As was stated in the paper, since our method is optimization-based, and the objective function is highly non-convex, the strokes converge to different local minimas. Therefore, the initialization of the strokes has a significant impact on the final result of the optimization.
In Figures \ref{fig:init_analysis_cat}, \ref{fig:init_analysis_dog}, and \ref{fig:init_analysis_woman}, we compare different approaches to initialization and demonstrate the synthesized sketches on images of cat, dog, and human face.
In particular, we investigate the performance of the following stroke initialization techniques: (1) ViT\footnote{Here ViT refers to CLIP ViT-B/32 model.} and XDoG, (2) only ViT, and (3) random initialization. All experiments were conducted using 16 strokes.
The figures show that when using random initialization, strokes converge to a sub-optimal solution, often missing the semantic components of the object such as the eyes.
We also provide the loss graphs for each image and initialization technique, in Figures \ref{fig:init_cat_conv}, \ref{fig:init_dog_conv}, and \ref{fig:init_woman_conv}. 

\pagebreak
\null\newpage
\subsection{CLIP Attention}
\begin{figure*}[h]
    \centering
        \begin{tabular}{c:c}
        \includegraphics[trim={0cm 0 65.8cm 0cm},clip, width=0.14\textwidth]{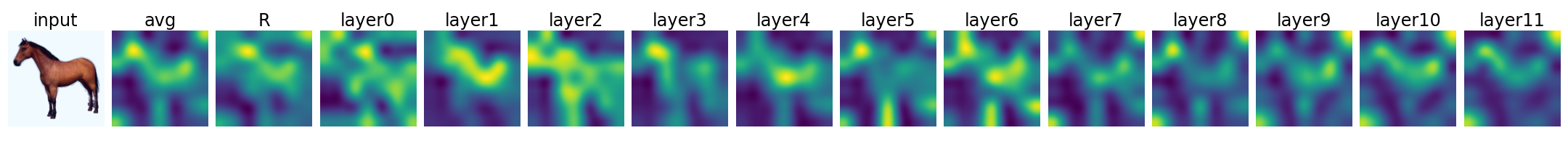} & \includegraphics[trim={15.5cm 0 0 0},clip, width=0.84\textwidth]{figs/clip_attention/horse_horse_132.png} \\
        \includegraphics[trim={0cm 0 65.8cm 0cm},clip, width=0.14\textwidth]{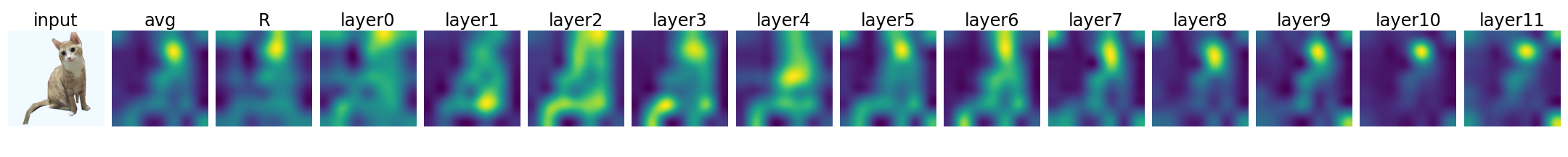} & \includegraphics[trim={15.5cm 0 0 0},clip, width=0.84\textwidth]{figs/clip_attention/cat_47654.png} \\
        \includegraphics[trim={0cm 0 65.8cm 0cm},clip, width=0.14\textwidth]{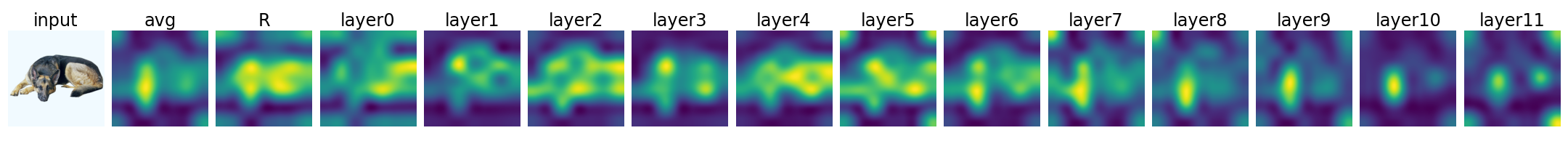} & \includegraphics[trim={15.5cm 0 0 0},clip, width=0.84\textwidth]{figs/clip_attention/dog_3626.png} \\
        \includegraphics[trim={0cm 0 65.8cm 0cm},clip, width=0.14\textwidth]{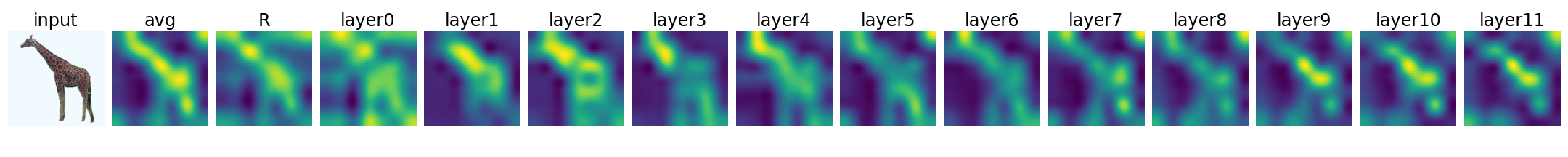} & \includegraphics[trim={15.5cm 0 0 0},clip, width=0.84\textwidth]{figs/clip_attention/giraffe_595745.png} \\
        \includegraphics[trim={0cm 0 65.8cm 0cm},clip, width=0.14\textwidth]{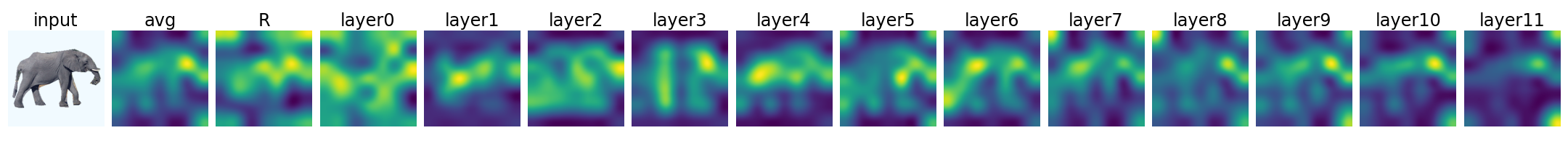} & \includegraphics[trim={15.5cm 0 0 0},clip, width=0.84\textwidth]{figs/clip_attention/elephant_580017.png} \\
        \includegraphics[trim={0cm 0 65.8cm 0cm},clip, width=0.14\textwidth]{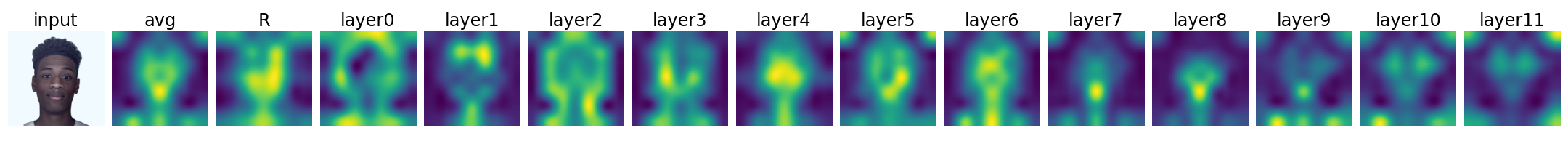} & \includegraphics[trim={15.5cm 0 0 0},clip, width=0.84\textwidth]{figs/clip_attention/men_05-34783750806_954ea83724_o.png} \\
        \includegraphics[trim={0cm 0 65.8cm 0cm},clip, width=0.14\textwidth]{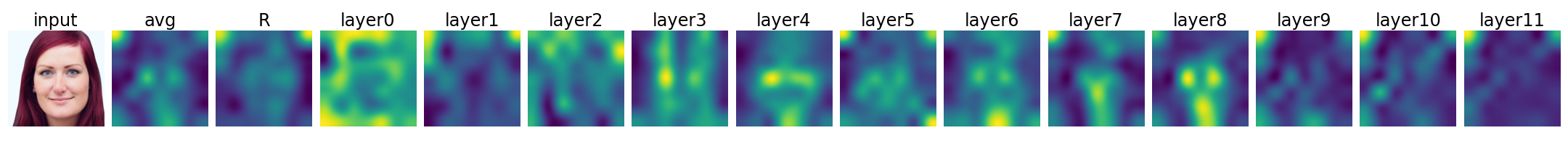} & \includegraphics[trim={15.5cm 0 0 0},clip, width=0.84\textwidth]{figs/clip_attention/women_01-12866391563_7312cae13a_o_w.png} 
        \end{tabular}
    \caption{Attention heads of CLIP among all layers.}
    \label{fig:clip_attention}
\end{figure*}

The transformer encoder is comprised of alternate layers of self-attention blocks and MLP blocks. For saliency visualization, it is common to use the attention probabilities of each layer, and aggregate them to get the final relevancy map.
In Figure \ref{fig:clip_attention}, we visualize the attention maps of each layer in ViT-32/B architecture on several image classes.
In order to produce the attention map for each layer, we take the average of 12 attention heads in that layer.
The final relevancy map is simply the average of the attention maps of all layers. It can be seen that the average attention map highlights the salient regions of the input image, such as the head of the horse, cat, and dog, as well as the body.
\hfill

\begin{figure*}[!b]
\centering
\begin{tabular}{@{\hskip2pt}c|@{\hskip2pt}c@{\hskip2pt}c@{\hskip2pt}c@{\hskip2pt}c@{\hskip2pt}c@{\hskip2pt}c@{\hskip2pt}c}
    \midrule
     & Input & Layer1 & Layer2 & Layer3 & Layer4 & Layer5 & Fully Connected \\
    
    \hline
    \parbox[t]{5mm}{\multirow{3}{*}{\rotatebox{90}{ResNet101}}} &
    \includegraphics[width=\widthresnet\linewidth]{figs/abstraction_levels/input/horse/horse_132.png} &
    \includegraphics[width=\widthresnet\linewidth]{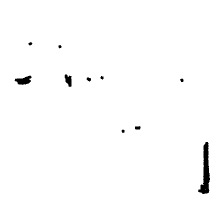} &
    \includegraphics[width=\widthresnet\linewidth]{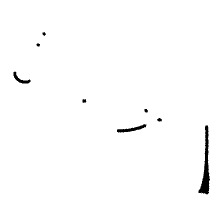} &
    \includegraphics[width=\widthresnet\linewidth]{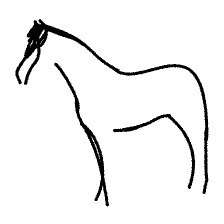} &
    \includegraphics[width=\widthresnet\linewidth]{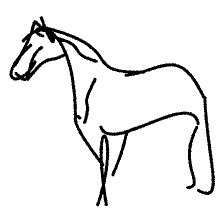} &
    \includegraphics[width=\widthresnet\linewidth]{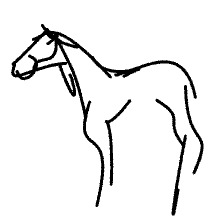} &
    \includegraphics[width=\widthresnet\linewidth]{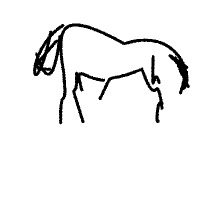} \\
    
    \hhline{~|-|-|-|-|-|-|-|}
    &
    \includegraphics[width=\widthresnet\linewidth]{figs/initialisation_comp/cat/xdog_clip/cat_10544.jpg} &
    \includegraphics[width=\widthresnet\linewidth]{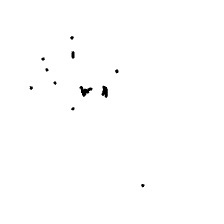} &
    \includegraphics[width=\widthresnet\linewidth]{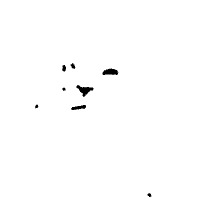} &
    \includegraphics[width=\widthresnet\linewidth]{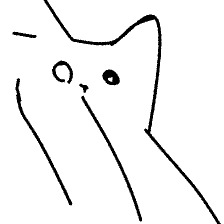} &
    \includegraphics[width=\widthresnet\linewidth]{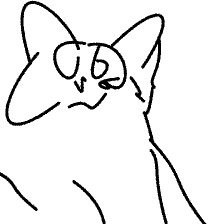} &
    \includegraphics[width=\widthresnet\linewidth]{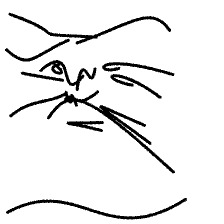} &
    \includegraphics[width=\widthresnet\linewidth]{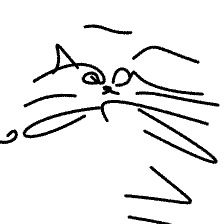} \\
    
    \hhline{~|-|-|-|-|-|-|-|}
    &
    \includegraphics[width=\widthresnet\linewidth]{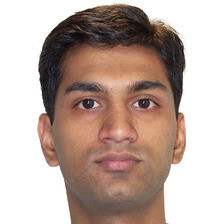} &
    \includegraphics[width=\widthresnet\linewidth]{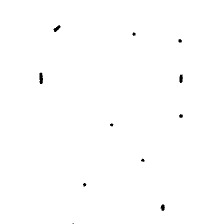} &
    \includegraphics[width=\widthresnet\linewidth]{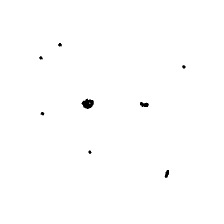} &
    \includegraphics[width=\widthresnet\linewidth]{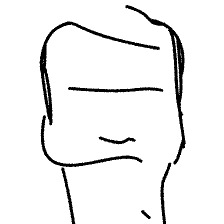} &
    \includegraphics[width=\widthresnet\linewidth]{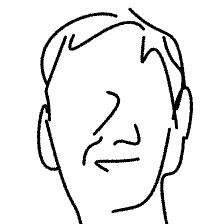} &
    \includegraphics[width=\widthresnet\linewidth]{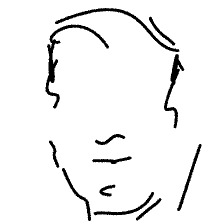} &
    \includegraphics[width=\widthresnet\linewidth]{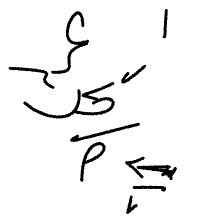} \\
    
    \midrule
    \midrule
    \parbox[t]{5mm}{\multirow{4}{*}{\rotatebox{90}{ResNet50}}} &
    \includegraphics[width=\widthresnet\linewidth]{figs/abstraction_levels/input/horse/horse_132.png} &
    \includegraphics[width=\widthresnet\linewidth]{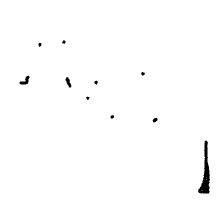} &
    \includegraphics[width=\widthresnet\linewidth]{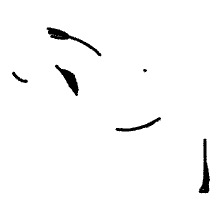} &
    \includegraphics[width=\widthresnet\linewidth]{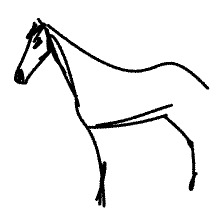} &
    \includegraphics[width=\widthresnet\linewidth]{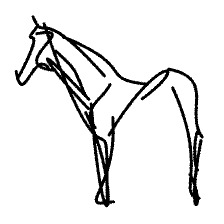} &
    \includegraphics[width=\widthresnet\linewidth]{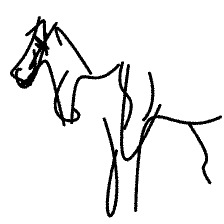} &
    \includegraphics[width=\widthresnet\linewidth]{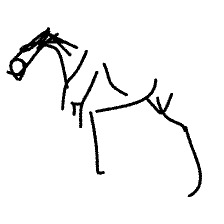} \\
    
    \hhline{~|-|-|-|-|-|-|-|}
    &
    \includegraphics[width=\widthresnet\linewidth]{figs/initialisation_comp/cat/xdog_clip/cat_10544.jpg} &
    \includegraphics[width=\widthresnet\linewidth]{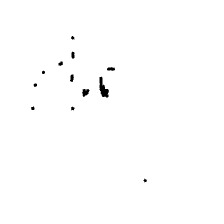} &
    \includegraphics[width=\widthresnet\linewidth]{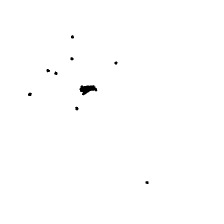} &
    \includegraphics[width=\widthresnet\linewidth]{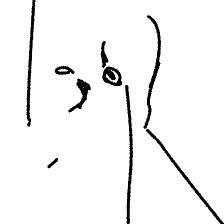} &
    \includegraphics[width=\widthresnet\linewidth]{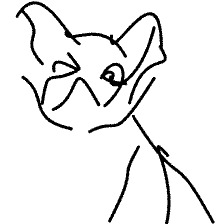} &
    \includegraphics[width=\widthresnet\linewidth]{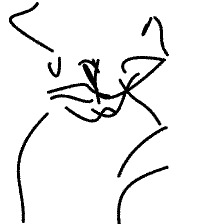} &
    \includegraphics[width=\widthresnet\linewidth]{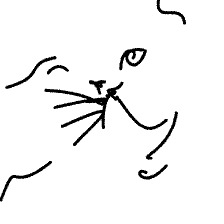} \\
    
    \hhline{~|-|-|-|-|-|-|-|}
    &
    \includegraphics[width=\widthresnet\linewidth]{figs/comp_objects/input/men/face_4_edge.jpg} &
    \includegraphics[width=\widthresnet\linewidth]{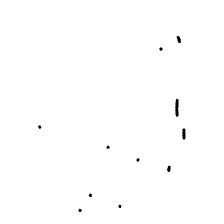} &
    \includegraphics[width=\widthresnet\linewidth]{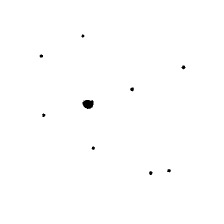} &
    \includegraphics[width=\widthresnet\linewidth]{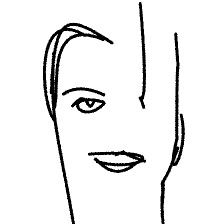} &
    \includegraphics[width=\widthresnet\linewidth]{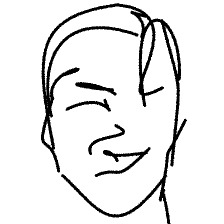} &
    \includegraphics[width=\widthresnet\linewidth]{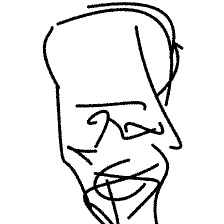} &
    \includegraphics[width=\widthresnet\linewidth]{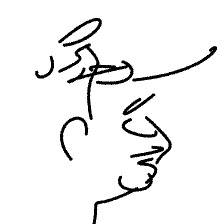} \\
    
\end{tabular}
 \caption{Synthesized sketches by optimizing w.r.t the different layers of RN101 and RN50 architectures}
\label{fig:arch_rn101_rn50}
\end{figure*}

\pagebreak
\null\newpage
\pagebreak
\null\newpage

\section{Loss Ablation and Analysis}
As stated in the paper, we chose to define our semantic and geometric loss according to CLIP's ResNet101 architecture. 
In section \ref{sec:clip_models_layers} we examine this choice by analyzing the results of optimization when different CLIP architectures and intermediate layers are chosen as the basis for the loss function.
In section \ref{sec:loss_term_weights} we provide an ablation study of the different components and weights in the chosen loss function.
In section \ref{sec:compare_l2_lpips} we further demonstrate the contribution of our loss compared to L2 and LPIPS.

It is important to note that throughout this analysis our goal is to select the objective which best encodes the visual representations of images for extracting a balanced combination of both geometric and semantic features. Enforcing too much geometrical fidelity would result in sketches being very close to simple contours. On the other hand, only focusing on the semantic features can produce uncoherent object depictions or reduce the instance-level recognizability of the synthesized sketches. With a lack of a good automatic metric for this objective, we use visual judgment to select the best configurations.

\begin{figure*}[!b]
\centering
\begin{tabular}{@{\hskip2pt}c|@{\hskip2pt}c@{\hskip2pt}c@{\hskip2pt}c@{\hskip2pt}c@{\hskip2pt}c@{\hskip2pt}c@{\hskip2pt}c}

    \hline
    \multicolumn{1}{l}{} & Input & Layer1 & Layer2 & Layer3 & Layer4 & Layer5 & Layer6 \\
    
    \hline
    \parbox[t]{5mm}{\multirow{3}{*}{\rotatebox{90}{ViT-32/B}}} &
    \includegraphics[width=\widthvit\linewidth]{figs/abstraction_levels/input/horse/horse_132.png} &
    \includegraphics[width=\widthvit\linewidth]{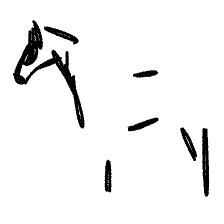} &
    \includegraphics[width=\widthvit\linewidth]{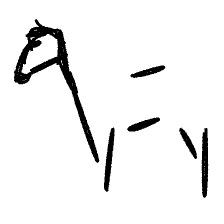} &
    \includegraphics[width=\widthvit\linewidth]{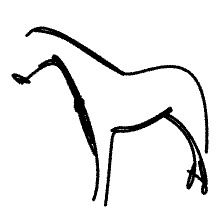} &
    \includegraphics[width=\widthvit\linewidth]{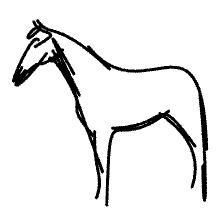} &
    \includegraphics[width=\widthvit\linewidth]{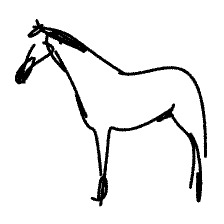} &
    \includegraphics[width=\widthvit\linewidth]{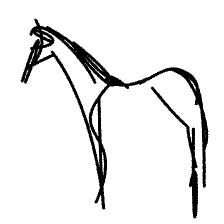} \\
    
    \hhline{~|-|-|-|-|-|-|-|}
    &
    \includegraphics[width=\widthvit\linewidth]{figs/initialisation_comp/cat/xdog_clip/cat_10544.jpg} &
    \includegraphics[width=\widthvit\linewidth]{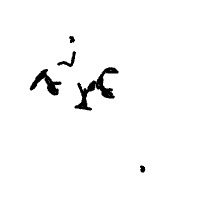} &
    \includegraphics[width=\widthvit\linewidth]{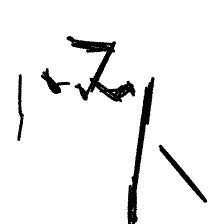} &
    \includegraphics[width=\widthvit\linewidth]{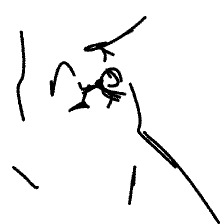} &
    \includegraphics[width=\widthvit\linewidth]{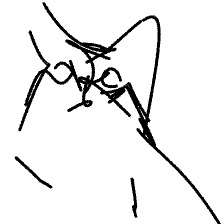} &
    \includegraphics[width=\widthvit\linewidth]{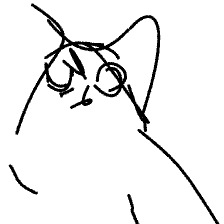} &
    \includegraphics[width=\widthvit\linewidth]{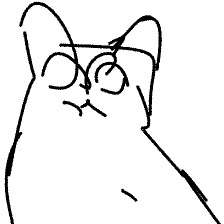} \\
    
    \hhline{~|-|-|-|-|-|-|-|}
    &
    \includegraphics[width=\widthvit\linewidth]{figs/comp_objects/input/men/face_4_edge.jpg} &
    \includegraphics[width=\widthvit\linewidth]{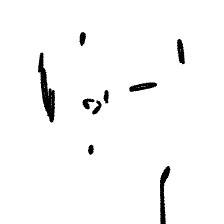} &
    \includegraphics[width=\widthvit\linewidth]{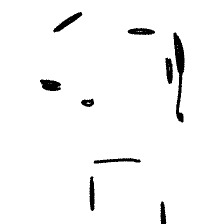} &
    \includegraphics[width=\widthvit\linewidth]{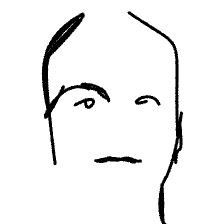} &
    \includegraphics[width=\widthvit\linewidth]{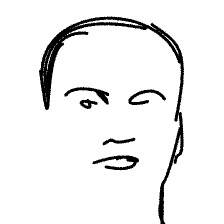} &
    \includegraphics[width=\widthvit\linewidth]{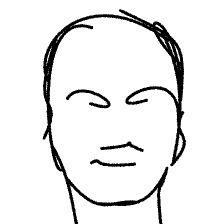} &
    \includegraphics[width=\widthvit\linewidth]{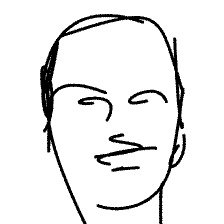} \\
    
    \midrule
    \midrule
    \parbox[t]{2mm}{\multirow{3}{*}{\rotatebox{90}{ViT-16/B}}} &
    \includegraphics[width=\widthvit\linewidth]{figs/abstraction_levels/input/horse/horse_132.png} &
    \includegraphics[width=\widthvit\linewidth]{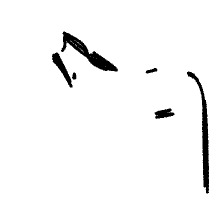} &
    \includegraphics[width=\widthvit\linewidth]{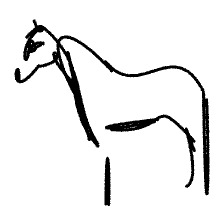} &
    \includegraphics[width=\widthvit\linewidth]{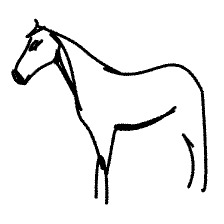} &
    \includegraphics[width=\widthvit\linewidth]{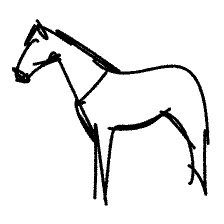} &
    \includegraphics[width=\widthvit\linewidth]{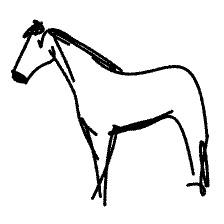} &
    \includegraphics[width=\widthvit\linewidth]{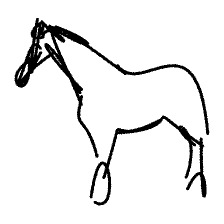} \\
    
    \hhline{~|-|-|-|-|-|-|-|}
    &
    \includegraphics[width=\widthvit\linewidth]{figs/initialisation_comp/cat/xdog_clip/cat_10544.jpg} &
    \includegraphics[width=\widthvit\linewidth]{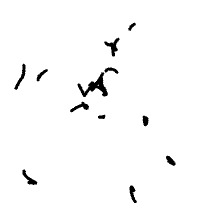} &
    \includegraphics[width=\widthvit\linewidth]{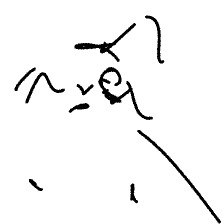} &
    \includegraphics[width=\widthvit\linewidth]{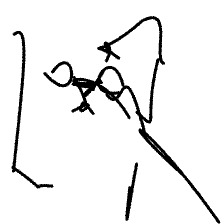} &
    \includegraphics[width=\widthvit\linewidth]{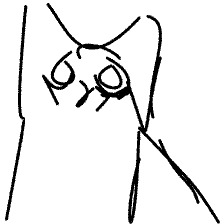} &
    \includegraphics[width=\widthvit\linewidth]{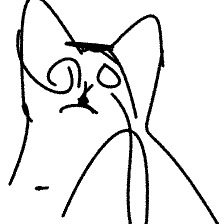} &
    \includegraphics[width=\widthvit\linewidth]{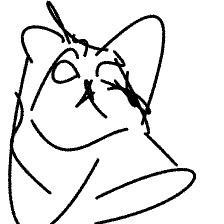} \\
    
    \hhline{~|-|-|-|-|-|-|-|}
    &
    \includegraphics[width=\widthvit\linewidth]{figs/comp_objects/input/men/face_4_edge.jpg} &
    \includegraphics[width=\widthvit\linewidth]{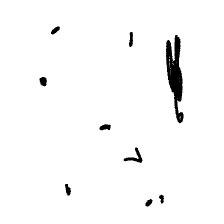} &
    \includegraphics[width=\widthvit\linewidth]{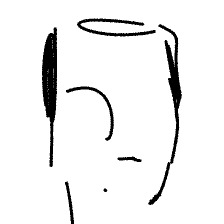} &
    \includegraphics[width=\widthvit\linewidth]{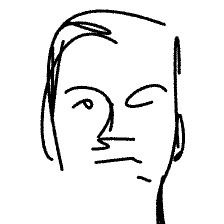} &
    \includegraphics[width=\widthvit\linewidth]{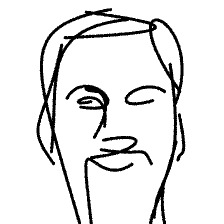} &
    \includegraphics[width=\widthvit\linewidth]{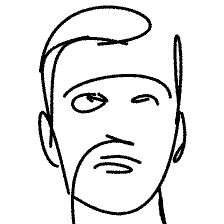} &
    \includegraphics[width=\widthvit\linewidth]{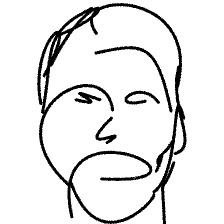}
\end{tabular}
 \caption{Synthesized sketches by optimizing w.r.t the different layers of ViT32 and ViT16 architectures}
\label{fig:arch_vit_LAYERS1-6}
\end{figure*}

\begin{figure*}[h]
\centering
\begin{tabular}{@{\hskip2pt}c|@{\hskip2pt}c@{\hskip2pt}c@{\hskip2pt}c@{\hskip2pt}c@{\hskip2pt}c@{\hskip2pt}c@{\hskip2pt}c@{\hskip2pt}c}

    \midrule
    \multicolumn{1}{l}{} & Input & Layer7 & Layer8 & Layer9 & Layer10 & Layer11 & Layer12 & Fully Connected \\
    
    \hline
    \parbox[t]{5mm}{\multirow{3}{*}{\rotatebox{90}{ViT-32/B}}} &
    \includegraphics[width=\widthvit\linewidth]{figs/abstraction_levels/input/horse/horse_132.png} &
    \includegraphics[width=\widthvit\linewidth]{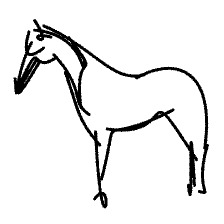} &
    \includegraphics[width=\widthvit\linewidth]{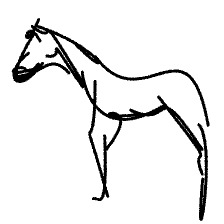} &
    \includegraphics[width=\widthvit\linewidth]{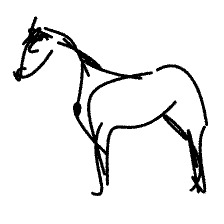} &
    \includegraphics[width=\widthvit\linewidth]{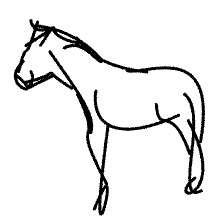} &
    \includegraphics[width=\widthvit\linewidth]{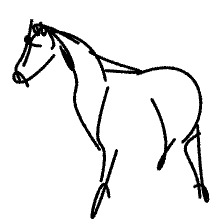} &
    \includegraphics[width=\widthvit\linewidth]{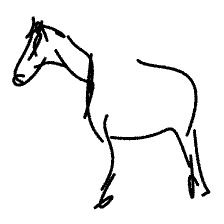} &
    \includegraphics[width=\widthvit\linewidth]{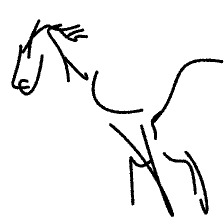} \\
    
    \hhline{~|-|-|-|-|-|-|-|-|}
    &
    \includegraphics[width=\widthvit\linewidth]{figs/initialisation_comp/cat/xdog_clip/cat_10544.jpg} &
    \includegraphics[width=\widthvit\linewidth]{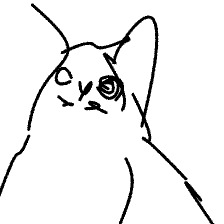} &
    \includegraphics[width=\widthvit\linewidth]{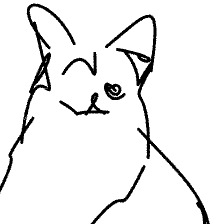} &
    \includegraphics[width=\widthvit\linewidth]{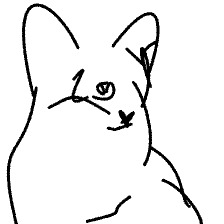} &
    \includegraphics[width=\widthvit\linewidth]{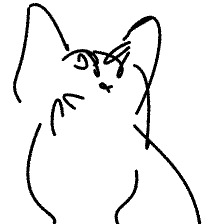} &
    \includegraphics[width=\widthvit\linewidth]{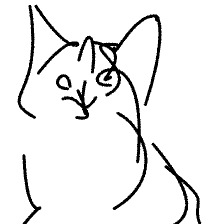} &
    \includegraphics[width=\widthvit\linewidth]{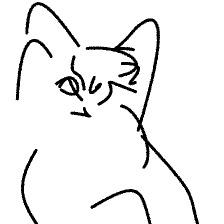} &
    \includegraphics[width=\widthvit\linewidth]{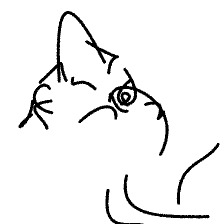} \\
    
    \hhline{~|-|-|-|-|-|-|-|-|}
    &
    \includegraphics[width=\widthvit\linewidth]{figs/comp_objects/input/men/face_4_edge.jpg} &
    \includegraphics[width=\widthvit\linewidth]{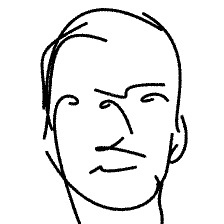} &
    \includegraphics[width=\widthvit\linewidth]{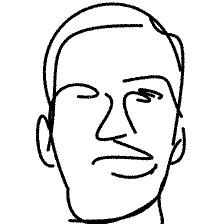} &
    \includegraphics[width=\widthvit\linewidth]{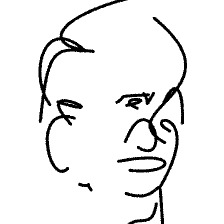} &
    \includegraphics[width=\widthvit\linewidth]{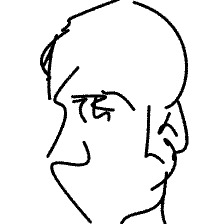} &
    \includegraphics[width=\widthvit\linewidth]{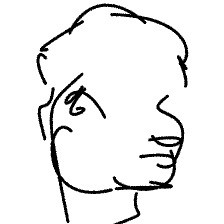} &
    \includegraphics[width=\widthvit\linewidth]{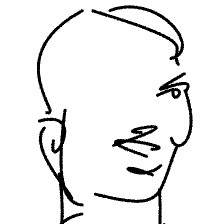} &
    \includegraphics[width=\widthvit\linewidth]{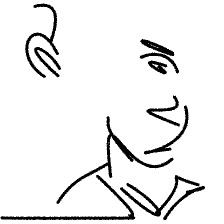} \\
    
    \midrule
    \midrule 
    \parbox[t]{5mm}{\multirow{3}{*}{\rotatebox{90}{ViT-16/B}}} &
    \includegraphics[width=\widthvit\linewidth]{figs/abstraction_levels/input/horse/horse_132.png} &
    \includegraphics[width=\widthvit\linewidth]{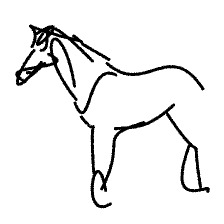} &
    \includegraphics[width=\widthvit\linewidth]{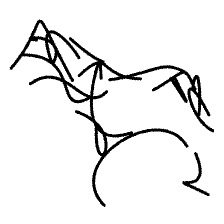} &
    \includegraphics[width=\widthvit\linewidth]{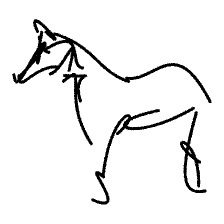} &
    \includegraphics[width=\widthvit\linewidth]{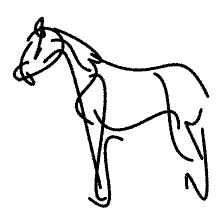} &
    \includegraphics[width=\widthvit\linewidth]{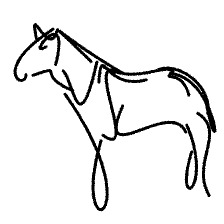} &
    \includegraphics[width=\widthvit\linewidth]{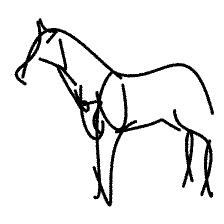} &
    \includegraphics[width=\widthvit\linewidth]{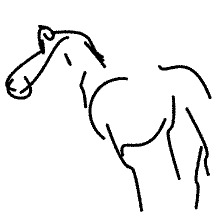} \\
    
    \hhline{~|-|-|-|-|-|-|-|-|}
    &
    \includegraphics[width=\widthvit\linewidth]{figs/initialisation_comp/cat/xdog_clip/cat_10544.jpg} &
    \includegraphics[width=\widthvit\linewidth]{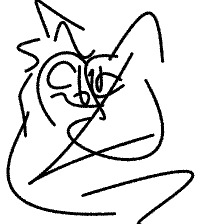} &
    \includegraphics[width=\widthvit\linewidth]{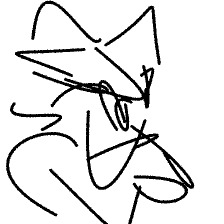} &
    \includegraphics[width=\widthvit\linewidth]{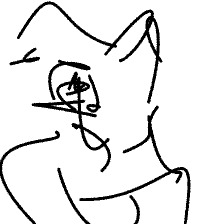} &
    \includegraphics[width=\widthvit\linewidth]{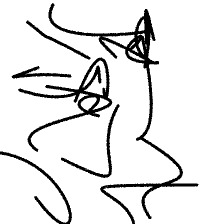} &
    \includegraphics[width=\widthvit\linewidth]{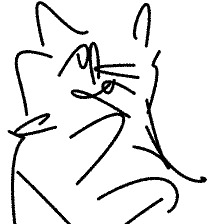} &
    \includegraphics[width=\widthvit\linewidth]{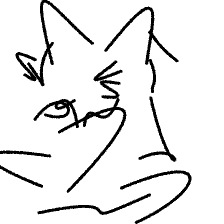} &
    \includegraphics[width=\widthvit\linewidth]{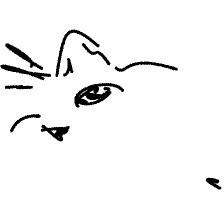} \\
    
    \hhline{~|-|-|-|-|-|-|-|-|}
    &
    \includegraphics[width=\widthvit\linewidth]{figs/comp_objects/input/men/face_4_edge.jpg} &
    \includegraphics[width=\widthvit\linewidth]{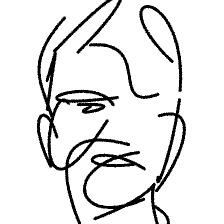} &
    \includegraphics[width=\widthvit\linewidth]{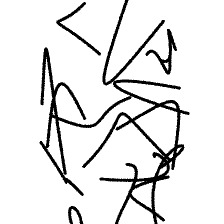} &
    \includegraphics[width=\widthvit\linewidth]{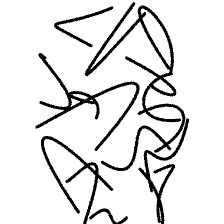} &
    \includegraphics[width=\widthvit\linewidth]{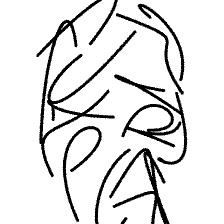} &
    \includegraphics[width=\widthvit\linewidth]{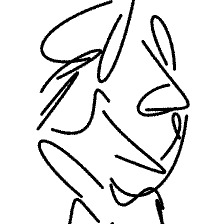} &
    \includegraphics[width=\widthvit\linewidth]{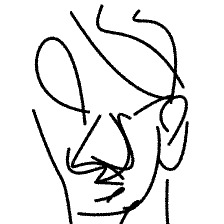} &
    \includegraphics[width=\widthvit\linewidth]{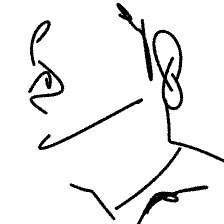} \\
\end{tabular}
 \caption{Synthesized sketches by optimizing w.r.t the different layers of ViT32 and ViT16 architectures}
\label{fig:arch_vit_LAYERS6-12}
\end{figure*}

\subsection{CLIP Layers and Architecture Analysis}
\label{sec:clip_models_layers}
The authors of CLIP published several pretrained models including ResNet50, ResNet101, and two additional vision transformers namely ViT-B/32 and ViT-B/16. 
In the following experiments, we used the same initialization method and general settings to obtain the sketches. We only change the layer or architecture used, and we apply the optimization with respect to the loss determined by this layer.
We use the cosine similarity to compute the loss for the fully connected layer (final output) and the L2 distance for the loss determined by intermediate layers.
In the comparisons, images from three different classes were examined - horse, cat, and human face.

Figure \ref{fig:arch_rn101_rn50} shows the results obtained by using ResNet101 and ResNet50 as the base model. The layer1 for the ResNet architecture is defined as the input to the first Bottleneck block, and the other four layers are simply the output of each Bottleneck blocks in the network. Each column in the figure shows the resulting sketches when the loss is defined based on the activations of the respective layer.
We can see that layers 1 and 2 produced non-recognizable results\footnote{This can be because of the small receptive field of the feature-maps at this layers.}.
In contrast, layers 3, 4, and 5 result in reasonable sketches that combine both geometry and semantics, leaning towards more semantic fidelity as we go deeper. The rightmost column contains the results obtained by using only the fully connected layer. This is equivalent to using text as input. We can see that spatial coherence has disappeared completely, and that the model has produced some recognizable class-based features, such as whiskers for the cat.
Overall, we observe that ResNet101 produced cleaner and more stable sketches than ResNet50. Layer 3 and 4 of the ResNet101 contain a desirable combination of globally coherent geometry and semantically recognizable features, which led us to use them for defining our geometric-fidelity loss $L_{geometric}$. 

In Figures \ref{fig:arch_vit_LAYERS1-6} and \ref{fig:arch_vit_LAYERS6-12} we present the results of a similar experiment using the vision transformer architectures ViT-16/B and ViT-32/B. To perform the analysis, we utilized the 12 self-attention layers and the last fully connected layer. In the results, we observe that both ViT models encode a higher bias towards shape compared to the ResNet models, especially in layers 5 and lower. The strokes attained using ViT16 are observably messier compared to ViT32 and do not have well-defined start and end points. 
Comparing the ViT vs. CNN architectures, we favor the latter, and specifically ResNet-101 as it seems to strike a balance between geometric fidelity and semantics. 

\null\newpage
\subsection{Loss Term Weights}
\label{sec:loss_term_weights}
\begin{figure}[h]
\centering
\begin{tabular}{@{\hskip2pt}c@{\hskip2pt}c@{\hskip2pt}c@{\hskip2pt}c@{\hskip2pt}c@{\hskip2pt}c}
    \midrule
    Input & $-L{s}$ & $-L_{g}$ & -Layer3 & -Layer2 &\begin{tabular}[c]{@{}c@{}}All \\ Together\end{tabular}  \\
    
    \midrule
    \includegraphics[width=\widthattn\linewidth]{figs/initialisation_comp/cat/xdog_clip/cat_10544.jpg} &
    \includegraphics[width=\widthattn\linewidth]{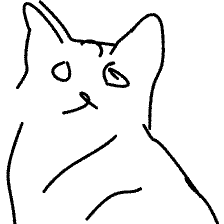} &
    \includegraphics[width=\widthattn\linewidth]{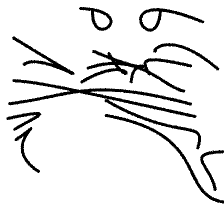} &
    \includegraphics[width=\widthattn\linewidth]{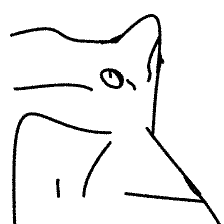} &
    \includegraphics[width=\widthattn\linewidth]{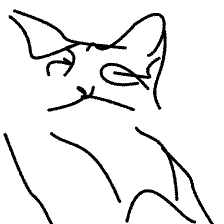} &
    \includegraphics[width=\widthattn\linewidth]{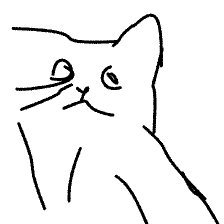} \\
    
    \midrule
    \includegraphics[width=\widthattn\linewidth]{figs/abstraction_levels/input/horse/horse_132.png} &
    \includegraphics[width=\widthattn\linewidth]{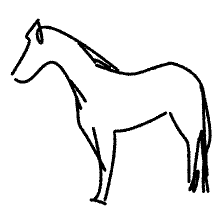} &
    \includegraphics[width=\widthattn\linewidth]{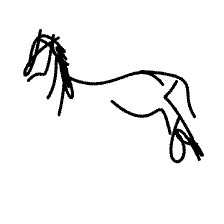} &
    \includegraphics[width=\widthattn\linewidth]{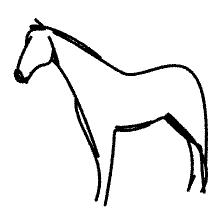} &
    \includegraphics[width=\widthattn\linewidth]{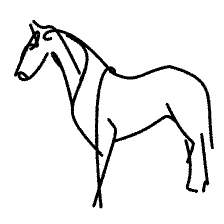} &
    \includegraphics[width=\widthattn\linewidth]{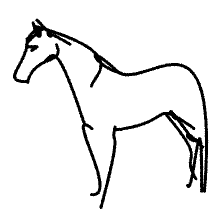} \\
    
    \midrule
    \includegraphics[width=\widthattn\linewidth]{figs/comp_objects/input/men/face_4_edge.jpg} &
    \includegraphics[width=\widthattn\linewidth]{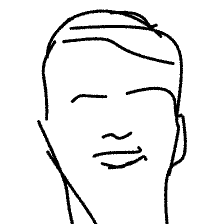} &
    \includegraphics[width=\widthattn\linewidth]{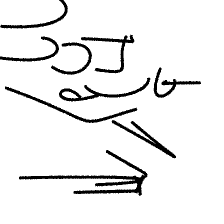} &
    \includegraphics[width=\widthattn\linewidth]{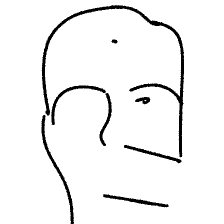} &
    \includegraphics[width=\widthattn\linewidth]{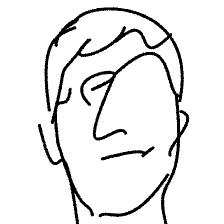} &
    \includegraphics[width=\widthattn\linewidth]{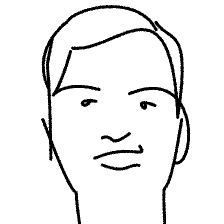} \\
    
\end{tabular}
 \caption{Excluding terms from the loss function - in the left column are the input images, and from left to right are the sketches produced when omitting - $L_{semantic}$, $L_{geometric}$, layer3 from $L_{geometric}$, and layer2 from $L_{geometric}$. In the rightmost column are the results of the proposed method when all parts are included. }
\label{fig:ablation_loss_remove_parts}
\end{figure}

Figure \ref{fig:ablation_loss_remove_parts} illustrates the effect of excluding each term from the loss formula, the term  excluded is listed on top of each column.
Omitting $L_{semantic}$ (column $L_s$) results in sketches that fit well with the geometry, but may not be semantic enough, whereas when omitting $L_{geometric}$ (column $L_g$), we lose instance-level recognition. For animal subjects, the sketches' class is recognizable, but they do not fit the specific object's geometry well, while for the face image, they are unrecognizable.
In columns 4 and 5, layers 3 and 2 have been omitted from $L_{geometric}$. It can be seen that when layer 3 is removed, the results become less semantic and more biased towards geometry, whereas the opposite happens when layer 2 is removed.

\begin{figure}
\centering
\begin{tabular}{@{\hskip2pt}c@{\hskip2pt}c@{\hskip2pt}c@{\hskip2pt}c@{\hskip2pt}c}
    \midrule
    Input & fc0 & fc0.1 & fc0.5 & fc1 \\
    
    \midrule
    \includegraphics[width=\widthattn\linewidth]{figs/initialisation_comp/cat/xdog_clip/cat_10544.jpg} &
    \includegraphics[width=\widthattn\linewidth]{figs/ablation_loss_weight/cat_l2-1_l3-1_fc0_p16.png} &
    \includegraphics[width=\widthattn\linewidth]{figs/ablation_loss_weight/cat_l2-1_l3-1_fc0.1_p16.png} &
    \includegraphics[width=\widthattn\linewidth]{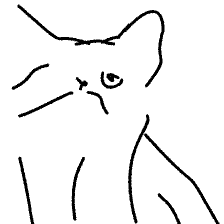} &
    \includegraphics[width=\widthattn\linewidth]{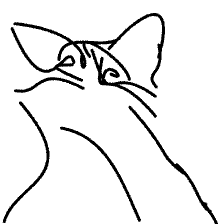} \\
    
    \midrule
    \includegraphics[width=\widthattn\linewidth]{figs/abstraction_levels/input/horse/horse_132.png} &
    \includegraphics[width=\widthattn\linewidth]{figs/ablation_loss_weight/horse_l2-1_l3-1_fc0_p16.png} &
    \includegraphics[width=\widthattn\linewidth]{figs/ablation_loss_weight/horse_l2-1_l3-1_fc0.1_p16.png} &
    \includegraphics[width=\widthattn\linewidth]{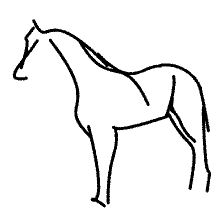} &
    \includegraphics[width=\widthattn\linewidth]{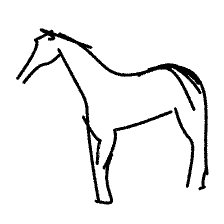} \\
    
    \midrule
    \includegraphics[width=\widthattn\linewidth]{figs/comp_objects/input/men/face_4_edge.jpg} &
    \includegraphics[width=\widthattn\linewidth]{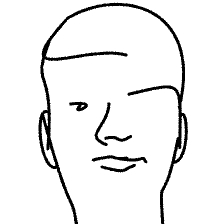} &
    \includegraphics[width=\widthattn\linewidth]{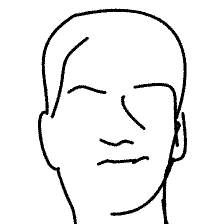} &
    \includegraphics[width=\widthattn\linewidth]{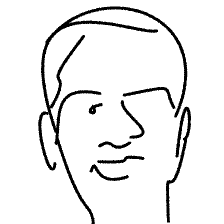} &
    \includegraphics[width=\widthattn\linewidth]{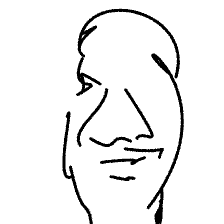} \\
\end{tabular}
 \caption{Weights intervals of semantic loss, columns 2 - 5 shows the optimization result when $L_{semantic}$ is weighted by 0, 0.1, 0.5, and 1 correspondingly.}
 \vspace{-1em}
\label{fig:ablation_loss_fc}
\end{figure}

Figure \ref{fig:ablation_loss_fc} illustrates the impact of applying different weights to $L_{semantic}$ - ranging from 0 to 1 (left to right).
When applied with weights of 0.5 and 1 (columns 4 and 5), the resulting sketches exhibit geometric changes that may be too extreme, however when we cancel $L_{semantic}$ completely (column 2) the results tend towards geometry. We find that weight 0.1 provides a good balance between geometry and semantics. For example, some whiskers are added to the cat while the general geometry is also maintained. Additionally, the geometry of the horse's head varies slightly from that in the input image, but the pose is correct.

\subsection{Comparison to Other Loss Functions}
\label{sec:compare_l2_lpips}
\begin{figure}[h]
\centering
\begin{tabular}{@{\hskip2pt}c@{\hskip2pt}c@{\hskip2pt}c@{\hskip2pt}c@{\hskip2pt}c}
    \midrule
    Input & XDoG & L2 & LPIPS & Ours \\
    
    \midrule
    \includegraphics[width=\widthattn\linewidth]{figs/initialisation_comp/cat/xdog_clip/cat_10544.jpg} &
    \includegraphics[width=\widthattn\linewidth]{figs/ablation_l_geometric/cat_xdog.png} &
    \includegraphics[width=\widthattn\linewidth]{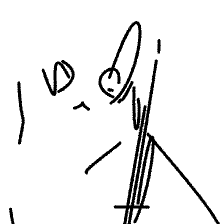} &
    \includegraphics[width=\widthattn\linewidth]{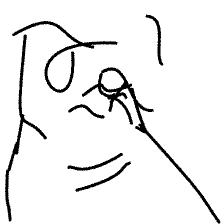} &
    \includegraphics[width=\widthattn\linewidth]{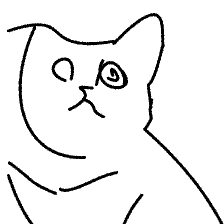} \\
    \hline

    \includegraphics[width=\widthattn\linewidth]{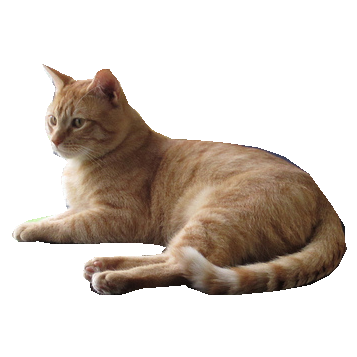} &
    \includegraphics[width=\widthattn\linewidth]{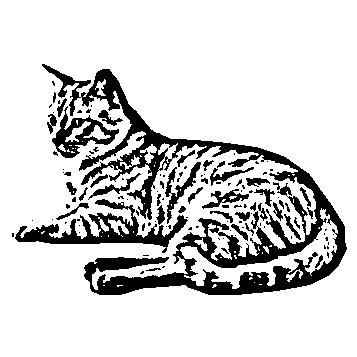} &
    \includegraphics[width=\widthattn\linewidth]{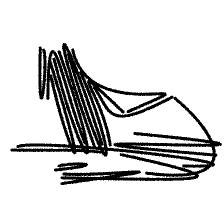} &
    \includegraphics[width=\widthattn\linewidth]{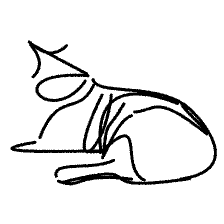} &
    \includegraphics[width=\widthattn\linewidth]{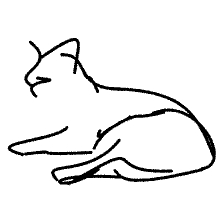} \\
    
    \hline
    \includegraphics[width=\widthattn\linewidth]{figs/abstraction_levels/input/horse/horse_132.png} &
    \includegraphics[width=\widthattn\linewidth]{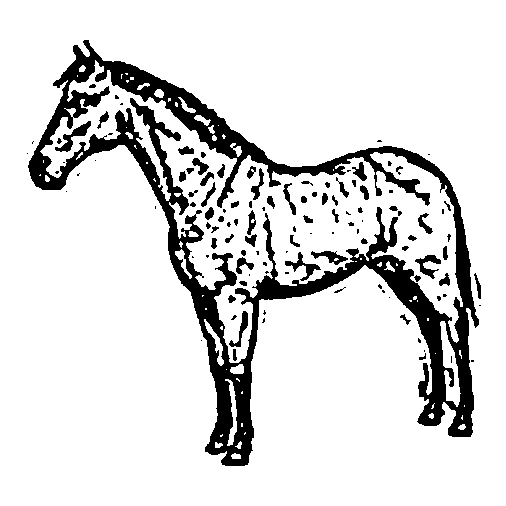} &
    \includegraphics[width=\widthattn\linewidth]{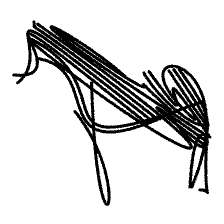} &
    \includegraphics[width=\widthattn\linewidth]{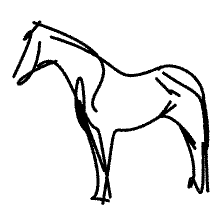} &
    \includegraphics[width=\widthattn\linewidth]{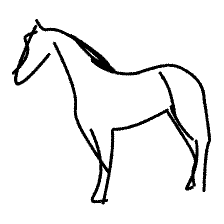} \\
    
    \hline
    \includegraphics[width=\widthattn\linewidth]{figs/comp_objects/input/men/face_4_edge.jpg} &
    \includegraphics[width=\widthattn\linewidth]{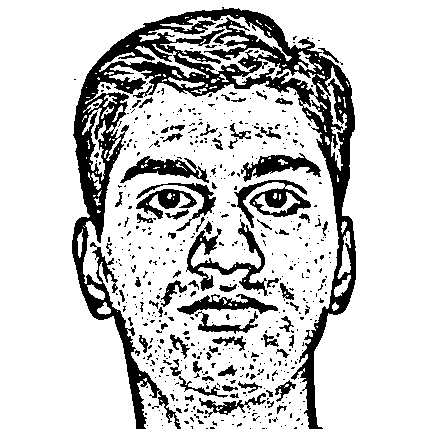} &
    \includegraphics[width=\widthattn\linewidth]{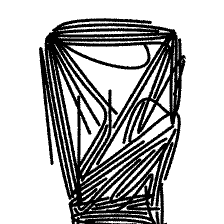} &
    \includegraphics[width=\widthattn\linewidth]{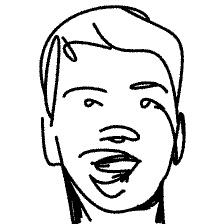} &
    \includegraphics[width=\widthattn\linewidth]{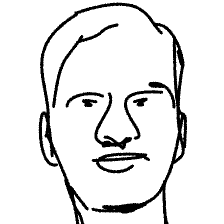} \\
\end{tabular}
 \caption{Replacing our CLIP-based geometric loss with LPIPS, and L2. The animal sketches are drawn using 16 strokes and the face sketch is drawn using 32 strokes.}
\label{fig:ablation_l_geomentry}
\end{figure}

Our proposed loss is a perceptual loss, providing a combination between geometry and semantics.
In this section we compare the proposed loss to L2 and LPIPS \cite{Zhang2018TheUE}.
While L2 is purely geometric in that it measures pixel-wise distance, LPIPS is considered more semantic, since it measures the distance between the intermediate activations of a pretrained vision network (trained for a classification task using ImageNet dataset and the VGG16 architecture).
In Figure \ref{fig:ablation_l_geomentry} we compare these losses with our loss on four images.
Each result displayed in the figure was obtained through the same initialization method (saliency guided), with the only difference being the loss used for optimization. The XDoG edge map is presented as an edge-map baseline.
It is evident that the results produced using L2 do not contain any semantic information. L2 encourage the optimization to "fill" the colored pixels in the input image.
The sketches improved significantly when using LPIPS, the strokes follow the semantics of the image. We observe, however, that LPIPS is too geometric in comparison with our loss. 
The results produced by our method are shown in the last column. Improvements can be seen clearly, especially for the face, where semantic visual attributes are introduced to the sketch.

\begin{figure*}[ht]
    \centering
    \includegraphics[width=\linewidth]{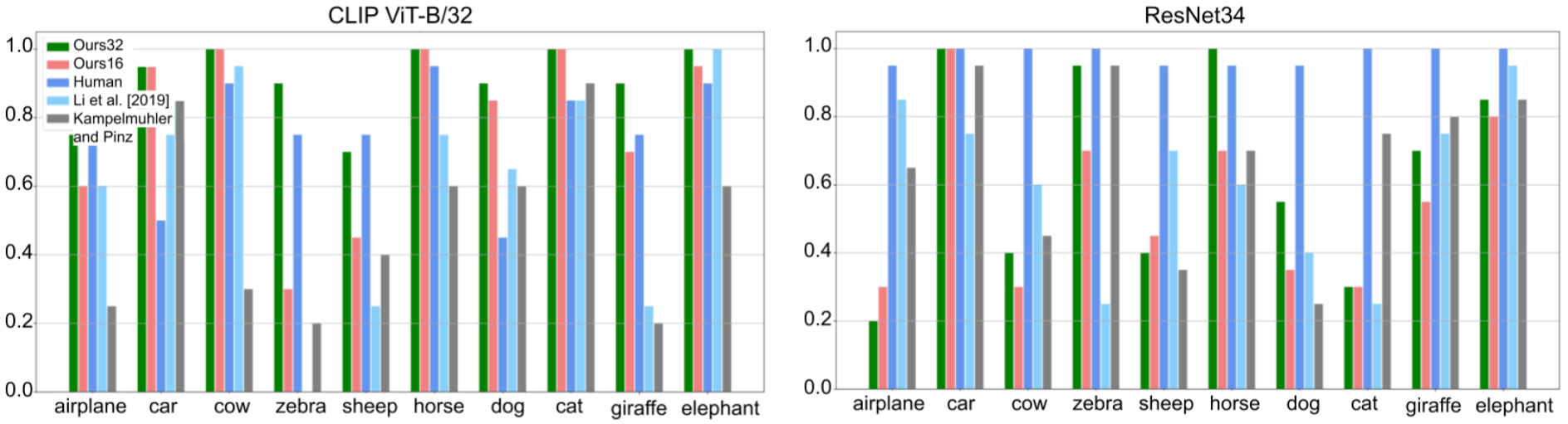}
    \caption{Per-class classification accuracy of the two classifiers we used in our quantitative analysis. The figure on top shows the results for CLIP ViT-B/32 and the figure below shows the results for ResNet34 classifier. Each color represent sketches by the corresponding sketch synthesis model.}
    \label{fig:sketchy_coco_detailed_classification}
\end{figure*}
\section{Quantitative Comparison}
\begin{figure}[h]
\centering
\begin{tabular}{@{\hskip2pt}c@{\hskip2pt}c@{\hskip2pt}c@{\hskip2pt}c}
    \midrule
    A & B & \begin{tabular}[c]{@{}c@{}}Human \\ Sketches\end{tabular} & Ours16 \\
    
    \midrule
    \includegraphics[width=\widthmiss\linewidth]{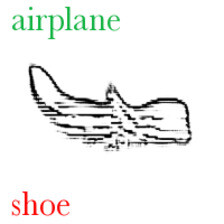} &
    \includegraphics[width=\widthmiss\linewidth]{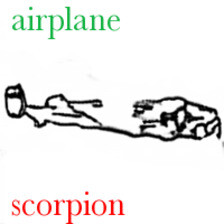} &
    \includegraphics[width=\widthmiss\linewidth]{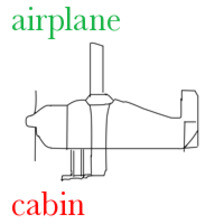} &
    \includegraphics[width=\widthmiss\linewidth]{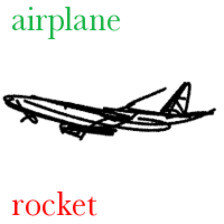} \\
    
    \midrule
    \includegraphics[width=\widthmiss\linewidth]{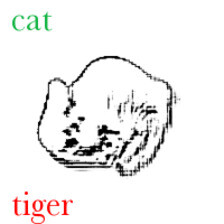} &
    \includegraphics[width=\widthmiss\linewidth]{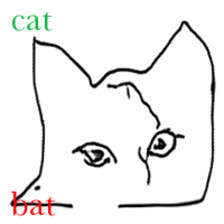} &
    &
    \includegraphics[width=\widthmiss\linewidth]{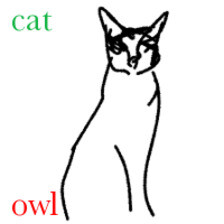} \\
    
    \midrule
    \includegraphics[width=\widthmiss\linewidth]{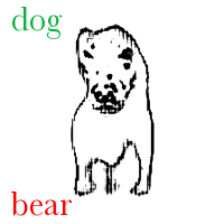} &
    \includegraphics[width=\widthmiss\linewidth]{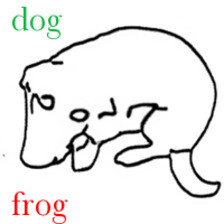} &
    \includegraphics[width=\widthmiss\linewidth]{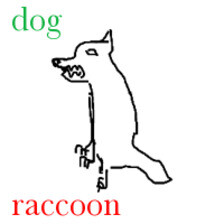} &
    \includegraphics[width=\widthmiss\linewidth]{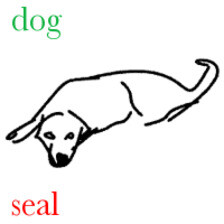} \\
    
    \midrule
    \includegraphics[width=\widthmiss\linewidth]{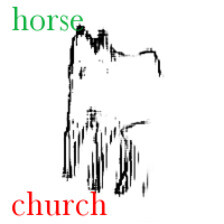} &
    \includegraphics[width=\widthmiss\linewidth]{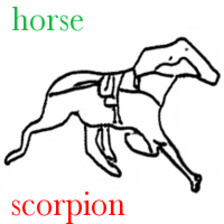} &
    \includegraphics[width=\widthmiss\linewidth]{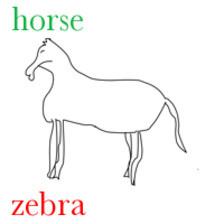} &
    \includegraphics[width=\widthmiss\linewidth]{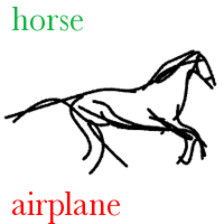} \\
    
    \midrule
    \includegraphics[width=\widthmiss\linewidth]{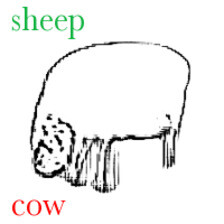} &
    \includegraphics[width=\widthmiss\linewidth]{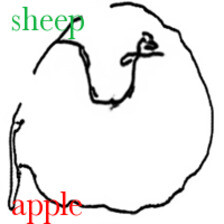} &
    \includegraphics[width=\widthmiss\linewidth]{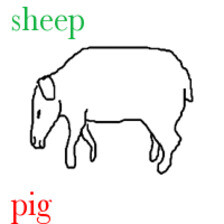} &
    \includegraphics[width=\widthmiss\linewidth]{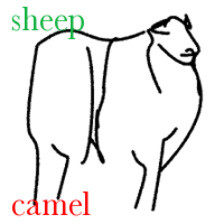} \\
    
\end{tabular}
 \caption{ResNet34 miss-classifications on the sketches generated by different methods (A) Kampelmühler and Pinz \cite{human-like-sketches}, (B) Li et al. \cite{li2019photosketching}. The green text on each sketch shows to correct class and the red text shows the predicted class.}
\label{fig:missclassification_rn34}
\end{figure}

\begin{figure}[h]
\centering
\begin{tabular}{@{\hskip2pt}c@{\hskip2pt}c@{\hskip2pt}c@{\hskip2pt}c}
    \midrule
    A & B & \begin{tabular}[c]{@{}c@{}}Human \\ Sketches\end{tabular} & Ours16 \\
    
    \midrule
    \includegraphics[width=\widthmiss\linewidth]{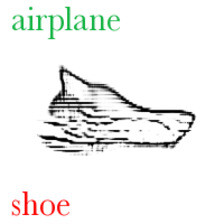} &
    \includegraphics[width=\widthmiss\linewidth]{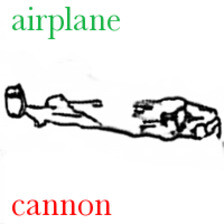} &
    \includegraphics[width=\widthmiss\linewidth]{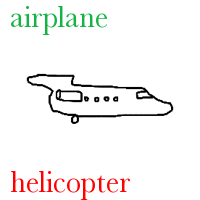} &
    \includegraphics[width=\widthmiss\linewidth]{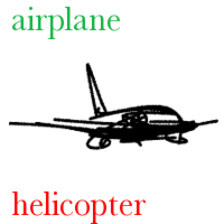} \\
    
    \midrule
    \includegraphics[width=\widthmiss\linewidth]{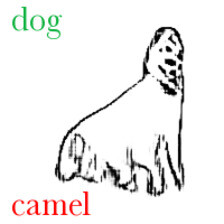} &
    \includegraphics[width=\widthmiss\linewidth]{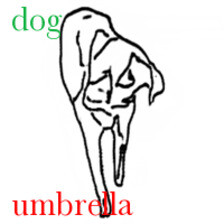} &
    \includegraphics[width=\widthmiss\linewidth]{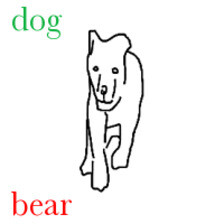} &
    \includegraphics[width=\widthmiss\linewidth]{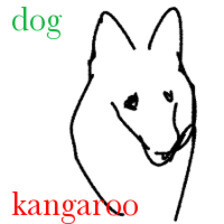} \\
    
    \midrule
    \includegraphics[width=\widthmiss\linewidth]{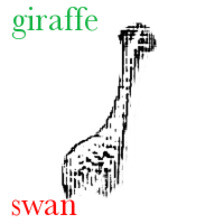} &
    \includegraphics[width=\widthmiss\linewidth]{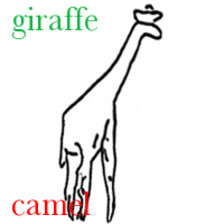} &
    \includegraphics[width=\widthmiss\linewidth]{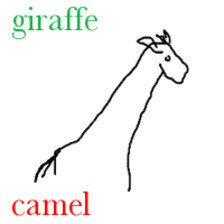} &
    \includegraphics[width=\widthmiss\linewidth]{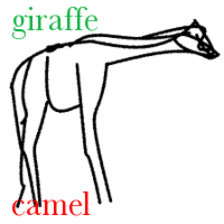} \\
    
    \midrule
    \includegraphics[width=\widthmiss\linewidth]{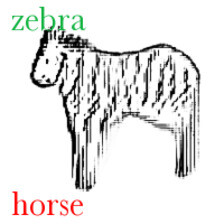} &
    \includegraphics[width=\widthmiss\linewidth]{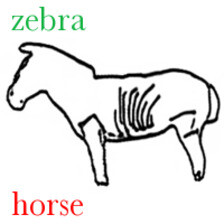} &
    \includegraphics[width=\widthmiss\linewidth]{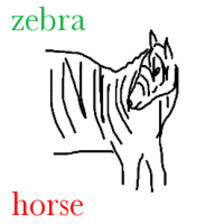} &
    \includegraphics[width=\widthmiss\linewidth]{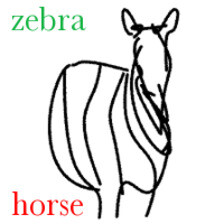} \\
    
    \midrule
    \includegraphics[width=\widthmiss\linewidth]{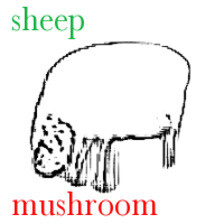} &
    \includegraphics[width=\widthmiss\linewidth]{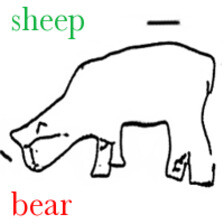} &
    \includegraphics[width=\widthmiss\linewidth]{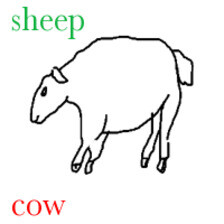} &
    \includegraphics[width=\widthmiss\linewidth]{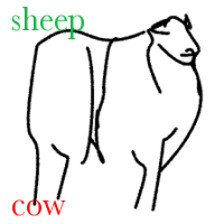} \\
    
\end{tabular}
 \caption{CLIP ViT-B/32 miss-classifications on the sketches generated by different methods (A) Kampelmühler and Pinz \cite{human-like-sketches}, (B) Li et al. \cite{li2019photosketching}. The green text on each sketch shows to correct class and the red text show the predicted class.}
\label{fig:missclassification_clip}
\end{figure}

Figure~\ref{fig:sketchy_coco_detailed_classification} provides the per-class classification accuracies of the CLIP ViT-B/32 and ResNet34 classifiers on the sketches synthesized by different methods using images from SketchyCOCO dataset.  
Figure~\ref{fig:missclassification_rn34}, \ref{fig:missclassification_clip} show sample sketches that are missclassified by the pretrained classifiers that we used for our quantitative evaluation, namely ResNet34, and CLIP ViT-B/32. Each column in the figure shows the sketches synthesized by the respective method, while the text in green showing the correct class and the text in red showing the prediction of the model.
\vfill

\pagebreak
\section{Additional Qualitative Comparisons}
\begin{figure}[h]
\centering
    \begin{tabular}{@{\hskip2pt}c@{\hskip2pt}c@{\hskip2pt}c@{\hskip2pt}c@{\hskip2pt}c@{\hskip2pt}c}
    \midrule
         Input & Text & Image & \begin{tabular}[c]{@{}c@{}}Text and\\ Image\end{tabular}  & \begin{tabular}[c]{@{}c@{}}Sketch \\ Style\end{tabular}  & Ours  \\
         \midrule
         \includegraphics[width=0.16\linewidth]{figs/comp_objects/input/men/face_7_edge.jpg} &
         \includegraphics[width=0.16\linewidth]{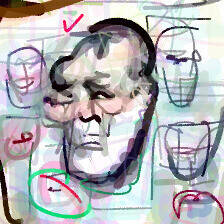} &
         \includegraphics[width=0.16\linewidth]{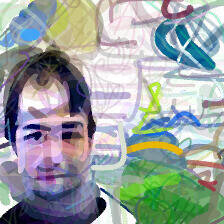} &
         \includegraphics[width=0.16\linewidth]{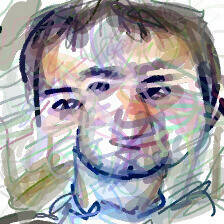} &
         \includegraphics[width=0.16\linewidth]{figs/comp_objects/clipdraw/men/face_7_edge_.png} &
         \includegraphics[width=0.16\linewidth]{figs/comp_objects/ours32/men/face_7_edge.png} 
         \\
        
         \hline
         \includegraphics[width=0.16\linewidth]{figs/comp_objects/input/horse/horse_1156303540_bd2d0bce80.jpg} &
         \includegraphics[width=0.16\linewidth]{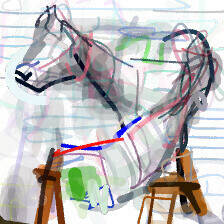} &
         \includegraphics[width=0.16\linewidth]{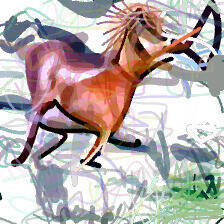} &
         \includegraphics[width=0.16\linewidth]{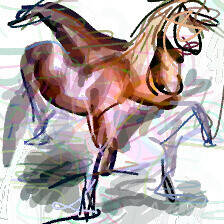} &
         \includegraphics[width=0.16\linewidth]{figs/comp_objects/clipdraw/horse/horse_1156303540_bd2d0bce80.png} &
         \includegraphics[width=0.16\linewidth]{figs/comp_objects/ours/horse/horse_1156303540_bd2d0bce80.png} 
    
    \end{tabular}
    \caption{Comparison to CLIPDraw \cite{CLIPDraw}.}
    \label{fig:clipdraw_comp_}
\end{figure}

In this section, we provide additional comparisons to competitor methods.
In Figure \ref{fig:clipdraw_comp_} we show the results generated by CLIPDraw when applied with different settings. Based on the authors' standard practice, the colored drawings were generated with 256 color strokes. In the second column, the input to the optimization is the text "A sketch of a(n) \textit{class-name}". Under these circumstances, it is not possible to control the geometry or appearance of the output drawing to produce a sketch that closely resembles an input image.
In the third column, we present the output of CLIPDraw when the image shown in the first column is provided as input instead of text. Under these settings, the output drawing is still not consistent with the geometry of the input image.
When using both text and image to guide the optimization of CLIPDraw (fourth column), the output drawing still lacks the geometric grounding. Results presented in column 5 are from the main paper, where we limit the output domain of CLIPDraw to sketch styles and utilize the input image to guide the optimization process. The sketches produced by our method are presented in the right column. Our method allows control over the visual appearance of the sketch, which is a valuable feature in a variety of art and design settings. Additionally, since we restrict the primitives to a small set of thin, black curves, we encourage the depiction of critical semantic and structural details from the input image.

\begin{figure}[ht]
    \centering
    \includegraphics[width=0.9\linewidth]{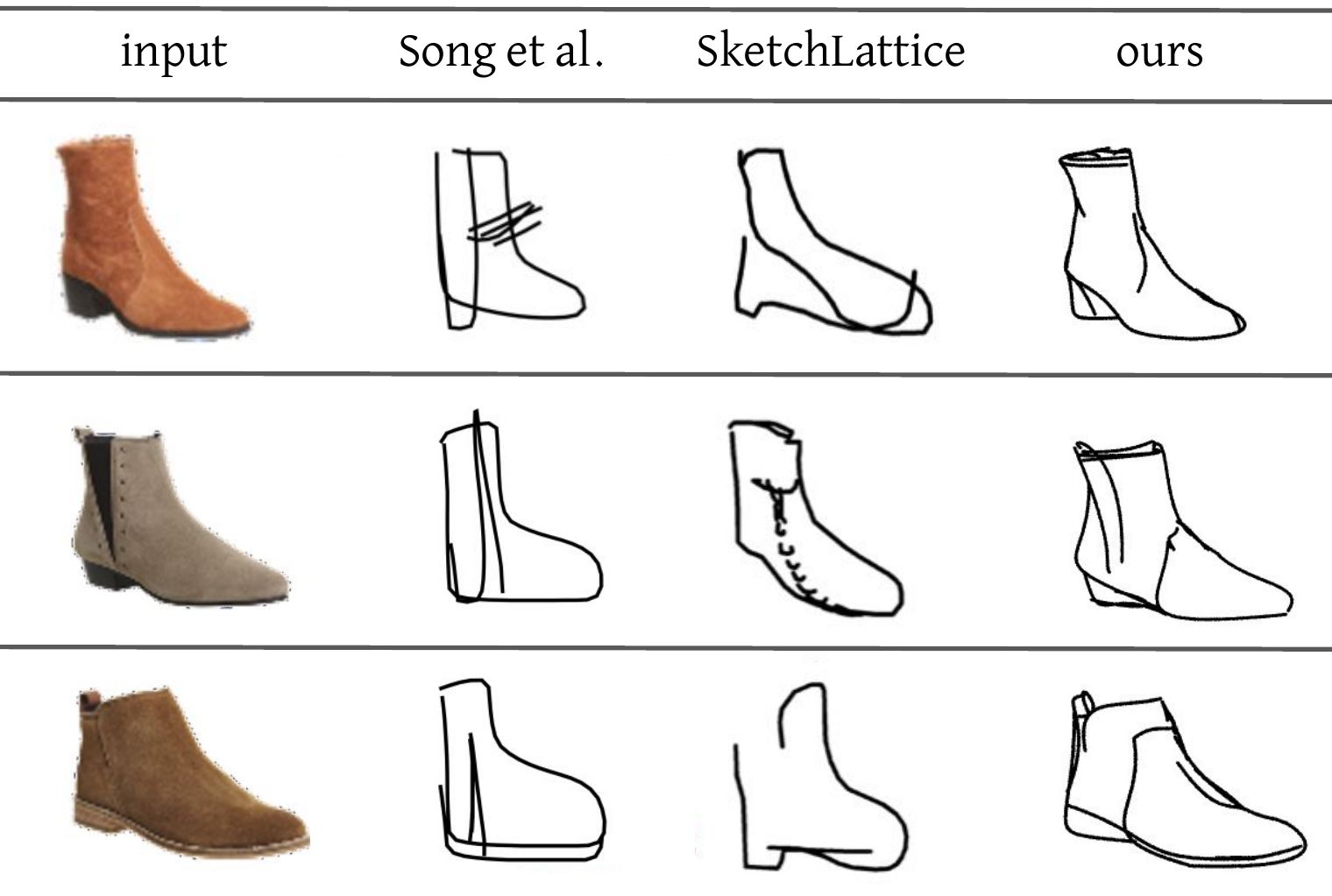}
    \caption{Comparison to Song et al. \cite{song2018learning} and SketchLattice \cite{Qi2021SketchLatticeLR}}
    \label{fig:comp-shoe-lattice-son}
\end{figure}

\begin{figure}[ht]
    \centering
\begin{tabular}{cccc}
    \midrule
    Input & Song et al. & Ours8 & Ours4 \\
    \midrule
    \includegraphics[width=0.16\linewidth]{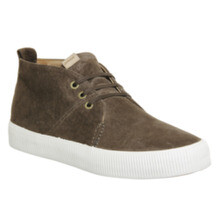} &
    \includegraphics[trim={0.7cm 0.7cm 0.7cm  0.7cm},clip,width=0.16\linewidth]{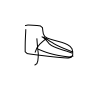}&
    \includegraphics[width=0.16\linewidth]{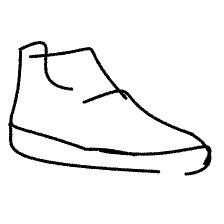} &
    \includegraphics[width=0.16\linewidth]{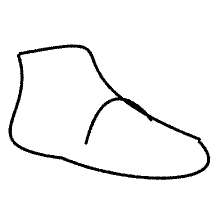}\\
    \midrule
    \includegraphics[width=0.16\linewidth]{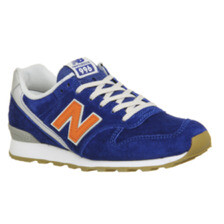} &
    \includegraphics[trim={0.7cm 0.7cm 0.7cm  0.7cm},clip,width=0.16\linewidth]{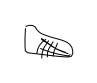} &
    \includegraphics[width=0.16\linewidth]{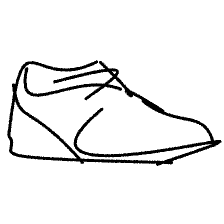} &
    \includegraphics[width=0.16\linewidth]{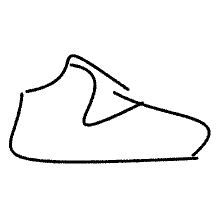}\\
    \midrule
    \includegraphics[width=0.16\linewidth]{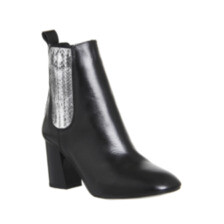} &
    \includegraphics[trim={0.7cm 0.7cm 0.7cm  0.7cm},clip,width=0.13\linewidth]{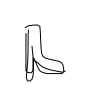} &
    \includegraphics[width=0.16\linewidth]{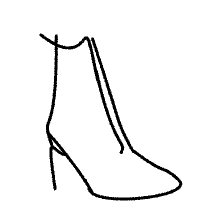} &
    \includegraphics[width=0.16\linewidth]{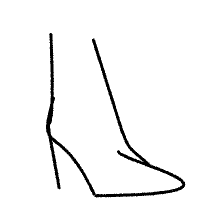}\\
    \end{tabular}
    \caption{Comparison to Song et al.}
    \label{fig:comp_cycle_consistency}
\end{figure}

In Figures \ref{fig:comp-shoe-lattice-son} and \ref{fig:comp_cycle_consistency} we provide additional comparison to SketchLattice \cite{Qi2021SketchLatticeLR} and Song et al.\cite{song2018learning} on shoes images. 

\begin{figure*}[ht]
\centering
\begin{tabular}{@{\hskip2pt}c@{\hskip2pt}c@{\hskip2pt}c@{\hskip2pt}c@{\hskip2pt}c@{\hskip2pt}c@{\hskip2pt}c}
    \midrule
    Input & A & B & C & CLIPDraw & Ours16 & Ours8 \\
    
    \midrule
    \includegraphics[width=\widthteapot\linewidth]{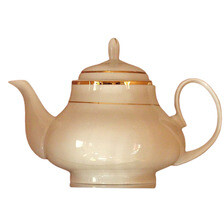} &
    \includegraphics[width=\widthteapot\linewidth]{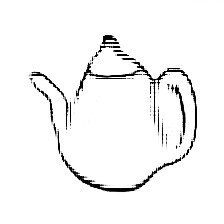} &
    \raisebox{.2\height}{\includegraphics[width=\widthteapot\linewidth]{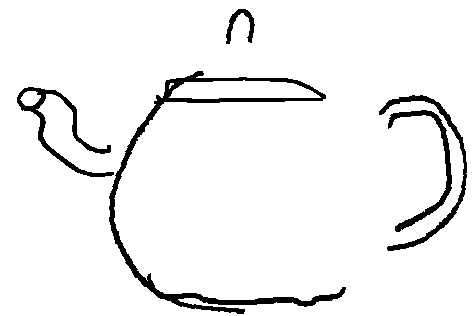}} &
    \includegraphics[width=\widthteapot\linewidth]{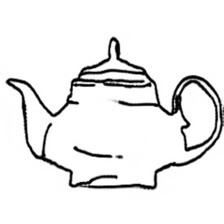} &
    \includegraphics[width=\widthteapot\linewidth]{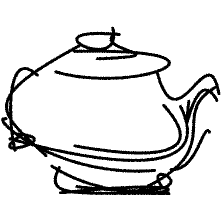} &
    \includegraphics[width=\widthteapot\linewidth]{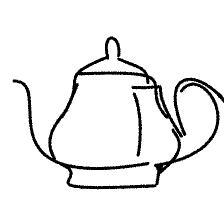} &
    \includegraphics[width=\widthteapot\linewidth]{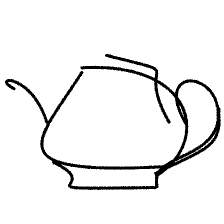} \\
    
    \midrule
    \includegraphics[width=\widthteapot\linewidth]{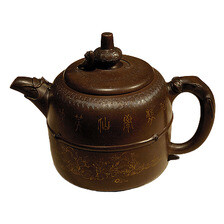} &
    \includegraphics[width=\widthteapot\linewidth]{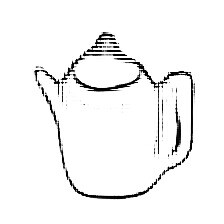} &
    \raisebox{.2\height}{\includegraphics[width=\widthteapot\linewidth]{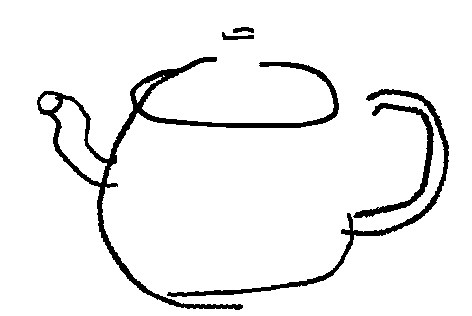}} &
    \includegraphics[width=\widthteapot\linewidth]{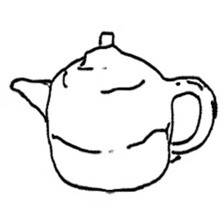} &
    \includegraphics[width=\widthteapot\linewidth]{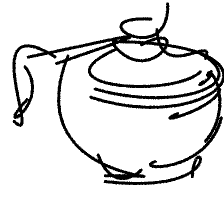} &
    \includegraphics[width=\widthteapot\linewidth]{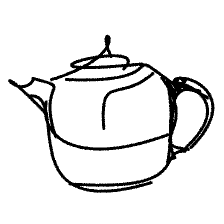} &
    \includegraphics[width=\widthteapot\linewidth]{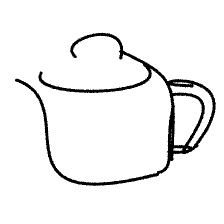}\\
    
    \midrule
    \includegraphics[width=\widthteapot\linewidth]{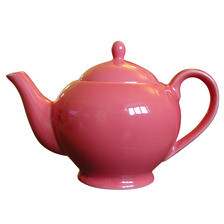} &
    \includegraphics[width=\widthteapot\linewidth]{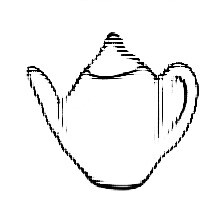} &
    \raisebox{.2\height}{\includegraphics[width=\widthteapot\linewidth]{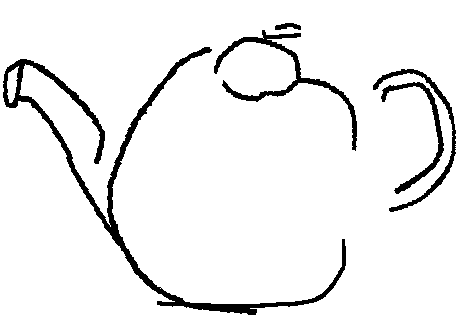}} &
    \includegraphics[width=\widthteapot\linewidth]{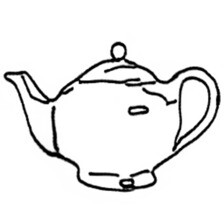} &
    \includegraphics[width=\widthteapot\linewidth]{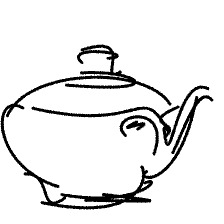} &
    \includegraphics[width=\widthteapot\linewidth]{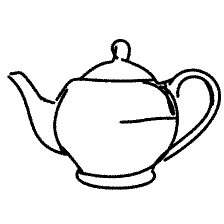} &
     \includegraphics[width=\widthteapot\linewidth]{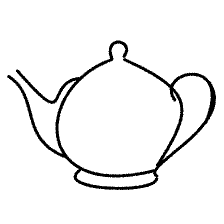} \\
    
    \midrule
    \includegraphics[width=\widthteapot\linewidth]{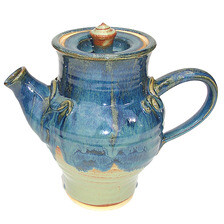} &
    \includegraphics[width=\widthteapot\linewidth]{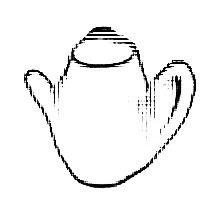} &
    \raisebox{.2\height}{\includegraphics[width=\widthteapot\linewidth]{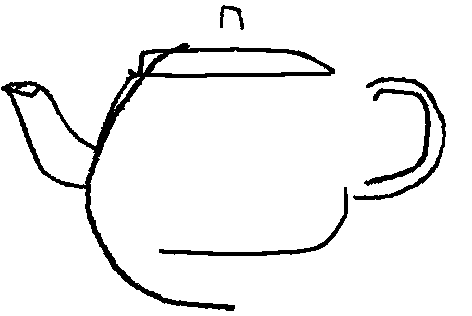}} &
    \includegraphics[width=\widthteapot\linewidth]{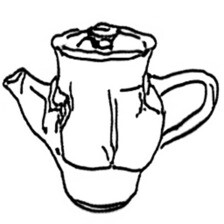} &
    \includegraphics[width=\widthteapot\linewidth]{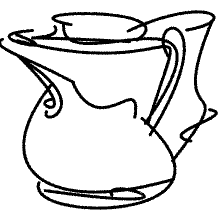} &
    \includegraphics[width=\widthteapot\linewidth]{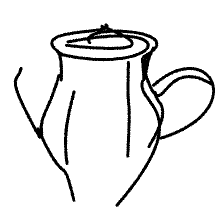} &
    \includegraphics[width=\widthteapot\linewidth]{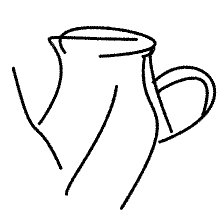}\\
    
    \midrule
    \includegraphics[width=\widthteapot\linewidth]{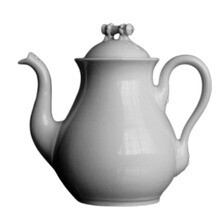} &
    \includegraphics[width=\widthteapot\linewidth]{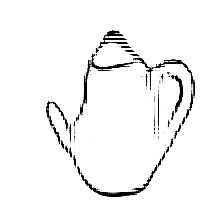} &
    \raisebox{.2\height}{\includegraphics[width=\widthteapot\linewidth]{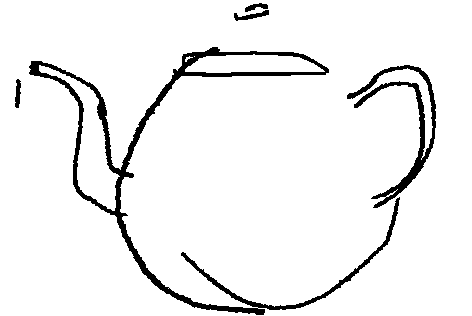}} &
    \includegraphics[width=\widthteapot\linewidth]{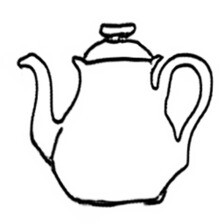} &
    \includegraphics[width=\widthteapot\linewidth]{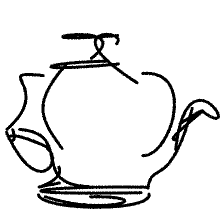} &
    \includegraphics[width=\widthteapot\linewidth]{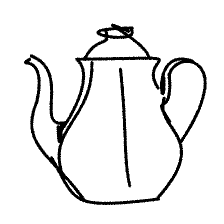} &
    \includegraphics[width=\widthteapot\linewidth]{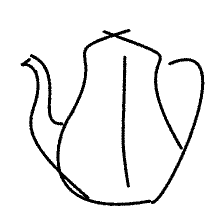}\\
    
    \midrule
    \includegraphics[width=\widthteapot\linewidth]{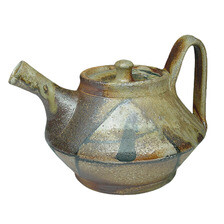} &
    \includegraphics[width=\widthteapot\linewidth]{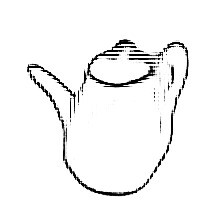} &
    \raisebox{.2\height}{\includegraphics[width=\widthteapot\linewidth]{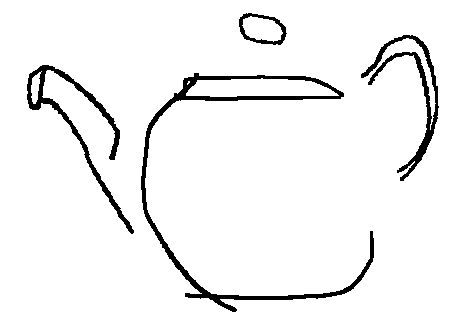}} &
    \includegraphics[width=\widthteapot\linewidth]{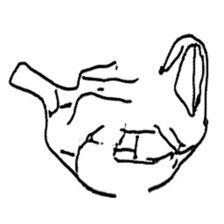} &
    \includegraphics[width=\widthteapot\linewidth]{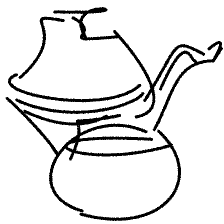} &
    \includegraphics[width=\widthteapot\linewidth]{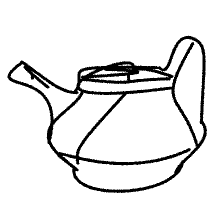} &
    \includegraphics[width=\widthteapot\linewidth]{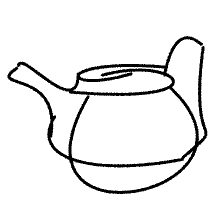}\\
    
    \midrule
    \includegraphics[width=\widthteapot\linewidth]{figs/comp_objects/input/teapot/teapot_2252638111_852bafb6e8.jpg} &
    \includegraphics[width=\widthteapot\linewidth]{figs/comp_objects/human_like/teapot/teapot_2252638111_852bafb6e8_processed.jpg} &
    \raisebox{.2\height}{\includegraphics[width=\widthteapot\linewidth]{figs/comp_objects/deformable/teapot/teapot_2252638111_852bafb6e8_def.png}} &
    \includegraphics[width=\widthteapot\linewidth]{figs/comp_objects/contour/teapot/teapot_2252638111_852bafb6e8_cont.jpg} &
    \includegraphics[width=\widthteapot\linewidth]{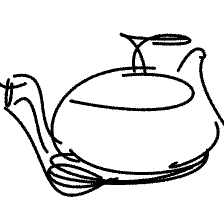} &
    \includegraphics[width=\widthteapot\linewidth]{figs/comp_objects/ours/teapot/teapot_2252638111_852bafb6e8.png} &
    \includegraphics[width=\widthteapot\linewidth]{figs/comp_objects/ours8/teapot/teapot_2252638111_852bafb6e8.png}\\
    
    \midrule
    \includegraphics[width=\widthteapot\linewidth]{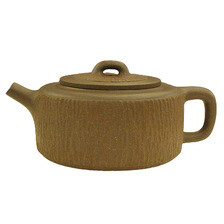} &
    \includegraphics[width=\widthteapot\linewidth]{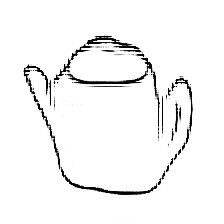} &
    \raisebox{.2\height}{\includegraphics[width=\widthteapot\linewidth]{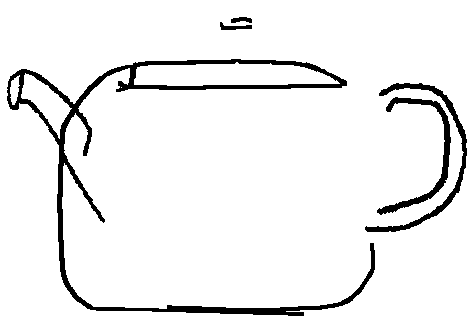}} &
    \includegraphics[width=\widthteapot\linewidth]{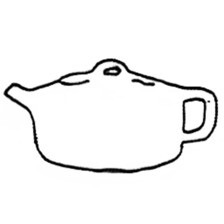} &
    \includegraphics[width=\widthteapot\linewidth]{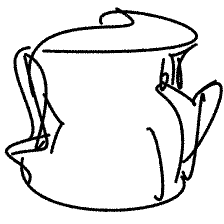} &
    \includegraphics[width=\widthteapot\linewidth]{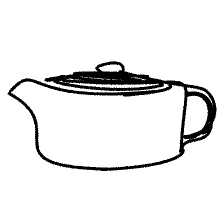} &
    \includegraphics[width=\widthteapot\linewidth]{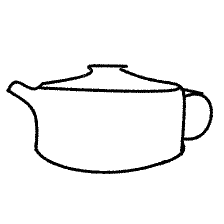}\\
    
    \midrule
    \includegraphics[width=\widthteapot\linewidth]{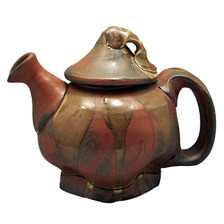} &
    \includegraphics[width=\widthteapot\linewidth]{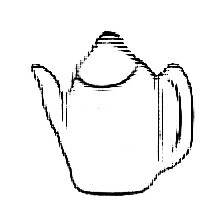} &
    \raisebox{.2\height}{\includegraphics[width=\widthteapot\linewidth]{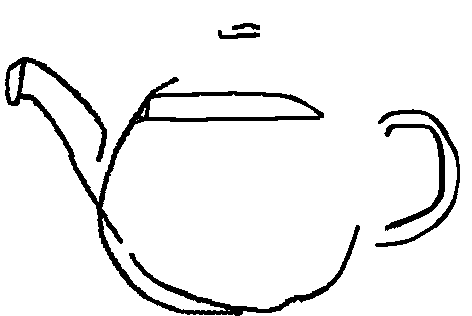}} &
    \includegraphics[width=\widthteapot\linewidth]{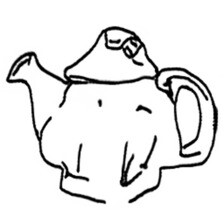} &
    \includegraphics[width=\widthteapot\linewidth]{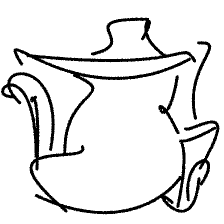} &
    \includegraphics[width=\widthteapot\linewidth]{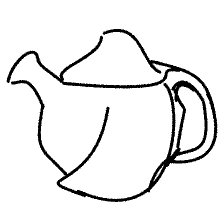} &
    \includegraphics[width=\widthteapot\linewidth]{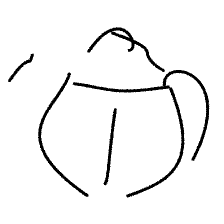}\\
    
\end{tabular}
 \caption{Comparison to competitor methods on images from the "teapot" class. Left to right : (A) Kampelm{\"{u}}hler and Pinz \cite{human-like-sketches}, (B) Li et al. \cite{Deformable_Stroke}, (C) Li et al. \cite{li2019photosketching}, and CLIPDraw \cite{CLIPDraw}.}
\label{fig:teapot}
\end{figure*}

\begin{figure*}[ht!]
\centering
\begin{tabular}{@{\hskip2pt}c@{\hskip2pt}c@{\hskip2pt}c@{\hskip2pt}c@{\hskip2pt}c@{\hskip2pt}c@{\hskip2pt}c}
    \midrule
    Input & A & B & C & CLIPDraw & Ours16 & Ours8 \\
    
    \midrule
    \includegraphics[width=\widthteapot\linewidth]{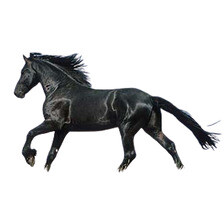} &
    \includegraphics[width=\widthteapot\linewidth]{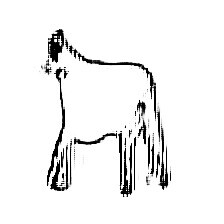} &
    \raisebox{.2\height}{\includegraphics[width=\widthteapot\linewidth]{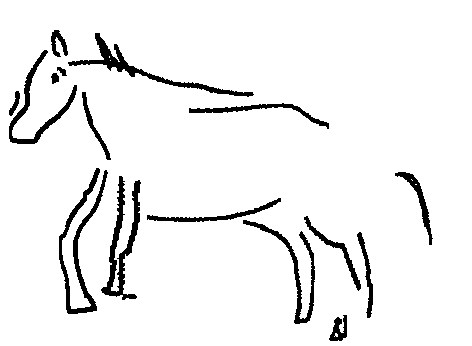}} &
    \includegraphics[width=\widthteapot\linewidth]{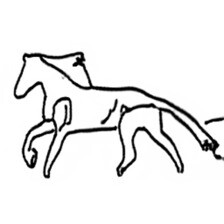} &
    \includegraphics[width=\widthteapot\linewidth]{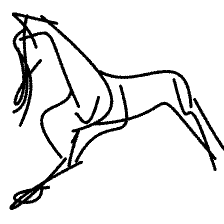} &
    \includegraphics[width=\widthteapot\linewidth]{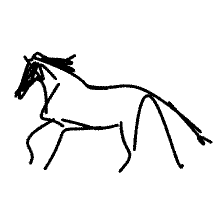} &
    \includegraphics[width=\widthteapot\linewidth]{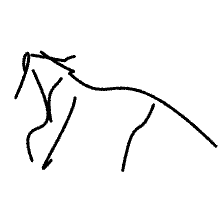} \\
    
    
    \midrule
    \includegraphics[width=\widthteapot\linewidth]{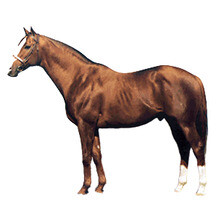} &
    \includegraphics[width=\widthteapot\linewidth]{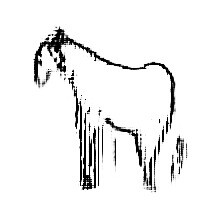} &
    \raisebox{.2\height}{\includegraphics[width=\widthteapot\linewidth]{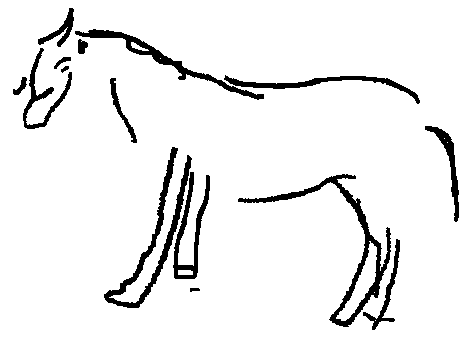}} &
    \includegraphics[width=\widthteapot\linewidth]{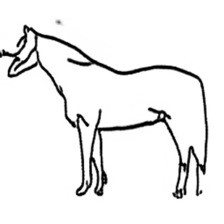} &
    \includegraphics[width=\widthteapot\linewidth]{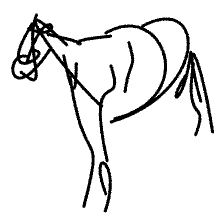} &
    \includegraphics[width=\widthteapot\linewidth]{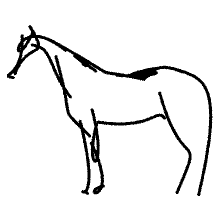} &
    \includegraphics[width=\widthteapot\linewidth]{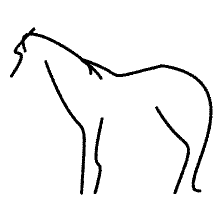}\\
    
    \midrule
    \includegraphics[width=\widthteapot\linewidth]{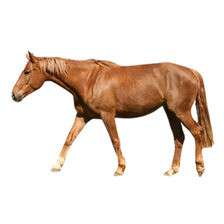} &
    \includegraphics[width=\widthteapot\linewidth]{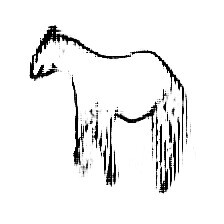} &
    \raisebox{.2\height}{\includegraphics[width=\widthteapot\linewidth]{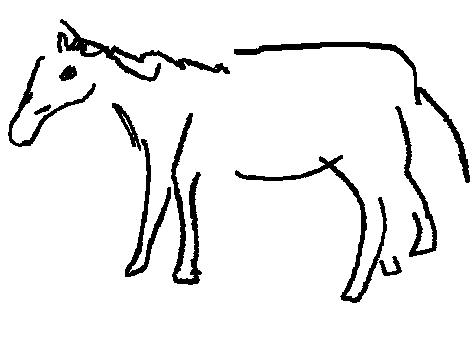}} &
    \includegraphics[width=\widthteapot\linewidth]{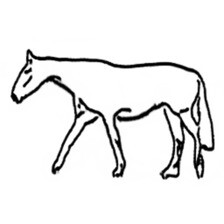} &
    \includegraphics[width=\widthteapot\linewidth]{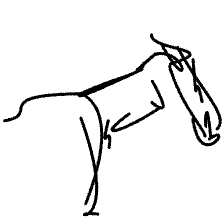} &
    \includegraphics[width=\widthteapot\linewidth]{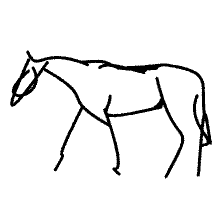} &
    \includegraphics[width=\widthteapot\linewidth]{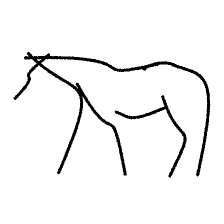}\\
    
    \midrule
    \includegraphics[width=\widthteapot\linewidth]{figs/comp_objects/input/horse/horse_1156303540_bd2d0bce80.jpg} &
    \includegraphics[width=\widthteapot\linewidth]{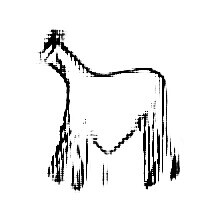} &
    \raisebox{.2\height}{\includegraphics[width=\widthteapot\linewidth]{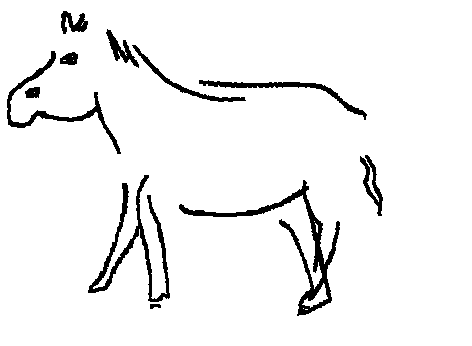}} &
    \includegraphics[width=\widthteapot\linewidth]{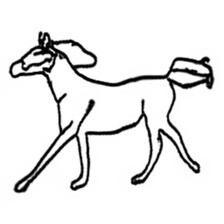} &
    \includegraphics[width=\widthteapot\linewidth]{figs/comp_objects/clipdraw/horse/horse_1156303540_bd2d0bce80.png} &
    \includegraphics[width=\widthteapot\linewidth]{figs/comp_objects/ours/horse/horse_1156303540_bd2d0bce80.png} &
    \includegraphics[width=\widthteapot\linewidth]{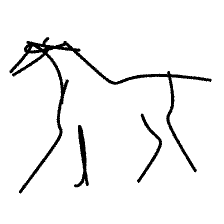} \\
    
    \midrule
    \includegraphics[width=\widthteapot\linewidth]{figs/comp_objects/input/horse/horse_2143679822_711e808c54.jpg} &
    \includegraphics[width=\widthteapot\linewidth]{figs/comp_objects/human_like/horse/horse_2143679822_711e808c54_processed.jpg} &
    \raisebox{.2\height}{\includegraphics[width=\widthteapot\linewidth]{figs/comp_objects/deformable/horse/horse_2143679822_711e808c54.png}} &
    \includegraphics[width=\widthteapot\linewidth]{figs/comp_objects/contour/horse/horse_2143679822_711e808c54.jpg} &
    \includegraphics[width=\widthteapot\linewidth]{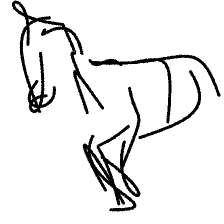} &
    \includegraphics[width=\widthteapot\linewidth]{figs/comp_objects/ours/horse/horse_2143679822_711e808c54.png} &
    \includegraphics[width=\widthteapot\linewidth]{figs/comp_objects/ours8/horse/horse_2143679822_711e808c54.png} \\
    
    \midrule
    \includegraphics[width=\widthteapot\linewidth]{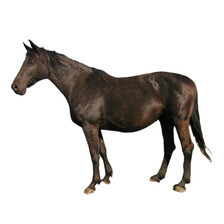} &
    \includegraphics[width=\widthteapot\linewidth]{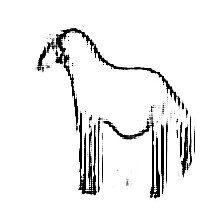} &
    \raisebox{.2\height}{\includegraphics[width=\widthteapot\linewidth]{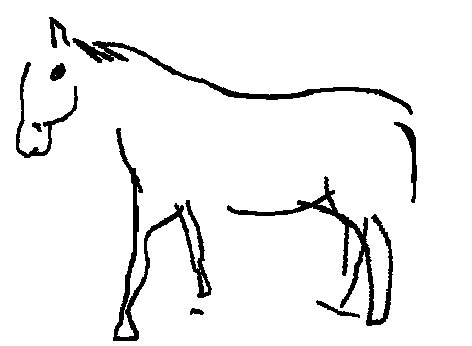}} &
    \includegraphics[width=\widthteapot\linewidth]{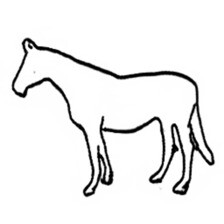} &
    \includegraphics[width=\widthteapot\linewidth]{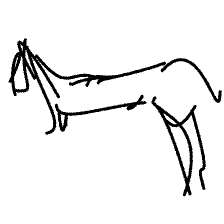} &
    \includegraphics[width=\widthteapot\linewidth]{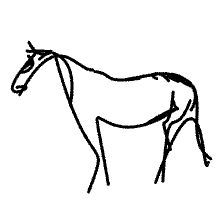} &
    \includegraphics[width=\widthteapot\linewidth]{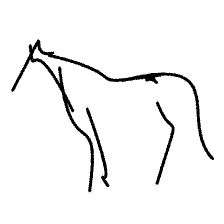}\\
    
    \midrule
    \includegraphics[width=\widthteapot\linewidth]{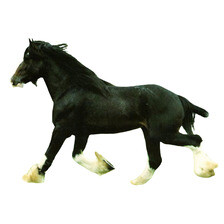} &
    \includegraphics[width=\widthteapot\linewidth]{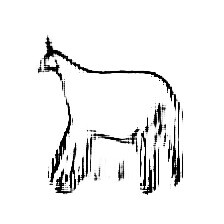} &
    \raisebox{.2\height}{\includegraphics[width=\widthteapot\linewidth]{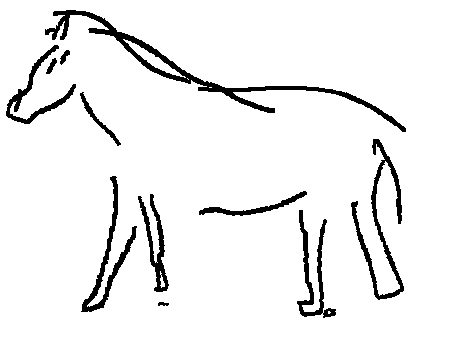}} &
    \includegraphics[width=\widthteapot\linewidth]{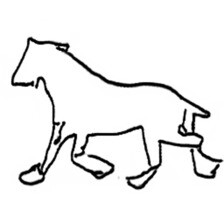} &
    \includegraphics[width=\widthteapot\linewidth]{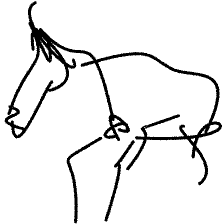} &
    \includegraphics[width=\widthteapot\linewidth]{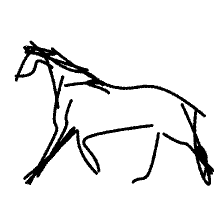} &
    \includegraphics[width=\widthteapot\linewidth]{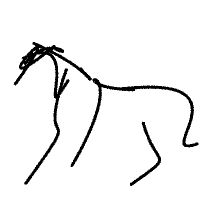}\\
    
    \midrule
    \includegraphics[width=\widthteapot\linewidth]{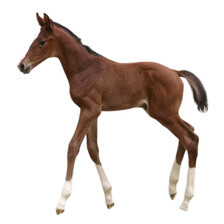} &
    \includegraphics[width=\widthteapot\linewidth]{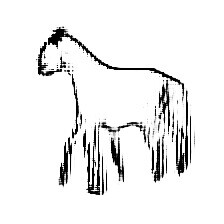} &
    \raisebox{.2\height}{\includegraphics[width=\widthteapot\linewidth]{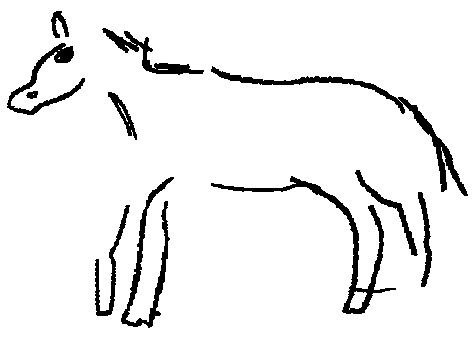}} &
    \includegraphics[width=\widthteapot\linewidth]{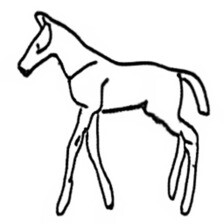} &
    \includegraphics[width=\widthteapot\linewidth]{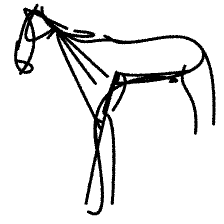} &
    \includegraphics[width=\widthteapot\linewidth]{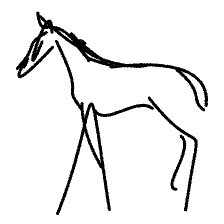} &
    \includegraphics[width=\widthteapot\linewidth]{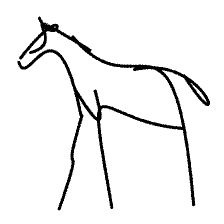} \\

\end{tabular}
 \caption{Comparison to competitor methods on images from the "horse" class. Left to right : (A) Kampelm{\"{u}}hler and Pinz \cite{human-like-sketches}, (B) Li et al. \cite{Deformable_Stroke}, (C) Li et al. \cite{li2019photosketching}, and CLIPDraw \cite{CLIPDraw}.}
\label{fig:horse}
\end{figure*}

\begin{figure*}[ht!]
\centering
\begin{tabular}{@{\hskip1pt}c@{\hskip1pt}c@{\hskip1pt}c@{\hskip1pt}c@{\hskip1pt}c@{\hskip1pt}c@{\hskip1pt}c}
    \midrule
    Input & A & B & C & CLIPDraw & Ours16 & Ours8 \\
    
    \midrule
    \includegraphics[width=\widthteapot\linewidth]{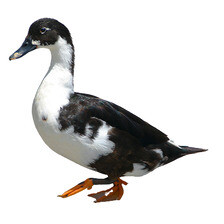} &
    \includegraphics[width=\widthteapot\linewidth]{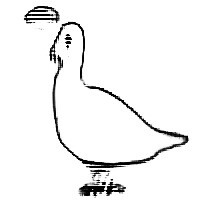} &
    \raisebox{.1\height}{\includegraphics[width=\widthteapot\linewidth]{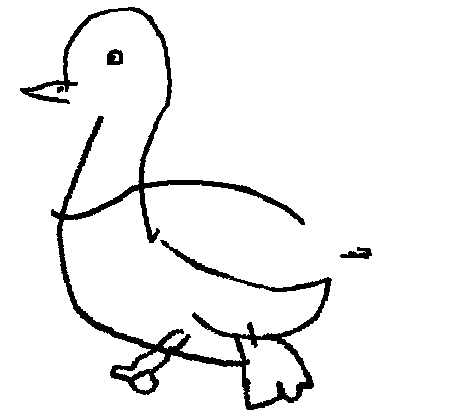}} &
    \includegraphics[width=\widthteapot\linewidth]{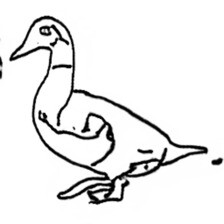} &
    \includegraphics[width=\widthteapot\linewidth]{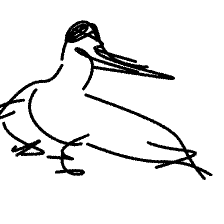} &
    \includegraphics[width=\widthteapot\linewidth]{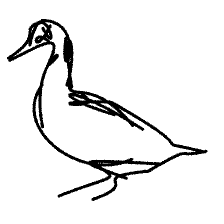} &
    \includegraphics[width=\widthteapot\linewidth]{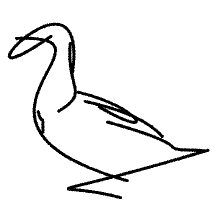} \\
    
    \midrule
    \includegraphics[width=\widthteapot\linewidth]{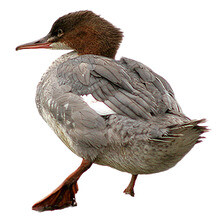} &
    \includegraphics[width=\widthteapot\linewidth]{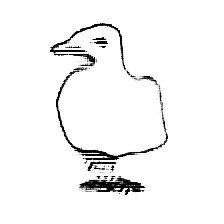} &
    \raisebox{.1\height}{\includegraphics[width=\widthteapot\linewidth]{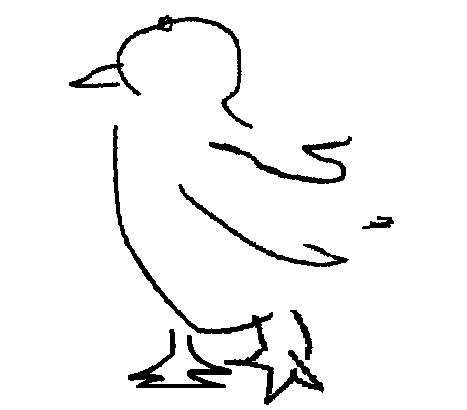}} &
    \includegraphics[width=\widthteapot\linewidth]{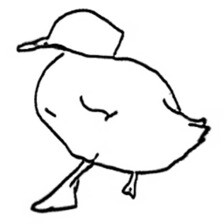} &
    \includegraphics[width=\widthteapot\linewidth]{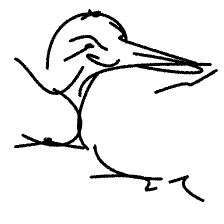} &
    \includegraphics[width=\widthteapot\linewidth]{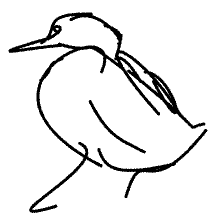} &
    \includegraphics[width=\widthteapot\linewidth]{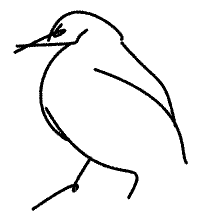} \\
    
    \midrule
    \includegraphics[width=\widthteapot\linewidth]{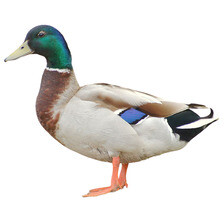} &
    \includegraphics[width=\widthteapot\linewidth]{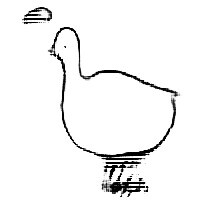} &
    \raisebox{.1\height}{\includegraphics[width=\widthteapot\linewidth]{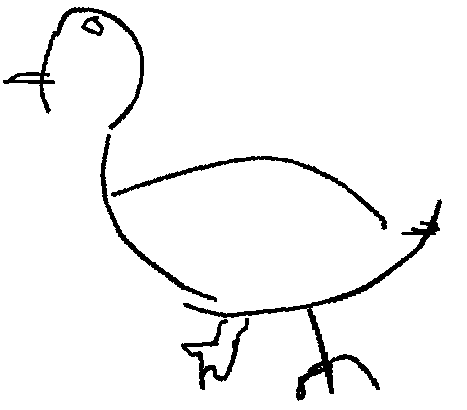}} &
    \includegraphics[width=\widthteapot\linewidth]{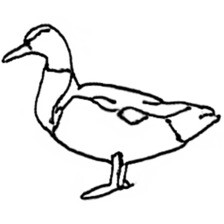} &
    \includegraphics[width=\widthteapot\linewidth]{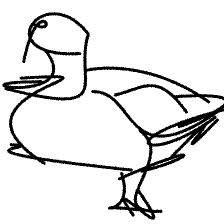} &
    \includegraphics[width=\widthteapot\linewidth]{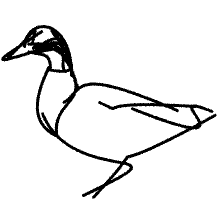} &
    \includegraphics[width=\widthteapot\linewidth]{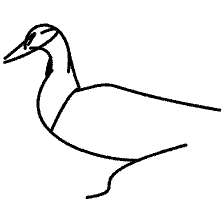}\\
    
    \midrule
    \includegraphics[width=\widthteapot\linewidth]{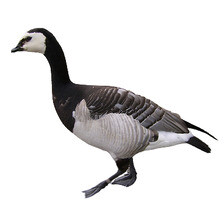} &
    \includegraphics[width=\widthteapot\linewidth]{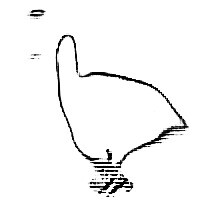} &
    \raisebox{.1\height}{\includegraphics[width=\widthteapot\linewidth]{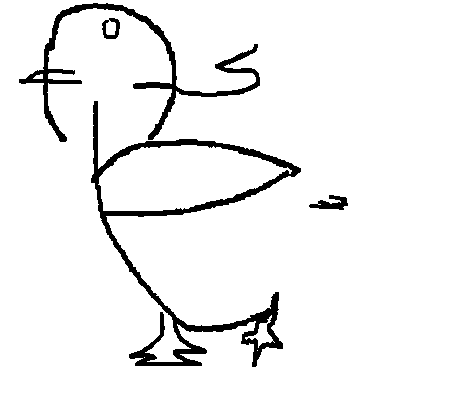}} &
    \includegraphics[width=\widthteapot\linewidth]{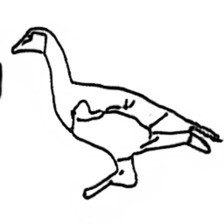} &
    \includegraphics[width=\widthteapot\linewidth]{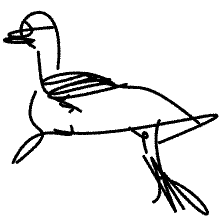} & 
    \includegraphics[width=\widthteapot\linewidth]{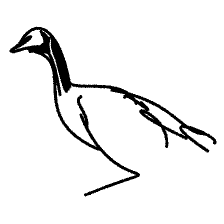} & 
    \includegraphics[width=\widthteapot\linewidth]{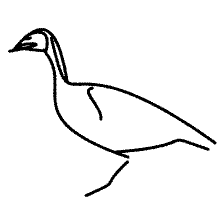}\\
    
    \midrule
    \includegraphics[width=\widthteapot\linewidth]{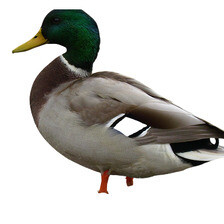} &
    \includegraphics[width=\widthteapot\linewidth]{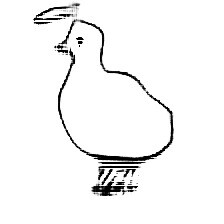} &
    \raisebox{.1\height}{\includegraphics[width=\widthteapot\linewidth]{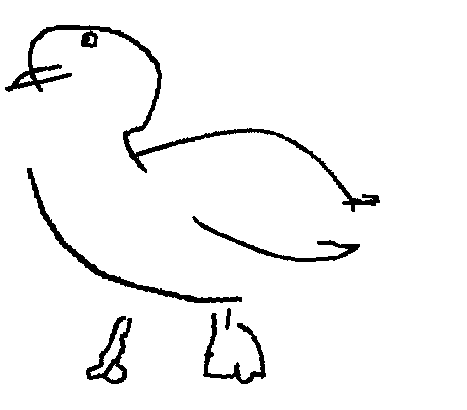}} &
    \includegraphics[width=\widthteapot\linewidth]{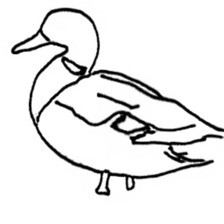} &
    \includegraphics[width=\widthteapot\linewidth]{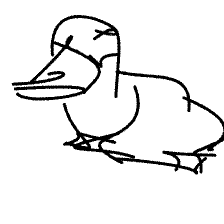} &
    \includegraphics[width=\widthteapot\linewidth]{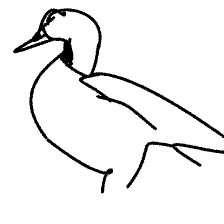} &
    \includegraphics[width=\widthteapot\linewidth]{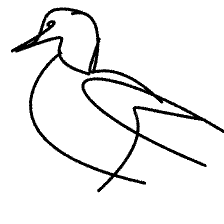}\\
    
    \midrule
    \includegraphics[width=\widthteapot\linewidth]{figs/comp_objects/input/duck/duck_1497314383_b36e13d3e1.jpg} &
    \includegraphics[width=\widthteapot\linewidth]{figs/comp_objects/human_like/duck/duck_1497314383_b36e13d3e1_processed.jpg} &
    \raisebox{.1\height}{\includegraphics[width=\widthteapot\linewidth]{figs/comp_objects/deformable/duck/duck_1497314383_b36e13d3e1.png}} &
    \includegraphics[width=\widthteapot\linewidth]{figs/comp_objects/contour/duck/duck_1497314383_b36e13d3e1.jpg} &
    \includegraphics[width=\widthteapot\linewidth]{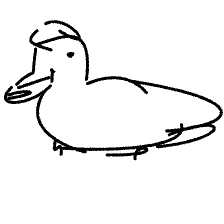} &
    \includegraphics[width=\widthteapot\linewidth]{figs/comp_objects/ours/duck/duck_1497314383_b36e13d3e1.png} &
    \includegraphics[width=\widthteapot\linewidth]{figs/comp_objects/ours8/duck/duck_1497314383_b36e13d3e1.png}\\
    
    \midrule
    \includegraphics[width=\widthteapot\linewidth]{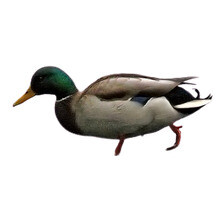} &
    \includegraphics[width=\widthteapot\linewidth]{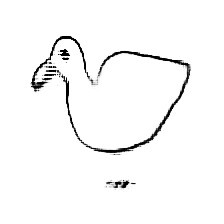} &
    \raisebox{.1\height}{\includegraphics[width=\widthteapot\linewidth]{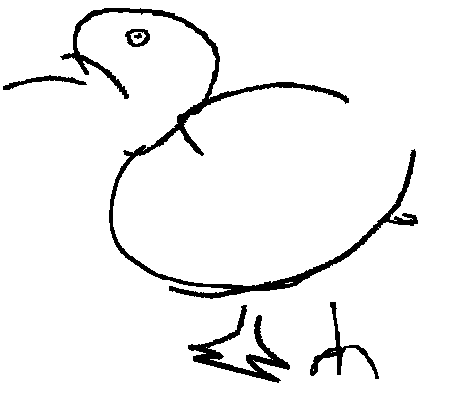}} &
    \includegraphics[width=\widthteapot\linewidth]{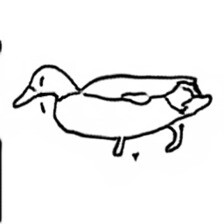} &
    \includegraphics[width=\widthteapot\linewidth]{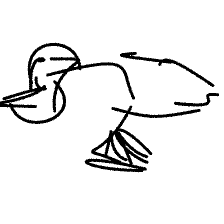} &
    \includegraphics[width=\widthteapot\linewidth]{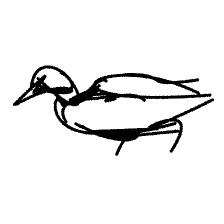} &
    \includegraphics[width=\widthteapot\linewidth]{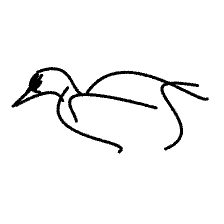}\\
    
    \midrule
    \includegraphics[width=\widthteapot\linewidth]{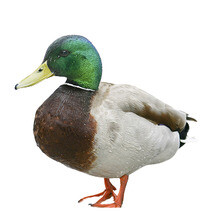} &
    \includegraphics[width=\widthteapot\linewidth]{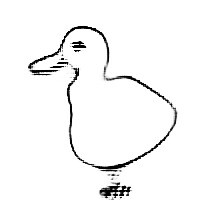} &
    \raisebox{.1\height}{\includegraphics[width=\widthteapot\linewidth]{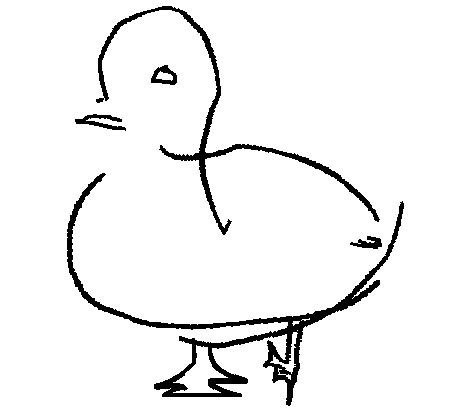}} &
    \includegraphics[width=\widthteapot\linewidth]{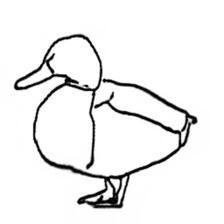} &
    \includegraphics[width=\widthteapot\linewidth]{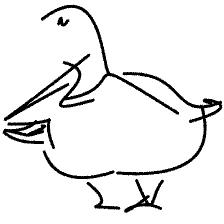} &
    \includegraphics[width=\widthteapot\linewidth]{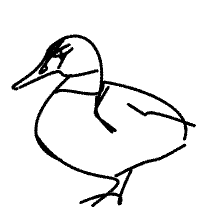} &
    \includegraphics[width=\widthteapot\linewidth]{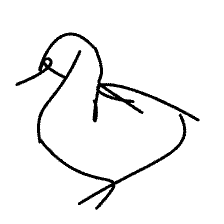}\\
    
    \midrule
    \includegraphics[width=\widthteapot\linewidth]{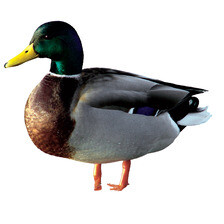} &
    \includegraphics[width=\widthteapot\linewidth]{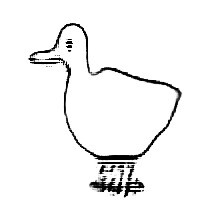} &
    \raisebox{.1\height}{\includegraphics[width=\widthteapot\linewidth]{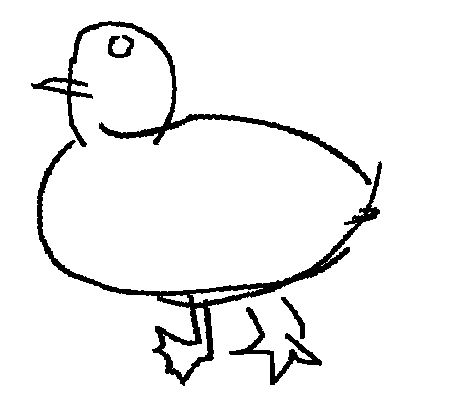}} &
    \includegraphics[width=\widthteapot\linewidth]{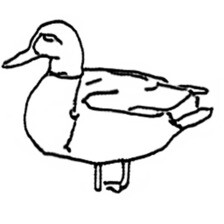} &
    \includegraphics[width=\widthteapot\linewidth]{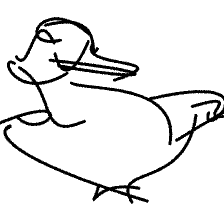} &
    \includegraphics[width=\widthteapot\linewidth]{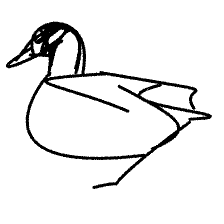} &
    \includegraphics[width=\widthteapot\linewidth]{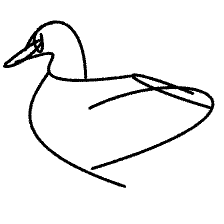}\\
    
\end{tabular}
 \caption{Comparison to competitor methods on images from the "duck" class. Left to right : (A) Kampelm{\"{u}}hler and Pinz \cite{human-like-sketches}, (B) Li et al. \cite{Deformable_Stroke}, (C) Li et al. \cite{li2019photosketching}, and CLIPDraw \cite{CLIPDraw}.}
\label{fig:duck}
\end{figure*}

\begin{figure*}[ht!]
\centering
\begin{tabular}{@{\hskip2pt}c@{\hskip2pt}c@{\hskip2pt}c@{\hskip2pt}c@{\hskip2pt}c@{\hskip2pt}c@{\hskip2pt}c}
    \midrule
    Input & A & B & C & CLIPDraw & Ours16 & Ours8 \\
    
    \midrule
    \includegraphics[width=\widthteapot\linewidth]{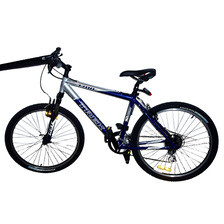} &
    \includegraphics[width=\widthteapot\linewidth]{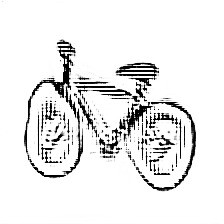} &
    \raisebox{.2\height}{\includegraphics[width=\widthteapot\linewidth]{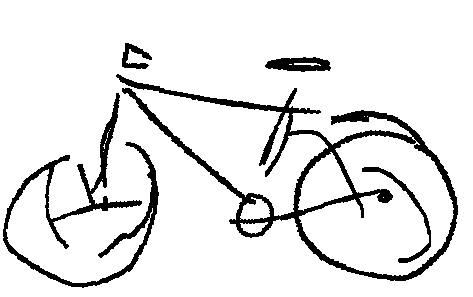}} &
    \includegraphics[width=\widthteapot\linewidth]{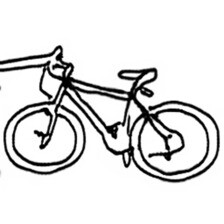} &
    \includegraphics[width=\widthteapot\linewidth]{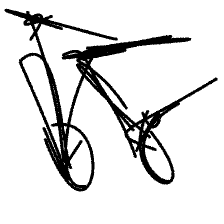} &
    \includegraphics[width=\widthteapot\linewidth]{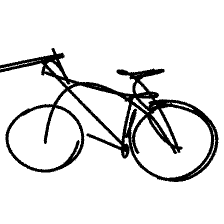} &
    \includegraphics[width=\widthteapot\linewidth]{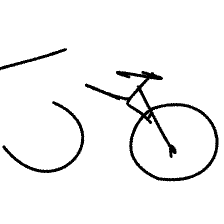}\\
    
    \midrule
    \includegraphics[width=\widthteapot\linewidth]{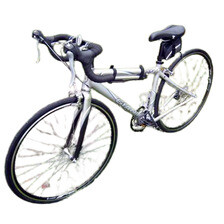} &
    \includegraphics[width=\widthteapot\linewidth]{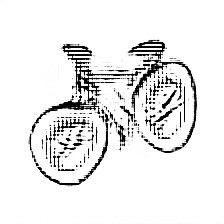} &
    \raisebox{.2\height}{\includegraphics[width=\widthteapot\linewidth]{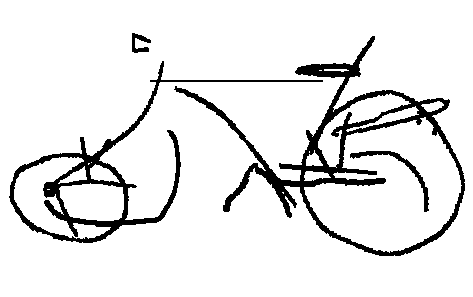}} &
    \includegraphics[width=\widthteapot\linewidth]{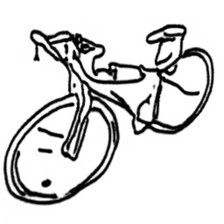} &
    \includegraphics[width=\widthteapot\linewidth]{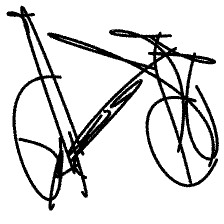} &
    \includegraphics[width=\widthteapot\linewidth]{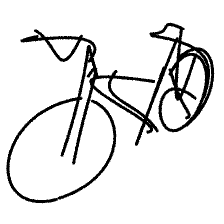} &
    \includegraphics[width=\widthteapot\linewidth]{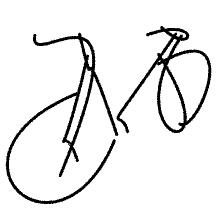}\\
    
    \midrule
    \includegraphics[width=\widthteapot\linewidth]{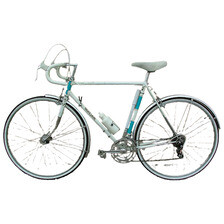} &
    \includegraphics[width=\widthteapot\linewidth]{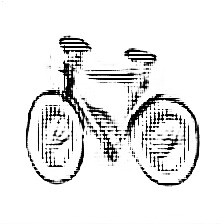} &
    \raisebox{.2\height}{\includegraphics[width=\widthteapot\linewidth]{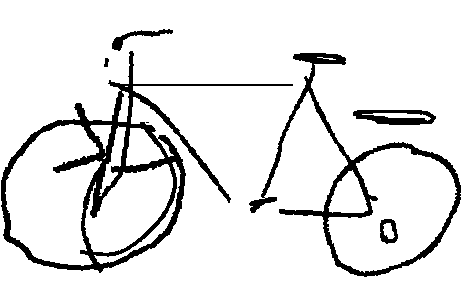}} &
    \includegraphics[width=\widthteapot\linewidth]{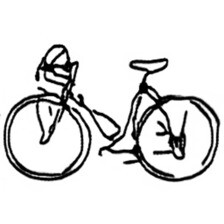} &
    \includegraphics[width=\widthteapot\linewidth]{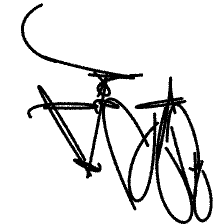} & 
    \includegraphics[width=\widthteapot\linewidth]{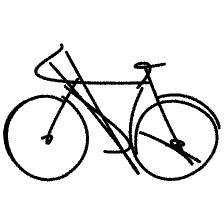} & 
    \includegraphics[width=\widthteapot\linewidth]{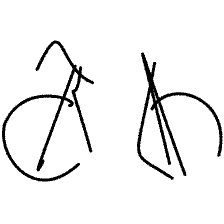}\\
    
    \midrule
    \includegraphics[width=\widthteapot\linewidth]{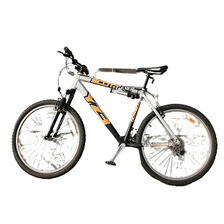} &
    \includegraphics[width=\widthteapot\linewidth]{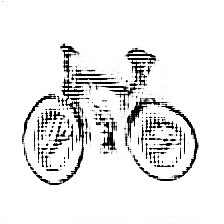} &
    \raisebox{.2\height}{\includegraphics[width=\widthteapot\linewidth]{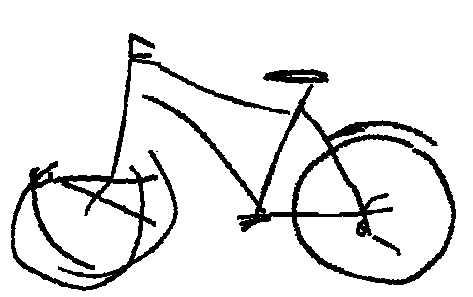}} &
    \includegraphics[width=\widthteapot\linewidth]{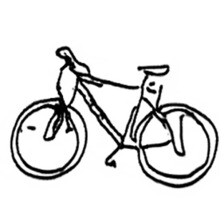} &
    \includegraphics[width=\widthteapot\linewidth]{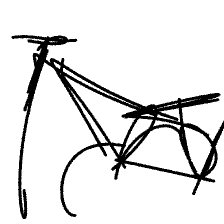} &
    \includegraphics[width=\widthteapot\linewidth]{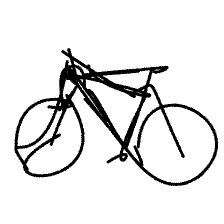} &
    \includegraphics[width=\widthteapot\linewidth]{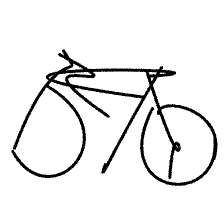}\\
    
    \midrule
    \includegraphics[width=\widthteapot\linewidth]{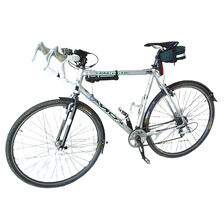} &
    \includegraphics[width=\widthteapot\linewidth]{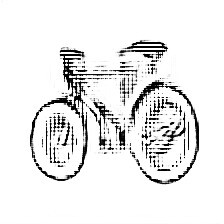} &
    \raisebox{.2\height}{\includegraphics[width=\widthteapot\linewidth]{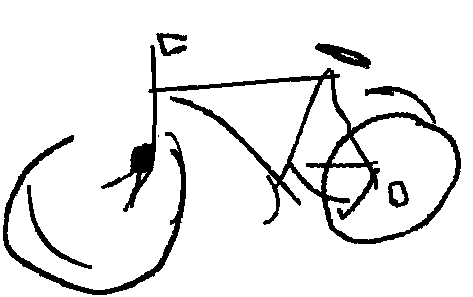}} &
    \includegraphics[width=\widthteapot\linewidth]{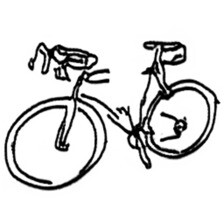} &
    \includegraphics[width=\widthteapot\linewidth]{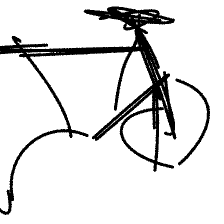} &
    \includegraphics[width=\widthteapot\linewidth]{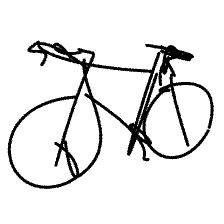} &
    \includegraphics[width=\widthteapot\linewidth]{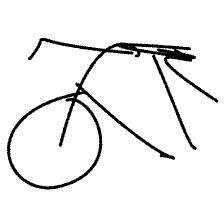}\\
    
    \midrule
    \includegraphics[width=\widthteapot\linewidth]{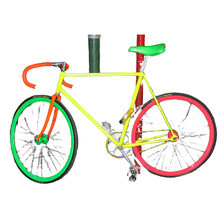} &
    \includegraphics[width=\widthteapot\linewidth]{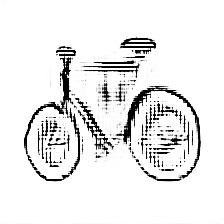} &
    \raisebox{.2\height}{\includegraphics[width=\widthteapot\linewidth]{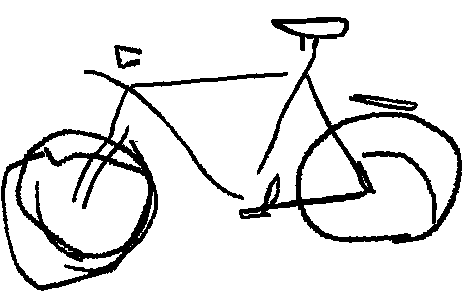}} &
    \includegraphics[width=\widthteapot\linewidth]{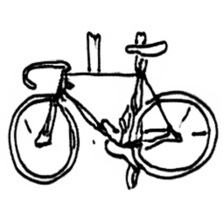} &
    \includegraphics[width=\widthteapot\linewidth]{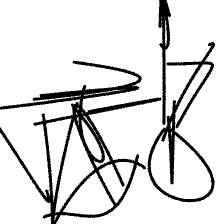} &
    \includegraphics[width=\widthteapot\linewidth]{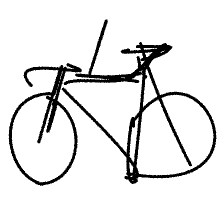} &
    \includegraphics[width=\widthteapot\linewidth]{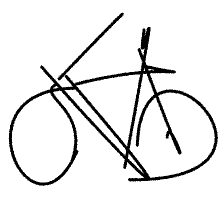}\\
    
    \midrule
    \includegraphics[width=\widthteapot\linewidth]{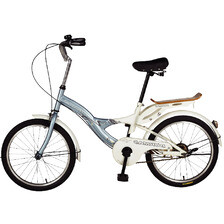} &
    \includegraphics[width=\widthteapot\linewidth]{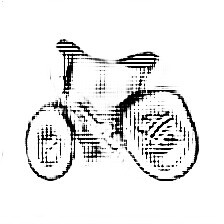} &
    \raisebox{.2\height}{\includegraphics[width=\widthteapot\linewidth]{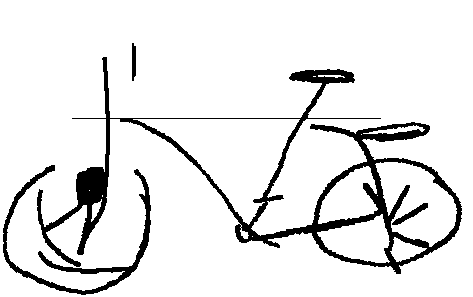}} &
    \includegraphics[width=\widthteapot\linewidth]{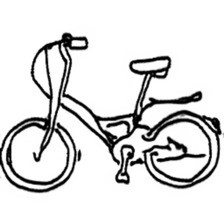} &
    \includegraphics[width=\widthteapot\linewidth]{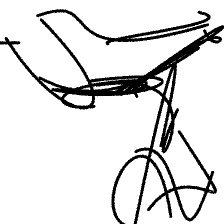} &
    \includegraphics[width=\widthteapot\linewidth]{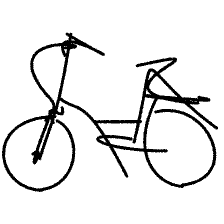} &
    \includegraphics[width=\widthteapot\linewidth]{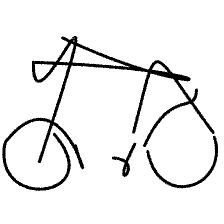}\\
    
    \midrule
    \includegraphics[width=\widthteapot\linewidth]{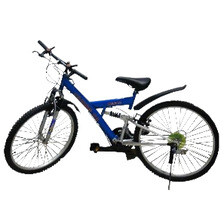} &
    \includegraphics[width=\widthteapot\linewidth]{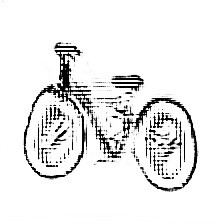} &
    \raisebox{.2\height}{\includegraphics[width=\widthteapot\linewidth]{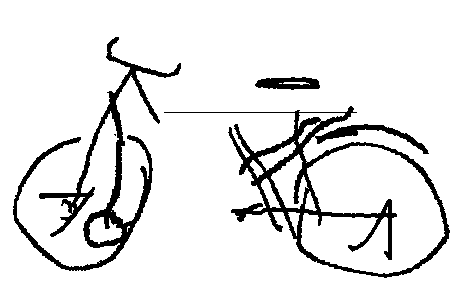}} &
    \includegraphics[width=\widthteapot\linewidth]{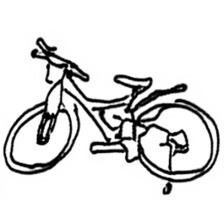} &
    \includegraphics[width=\widthteapot\linewidth]{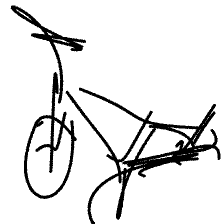} &
    \includegraphics[width=\widthteapot\linewidth]{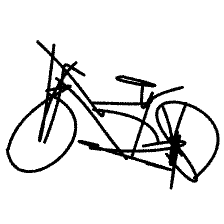} &
    \includegraphics[width=\widthteapot\linewidth]{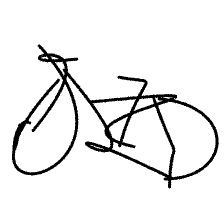}\\
    
    \midrule
    \includegraphics[width=\widthteapot\linewidth]{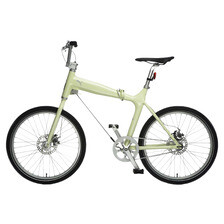} &
    \includegraphics[width=\widthteapot\linewidth]{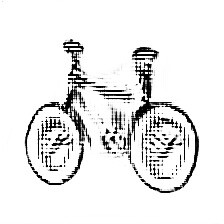} &
    \raisebox{.2\height}{\includegraphics[width=\widthteapot\linewidth]{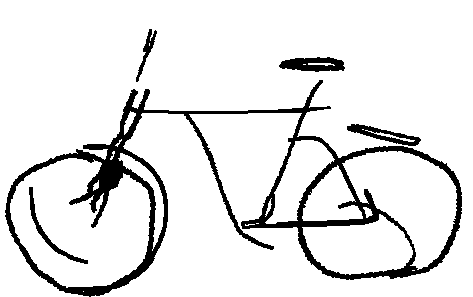}} &
    \includegraphics[width=\widthteapot\linewidth]{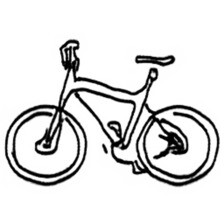} &
    \includegraphics[width=\widthteapot\linewidth]{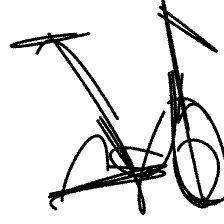} &
    \includegraphics[width=\widthteapot\linewidth]{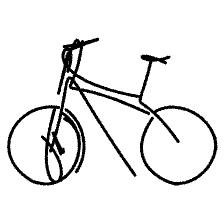} &
    \includegraphics[width=\widthteapot\linewidth]{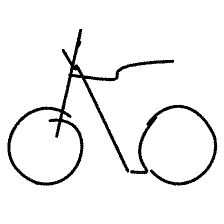}\\

\end{tabular}
 \caption{Comparison to competitor methods on images from the "bicycle" class. Left to right : (A) Kampelm{\"{u}}hler and Pinz \cite{human-like-sketches}, (B) Li et al. \cite{Deformable_Stroke}, (C) Li et al. \cite{li2019photosketching}, and CLIPDraw \cite{CLIPDraw}.}
\label{fig:bicycle}
\end{figure*}

In Figures \ref{fig:teapot}, \ref{fig:horse}, \ref{fig:duck}, and \ref{fig:bicycle} we provide additional comparison to (A) 
Kampelm{\"{u}}hler and Pinz \cite{human-like-sketches}, (B) Li et al. \cite{Deformable_Stroke}, (C) Li et al. \cite{li2019photosketching}, and CLIPDraw \cite{CLIPDraw} (using the sketch style and image as an input). The input images used are from the SketchyDatabase \cite{Sketchy-Database} with four classes - teapot, horse, duck, and bicycle. The classes and images were chosen according to the publicly available results provided by Li et al. \cite{Deformable_Stroke}.

\begin{figure}[ht!]
\centering
\begin{tabular}{@{\hskip2pt}c@{\hskip2pt}c@{\hskip2pt}c@{\hskip2pt}c@{\hskip2pt}c@{\hskip2pt}c@{\hskip2pt}c@{\hskip2pt}c}
    \midrule
    Input & A & B & C & \begin{tabular}[c]{@{}c@{}}CLIP \\ Draw\end{tabular} & Ours \\
    
    \midrule
    \includegraphics[width=\widthface\linewidth]{figs/comp_objects/input/men/face_9_edge.jpg} &
    \includegraphics[width=\widthdeformfaces\linewidth]{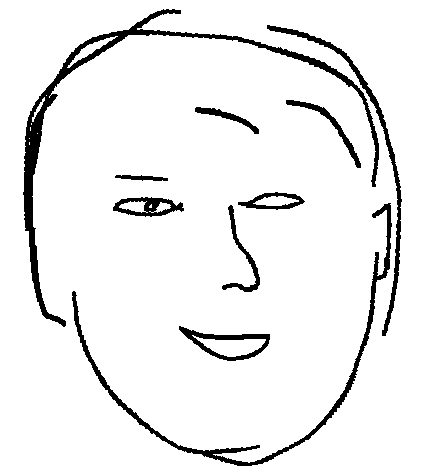} &
    \includegraphics[width=\widthface\linewidth]{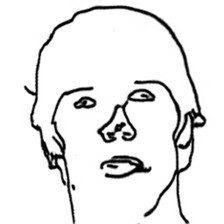} &
    \includegraphics[width=\widthface\linewidth]{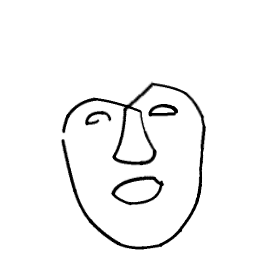} &
    \includegraphics[width=\widthface\linewidth]{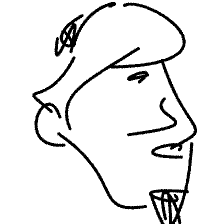} &
    \includegraphics[width=\widthface\linewidth]{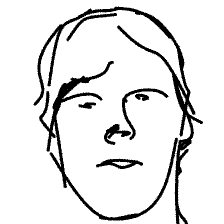} \\
    
    \midrule
    \includegraphics[width=\widthface\linewidth]{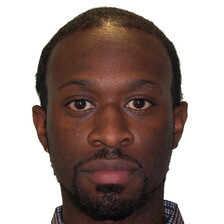} &
    \includegraphics[width=\widthdeformfaces\linewidth]{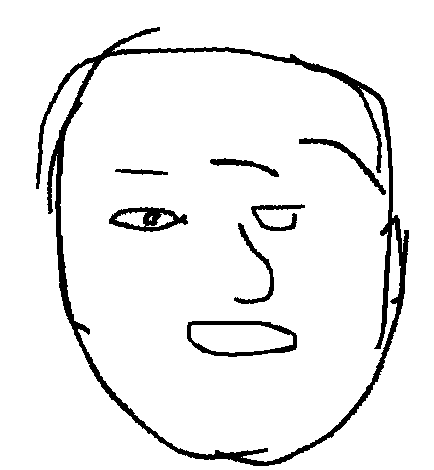} &
    \includegraphics[width=\widthface\linewidth]{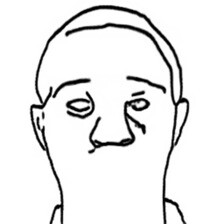} &
    \includegraphics[width=\widthface\linewidth]{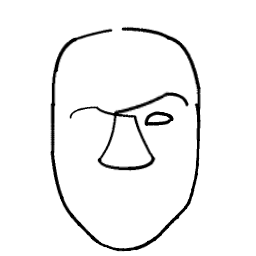} &
    \includegraphics[width=\widthface\linewidth]{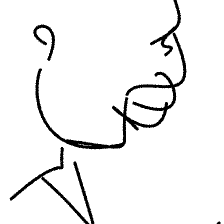} &
    \includegraphics[width=\widthface\linewidth]{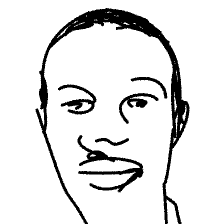} \\
    
    \midrule
    \includegraphics[width=\widthface\linewidth]{figs/comp_objects/input/men/face_4_edge.jpg} &
    \includegraphics[width=\widthdeformfaces\linewidth]{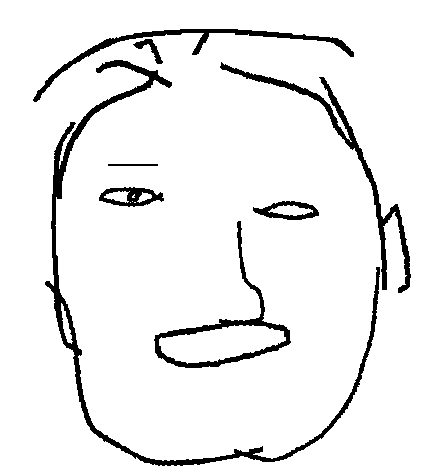} &
    \includegraphics[width=\widthface\linewidth]{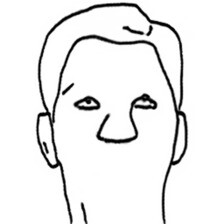} &
    \includegraphics[width=\widthface\linewidth]{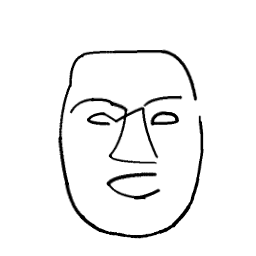} &
    \includegraphics[width=\widthface\linewidth]{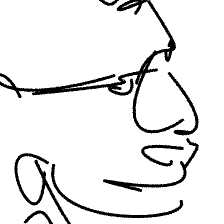} &
    \includegraphics[width=\widthface\linewidth]{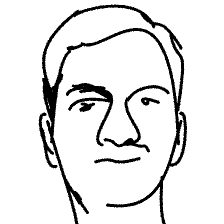} \\
    
    \midrule
    \includegraphics[width=\widthface\linewidth]{figs/comp_objects/input/men/face_7_edge.jpg} &
    \includegraphics[width=\widthdeformfaces\linewidth]{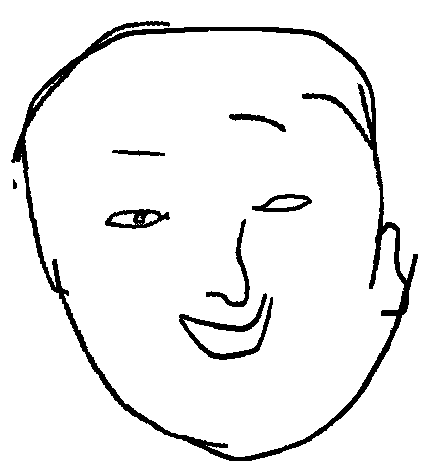} &
    \includegraphics[width=\widthface\linewidth]{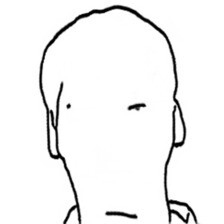} &
    \includegraphics[width=\widthface\linewidth]{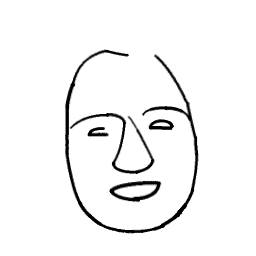} &
    \includegraphics[width=\widthface\linewidth]{figs/comp_objects/clipdraw/men/face_7_edge_.png} &
    \includegraphics[width=\widthface\linewidth]{figs/comp_objects/ours32/men/face_7_edge.png} \\
    
    \midrule
    \includegraphics[width=\widthface\linewidth]{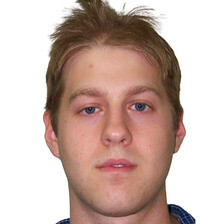} &
    \includegraphics[width=\widthdeformfaces\linewidth]{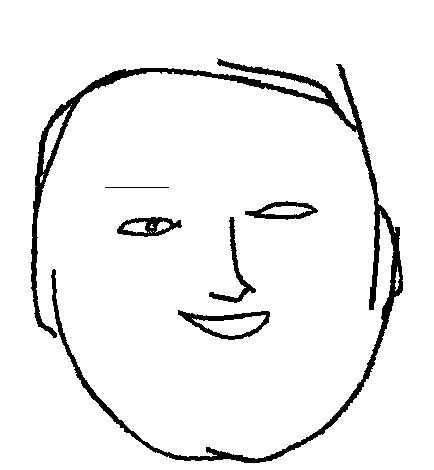} &
    \includegraphics[width=\widthface\linewidth]{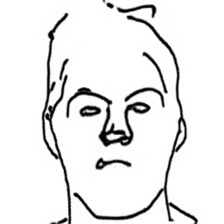} &
    \includegraphics[width=\widthface\linewidth]{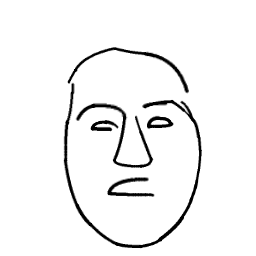} &
    \includegraphics[width=\widthface\linewidth]{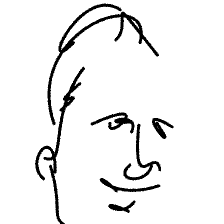} &
    \includegraphics[width=\widthface\linewidth]{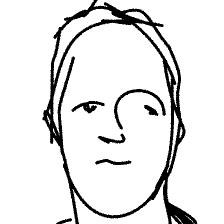} \\
    
    \midrule
    \includegraphics[width=\widthface\linewidth]{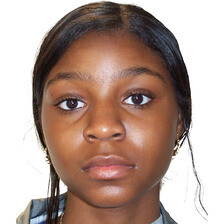} &
    \includegraphics[width=\widthdeformfaces\linewidth]{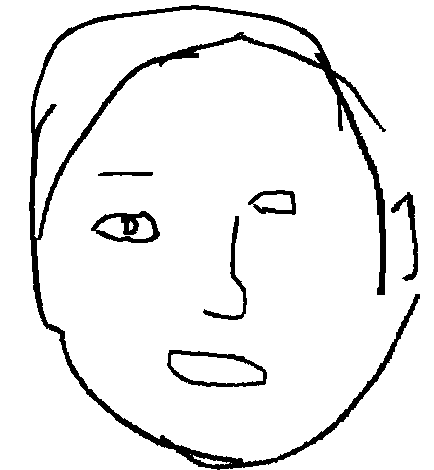} &
    \includegraphics[width=\widthface\linewidth]{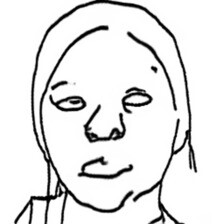} &
    \includegraphics[width=\widthface\linewidth]{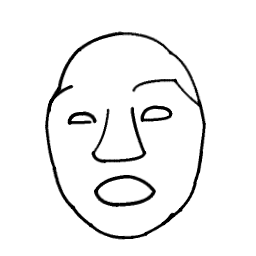} &
    \includegraphics[width=\widthface\linewidth]{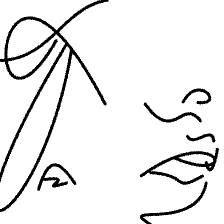} &
    \includegraphics[width=\widthface\linewidth]{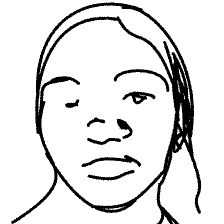} \\
    
      \midrule
    \includegraphics[width=\widthface\linewidth]{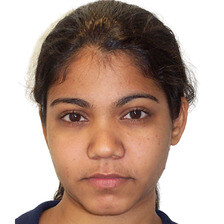} &
    \includegraphics[width=\widthdeformfaces\linewidth]{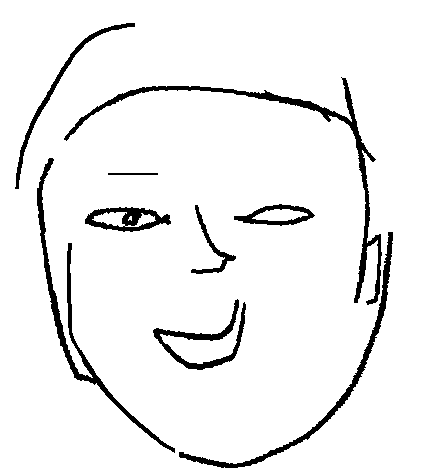} &
    \includegraphics[width=\widthface\linewidth]{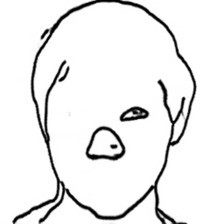} &
    \includegraphics[width=\widthface\linewidth]{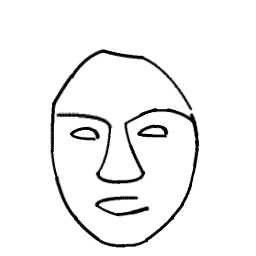} &
    \includegraphics[width=\widthface\linewidth]{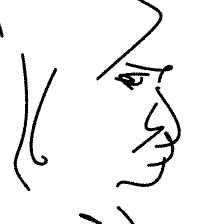} &
    \includegraphics[width=\widthface\linewidth]{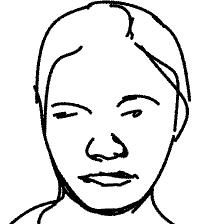} \\
    
    \midrule
    \includegraphics[width=\widthface\linewidth]{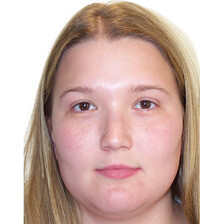} &
    \includegraphics[width=\widthdeformfaces\linewidth]{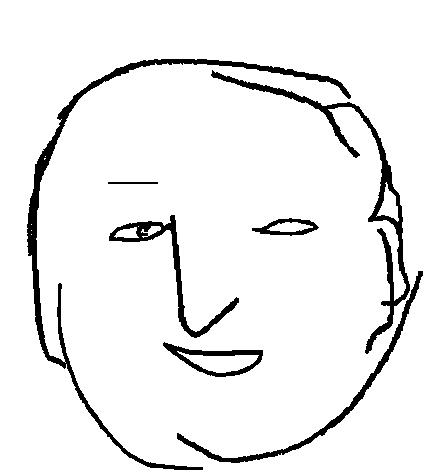} &
    \includegraphics[width=\widthface\linewidth]{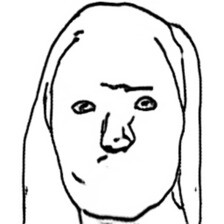} &
    \includegraphics[width=\widthface\linewidth]{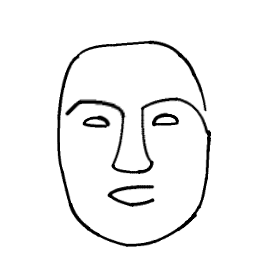} &
    \includegraphics[width=\widthface\linewidth]{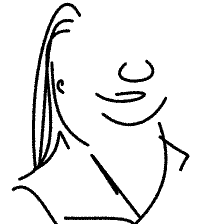} &
    \includegraphics[width=\widthface\linewidth]{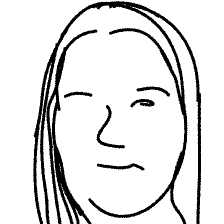} \\
    
    \midrule
    \includegraphics[width=\widthface\linewidth]{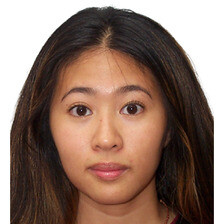} &
    \includegraphics[width=\widthdeformfaces\linewidth]{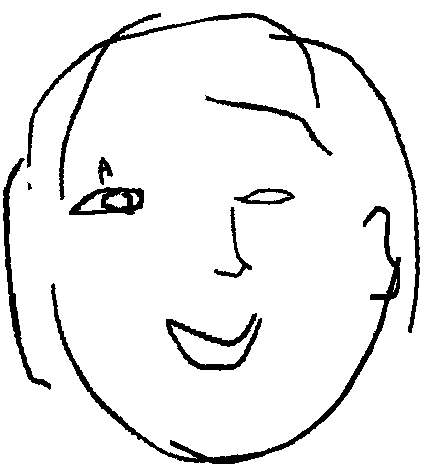} &
    \includegraphics[width=\widthface\linewidth]{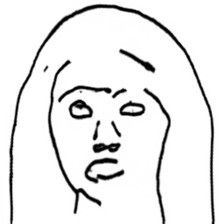} &
    \includegraphics[width=\widthface\linewidth]{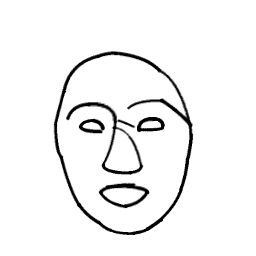} &
    \includegraphics[width=\widthface\linewidth]{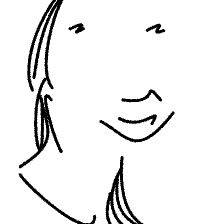} &
    \includegraphics[width=\widthface\linewidth]{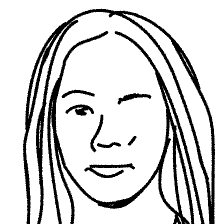} \\
    
    \midrule
    \includegraphics[width=\widthface\linewidth]{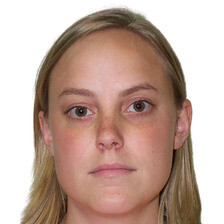} &
    \includegraphics[width=\widthdeformfaces\linewidth]{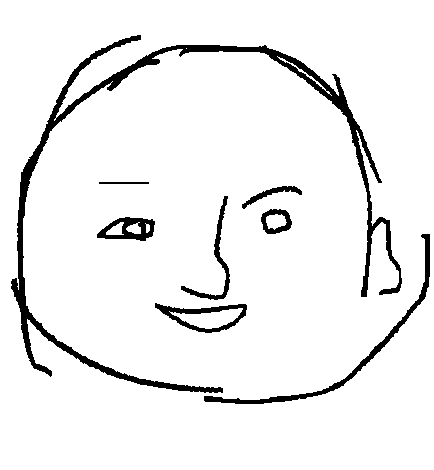} &
    \includegraphics[width=\widthface\linewidth]{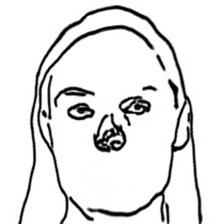} &
    \includegraphics[width=\widthface\linewidth]{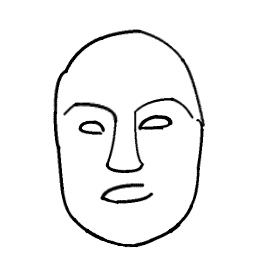} &
    \includegraphics[width=\widthface\linewidth]{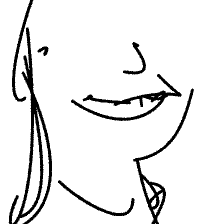} &
    \includegraphics[width=\widthface\linewidth]{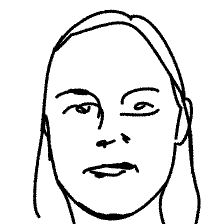} \\
    
\end{tabular}
 \caption{Comparison to competitor methods on human faces. Left to right: (A) Li et al. \cite{Deformable_Stroke}, (B) Li et al. \cite{li2019photosketching}, (C) Mo et al. \cite{mo2021virtualsketching}, and CLIPDraw \cite{CLIPDraw}.}
\label{fig:men}
\end{figure}

In Figure \ref{fig:men} we provide additional comparison on portrait-style human faces from the NPR benchmark \cite{Rosin2017BenchmarkingNR}. This comparison was conducted using the methods of (A) Li et al. \cite{Deformable_Stroke}, (B) Li et al. \cite{li2019photosketching} and (C) Mo et al. \cite{mo2021virtualsketching}. (C) is a recent method to produce vector line drawings from a rasterized input image. Photo-sketch generation is not the focus of this work; nonetheless, they demonstrate an application of converting a face image into a line drawing.
The figures illustrate that the sketches produced by Mo et al. are mostly geometric and do not reflect the semantic characteristics of the input faces.
\vfill


\pagebreak
\null\newpage
\pagebreak
\null\newpage
\section{Additional Qualitative Results}
As stated in the paper, we mainly use four control points to define each Bezier curve, however, it is also possible to change the degree of the curves to achieve different styles and levels of abstraction. 
In Figures \ref{fig:abstraction_levels2}, \ref{fig:abstraction_levels3}, and \ref{fig:abstraction_levels4} we provide additional results generated by our method, demonstrating three levels of abstraction for each image, and three different styles.
The numbers on top indicate the total number of control points used to generate the sketches (64, 24, and 16).
In each block, each row shows a different style: first row – 4 control points per stroke, second row – 3 control points per stroke, third row – 2 control points per stroke (i.e. straight lines). 

\begin{figure*}[ht]
\centering
\vspace{-3em}
\begin{tabular}{ccccc}
    \midrule
    Input & 64 & 32 & 16 & 8 \\
    
    \midrule
    \includegraphics[width=\widthfaceour\linewidth]{figs/comp_objects/input/men/face_9_edge.jpg} &
  \includegraphics[width=\widthfaceour\linewidth]{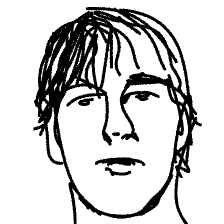} &
  \includegraphics[width=\widthfaceour\linewidth]{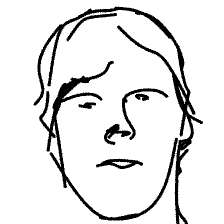} &
  \includegraphics[width=\widthfaceour\linewidth]{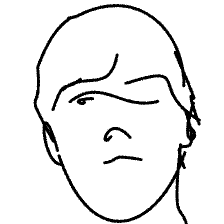} &
  \includegraphics[width=\widthfaceour\linewidth]{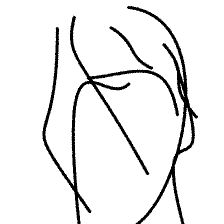} \\

    \midrule
    \includegraphics[width=\widthfaceour\linewidth]{figs/comp_objects/input/men/face_2_edge.jpg} &
    \includegraphics[width=\widthfaceour\linewidth]{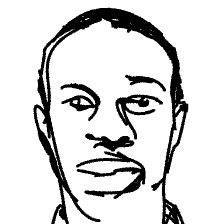} &
  \includegraphics[width=\widthfaceour\linewidth]{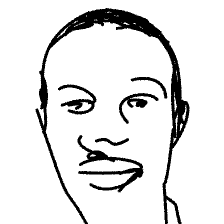} &
  \includegraphics[width=\widthfaceour\linewidth]{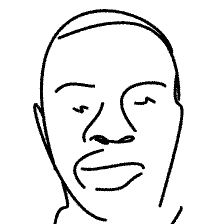} &
  \includegraphics[width=\widthfaceour\linewidth]{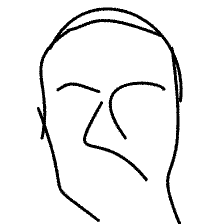} \\
    
    \midrule
    \includegraphics[width=\widthfaceour\linewidth]{figs/comp_objects/input/men/face_4_edge.jpg} &
      \includegraphics[width=\widthfaceour\linewidth]{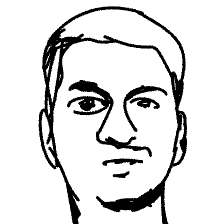} &
  \includegraphics[width=\widthfaceour\linewidth]{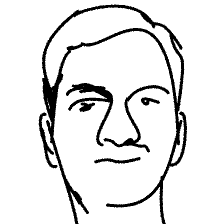} &
  \includegraphics[width=\widthfaceour\linewidth]{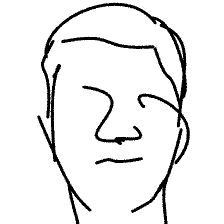} &
  \includegraphics[width=\widthfaceour\linewidth]{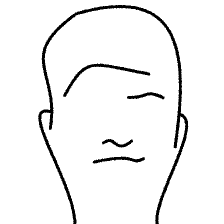} \\
   
    \midrule
    \includegraphics[width=\widthfaceour\linewidth]{figs/comp_objects/input/men/face_7_edge.jpg} &
  \includegraphics[width=\widthfaceour\linewidth]{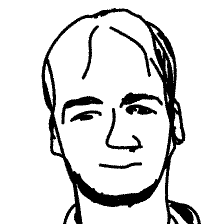} &
  \includegraphics[width=\widthfaceour\linewidth]{figs/abstraction_levels/32/men/face_7_edge.png} &
  \includegraphics[width=\widthfaceour\linewidth]{figs/abstraction_levels/16/men/face_7_edge.png} &
  \includegraphics[width=\widthfaceour\linewidth]{figs/abstraction_levels/8/men/face_7_edge.png} \\

    \midrule
    \includegraphics[width=\widthfaceour\linewidth]{figs/comp_objects/input/men/face_10_edge.jpg} &
  \includegraphics[width=\widthfaceour\linewidth]{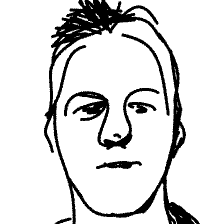} &
  \includegraphics[width=\widthfaceour\linewidth]{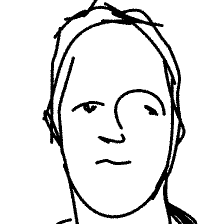} &
  \includegraphics[width=\widthfaceour\linewidth]{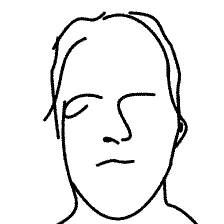} &
  \includegraphics[width=\widthfaceour\linewidth]{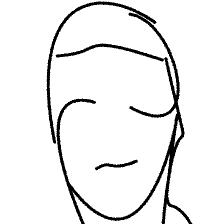} \\
  
  \midrule
    \includegraphics[width=\widthfaceour\linewidth]{figs/comp_objects/input/women/face_1_edge.jpg} &
  \includegraphics[width=\widthfaceour\linewidth]{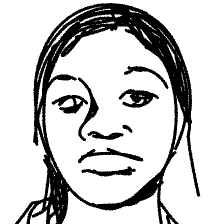} &
  \includegraphics[width=\widthfaceour\linewidth]{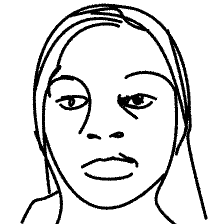} &
  \includegraphics[width=\widthfaceour\linewidth]{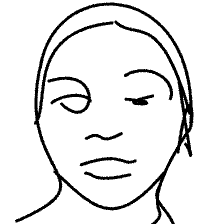} &
  \includegraphics[width=\widthfaceour\linewidth]{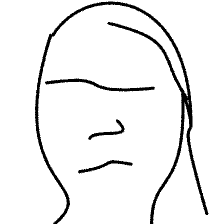} \\
   
  \midrule
    \includegraphics[width=\widthfaceour\linewidth]{figs/comp_objects/input/women/face_3_edge.jpg} &
  \includegraphics[width=\widthfaceour\linewidth]{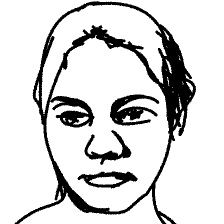} &
  \includegraphics[width=\widthfaceour\linewidth]{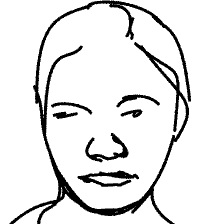} &
  \includegraphics[width=\widthfaceour\linewidth]{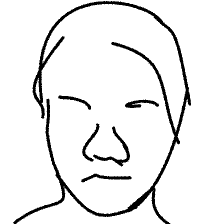} &
  \includegraphics[width=\widthfaceour\linewidth]{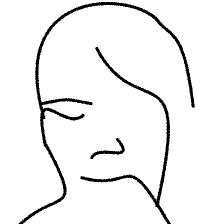} \\
   
  \midrule
    \includegraphics[width=\widthfaceour\linewidth]{figs/comp_objects/input/women/face_5_edge.jpg} &
  \includegraphics[width=\widthfaceour\linewidth]{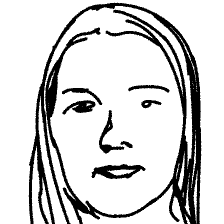} &
  \includegraphics[width=\widthfaceour\linewidth]{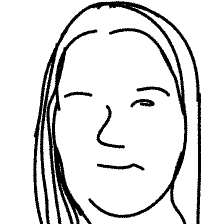} &
  \includegraphics[width=\widthfaceour\linewidth]{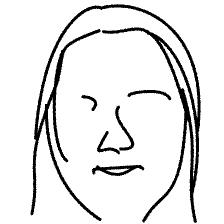} &
  \includegraphics[width=\widthfaceour\linewidth]{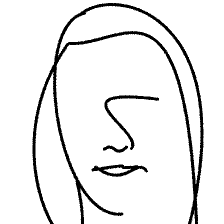} \\
   
  \midrule
    \includegraphics[width=\widthfaceour\linewidth]{figs/comp_objects/input/women/face_6_edge.jpg} &
  \includegraphics[width=\widthfaceour\linewidth]{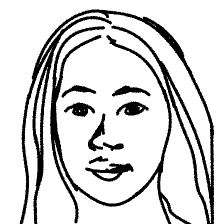} &
  \includegraphics[width=\widthfaceour\linewidth]{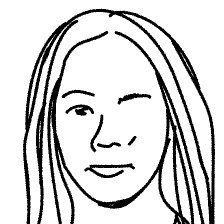} &
  \includegraphics[width=\widthfaceour\linewidth]{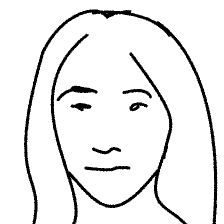} &
  \includegraphics[width=\widthfaceour\linewidth]{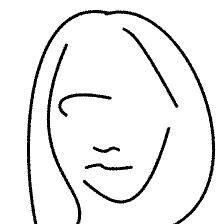} \\
   
  \midrule
    \includegraphics[width=\widthfaceour\linewidth]{figs/comp_objects/input/women/face_8_edge.jpg} &
  \includegraphics[width=\widthfaceour\linewidth]{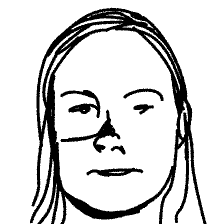} &
  \includegraphics[width=\widthfaceour\linewidth]{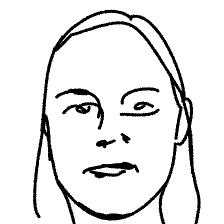} &
  \includegraphics[width=\widthfaceour\linewidth]{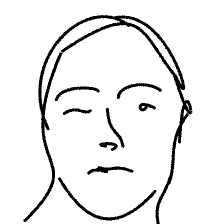} &
  \includegraphics[width=\widthfaceour\linewidth]{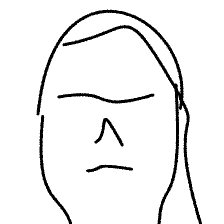} \\
   
\end{tabular}
 \caption{Sketches of human faces produced by our method with different levels of abstraction.}
\label{fig:men_abstractions_level}
\end{figure*}

In Figure \ref{fig:men_abstractions_level} we show additional results on portrait-style human faces from the NPR benchmark \cite{Rosin2017BenchmarkingNR}. The sketches produced using 64, 32, 16, and 8 strokes, with 4 control points each.

\newcommand{\widthabsup}{0.1}

\begin{figure*}[ht]
\centering
\begin{tabular}{@{\hskip2pt}c@{\hskip2pt}c@{\hskip2pt}c@{\hskip2pt}c|@{\hskip2pt}c@{\hskip2pt}c@{\hskip2pt}c@{\hskip2pt}c@{\hskip2pt}c}
    \midrule
    Input & 64 & 24 & 16 & Input & 64 & 24 & 16 \\
    \midrule
    \includegraphics[width=\widthabsup\linewidth]{figs/abstraction_levels/input/camel/camel.png} &
    \includegraphics[width=\widthabsup\linewidth]{figs/abstraction_levels/camel/camel_p16_cp4.png} &
    \includegraphics[width=\widthabsup\linewidth]{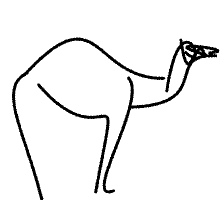} & 
    \includegraphics[width=\widthabsup\linewidth]{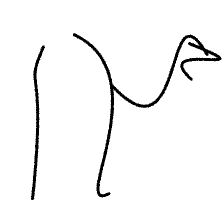} &
    \includegraphics[width=\widthabsup\linewidth]{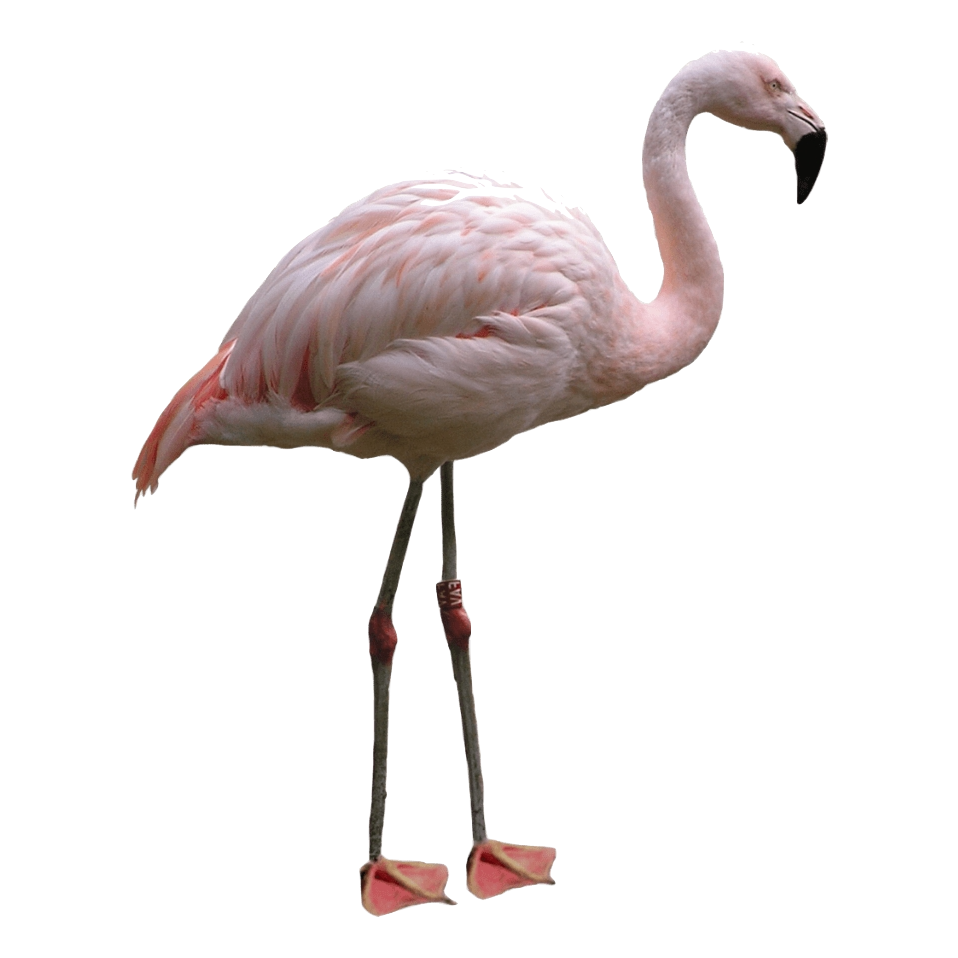} &
    \includegraphics[width=\widthabsup\linewidth]{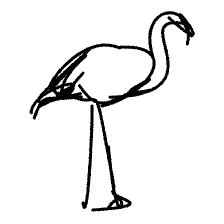} &
    \includegraphics[width=\widthabsup\linewidth]{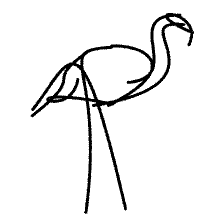} & 
    \includegraphics[width=\widthabsup\linewidth]{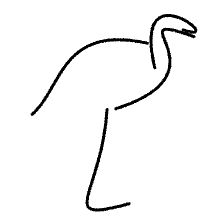} \\
    &
    \includegraphics[width=\widthabsup\linewidth]{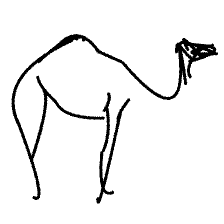} &
    \includegraphics[width=\widthabsup\linewidth]{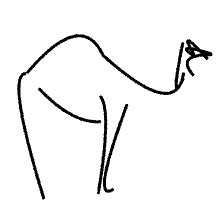} &
    \includegraphics[width=\widthabsup\linewidth]{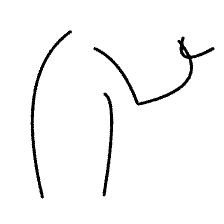} & &
    \includegraphics[width=\widthabsup\linewidth]{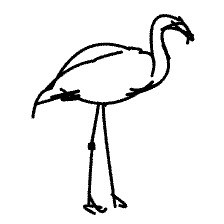} &
    \includegraphics[width=\widthabsup\linewidth]{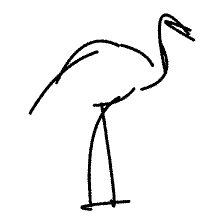} &
    \includegraphics[width=\widthabsup\linewidth]{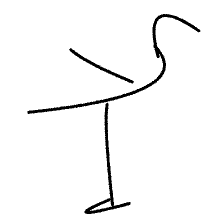} \\
    &
    \includegraphics[width=\widthabsup\linewidth]{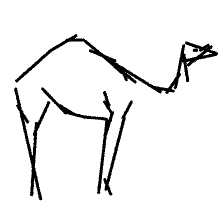} &
    \includegraphics[width=\widthabsup\linewidth]{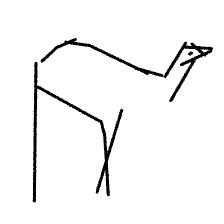} &
    \includegraphics[width=\widthabsup\linewidth]{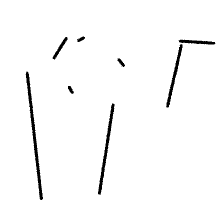} & &
    \includegraphics[width=\widthabsup\linewidth]{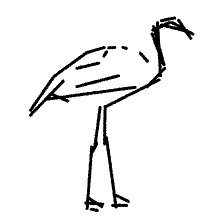} &
    \includegraphics[width=\widthabsup\linewidth]{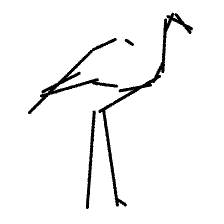} &
    \includegraphics[width=\widthabsup\linewidth]{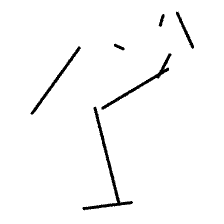} \\
    
    \midrule
    \includegraphics[width=\widthabsup\linewidth]{figs/abstraction_levels/input/rose/rose1.jpeg} &
    \includegraphics[width=\widthabsup\linewidth]{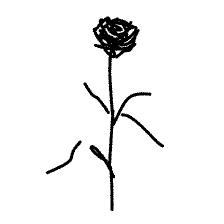} &
    \includegraphics[width=\widthabsup\linewidth]{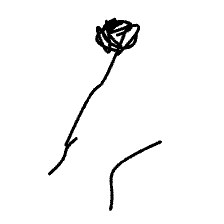} & 
    \includegraphics[width=\widthabsup\linewidth]{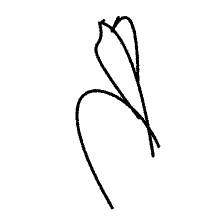} &
    \includegraphics[width=\widthabsup\linewidth]{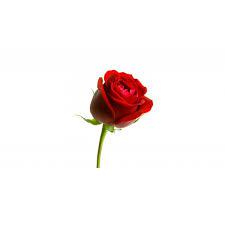} &
    \includegraphics[width=\widthabsup\linewidth]{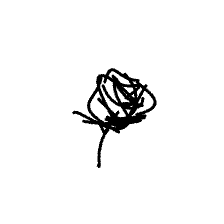} &
    \includegraphics[width=\widthabsup\linewidth]{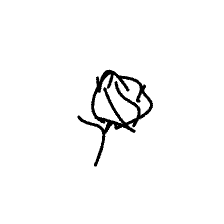} & 
    \includegraphics[width=\widthabsup\linewidth]{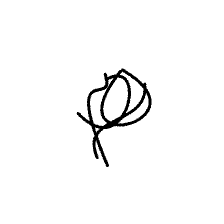} \\
    &
    \includegraphics[width=\widthabsup\linewidth]{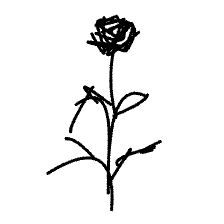} &
    \includegraphics[width=\widthabsup\linewidth]{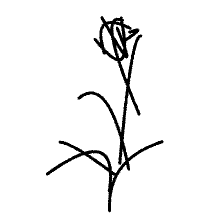} &
    \includegraphics[width=\widthabsup\linewidth]{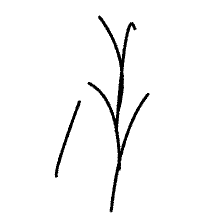} & &
    \includegraphics[width=\widthabsup\linewidth]{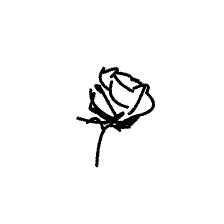} &
    \includegraphics[width=\widthabsup\linewidth]{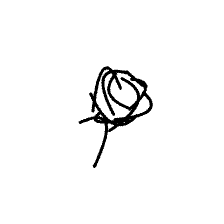} &
    \includegraphics[width=\widthabsup\linewidth]{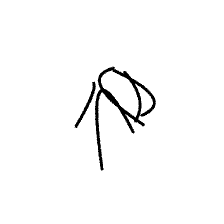} \\
    &
    \includegraphics[width=\widthabsup\linewidth]{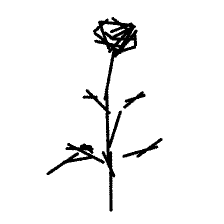} &
    \includegraphics[width=\widthabsup\linewidth]{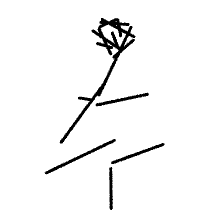} &
    \includegraphics[width=\widthabsup\linewidth]{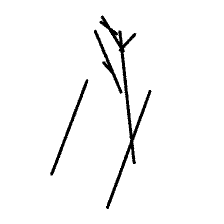} & &
    \includegraphics[width=\widthabsup\linewidth]{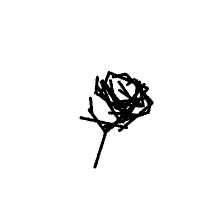} &
    \includegraphics[width=\widthabsup\linewidth]{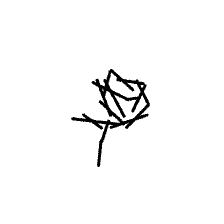} &
    \includegraphics[width=\widthabsup\linewidth]{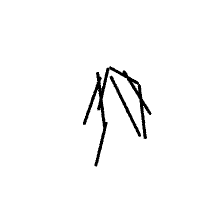} \\
    
    \midrule
    \includegraphics[width=\widthabsup\linewidth]{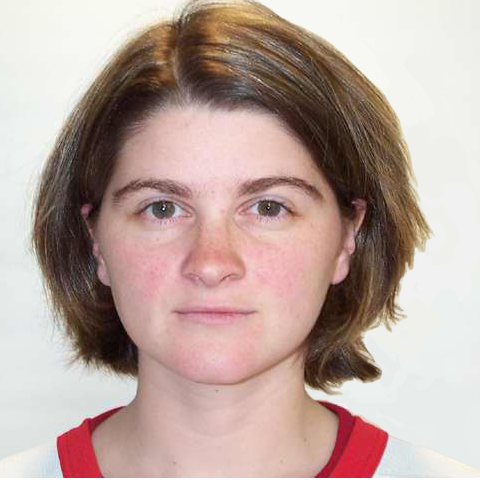} &
    \includegraphics[width=\widthabsup\linewidth]{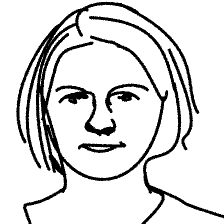} &
    \includegraphics[width=\widthabsup\linewidth]{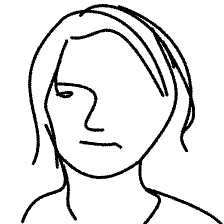} &
    \includegraphics[width=\widthabsup\linewidth]{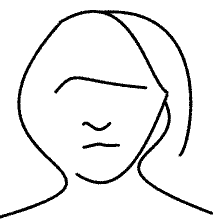} & 
    \includegraphics[width=\widthabsup\linewidth]{figs/comp_objects/input/men/face_7_edge.jpg} &
    \includegraphics[width=\widthabsup\linewidth]{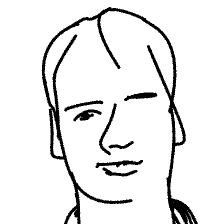} &
    \includegraphics[width=\widthabsup\linewidth]{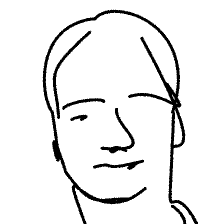} &
    \includegraphics[width=\widthabsup\linewidth]{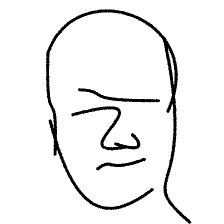} \\
    &
    \includegraphics[width=\widthabsup\linewidth]{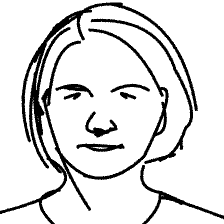} &
    \includegraphics[width=\widthabsup\linewidth]{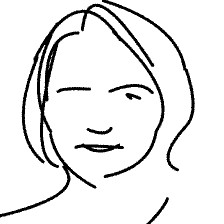} &
    \includegraphics[width=\widthabsup\linewidth]{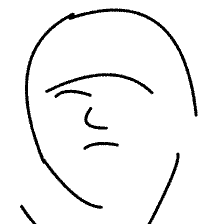} & &
    \includegraphics[width=\widthabsup\linewidth]{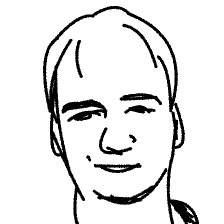} &
    \includegraphics[width=\widthabsup\linewidth]{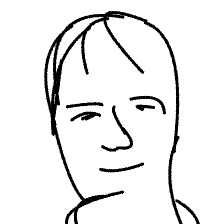} &
    \includegraphics[width=\widthabsup\linewidth]{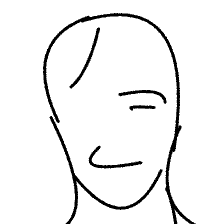} \\
    &
    \includegraphics[width=\widthabsup\linewidth]{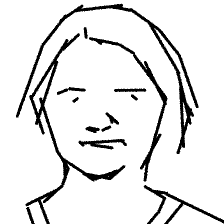} &
    \includegraphics[width=\widthabsup\linewidth]{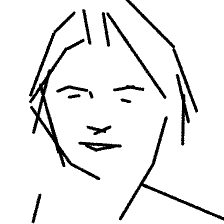} &
    \includegraphics[width=\widthabsup\linewidth]{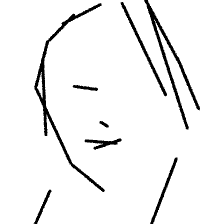} & &
    \includegraphics[width=\widthabsup\linewidth]{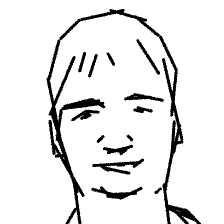} &
    \includegraphics[width=\widthabsup\linewidth]{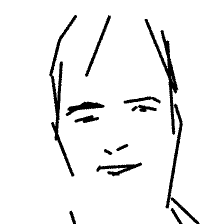} &
    \includegraphics[width=\widthabsup\linewidth]{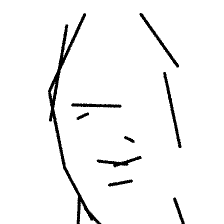} \\

\end{tabular}
 \caption{Sketches produced by our method with different styles and levels of abstraction. The columns indicate the total number of control points used to generate the sketch - 64, 24, and 16. The rows indicate the number of control points used for each stroke, which defines the style - 4, 3, and 2 control points respectively.}
\label{fig:abstraction_levels2}
\end{figure*}

\begin{figure*}[ht]
\centering
\begin{tabular}{@{\hskip2pt}c@{\hskip2pt}c@{\hskip2pt}c@{\hskip2pt}c|@{\hskip2pt}c@{\hskip2pt}c@{\hskip2pt}c@{\hskip2pt}c@{\hskip2pt}c}
    \midrule
    Input & 64 & 24 & 16 & Input & 64 & 24 & 16 \\

    \midrule
    \includegraphics[width=\widthabsup\linewidth]{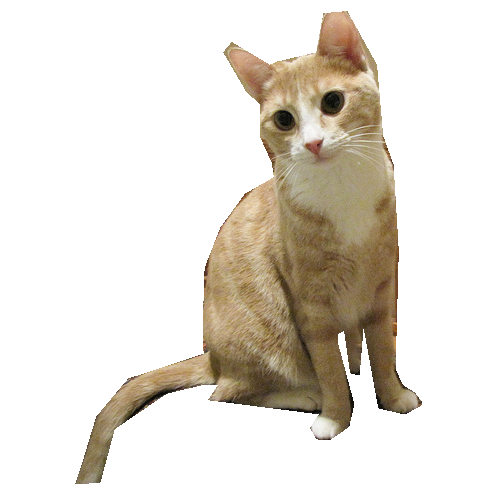} &
    \includegraphics[width=\widthabsup\linewidth]{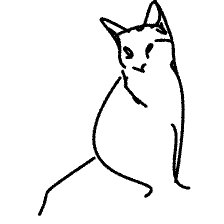} &
    \includegraphics[width=\widthabsup\linewidth]{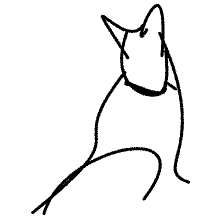} & 
    \includegraphics[width=\widthabsup\linewidth]{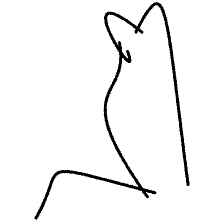} &
    \includegraphics[width=\widthabsup\linewidth]{figs/abstraction_levels/cat_49281/49281.png} &
    \includegraphics[width=\widthabsup\linewidth]{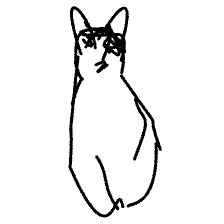} &
    \includegraphics[width=\widthabsup\linewidth]{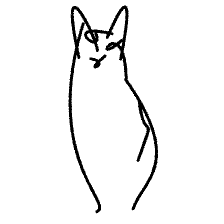} & 
    \includegraphics[width=\widthabsup\linewidth]{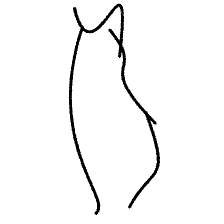} \\
    &
    \includegraphics[width=\widthabsup\linewidth]{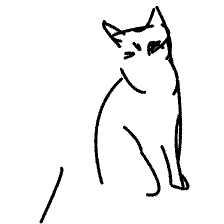} &
    \includegraphics[width=\widthabsup\linewidth]{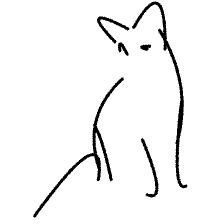} &
    \includegraphics[width=\widthabsup\linewidth]{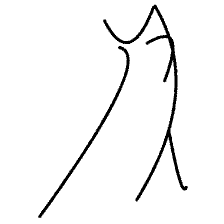} & &
    \includegraphics[width=\widthabsup\linewidth]{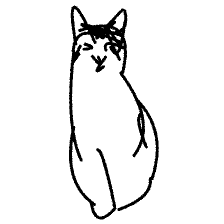} &
    \includegraphics[width=\widthabsup\linewidth]{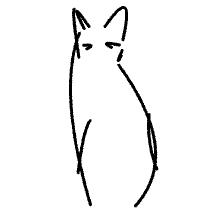} &
    \includegraphics[width=\widthabsup\linewidth]{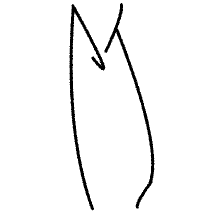} \\
    &
    \includegraphics[width=\widthabsup\linewidth]{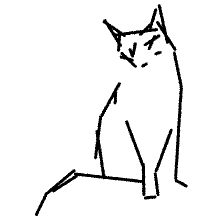} &
    \includegraphics[width=\widthabsup\linewidth]{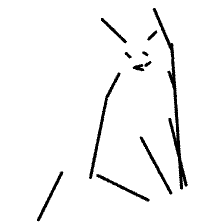} &
    \includegraphics[width=\widthabsup\linewidth]{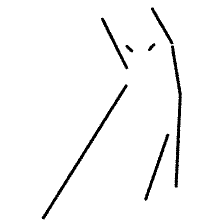} & &
    \includegraphics[width=\widthabsup\linewidth]{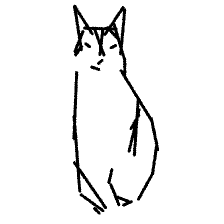} &
    \includegraphics[width=\widthabsup\linewidth]{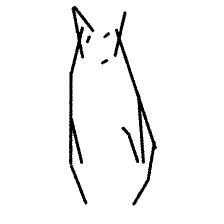} &
    \includegraphics[width=\widthabsup\linewidth]{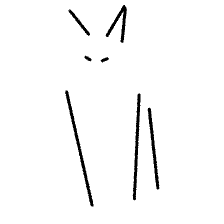} \\
    
    \midrule
    \includegraphics[width=\widthabsup\linewidth]{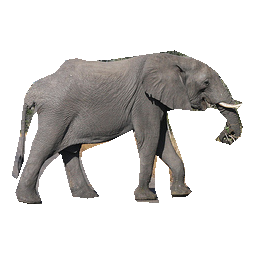} &
    \includegraphics[width=\widthabsup\linewidth]{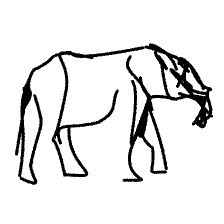} &
    \includegraphics[width=\widthabsup\linewidth]{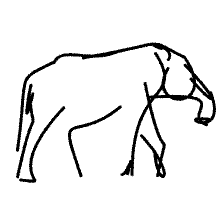} &
    \includegraphics[width=\widthabsup\linewidth]{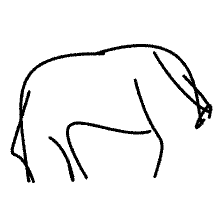} & 
    \includegraphics[width=\widthabsup\linewidth]{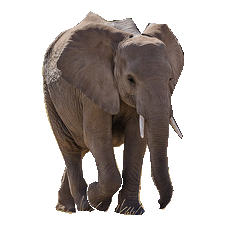} &
    \includegraphics[width=\widthabsup\linewidth]{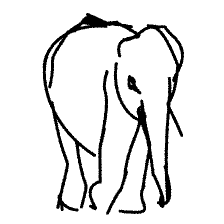} &
    \includegraphics[width=\widthabsup\linewidth]{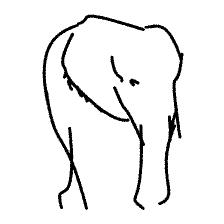} &
    \includegraphics[width=\widthabsup\linewidth]{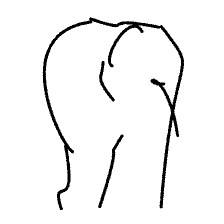} \\
    &
    \includegraphics[width=\widthabsup\linewidth]{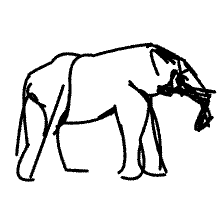} &
    \includegraphics[width=\widthabsup\linewidth]{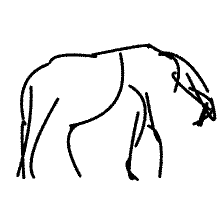} &
    \includegraphics[width=\widthabsup\linewidth]{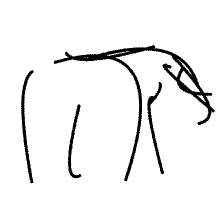} & &
    \includegraphics[width=\widthabsup\linewidth]{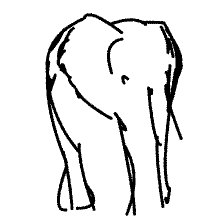} &
    \includegraphics[width=\widthabsup\linewidth]{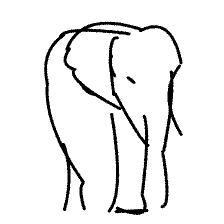} &
    \includegraphics[width=\widthabsup\linewidth]{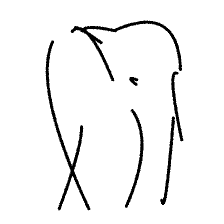} \\
    &
    \includegraphics[width=\widthabsup\linewidth]{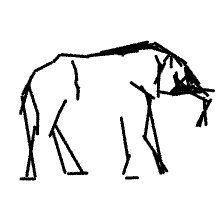} &
    \includegraphics[width=\widthabsup\linewidth]{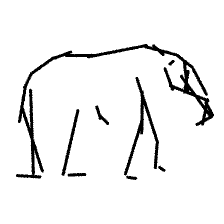} &
    \includegraphics[width=\widthabsup\linewidth]{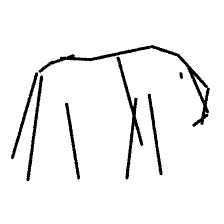} & &
    \includegraphics[width=\widthabsup\linewidth]{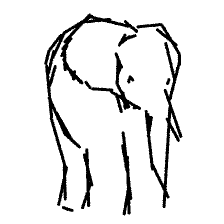} &
    \includegraphics[width=\widthabsup\linewidth]{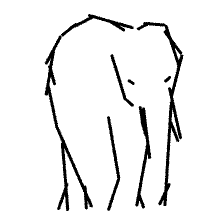} &
    \includegraphics[width=\widthabsup\linewidth]{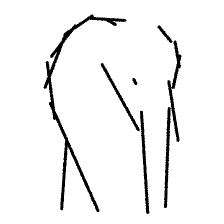} \\
    
\end{tabular}
 \caption{Sketches produced by our method with different styles and levels of abstraction. The columns indicate the total number of control points used to generate the sketch - 64, 24, and 16. The rows indicate the number of control points used for each stroke, which defines the style - 4, 3, and 2 control points respectively.}
\label{fig:abstraction_levels3}
\end{figure*}

\begin{figure*}[ht]
\centering
\begin{tabular}{@{\hskip2pt}c@{\hskip2pt}c@{\hskip2pt}c@{\hskip2pt}c|@{\hskip2pt}c@{\hskip2pt}c@{\hskip2pt}c@{\hskip2pt}c@{\hskip2pt}c}
    \midrule
    Input & 64 & 24 & 16 & Input & 64 & 24 & 16 \\
    \midrule
    \includegraphics[width=\widthabsup\linewidth]{figs/abstraction_levels/input/horse/horse_132.png} &
    \includegraphics[width=\widthabsup\linewidth]{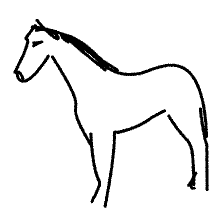} &
    \includegraphics[width=\widthabsup\linewidth]{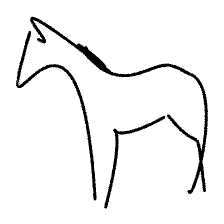} & 
    \includegraphics[width=\widthabsup\linewidth]{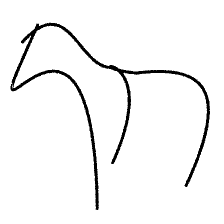} &
    \includegraphics[width=\widthabsup\linewidth]{figs/abstraction_levels/elephant_1096/elephant_1096.png} &
    \includegraphics[width=\widthabsup\linewidth]{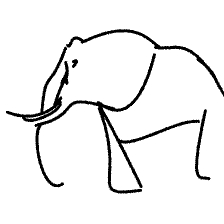} &
    \includegraphics[width=\widthabsup\linewidth]{figs/abstraction_levels/elephant_1096/elephant_1096_p8_cp4.png} & 
    \includegraphics[width=\widthabsup\linewidth]{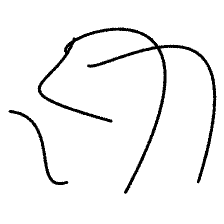} \\
    &
    \includegraphics[width=\widthabsup\linewidth]{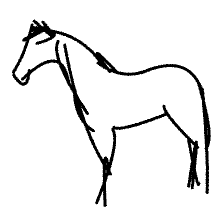} &
    \includegraphics[width=\widthabsup\linewidth]{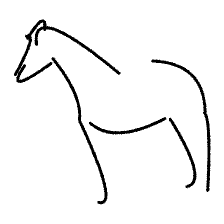} &
    \includegraphics[width=\widthabsup\linewidth]{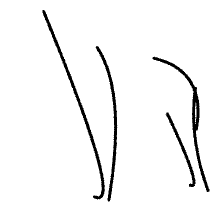} & &
    \includegraphics[width=\widthabsup\linewidth]{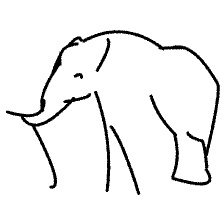} &
    \includegraphics[width=\widthabsup\linewidth]{figs/abstraction_levels/elephant_1096/elephant_1096_p10_cp3.png} &
    \includegraphics[width=\widthabsup\linewidth]{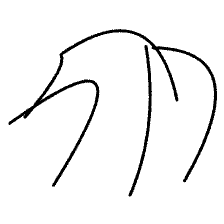} \\
    &
    \includegraphics[width=\widthabsup\linewidth]{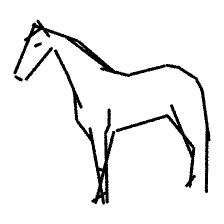} &
    \includegraphics[width=\widthabsup\linewidth]{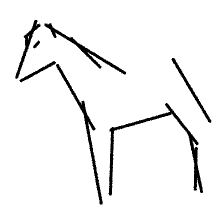} &
    \includegraphics[width=\widthabsup\linewidth]{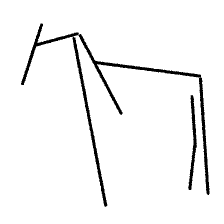} & &
    \includegraphics[width=\widthabsup\linewidth]{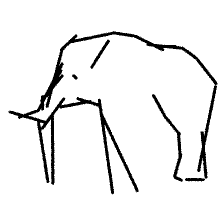} &
    \includegraphics[width=\widthabsup\linewidth]{figs/abstraction_levels/elephant_1096/elephant_1096_p16_cp2.png} &
    \includegraphics[width=\widthabsup\linewidth]{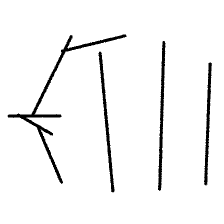} \\
    
    \midrule
    \includegraphics[width=\widthabsup\linewidth]{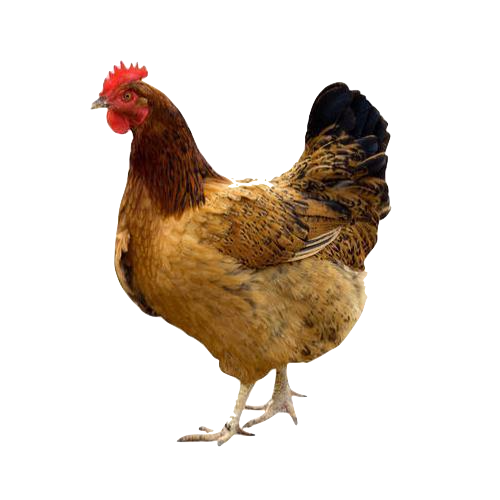} &
    \includegraphics[width=\widthabsup\linewidth]{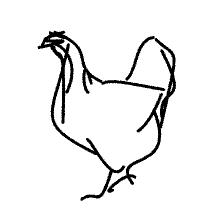} &
    \includegraphics[width=\widthabsup\linewidth]{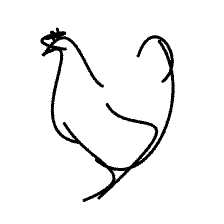} & 
    \includegraphics[width=\widthabsup\linewidth]{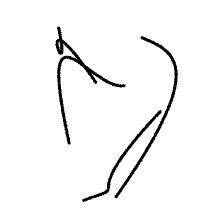} &
    \includegraphics[width=\widthabsup\linewidth]{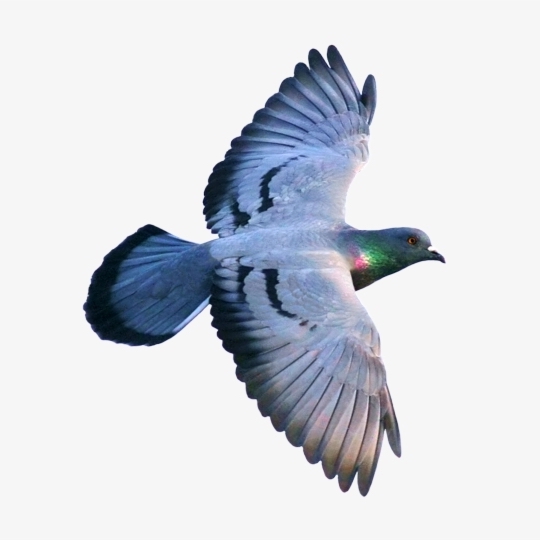} &
    \includegraphics[width=\widthabsup\linewidth]{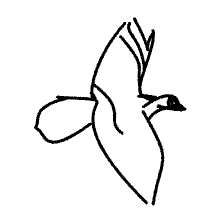} &
    \includegraphics[width=\widthabsup\linewidth]{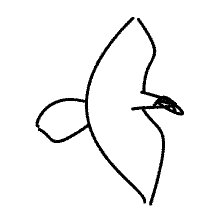} & 
    \includegraphics[width=\widthabsup\linewidth]{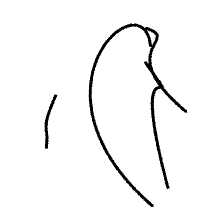} \\
    &
    \includegraphics[width=\widthabsup\linewidth]{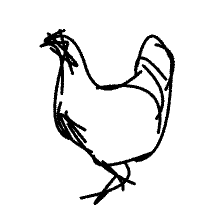} &
    \includegraphics[width=\widthabsup\linewidth]{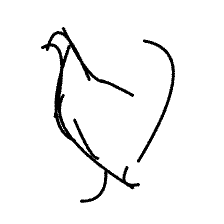} &
    \includegraphics[width=\widthabsup\linewidth]{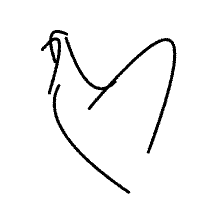} & &
    \includegraphics[width=\widthabsup\linewidth]{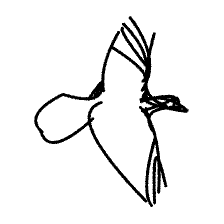} &
    \includegraphics[width=\widthabsup\linewidth]{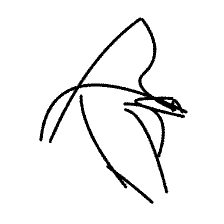} &
    \includegraphics[width=\widthabsup\linewidth]{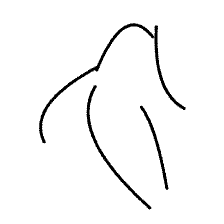} \\
    &
    \includegraphics[width=\widthabsup\linewidth]{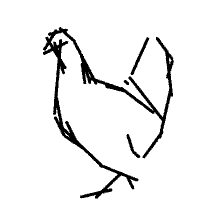} &
    \includegraphics[width=\widthabsup\linewidth]{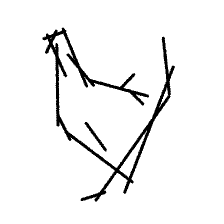} &
    \includegraphics[width=\widthabsup\linewidth]{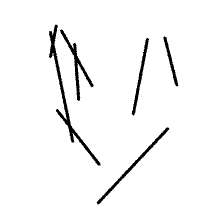} & &
    \includegraphics[width=\widthabsup\linewidth]{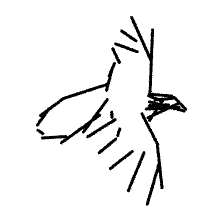} &
    \includegraphics[width=\widthabsup\linewidth]{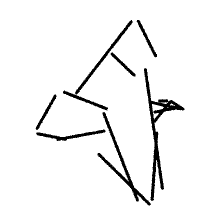} &
    \includegraphics[width=\widthabsup\linewidth]{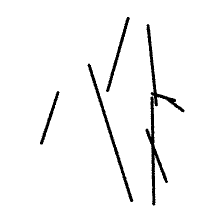} \\
    
    \midrule
    \includegraphics[width=\widthabsup\linewidth]{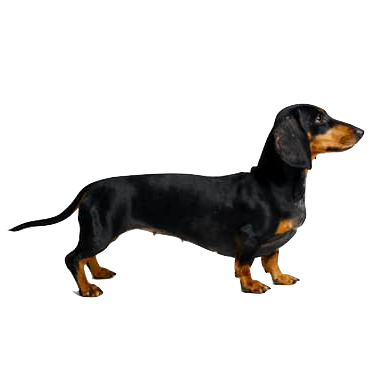} &
    \includegraphics[width=\widthabsup\linewidth]{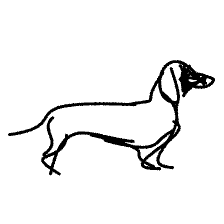} &
    \includegraphics[width=\widthabsup\linewidth]{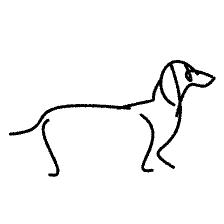} & 
    \includegraphics[width=\widthabsup\linewidth]{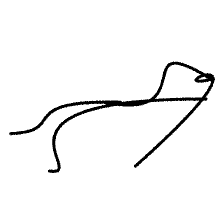} &
    \includegraphics[width=\widthabsup\linewidth]{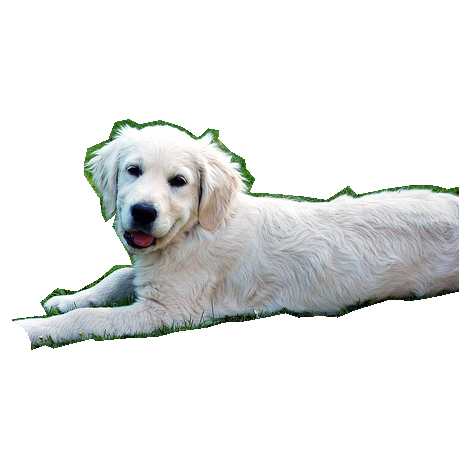} &
    \includegraphics[width=\widthabsup\linewidth]{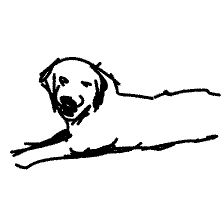} &
    \includegraphics[width=\widthabsup\linewidth]{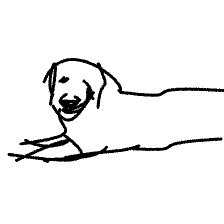} &
    \includegraphics[width=\widthabsup\linewidth]{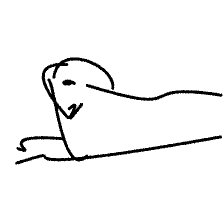} \\
    &
    \includegraphics[width=\widthabsup\linewidth]{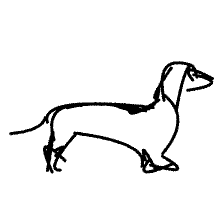} &
    \includegraphics[width=\widthabsup\linewidth]{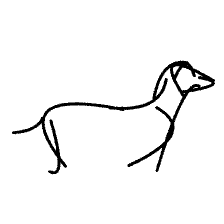} &
    \includegraphics[width=\widthabsup\linewidth]{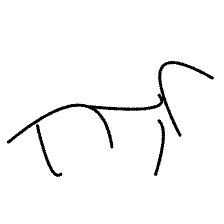} & &
    \includegraphics[width=\widthabsup\linewidth]{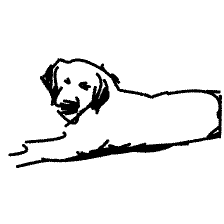} &
    \includegraphics[width=\widthabsup\linewidth]{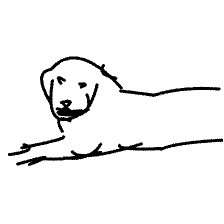} &
    \includegraphics[width=\widthabsup\linewidth]{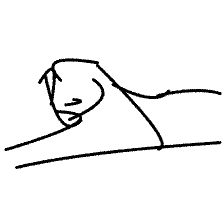} \\
    &
    \includegraphics[width=\widthabsup\linewidth]{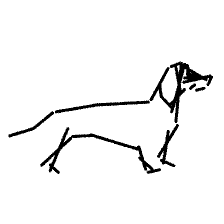} &
    \includegraphics[width=\widthabsup\linewidth]{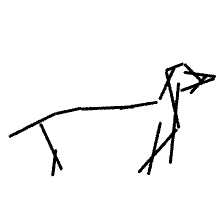} &
    \includegraphics[width=\widthabsup\linewidth]{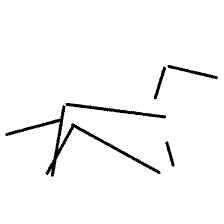} & &
    \includegraphics[width=\widthabsup\linewidth]{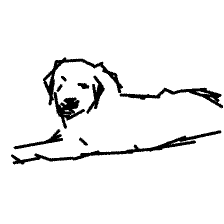} &
    \includegraphics[width=\widthabsup\linewidth]{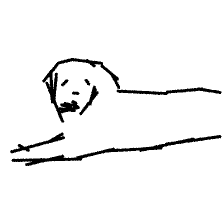} &
    \includegraphics[width=\widthabsup\linewidth]{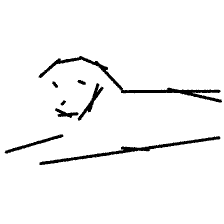} \\

\end{tabular}
 \caption{Sketches produced by our method with different styles and levels of abstraction. The columns indicate the total number of control points used to generate the sketch - 64, 24, and 16. The rows indicate the number of control points used for each stroke, which defines the style - 4, 3, and 2 control points respectively.}
\label{fig:abstraction_levels4}
\end{figure*}

\newpage

\begin{figure*}[ht]
\centering
\begin{tabular}{@{\hskip2pt}c@{\hskip2pt}c@{\hskip2pt}c@{\hskip2pt}c@{\hskip2pt}c@{\hskip2pt}c@{\hskip2pt}c@{\hskip2pt}c@{\hskip2pt}c@{\hskip2pt}c}
   
    \includegraphics[width=\widthmany\linewidth]{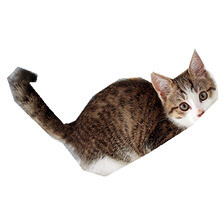} &
    \includegraphics[width=\widthmany\linewidth]{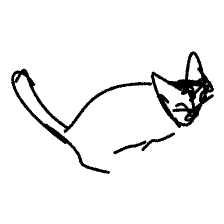} &
    \includegraphics[width=\widthmany\linewidth]{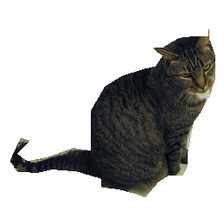} &
    \includegraphics[width=\widthmany\linewidth]{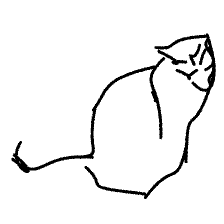} &
    \includegraphics[width=\widthmany\linewidth]{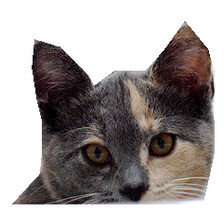} &
    \includegraphics[width=\widthmany\linewidth]{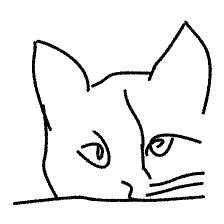} &
    \includegraphics[width=\widthmany\linewidth]{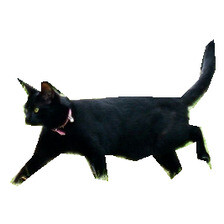} &
    \includegraphics[width=\widthmany\linewidth]{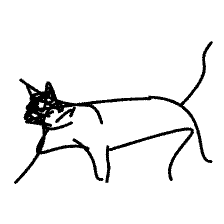} &
    \includegraphics[width=\widthmany\linewidth]{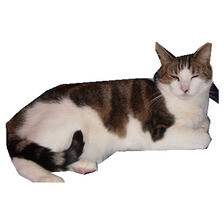} &
    \includegraphics[width=\widthmany\linewidth]{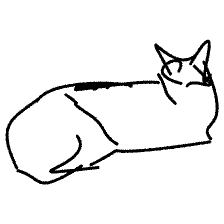} \\
    
    \includegraphics[width=\widthmany\linewidth]{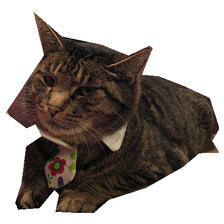} &
    \includegraphics[width=\widthmany\linewidth]{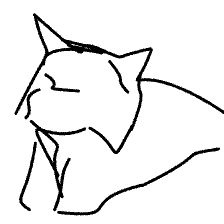} &
    \includegraphics[width=\widthmany\linewidth]{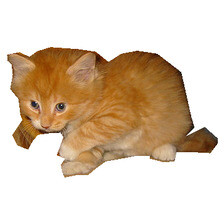} &
    \includegraphics[width=\widthmany\linewidth]{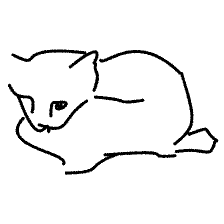} &
    \includegraphics[width=\widthmany\linewidth]{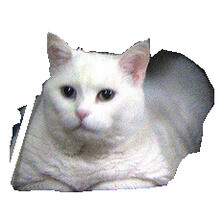} &
    \includegraphics[width=\widthmany\linewidth]{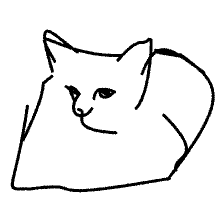} &
    \includegraphics[width=\widthmany\linewidth]{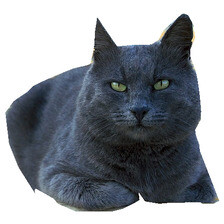} &
    \includegraphics[width=\widthmany\linewidth]{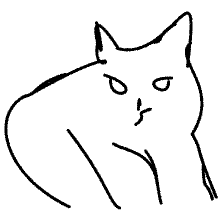} &
    \includegraphics[width=\widthmany\linewidth]{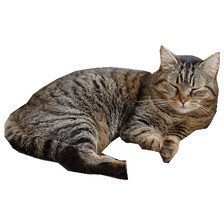} &
    \includegraphics[width=\widthmany\linewidth]{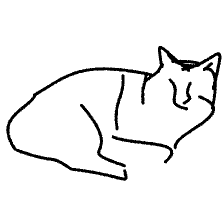} \\
    
    \includegraphics[width=\widthmany\linewidth]{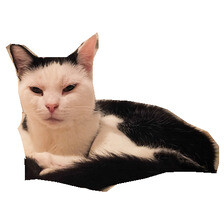} &
    \includegraphics[width=\widthmany\linewidth]{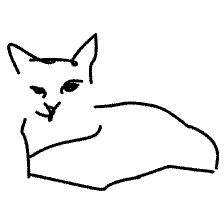} &
    \includegraphics[width=\widthmany\linewidth]{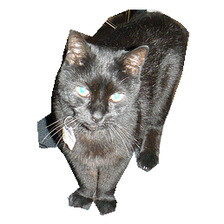} &
    \includegraphics[width=\widthmany\linewidth]{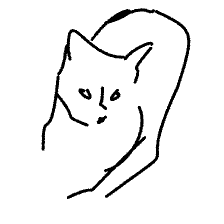} &
    \includegraphics[width=\widthmany\linewidth]{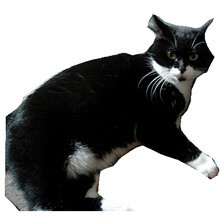} &
    \includegraphics[width=\widthmany\linewidth]{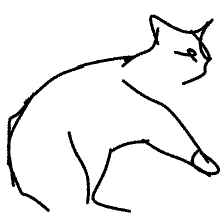} &
    \includegraphics[width=\widthmany\linewidth]{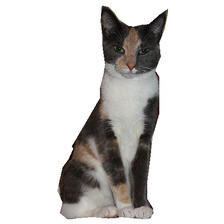} &
    \includegraphics[width=\widthmany\linewidth]{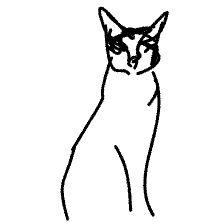} &
    \includegraphics[width=\widthmany\linewidth]{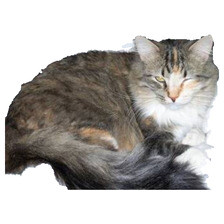} &
    \includegraphics[width=\widthmany\linewidth]{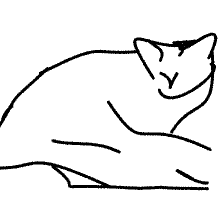} \\
    
    \includegraphics[width=\widthmany\linewidth]{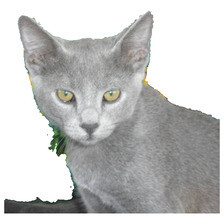} &
    \includegraphics[width=\widthmany\linewidth]{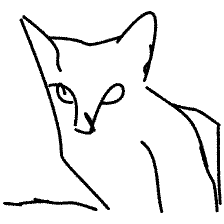} &
    \includegraphics[width=\widthmany\linewidth]{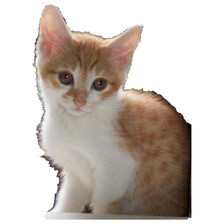} &
    \includegraphics[width=\widthmany\linewidth]{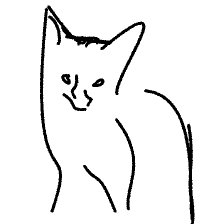} &
    \includegraphics[width=\widthmany\linewidth]{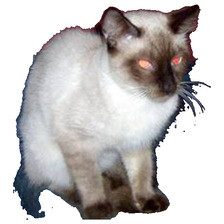} &
    \includegraphics[width=\widthmany\linewidth]{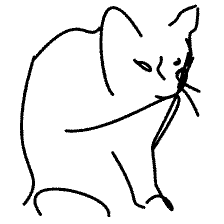} &
    \includegraphics[width=\widthmany\linewidth]{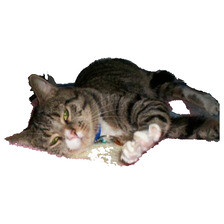} &
    \includegraphics[width=\widthmany\linewidth]{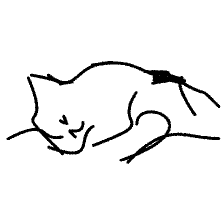} &
    \includegraphics[width=\widthmany\linewidth]{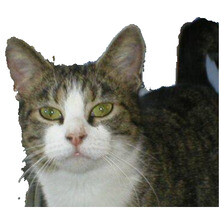} &
    \includegraphics[width=\widthmany\linewidth]{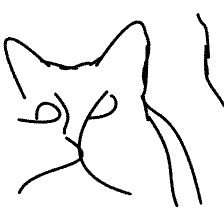} \\
    
    \includegraphics[width=\widthmany\linewidth]{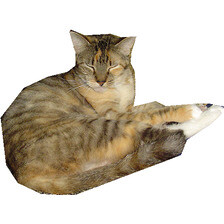} &
    \includegraphics[width=\widthmany\linewidth]{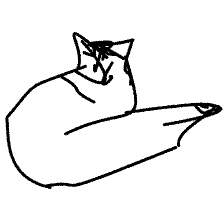} &
    \includegraphics[width=\widthmany\linewidth]{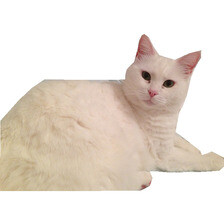} &
    \includegraphics[width=\widthmany\linewidth]{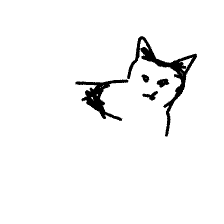} &
    \includegraphics[width=\widthmany\linewidth]{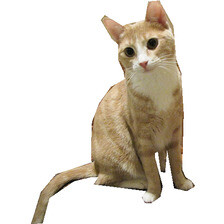} &
    \includegraphics[width=\widthmany\linewidth]{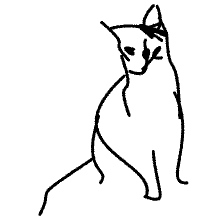} &
    \includegraphics[width=\widthmany\linewidth]{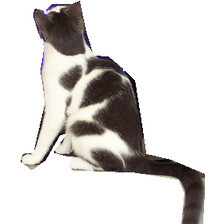} &
    \includegraphics[width=\widthmany\linewidth]{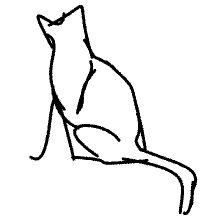} &
    \includegraphics[width=\widthmany\linewidth]{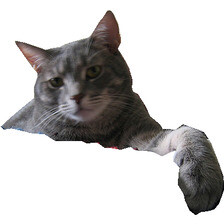} &
    \includegraphics[width=\widthmany\linewidth]{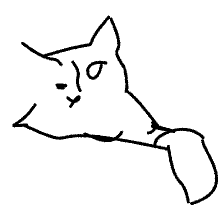}
    \\
    \includegraphics[width=\widthmany\linewidth]{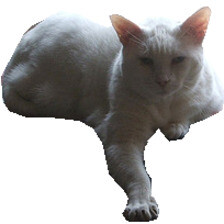} &
    \includegraphics[width=\widthmany\linewidth]{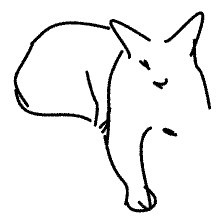} &
    \includegraphics[width=\widthmany\linewidth]{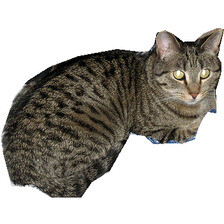} &
    \includegraphics[width=\widthmany\linewidth]{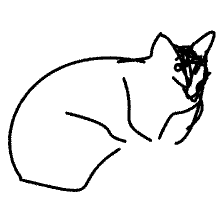} &
    \includegraphics[width=\widthmany\linewidth]{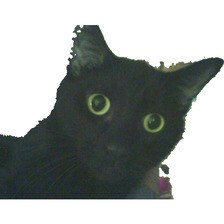} &
    \includegraphics[width=\widthmany\linewidth]{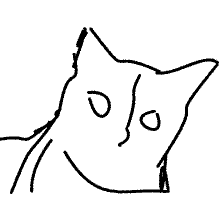} &
    \includegraphics[width=\widthmany\linewidth]{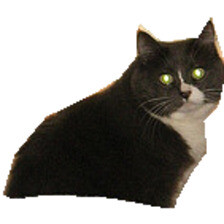} &
    \includegraphics[width=\widthmany\linewidth]{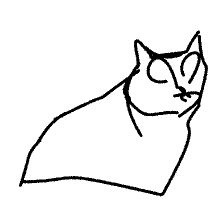} &
    \includegraphics[width=\widthmany\linewidth]{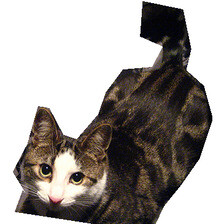} &
    \includegraphics[width=\widthmany\linewidth]{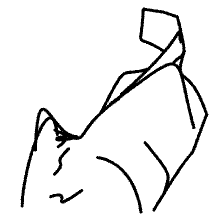} 
    \\
    \includegraphics[width=\widthmany\linewidth]{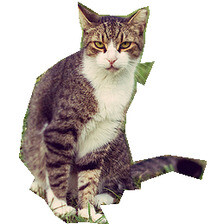} &
    \includegraphics[width=\widthmany\linewidth]{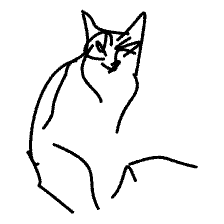} &
    \includegraphics[width=\widthmany\linewidth]{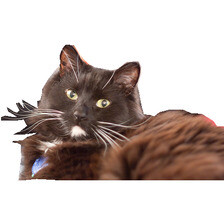} &
    \includegraphics[width=\widthmany\linewidth]{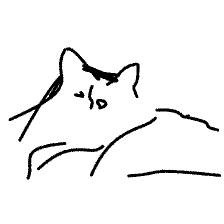} &
    \includegraphics[width=\widthmany\linewidth]{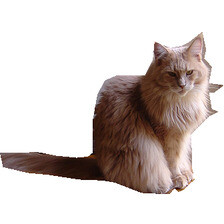} &
    \includegraphics[width=\widthmany\linewidth]{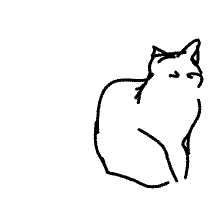} &
    \includegraphics[width=\widthmany\linewidth]{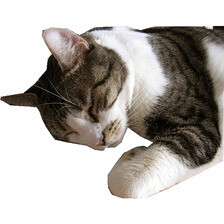} &
    \includegraphics[width=\widthmany\linewidth]{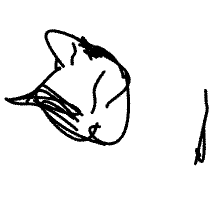} &
    \includegraphics[width=\widthmany\linewidth]{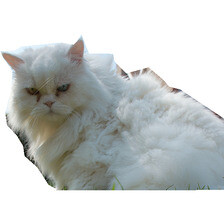} &
    \includegraphics[width=\widthmany\linewidth]{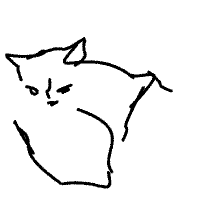} 
    \\
    \includegraphics[width=\widthmany\linewidth]{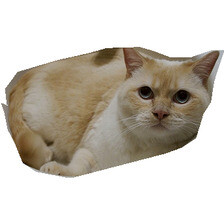} &
    \includegraphics[width=\widthmany\linewidth]{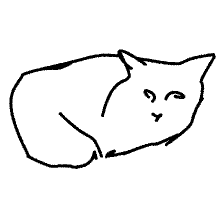} &
    \includegraphics[width=\widthmany\linewidth]{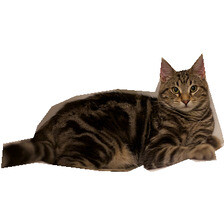} &
    \includegraphics[width=\widthmany\linewidth]{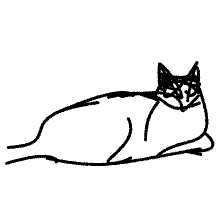} &
    \includegraphics[width=\widthmany\linewidth]{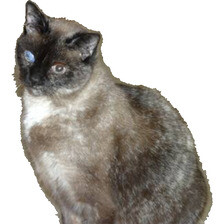} &
    \includegraphics[width=\widthmany\linewidth]{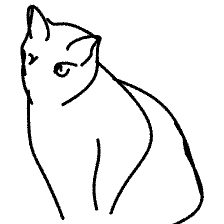} &
    \includegraphics[width=\widthmany\linewidth]{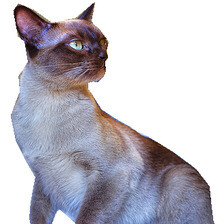} &
    \includegraphics[width=\widthmany\linewidth]{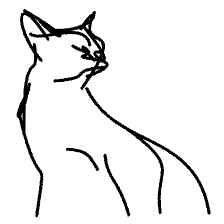} &
    \includegraphics[width=\widthmany\linewidth]{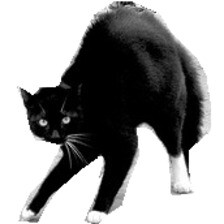} &
    \includegraphics[width=\widthmany\linewidth]{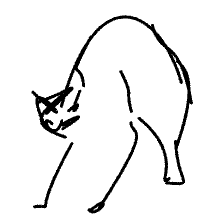}
    \\
    \includegraphics[width=\widthmany\linewidth]{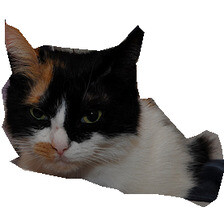} &
    \includegraphics[width=\widthmany\linewidth]{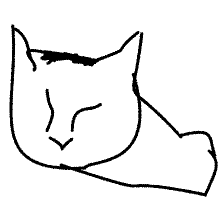} &
    \includegraphics[width=\widthmany\linewidth]{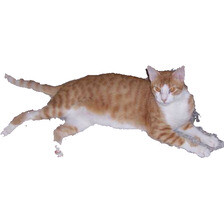} &
    \includegraphics[width=\widthmany\linewidth]{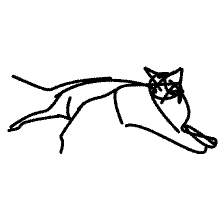} &
    \includegraphics[width=\widthmany\linewidth]{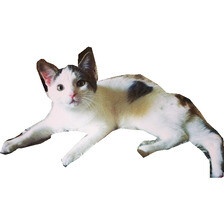} &
    \includegraphics[width=\widthmany\linewidth]{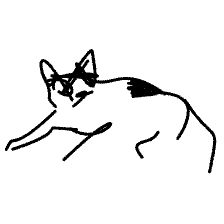} &
    \includegraphics[width=\widthmany\linewidth]{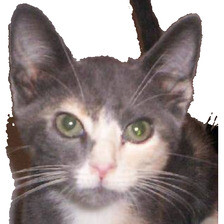} &
    \includegraphics[width=\widthmany\linewidth]{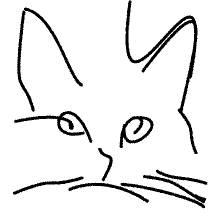} &
    \includegraphics[width=\widthmany\linewidth]{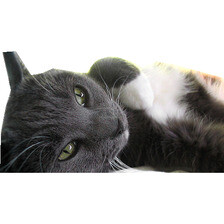} &
    \includegraphics[width=\widthmany\linewidth]{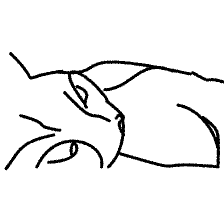}
    \\
    \includegraphics[width=\widthmany\linewidth]{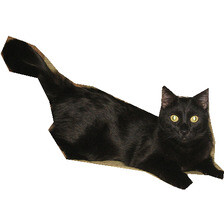} &
    \includegraphics[width=\widthmany\linewidth]{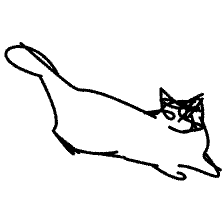} &
    \includegraphics[width=\widthmany\linewidth]{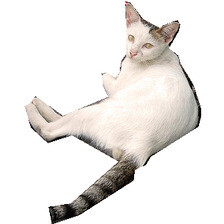} &
    \includegraphics[width=\widthmany\linewidth]{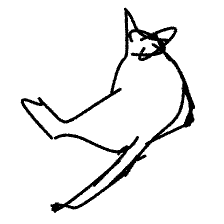} &
    \includegraphics[width=\widthmany\linewidth]{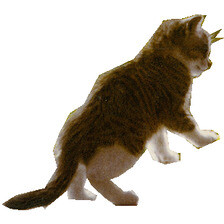} &
    \includegraphics[width=\widthmany\linewidth]{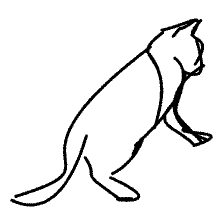} &
    \includegraphics[width=\widthmany\linewidth]{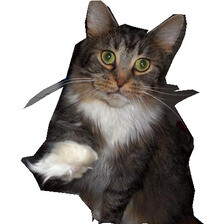} &
    \includegraphics[width=\widthmany\linewidth]{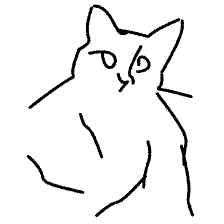} &
    \includegraphics[width=\widthmany\linewidth]{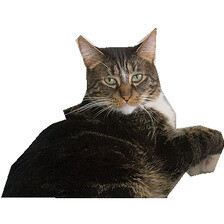} &
    \includegraphics[width=\widthmany\linewidth]{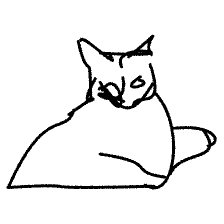}
    \\
    \includegraphics[width=\widthmany\linewidth]{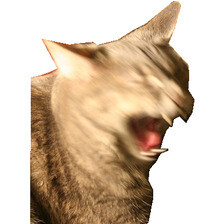} &
    \includegraphics[width=\widthmany\linewidth]{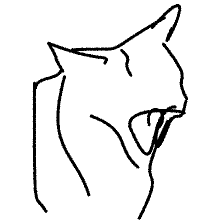} &
    \includegraphics[width=\widthmany\linewidth]{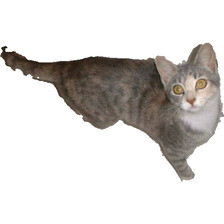} &
    \includegraphics[width=\widthmany\linewidth]{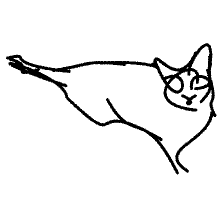} &
    \includegraphics[width=\widthmany\linewidth]{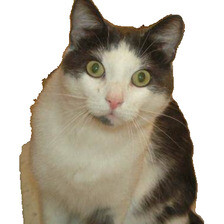} &
    \includegraphics[width=\widthmany\linewidth]{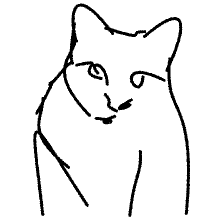} &
    \includegraphics[width=\widthmany\linewidth]{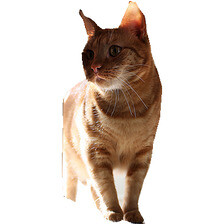} &
    \includegraphics[width=\widthmany\linewidth]{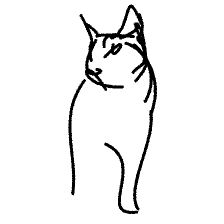} &
    \includegraphics[width=\widthmany\linewidth]{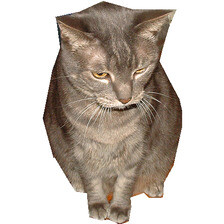} &
    \includegraphics[width=\widthmany\linewidth]{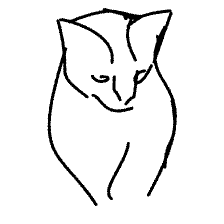}
    \\
    \includegraphics[width=\widthmany\linewidth]{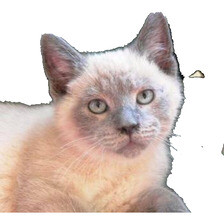} &
    \includegraphics[width=\widthmany\linewidth]{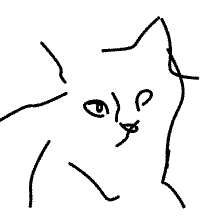} &
    \includegraphics[width=\widthmany\linewidth]{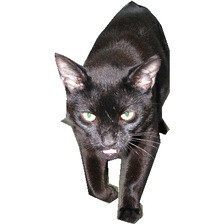} &
    \includegraphics[width=\widthmany\linewidth]{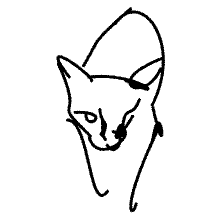} &
    \includegraphics[width=\widthmany\linewidth]{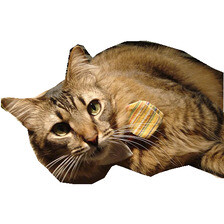} &
    \includegraphics[width=\widthmany\linewidth]{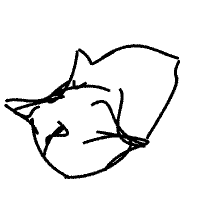} &
    \includegraphics[width=\widthmany\linewidth]{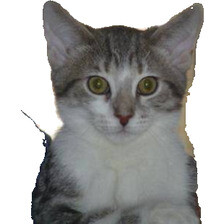} &
    \includegraphics[width=\widthmany\linewidth]{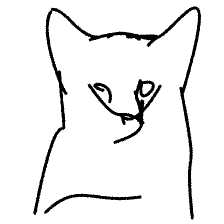} &
    \includegraphics[width=\widthmany\linewidth]{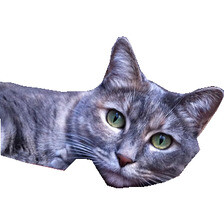} &
    \includegraphics[width=\widthmany\linewidth]{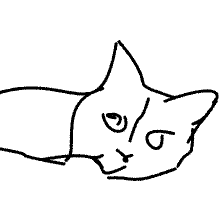}

\end{tabular}
 \caption{Sketching "in the wild": results of 100 random images of cats from SketchyCOCO \cite{gao2020sketchycoco}.}
\label{fig:many_from_same_class1}
\end{figure*}

\begin{figure*}[ht]
\centering
\begin{tabular}{@{\hskip2pt}c@{\hskip2pt}c@{\hskip2pt}c@{\hskip2pt}c@{\hskip2pt}c@{\hskip2pt}c@{\hskip2pt}c@{\hskip2pt}c@{\hskip2pt}c@{\hskip2pt}c}

    \includegraphics[width=\widthmany\linewidth]{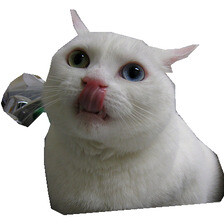} &
    \includegraphics[width=\widthmany\linewidth]{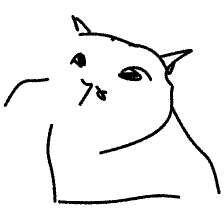} &
    \includegraphics[width=\widthmany\linewidth]{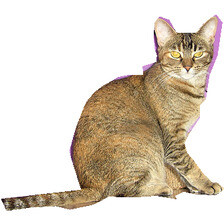} &
    \includegraphics[width=\widthmany\linewidth]{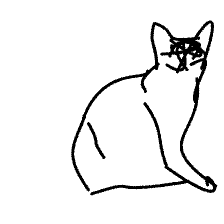} &
    \includegraphics[width=\widthmany\linewidth]{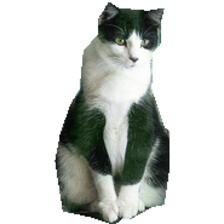} &
    \includegraphics[width=\widthmany\linewidth]{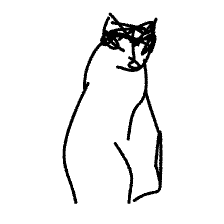} &
    \includegraphics[width=\widthmany\linewidth]{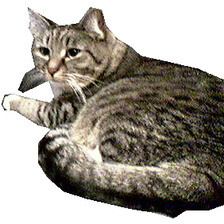} &
    \includegraphics[width=\widthmany\linewidth]{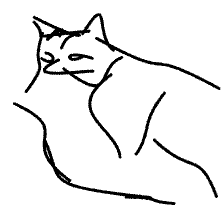} &
    \includegraphics[width=\widthmany\linewidth]{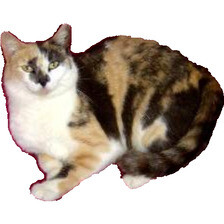} &
    \includegraphics[width=\widthmany\linewidth]{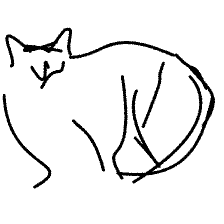}
    \\
    \includegraphics[width=\widthmany\linewidth]{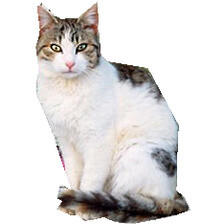} &
    \includegraphics[width=\widthmany\linewidth]{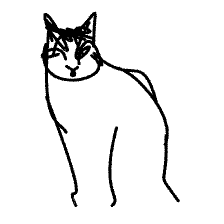} &
    \includegraphics[width=\widthmany\linewidth]{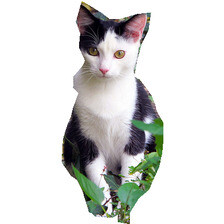} &
    \includegraphics[width=\widthmany\linewidth]{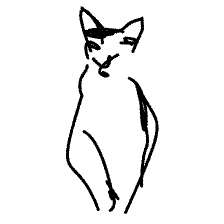} &
    \includegraphics[width=\widthmany\linewidth]{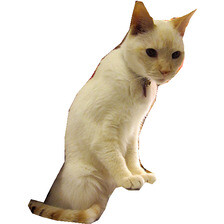} &
    \includegraphics[width=\widthmany\linewidth]{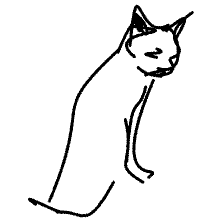} &
    \includegraphics[width=\widthmany\linewidth]{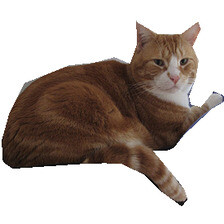} &
    \includegraphics[width=\widthmany\linewidth]{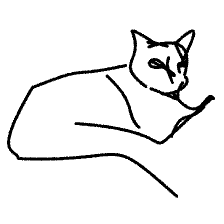} &
    \includegraphics[width=\widthmany\linewidth]{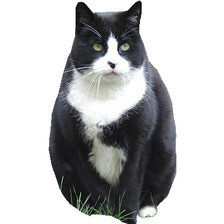} &
    \includegraphics[width=\widthmany\linewidth]{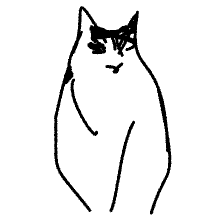}
    \\
    \includegraphics[width=\widthmany\linewidth]{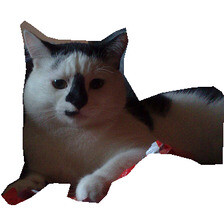} &
    \includegraphics[width=\widthmany\linewidth]{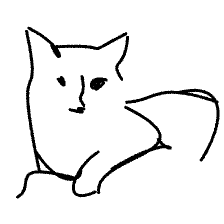} &
    \includegraphics[width=\widthmany\linewidth]{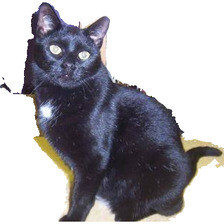} &
    \includegraphics[width=\widthmany\linewidth]{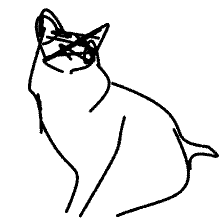} &
    \includegraphics[width=\widthmany\linewidth]{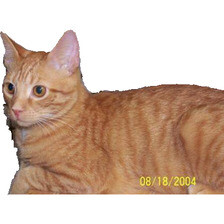} &
    \includegraphics[width=\widthmany\linewidth]{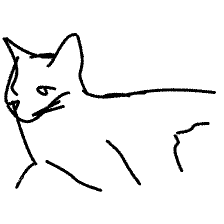} &
    \includegraphics[width=\widthmany\linewidth]{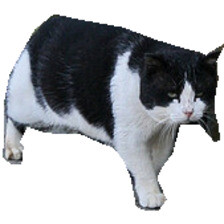} &
    \includegraphics[width=\widthmany\linewidth]{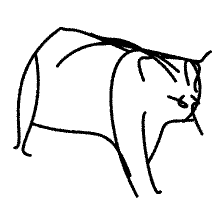} &
    \includegraphics[width=\widthmany\linewidth]{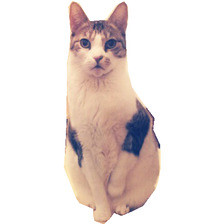} &
    \includegraphics[width=\widthmany\linewidth]{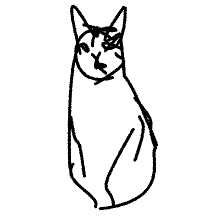}
    \\
    \includegraphics[width=\widthmany\linewidth]{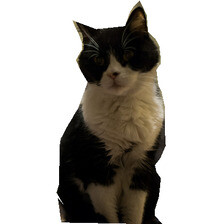} &
    \includegraphics[width=\widthmany\linewidth]{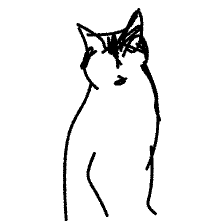} &
    \includegraphics[width=\widthmany\linewidth]{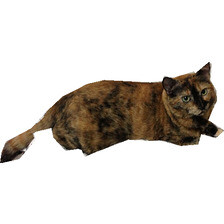} &
    \includegraphics[width=\widthmany\linewidth]{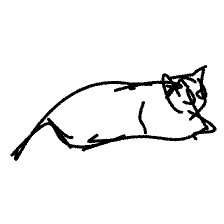} &
    \includegraphics[width=\widthmany\linewidth]{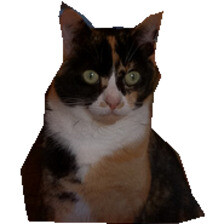} &
    \includegraphics[width=\widthmany\linewidth]{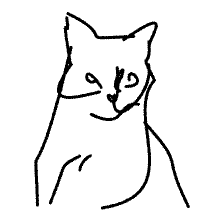} &
    \includegraphics[width=\widthmany\linewidth]{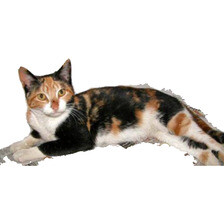} &
    \includegraphics[width=\widthmany\linewidth]{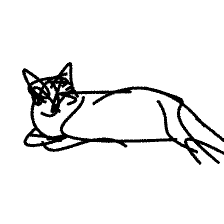} &
    \includegraphics[width=\widthmany\linewidth]{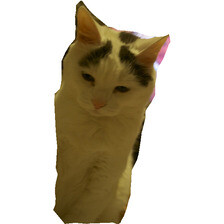} &
    \includegraphics[width=\widthmany\linewidth]{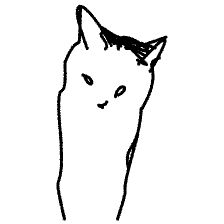}
    \\
    \includegraphics[width=\widthmany\linewidth]{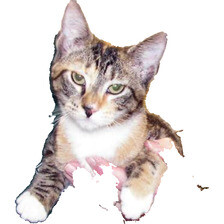} &
    \includegraphics[width=\widthmany\linewidth]{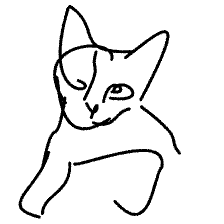} &
    \includegraphics[width=\widthmany\linewidth]{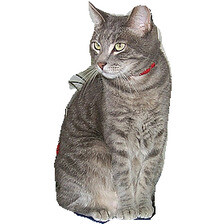} &
    \includegraphics[width=\widthmany\linewidth]{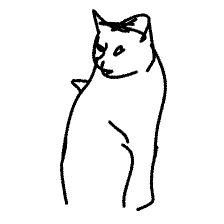} &
    \includegraphics[width=\widthmany\linewidth]{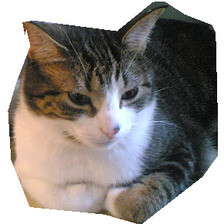} &
    \includegraphics[width=\widthmany\linewidth]{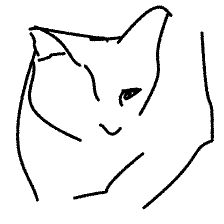} &
    \includegraphics[width=\widthmany\linewidth]{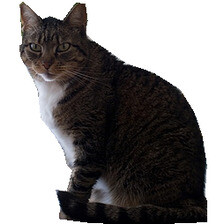} &
    \includegraphics[width=\widthmany\linewidth]{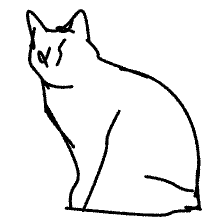} &
    \includegraphics[width=\widthmany\linewidth]{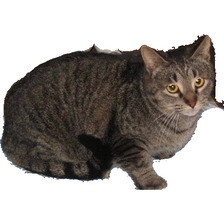} &
    \includegraphics[width=\widthmany\linewidth]{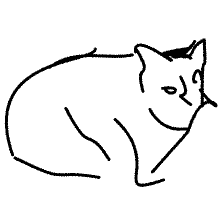}
    \\
    \includegraphics[width=\widthmany\linewidth]{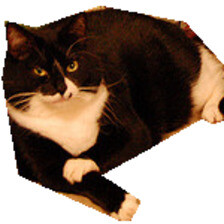} &
    \includegraphics[width=\widthmany\linewidth]{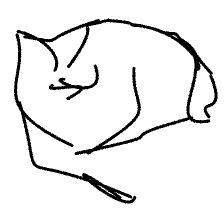} &
    \includegraphics[width=\widthmany\linewidth]{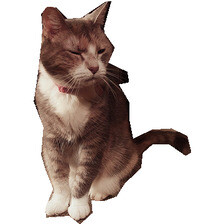} &
    \includegraphics[width=\widthmany\linewidth]{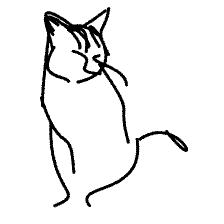} &
    \includegraphics[width=\widthmany\linewidth]{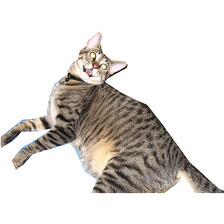} &
    \includegraphics[width=\widthmany\linewidth]{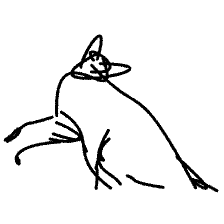} &
    \includegraphics[width=\widthmany\linewidth]{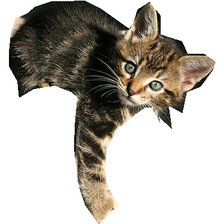} &
    \includegraphics[width=\widthmany\linewidth]{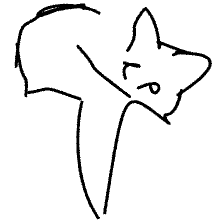} &
    \includegraphics[width=\widthmany\linewidth]{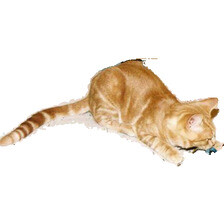} &
    \includegraphics[width=\widthmany\linewidth]{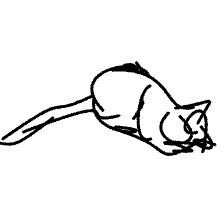}
    \\
    \includegraphics[width=\widthmany\linewidth]{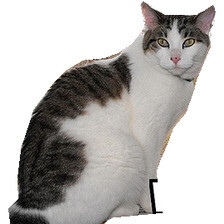} &
    \includegraphics[width=\widthmany\linewidth]{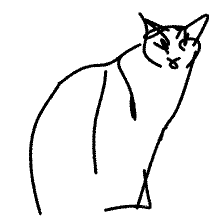} &
    \includegraphics[width=\widthmany\linewidth]{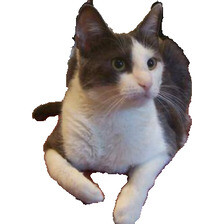} &
    \includegraphics[width=\widthmany\linewidth]{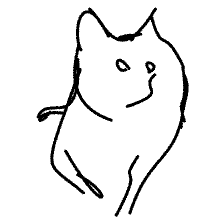} &
    \includegraphics[width=\widthmany\linewidth]{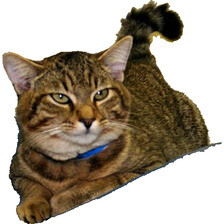} &
    \includegraphics[width=\widthmany\linewidth]{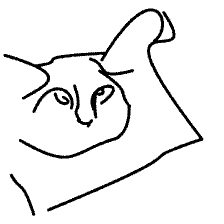} &
    \includegraphics[width=\widthmany\linewidth]{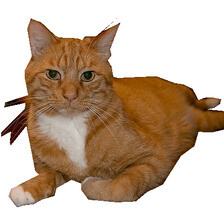} &
    \includegraphics[width=\widthmany\linewidth]{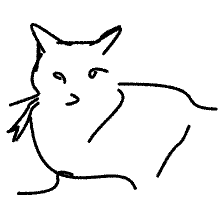} &
    \includegraphics[width=\widthmany\linewidth]{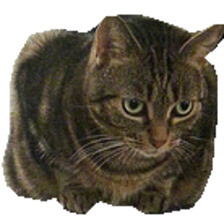} &
    \includegraphics[width=\widthmany\linewidth]{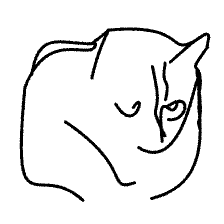}
    \\
    \includegraphics[width=\widthmany\linewidth]{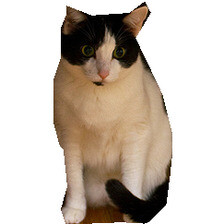} &
    \includegraphics[width=\widthmany\linewidth]{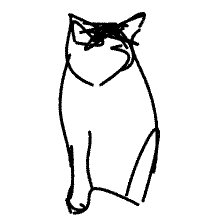} &
    \includegraphics[width=\widthmany\linewidth]{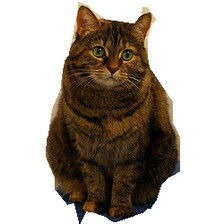} &
    \includegraphics[width=\widthmany\linewidth]{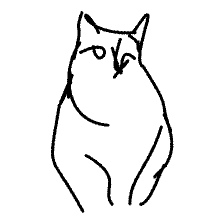} &
    \includegraphics[width=\widthmany\linewidth]{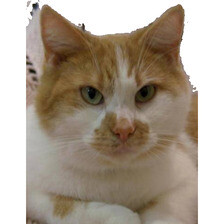} &
    \includegraphics[width=\widthmany\linewidth]{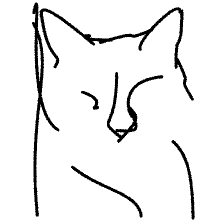} &
    \includegraphics[width=\widthmany\linewidth]{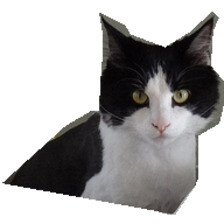} &
    \includegraphics[width=\widthmany\linewidth]{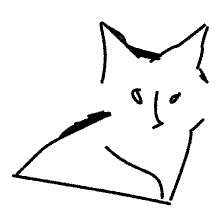} &
    \includegraphics[width=\widthmany\linewidth]{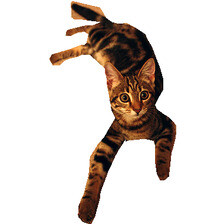} &
    \includegraphics[width=\widthmany\linewidth]{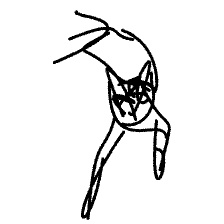}
    \\
    \includegraphics[width=\widthmany\linewidth]{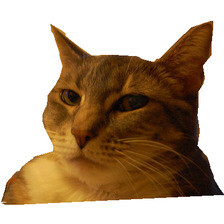} &
    \includegraphics[width=\widthmany\linewidth]{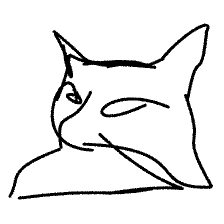} &
    \includegraphics[width=\widthmany\linewidth]{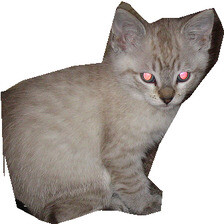} &
    \includegraphics[width=\widthmany\linewidth]{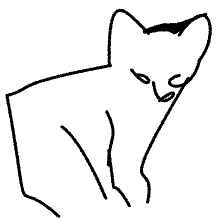} &
    \includegraphics[width=\widthmany\linewidth]{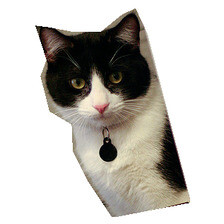} &
    \includegraphics[width=\widthmany\linewidth]{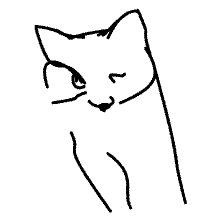} &
    \includegraphics[width=\widthmany\linewidth]{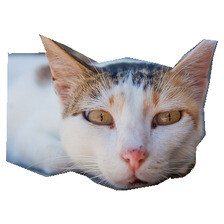} &
    \includegraphics[width=\widthmany\linewidth]{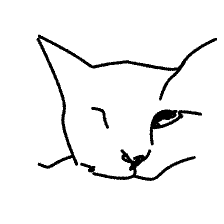} &
    \includegraphics[width=\widthmany\linewidth]{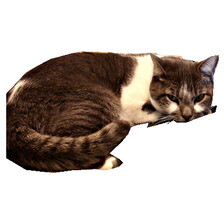} &
    \includegraphics[width=\widthmany\linewidth]{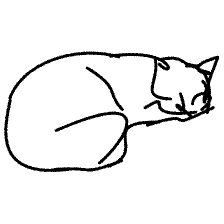} 
    \\
\end{tabular}
 \caption{Sketching "in the wild": results of 100 random images of cats from SketchyCOCO \cite{gao2020sketchycoco}.}
\label{fig:many_from_same_class}
\end{figure*}

\end{document}